\documentclass[12pt,preprint]{aastex}
\usepackage{pdflscape}

\usepackage{color}

\begin{document}

\title{A Comparative Study of Long and Short GRBs. I. Overlapping Properties}

\author{Ye Li, Bing Zhang, Hou-Jun L\"{u}\footnote{Current address: GXU-NAOC Center for Astrophysics and Space Sciences, Department of Physics, Guangxi University, and Guangxi Key Laboratory for Relativistic Astrophysics, Nanning 530004, China}}

\affil{Department of Physics and Astronomy, University of Nevada, Las Vegas, NV 89154, USA}

\begin{abstract}
Gamma ray bursts (GRBs) are classified into long and short categories based on their durations. 
Broad band studies suggest that these two categories of objects 
roughly correspond to two different classes of progenitor systems, i.e. compact star mergers
(Type I) vs. massive star core collapse (Type II).
However, the duration criterion sometimes leads to mis-identification of the progenitor systems. 
We perform a comprehensive multi-wavelength comparative study between duration-defined long 
GRBs and short GRBs as well as the so-called ``consensus" long GRBs and short GRBs (which
are believed to be more closely related to the two types of progenitor systems). 
The parameters we study include two parts: the prompt emission properties including
duration ($T_{90}$), spectral peak energy ($E_{\rm p}$), 
low energy photon index ($\alpha$), isotropic $\gamma$-ray energy ($E_{\rm \gamma, iso}$), 
isotropic peak luminosity ($L_{\rm p,iso}$), and the amplitude parameters ($f$ and $f_{\rm eff}$);
and the host galaxy properties including stellar mass ($M_*$), 
star formation rate (SFR), metallicity ([X/H]), half light radius ($R_{50}$), angular and physical ($R_{\rm off}$)
offset of the afterglow from the center of the host galaxy, the normalized offset
($r_{\rm off}=R_{\rm off}/R_{50}$), and the brightness fraction $F_{\rm light}$.
For most parameters, we find interesting overlapping properties between the two populations in both
1D and 2D distribution plots. The three best parameters for the classification purpose are
$T_{90}$, $f_{\rm eff}$, and $F_{\rm light}$. However, no single parameter alone is good
enough to place a particular burst into the right physical category, suggesting a need of
multiple criteria for physical classification.

\end{abstract}


\section{Introduction}
{The Gamma-Ray Burst (GRB) duration has a bimodal distribution. It had been seen in the early
GRB data \citep{1981Ap&SS..80....3M}, and was more clearly seen in
the GRB sample collected by Burst And Transient Source Experiment (BASTE) on board 
the Compton Gamma Ray Observatory (CGRO)  \citep{ 1993ApJ...413L.101K}.}
The division line between the long-duration GRBs (LGRBs) and 
short-duration GRBs (SGRBs) is around 2 seconds in the BATSE 50 $-$ 300 keV band.
Although the significance of this bimodality and the division line
depend on the sensitivity and energy band of the detectors
\citep{1996AIPC..384...87R, 2011ApJ...733...97B, 2012ApJ...750...88Z, 2013ApJ...763...15Q},
the authenticity of the bimodal $T_{90}$ distribution is confirmed 
not only with a larger BATSE sample \citep{1996ApJS..106...65M, 1999ApJS..122..465P},
but also by GRB data collected from other instruments such as BeppoSAX \citep{2009ApJS..180..192F}, 
INTEGRAL \citep{2008A&A...484..143F, 2012A&A...541A.122S, 2014A&A...561A..25B}, 
Swift \citep{2011ApJS..195....2S} 
and Fermi \citep{2012ApJS..199...18P, 2014ApJS..211...13V}.
The existence of two phenomenological classes of GRBs is firmly established.
The connection between these two phenomenological classes of GRBs with two physically
distinct progenitor systems are theoretical motivated and observationally confirmed through
observations of afterglow and host galaxies of both LGRBs and SGRBs.


LGRBs, typically with duration $T_{90} >$ 2 s, 
are supposed to originate from core-collapse of massive stars
\citep{1993ApJ...405..273W, 1998ApJ...494L..45P, 1999ApJ...524..262M}.
A direct observational support comes from the associations of some
LGRBs with Type Ic supernovae (SNe)
\citep{1998Natur.395..670G, 2003Natur.423..847H, 2003ApJ...591L..17S, 2006ARA&A..44..507W, 2012grbu.book..169H, 2013ApJ...776...98X}.
It strongly suggests that LGRBs are related to the death of massive stars,
$>$ 30 $M_{\odot}$ in general.
Also, host galaxies of LGRBs are generally dwarf star-forming galaxies
with low metallicity, sometimes interacting with other galaxies 
\citep{1997Natur.387..476S, 1998ApJ...507L..25B, 2002AJ....123.1111B, 
2002ApJ...566..229C, 2004A&A...425..913C, 2009ApJ...691..182S, 2015A&A...581A.125K}.
Within their host galaxies,
LGRBs are also located in bright star-forming regions with a small offset from the center of the host galaxy
\citep{2002AJ....123.1111B, 2006Natur.441..463F, 2016ApJ...817..144B}.
Both galactic and sub-galactic environments of LGRBs are consistent with an association
of LGRBs with recent star formation, supporting the massive star origin of LGRBs.

SGRBs, typically with duration $T_{90} <$ 2 s, 
are believed to be products of compact star mergers, 
i.e.,  neutron star - neutron star (NS-NS) 
or neutron star - black hole (NS-BH) mergers
(\cite{1986ApJ...308L..43P, 1989Natur.340..126E,1992ApJ...395L..83N}, see \cite{2014ARA&A..52...43B} for a review).
{ Contrary to LGRBs, so far
no SN was found associated with any SGRB and
the limits for the existence of a SN are 2-7 magnitudes         
deeper than typical SNe associated with LGRBs
\citep{2005Natur.437..845F, 2005ApJ...630L.117H, 2005Natur.437..859H, 2011ApJ...734...96K, 2013ApJ...774L..23B}. }
The absence of SN associations strongly disfavors a massive star origin,
but is consistent with the compact star origin of SGRBs.
Also, SGRBs reside in diverse types of galaxies, 
including both late-type galaxies and early-type galaxies,
e.g. GRB 050509B \citep{2005Natur.437..851G} and 
GRB 050724 \citep{2005Natur.438..988B}.
{ The offset of SGRBs from the center of their host galaxy is generally large, which is 
generally consistent with theoretical prediction of compact star mergers
\citep{2010ApJ...708....9F, 2011ApJ...734...96K, 2013ApJ...776...18F}.}
The compact star merger origin is also supported 
by the putative discovery of r-process-powered ``kilonovae/macronovae"
associated with SGRBs 130603B,  060614, and probably 080503, {050709} as well
\citep{1998ApJ...507L..59L, 2010MNRAS.406.2650M, 2013Natur.500..547T, 2013ApJ...774L..23B, 2015NatCo...6E7323Y, 2015ApJ...807..163G, 2016arXiv160307869J}.

However, {the} duration criterion alone is not always reliable 
to reveal the physical origin of individual GRBs, i.e., 
a massive star collapsar or a compact star merger.
GRB 060614 is classified as a LGRB by duration 
since its prompt emission shows 
4.5-s hard spikes followed by $\sim$190 s extended emission 
\citep{2006Natur.444.1044G, 2010ApJ...717..411N}.
However, no SN was found down to hundreds of times less luminous than SN 1998bw,
the Type Ic SN associated with GRB 980425
\citep{2006Natur.444.1053G, 2006Natur.444.1050D, 2006Natur.444.1047F}.
Moreover, its host galaxy is much more passive than normal LGRB host galaxies,
and its afterglow is located at a relatively faint position within the host galaxy
\citep{2006Natur.444.1053G, 2006Natur.444.1047F, 2016ApJ...817..144B}.
There is even a putative kilonova associated with it
\citep{2015NatCo...6E7323Y}.
The analogy of GRB 060614 with some SGRBs with extended emission allows
\cite{2007ApJ...655L..25Z} to suggest a compact star merger origin of this
apparently long duration GRB.
Another nearby long-duration GRB 060505 also showed a similar puzzle:
stringent SN limit, low sSFR host galaxy and large offset
\citep{2006Natur.444.1047F, 2016ApJ...817..144B}.
On the other hand, GRB 090426 is classified as a SGRB by duration
$T_{90}$ = 1.24 s.
However, it has a blue, star-forming and interacting host galaxy, its afterglow had a small offset 
with respect to the galaxy, and it is located in the LGRB region in $E_{\rm p,rest} - E_{\rm \gamma,iso}$ plot 
\citep{2009A&A...507L..45A, 2010MNRAS.401..963L}. All these
indicate that it might be of a collapsar origin. Indeed, it can be understood as
a long GRB with a short-duration ``tip-of-iceberg'' detected above the background level
\citep{2014MNRAS.442.1922L}.

Motivated by these observations, \cite{2006Natur.444.1010Z} suggested to separate the 
phenomenological classification scheme (short vs. long) from the physical
classification scheme (compact star origin or Type I vs. massive star origin or
Type II). \cite{2009ApJ...703.1696Z} presented a detailed study of observational
and theoretical motivations of connecting various observational properties with
progenitor systems, and suggested that one should apply multi-wavelength criteria
(including properties of prompt emission, afterglow emission and
host galaxy) to judge the physical category of individual GRBs.

In order to apply these multi-wavelength criteria, the first task is to investigate
how different/similar the two phenomenological types of GRBs are from each other
for each individual observational property. 
In previous papers, some individual properties of LGRBs and SGRBs have been compared,
such as prompt emission properties
\citep{2012ApJ...750...88Z},
afterglow properties
\citep{2008ApJ...689.1161G, 2009ARA&A..47..567G, 2010ApJ...720.1513K, 2011ApJ...734...96K},
and host galaxy properties
\citep{2013ApJ...776...18F}.
However, these studies mostly focus on one particular type of properties.
In order to get a global understanding of the differences and similarities 
between LGRBs and SGRBs, we need a comprehensive comparative study of
multiple criteria, especially between prompt emission properties and host galaxy 
properties, both carry important information to diagnose the physical origin of GRBs.

In this paper, we gather prompt emission and host galaxy properties for a large sample 
of LGRBs and SGRBs detected/observed before June 30th, 2014, and examine how much 
LGRBs and SGRBs properties overlap with each other.
We compare the properties of $T_{90}$-defined LGRBs and SGRBs,
and also the ``consensus'' LGRBs and SGRBs. The latter are based on the definition in 
Jochen Greiner's online catalog,\footnote{http://www.mpe.mpg.de/$\sim$jcg/grbgen.html}
with SGRBs labeled as `S'.\footnote{One exception is GRB 061210, which has a 0.13s spike 
with a 77s extended emission. 
It does not have a label `S' in Greiner's catalog, while other catalogs such as 
\cite{2011ApJS..195....2S} classify it as SGRB with extended emission. 
We use it as a SGRB hereafter.}
Some of the bursts in consensus SGRB sample have $T_{90}$ longer than 2 s, so 
the classification is not based on duration only. It reflects the consensus from the community,
which already takes into account multi-wavelength criteria (e.g. spectral lag, host galaxy
type, offset) in the definition. In a sense, the consensus classification of short vs. long
GRBs are more analogous to physical classification scheme of Type I vs. Type II by
\cite{2009ApJ...703.1696Z}.

The cosmological parameters $H_0=71\rm \ km\ s^{-1}\ Mpc^{-1}$,
$\Omega_m=0.27$, and $\Omega_{\Lambda}=0.73$ are adopted in this paper.
The sample and the observational properties we are interested in are presented in Section 2.
Section 3 shows the 1D distributions of each parameter for both LGRBs and SGRBs, which show
overlapping behaviors. In Section 4, we show a series of 2-D distribution plots, each with a pair
of prompt emission vs. host galaxy parameters, and quantify their overlapping properties.
We conclude and discuss the implications of these distributions for GRB classification schemes
in Section 5. 

\section{Samples}
Our main sample includes 375 GRBs with spectroscopic redshift measurements
in the literature before June 30th, 2014.  
Also included are 32 GRBs with host galaxy information, even though
no spectroscopic redshifts have been reported for these bursts.
Altogether we have 407 GRBs in total.

Column $2-4$ of Table \ref{tbbasic} show the redshift, the
method of redshift measurement/estimate, and the reference for each GRB in our sample.
They are obtained from refereed papers when possible, otherwise GCN circulars.
The redshifts of GRBs are usually measured/estimated via host galaxy emission lines (E),
afterglow absorption lines (A), 
or broad band SED fitting based on photometric properties (P).
{For a few objects such as GRB 050509B, host galaxies spectra are obtained 
and only absorption lines are detected. They are indicated by `HA'.}
Redshifts measured with emission lines are favored when possible,
since absorption lines strictly speaking only give the lower limit of the redshift.
When no emission line is detected,
we adopt the highest redshift in absorption line systems 
if no conflict with photometric redshift is claimed.
There are 187 GRBs with emission line redshifts, 
and 188 GRBs with absorption line redshifts.
For those GRBs whose spectroscopic redshift is not available but are included in our sample 
due to their host galaxy information, we list their photometric redshifts (16 GRBs) if available.
One object, GRB 080123, has a redshift reported in \cite{2010ApJ...725.1202L},
but no measurement/estimate method is given.
There are 15 GRBs included in our sample that do not have any redshift information.

\subsection{Prompt Emission properties}

\subsubsection{Duration}

The most basic prompt emission property of GRBs
is duration $T_{90}$,
the timescale during which 5\% to 95\% of its $\gamma$-ray fluence is detected. 
It shows a clear bimodal distribution in the BATSE 50 - 300 keV Band 
\citep{1993ApJ...413L.101K},
which is used as the criterion to classify SGRBs and LGRBs.
Column 6 and 7 of Table \ref{tbbasic} present 
the observed duration the $T_{90}$ and related reference for each GRB.
Also shown in Column 5 is the GRB detector from which the $T_{90}$ is derived.
For Swift GRBs, $T_{90}$ values are derived in the $15-150$ keV band, which are 
obtained from \cite{2011ApJS..195....2S} when possible, 
otherwise from the Swift GRB table\footnote{http://swift.gsfc.nasa.gov/archive/grb\_table/}.
For Fermi GRBs without Swift detections,  the 
$T_{90}$ ($10-1000$ keV) values from the Fermi GRB table \citep{2014ApJS..211...13V} 
are presented. For GRBs before the Swift era, the $T_{90}$ values from BeppoSAX 
\citep{2009ApJS..180..192F} and HETE-2 \citep{2008A&A...491..157P} are adopted 
when possible. Some GRBs were detected from other detectors, e.g.
Konus-Wind, INTEGRAL and Suzaku. Their $T_{90}$ values
are obtained from GRB GCN circulars or related publications when available.

\subsubsection{Spectral parameters, fluence and flux}

The broad band spectra of GRB prompt emission are usually fitted with the so-called ``Band function''
\citep{1993ApJ...413..281B}, which is a smoothly joint broken power law defined by
\begin{equation}
\label{eqBand}
N(E)=
\left.
\Big \{
\begin{array}{lr}
A\left(\frac{E}{100\ {\rm keV}}\right)^{\alpha} {\rm exp} \left(-\frac{E}{E_0}\right),& E < (\alpha-\beta)E_0\\
A\left(\frac{E}{100\ {\rm keV}}\right)^{\beta} \left[ \frac{(\alpha-\beta)E_0}{100\ {\rm keV}}\right]^{\alpha-\beta} {\rm exp} (\beta-\alpha),& E \ge (\alpha-\beta)E_0\\
\end{array}
\right.
\end{equation}
where $\alpha$ is low energy photon index, $\beta$ is high energy photon index 
and $E_0$ is the break energy.
Instead of $E_0$, more frequently quoted is $E_{\rm p}$,
the peak energy in the energy spectrum $E^2N$,
where $E_{\rm p}=(2+\alpha)E_0$. The spectral parameters
$\alpha$, $\beta$, and $E_{\rm p}$ in time-integrated spectra
are provided in columns $4-6$ of Table \ref{tbprompt}
when Band function fitting is available 
\citep{2002A&A...390...81A, 2013ApJS..208...21G}.
Sometimes not all of these parameters are well constrained,
due to the narrowness of the detector's energy band (Swift/BAT, \cite{2011ApJS..195....2S}),
and the low fluence of the bursts. 
In these cases, a cutoff power law (CPL) model, which is essentially the first half of Eq.(\ref{eqBand}), 
or a simple power law (PL) model
$$N(E) = A E^{-\Gamma}$$
are used for spectral fitting instead.
If a CPL fitting is available,
the parameters  $\alpha$ and $E_{\rm p}$ are recorded.
If the spectrum can be fitted with a PL model without an $E_{\rm p}$ estimation, 
the PL index $\Gamma$ is recorded in the third column of Table \ref{tbprompt}.
The PL $\Gamma$ shows a systematic difference from Band $\alpha$
\citep{2012MNRAS.424.2821V}. For sake of fair comparison,
$\Gamma$ is presented whenever available, regardless of whether
$\alpha$ is provided or not.

Table \ref{tbprompt} also shows fluence $S_{\gamma}$ and peak flux $F_{\rm p}$ 
of each GRB, as well as the respective energy band and the detector used to derive them.
The fluence $S_{\rm \gamma}$ is given as energy fluence, in units of $\rm erg\ cm^{-2} $.
If not specified, the peak flux is energy flux at the peak time $t_{\rm p}$,
in units of erg s$^{-1}$ cm$^{-2}$, with a time bin of 1 second.
The photon peak flux $P_{\rm p}$, in units of photons s$^{-1}$ cm$^{-2}$, is shown
with superscript `P'.
Peak flux from Konus-Wind is usually not  binned in 1 second.
The binning timescale is labelled in Table \ref{tbprompt} using the following convention:
1) 0.004 s; 2) 0.016 s; 3) 0.064 s; 4) 0.128 s; 5) 0.256 s; 6) 2.944 s; 7) 3 s, respectively.

Parameters of BeppoSAX are obtained from \cite{2002A&A...390...81A} and 
\cite{2009ApJS..180..192F} in general.
Parameters from BASTE are obtained
from \cite{2004ApJ...609..935Y} and 5B BASTE catalogue \citep{2013ApJS..208...21G}.
The 5B BASTE catalogue does not come with traditional GRB names, so
we match 5B BASTE catalogue burst IDs with traditional GRB names by 
requiring the positional difference smaller than 10 degree and a close
temporal match (typically the difference less than a few seconds).
Spectral parameters from HETE-II are obtained from
\cite{2005ApJ...629..311S} and \cite{2008A&A...491..157P}, 
covering 2-400 keV in general.
Spectral parameters from INTEGRAL are obtained from
\cite{2014A&A...561A..25B} and \cite{2008A&A...484..143F}, 
covering a time span from 2002 to 2012.
While \cite{2014A&A...561A..25B} makes a joint IBIS/SPI spectral fit
covering 20-1000 keV, 
spectral fitting in \cite{2008A&A...484..143F} uses data from 
IBIS only, covering 20-200 keV.
So data from \cite{2014A&A...561A..25B} are favored 
if parameters of a same burst are provided in both catalogs.
Spectral parameters from Fermi/GBM are obtained from the
Fermi GRB catalog \citep{2014ApJS..211...12G},
typically covering 10 - 1000 keV.
Sometimes, parameters from Konus-Wind, RHESSI or Suzaku/WAM are used,
obtained from GCN circulars in general.
Otherwise, Swift BAT parameters are presented, 
including $\Gamma$, fluence and flux for those with PL as the best fit model,
and $\alpha$, $E_{\rm p}$, fluence and flux for those with CPL as the best fit model.

\subsubsection{$E_{\rm \gamma,iso}$ and $L_{\rm p,iso}$} \label{sec:Eiso}

The isotropic gamma-ray energy $E_{\rm \gamma,iso}$ 
and peak luminosity $L_{\rm p,iso}$
in the cosmological rest frame $1-10^4$ keV are 
estimated with parameters presented in the previous sections.
The estimated value are shown in Column 8 and 9 of Table \ref{tbbasic}.

The isotropic energy $E_{\rm \gamma,iso}$ is estimated as
\begin{equation}
E_{\rm \gamma,iso} = 4 \pi D_{\rm L}^2 S_{\gamma} k/(1+z),
\end{equation}
where $S_{\gamma}$ is the $\gamma$-ray fluence, in units of erg cm$^{-2}$,
$D_{\rm L}$ is the luminosity distance estimated with redshift, 
and $k$ is a $k$-correction factor
from the lab frame to the bolometric rest frame, defined as
\begin{equation}
k=\frac{\int^{10^4/(1+z)}_{1/(1+z)}\ EN(E) dE }{\int^{e_{\rm max}}_{e_{\rm min}}\ EN(E) dE}.
\end{equation}
Here $e_{\rm max}$ and $e_{\rm min}$ are the observational energy range of fluence,
presented in Column 8 of Table \ref{tbprompt},
$N(E)$ denotes the photon spectrum of GRBs.
All the GRB spectra are assumed to be a Band function 
as shown in Eq. \ref{eqBand},
with the spectral parameters listed in Table \ref{tbprompt}.
For those GRBs fitted with a CPL, $\beta = -2.3$ is assumed.
For the GRBs with PL fitting only, 
the rough correlation between power law index $\Gamma$ 
and the peak energy $E_{\rm p}$, i.e.
$${\rm log}\ E_{\rm p} = (4.34 \pm 0.48) - (1.32 \pm 0.13) \Gamma$$
is used to estimate $E_{\rm p}$ \citep{2007ApJ...655L..25Z, 2009ApJ...693..922S, 2012MNRAS.424.2821V}.
For bursts without $\alpha$ and $\beta$, 
$\alpha = -1.0$ and $\beta = -2.3$ are assumed for LGRBs, 
and $\alpha = -0.5$, $\beta=-2.3$ for SGRBs 
\citep{1993ApJ...413..281B, 2000ApJS..126...19P}.
For the bursts without redshift estimation, $z=2$ is assumed for LGRBs,
and $z=0.5$ for SGRBs.
{One exception is GRB 020410, for which $z=0.5$ is assumed 
according to the redshift estimation based on the possible SN detection
in \cite{2005ApJ...624..880L}.}

The peak luminosity $L_{\rm p,iso}$ is estimated as
\begin{equation}
L_{\rm p,iso} = 4 \pi D_{\rm L}^2 F_{\rm p} k,
\end{equation}
with the same $k$ correction as $E_{\rm \gamma,iso}$ estimation, 
and the peak flux $F_{\rm p}$ in units of erg s$^{-1}$ cm$^{-2}$.
For GRBs with photon peak flux $P_{\rm p}$ (in units of photon s$^{-1}$ cm$^{-2}$) reported only, 
$F_{\rm p}$ is estimated from $P_{\rm p}$
\begin{equation}
F_{\rm p}=P_{\rm p} \frac{\int^{e_{\rm max}}_{e_{\rm min}} EN(E) dE}
{\int^{e_{\rm max}}_{e_{\rm min}} N(E) dE},
\end{equation}
where $e_{\rm max}$ and $e_{\rm min}$ define the observational energy range of flux, 
presented in Column 10 of Table \ref{tbprompt}.

\subsubsection{Amplitude $f$ and $f_{\rm eff}$}

\cite{2014MNRAS.442.1922L} introduced the amplitude parameters $f$ and $f_{\rm eff}$ to assist
classification of GRBs. The $f$ parameter is defined as the ratio between 1-s peak flux
and background flux $f=\frac{F_{\rm p}}{F_{\rm B}}$, which measures how bright the brightest peak 
of a burst is above the background level. The effective amplitude parameter is defined as 
$f_{\rm eff}=\frac{F'_{\rm p}}{F_{\rm B}}$, which is the amplitude of a pseudo GRB which was scaled
down from the original burst until the new duration $T_{90}$ is shorter than 2 s. It reflects the 
measured $f$ value for an intrinsically long GRB to be confused as a short GRB when the bulk of
the emission is buried below the background. Since short GRBs already have $T_{90} < 2$ s, 
their $f_{\rm eff}$ parameter is the same as the $f$ parameter. \cite{2014MNRAS.442.1922L} 
showed that the $f_{\rm eff}$ values of long GRBs are typically smaller than 2, which means that the
``tip-of-iceberg'' effect cannot give very high-amplitude short GRBs. In contrast, short GRBs typically
have $f_{\rm eff} = f$ greater than 2. As a result, the $f$ and $f_{\rm eff}$ parameters are useful 
parameters to diagnose the physical origin of a burst. We include all the $f$ and $f_{\rm eff}$
parameters published in \cite{2014MNRAS.442.1922L}\footnote{Available at http://grb.physics.unlv.edu/f/data.txt}
in our analysis. 
{The $f$ and $f_{\rm eff}$ values of later Swift GRBs are also calculated 
using the same method of \cite{2014MNRAS.442.1922L}.}
They are presented in the last two columns of Table \ref{tbbasic}.


\subsection{Host Galaxy Properties}
In general, GRB host galaxies can be detected with deep observations
after the GRB afterglows fade away.
With images of the host galaxies, morphological properties 
such as galaxy size $R_{\rm 50}$,
angular and physical offsets of GRBs from the center of host galaxies,
in units of arcsec $\theta_{\rm off}$ and kpc $R_{\rm off}$,
as well as normalized offset $r_{\rm off}=R_{\rm off}/R_{\rm 50}$,
 can be obtained.
If multi color photometric properties are available,
especially if the rest frame 4000 $\rm \AA$ is covered,
the host galaxy stellar mass $M_*$ may be estimated through
stellar population syntheses.
With emission lines, which are quite common for GRB host galaxies, physical properties
such as SFR and metallicity [X/H] can be studied.
Together with the stellar mass information,
one can estimate specific SFR (sSFR), average SFR per unit stellar mass.
We go through the {refereed} papers and GCN reports related to each GRB 
to gather host galaxy property information and present them in this section.
For each GRB with redshift, we use ADS\footnote{http://adsabs.harvard.edu/abstract\_service.html}
to search for papers and reports with the GRB name in the title and abstract, 
and use SIMBAD\footnote{http://simbad.u-strasbg.fr/simbad/}
to search for papers and reports that refer to the burst, 
regardless of in which parts of the paper and reports it is mentioned.

\subsubsection{Stellar mass, Star formation rate, and Metallicity}

Stellar mass, $M_*$,
which is the main control of luminosity, SFR and metallicity of a galaxy,
is the most important host galaxy parameter.
It is also used to estimate the specific SFR (sSFR), defined as SFR per unit
stellar mass  (SFR/$M_*$), which shows the intrinsic star formation status of a galaxy.
Broad band spectral energy distribution (SED)
fitting to stellar population synthesis models is the most common method
to estimate $M_*$.
For most of the bursts in our sample, the SED-fitted $M_{*}$ 
is collected from the catalogs of \cite{2009ApJ...691..182S} and \cite{2010ApJ...725.1202L}.
Others are obtained from individual papers.
When SED estimated $M_*$ is not available, {a} single band luminosity such as
the K band magnitude \citep{2010MNRAS.405...57S} or infrared magnitude \citep{2011ApJ...739....1L}
is used as the indicator of stellar mass.
In these cases, the uncertainty is larger than one order of magnitude.
For GRBs from \cite{2011ApJ...739....1L},
upper limits of $M_{*}$ are used when only upper limits are available.
For those with detections, $M_{\rm 70Myr}$ are used 
since 70 Myr is a typical age of LGRB hosts at $z \sim 1$
\citep{2010ApJ...725.1202L}.
The values of $M_{*}$ and the method to estimate them are presented in
column 4 and 5 of Table \ref{tbsfr}.

SFR indicates average rate of star formation in a ``recent'' time range.
It can be estimated with emission lines, ultra-violet (UV) light, 
infared (IR), radio, and X-rays 
(see \cite{1998ARA&A..36..189K} and \cite{2012ARA&A..50..531K} for reviews).
Among different diagnostics, emission lines 
such as H$\alpha$, H$\beta$, and [OII]
represent the most recent 0-10 Myr SFR, 
best matching the life of LGRB progenitor, stars with $>$ 30 $M_{\sun}$
\citep{2012ARA&A..50..531K}. 
Among different emission lines, H$\alpha$ is the best indicator of SFR,
due to its relative strength, small dust extinction 
and independence of metallicity.
However, for objects with redshift larger than 0.4, 
H$\alpha$ shifts out of the optical band
and requires an infared detection.
For these cases, H$\beta$ and [OII] emission lines become good indicators instead
in the optical band, which are applicable up to the redshift 0.9 and 1.4, respectively.
The benefit of H$\beta$ over [OII] is its independence of metallicity.
\cite{2006ApJ...642..775M} shows that 
the dependence of [OII] estimated SFR on metallicity is
weak in the range of 8.2 $<$ 12+log(O/H) $<$ 8.7,
but is significant in the range of 12+log(O/H) $<$ 8.2 and 12+log(O/H) $>$ 8.7.
Since a lot of GRB hosts show 12+log(O/H) $<$ 8.2
\citep{2009ApJ...691..182S, 2015A&A...581A.125K},
H$\beta$ is in general a better indicator than [OII].
However, [OII] is usually stronger than H$\beta$.
So [OII] is more frequently used as the SFR indicator for galaxies
with redshifts as high as 1.4.
For objects with redshifts higher than 2, the
Ly$\alpha$ emission line shifts into optical 
and may be used as an SFR indicator
\citep{2012ApJ...756...25M}.
The SFR and the method used to estimate it
are shown in column 6 and 7 of Table \ref{tbsfr}, 
with column 3 presenting the instrument offering the spectrum.
{For those without SFR information, 
column 3 gives the instruments of spectral observations that provide redshift information.}
The criteria mentioned above are used. 

The two largest LGRB SFR catalogs are from
\cite{2009ApJ...691..182S} and \cite{2015A&A...581A.125K}.
\cite{2009ApJ...691..182S} summarized emission line information
of GRBs before Dec 2006 and presented a systematic estimation of SFR
using H$\alpha$, H$\beta$, [OII] and UV, respectively.
We record the SFR of each GRB host according to the criteria mentioned above.
\cite{2015A&A...581A.125K} estimated the host galaxy SFR for GRBs later than April 2005,
with emission line luminosities obtained from the VLT/X-Shooter spectra.
Due to the infared coverage of VLT/X-Shooter, \cite{2015A&A...581A.125K}
extended H$\alpha$ detection to z $\sim$ 2.5.
It enables SFR estimation with H$\alpha$ 
and better dust extinction $A_{\rm V}$ estimation.
There are only two objects presented in both of these two catalgs,
GRB 050416A and GRB 051022A, with redshift 0.653 and 0.807 respectively.
\cite{2015A&A...581A.125K} showed that their Balmer decrement estimated $A_{\rm V}$ are 
$1.62^{+0.36}_{-0.36}$ mag and $1.86^{+0.17}_{-0.13}$ mag, respectively,
and made dust extinction correction with these values.
It turns out that the estimated SFR of these two bursts by \cite{2015A&A...581A.125K} 
are around two times larger than those estimated by
\cite{2009ApJ...691..182S}, who applied a mean $A_{\rm V}=0.53$ 
to their LGRB sample
due to the lack of $A_{\rm V}$ estimate for both of bursts.
As both of GRB 050416A and GRB 051022A have dust extinction $A_{\rm V}$ 
much larger than $A_{\rm V}=0.53$,
the diversity between these two papers can be easily accounted for by
the discrepancy of the $A_{\rm V}$ applied.
On the other hand, the average $A_{\rm V}$ and SFR for the same redshift range
are consistent with each other
between \cite{2015A&A...581A.125K} and \cite{2009ApJ...691..182S}, so that
these two GRBs do not indicate statistically inconsistency between these two catalogs.

Two largest SGRB SFR catalogs are from 
\cite{2009ApJ...691..182S} and \cite{2009ApJ...690..231B}, each presenting 5 bursts.
Three of their host galaxies, GRB 051221, GRB 050709 and GRB 061006, 
show active star formation with emission lines
and their emission-line-estimated SFRs show consistency
between these two papers.
The other two, GRB 050509B and GRB 050724, 
have passive hosts without emission lines.
While the emission-line-estimated SFR upper limits are
$<$ 0.1 and $<$ 0.05 $M_{\sun}\ \rm yr^{-1}$, respectively 
\citep{2009ApJ...690..231B},
their UV-estimated SFRs are as high as 16.87 and 18.76 $M_{\sun}\ \rm yr^{-1}$,
respectively \citep{2009ApJ...691..182S}.
The discrepancy could be understood by the difference of the age of stars that
emission lines and UV light trace, i.e.,
$0-10$ Myr for emission lines 
and $10-200$ Myr for UV light.
Since LGRBs originate from stars with mass $>$ 30 $M_{\sun}$ 
and age {$\sim$} 10 Myr, emission lines are better 
diagnostics than UV light.
As a result, we do not include UV SFRs in our analysis 
even though we still list them in Table \ref{tbsfr} for completeness.

Metallicity, abundance of elements other than hydrogen and helium, is
generally described by the number density ratio 
between a specific element and hydrogen.
It may be estimated with absorption lines or emission lines.
Although these two methods provide metallicity estimation 
for somewhat different regimes in GRB host galaxies, 
they show consisteny in GRB 121024A, 
which has both emission- and absorption-line estimated metallicities
\citep{2015MNRAS.451..167F}.
The two methods also cover complementary redshift ranges,
so we include both of them here.
We caution that one needs more objects with both absorption lines and emission lines
to provide metallicity estimates to confirm the consistency between the two methods.
Metallicities estimated by absorption lines are generally described by
[X/H]=log($N_{\rm X}$/$N_{\rm H}$)-log($N_{\rm X_{\odot}}$/$N_{\rm H_{\odot}}$),
where $N_{\rm X}$ indicates column density of element X.
Metallicities estimated by emission lines, on the other hand, are generally described by
$12+{\rm log(O/H)}$, with solar value $12+{\rm log(O/H)}_{\sun}$=8.69.
In order to be consistent with each other,
we convert $12+{\rm log(O/H)}$ to [X/H] with X = O 
by $12+{\rm log(O/H)}-8.69$ \citep{2009ARA&A..47..481A}.
The estimated values as well as corresponding methods are presented
in column 8 and 9 of Table \ref{tbsfr}.

Emission line ratios are the most common diagnostics 
for late type galaxy metallicity
\citep{2002ApJS..142...35K, 2004ApJ...617..240K, 2004MNRAS.348L..59P}.
This method estimates the metallicities in HII regions of the host galaxy.
It is based on photoionisation models
\citep{2002ApJS..142...35K}
and local HII region and galaxies observations
\citep{2004MNRAS.348L..59P}.
If the host galaxy redshift is larger than 0.2,
it is hard to obtain a {spatially} resolved spectrum of a specific point. 
Since most GRBs have redshifts greater than 0.2,
emission-line-estimated metallicity is generally
the luminosity-weighted metallicity of the hosts.

The largest two LGRB samples with metallicity measurements
are \cite{2009ApJ...691..182S} and \cite{2015A&A...581A.125K}.
\cite{2009ApJ...691..182S} used different emission line diagnostics
for different GRBs, due to different available emission lines.
A direct estimation comes from electron temperature $T_{\rm e}$,
which requires a comparison of different ionization lines with the same elements
\citep{2006A&A...448..955I}.
This is only valid for a few cases 
where both [OIII]$\lambda$4363 and [OIII]$\lambda$4959,5007 are available. 
For most cases, 
other indicators with higher uncertainties are generally used.
If H$\alpha$ is detected, for GRBs with z $<$ 0.4 in general,
$\rm O3N2=log\{([OIII]\lambda5007/H\beta)/([NII]\lambda6583/H\alpha)\}$
is used, with \citep{2004MNRAS.348L..59P}
$$12+{\rm log(O/H)}=8.73-0.32\times{\rm O3N2}.$$
If H$\alpha$ is not available, for most high redshift GRBs, 
$${\rm log}\ R_{23} =\rm log\{([OII]\lambda3727+[OIII]\lambda4959, 5007)/H\beta\}$$
are used for metallicity estimation, 
with $${\rm log}\ O_{32} =\rm log\{([OIII]\lambda4959, 5007)/([OII]\lambda3727)\}$$
as an indicator of the ionization parameter.
However, the relation between $R_{23}$ and 12+log(O/H) is double-valued.
Following \cite{2008ApJ...681.1183K},
equation (18) of \cite{2004ApJ...617..240K} is applied for the upper branch
and \cite{2002ApJS..142...35K} for the lower branch.
These $R_{23}$ metallicities are corrected to $\rm O3N2$ values
as suggested by \cite{2008ApJ...681.1183K}.
Due to the lack of [NII], 
which is usually needed to decouple the double value effect,
the two values are sometimes both listed \citep{2009ApJ...691..182S}.
\cite{2015A&A...581A.125K} used a combination of the methods
by estimating the probability density profile (PDF) of metallicities 
for each GRB.
{Benefiting} from IR {spectra} with H$\alpha$ lines and [NII] lines,
their values do not encounter the double value problem.

The largest SGRB sample is from \cite{2009ApJ...690..231B}.
The $R_{23}$ method is used and only the upper branch is presented,
as suggested by \cite{2004ApJ...617..240K}.
However, the event available for O3N2, i.e. GRB 061210, 
shows 12+log(O/H)=8.47 by the method applied in \cite{2009ApJ...691..182S},
which is 0.35 smaller than the value 12+log(O/H)=8.82 estimated with the upper $R_{23}$ branch.
It indicates that the upper $R_{23}$ branch method may overestimate 
the metallicities of SGRB host galaxies, 
and the diversity of SGRB and LGRB metallicities may not be as significant as shown.
However, due to the complexity of metallicity estimation
and the lack of H$\alpha$ and [NII] line information of
other three events, GRB 061006, GRB 070724 and GRB 051221A,
we still present the values of \cite{2009ApJ...690..231B} in Table \ref{tbsfr}. 
We notice here that the true metallicities may be 
a factor of 0.4 smaller than the listed values.
More observations, especially of the IR spectra, are required to 
verify the metallicities of these SGRBs.

With the absorption line equivalent width (EW) and line profile,
column densities of various elements $N_{\rm X}$ along the line of sight
can be estimated
\citep{2011piim.book.....D}.
By {comparing} with hydrogen column density $N_{\rm H}$ obtained from Ly$\alpha$,
metallicities
[X/H]=log($N_{\rm X}$/$N_{\rm H}$)-log($N_{\rm X_{\odot}}$/$N_{\rm H_{\odot}}$)
can be estimated
for various elements X, such as Oxygen.
The condition to produce absorption lines is that the probed regions are cooler
than those probed with emission lines.
These absorption line regions are estimated to be around 100 pc away from GRBs
\citep{2012MSAIS..21...14V, 2013A&A...557A..18K, 2014A&A...564A..38D}, so that
they can reveal the properties of local environment of GRBs.
Since generally the detection of Ly$\alpha$ absorption line is needed,
the absorption line metallicity
estimation is generally valid for high redshift GRBs, i.e.
z $>$ 1.8 in general.
If metallicity is estimated for more than one element,
the value for most abundant element is recorded,
e.g., in the order of O, C, N, Mg, Si, Fe, S
\citep{2009ARA&A..47..481A}.
The largest absorption line estimated metalllicity catalog is from 
\cite{2015ApJ...804...51C}, and other cases are obtained from individual papers.
These values are labelled as 'A' in the metallicity method column
and the specific elements used to estimate it is also recorded.
Lower limits for metallicity is usually due to saturation of the absorption lines,
e.g., in GRB 140515A.
Although these values are lower limits in definition,
they are generally used as the metallicity in the literature,
so we treat them as the measured metallicity in the rest of the paper.
The upper limits are usually due to non-detection of metal lines,
e.g. in GRB 140518A.

 
\subsubsection{Morphological properties: galaxy size and offset}

Morphological properties of GRB host galaxies 
are obtained from optical images.
Due to the faintness of GRB hosts, 
it usually requires deep and high angular resolution 
photometric observations, e.g. with
Hubble Space Telescope (HST).
The identification of a GRB host galaxy is not straightforward
{\citep{2002AJ....123.1111B, 2010ApJ...722.1946B, 2011MNRAS.413.2004C, 2014MNRAS.437.1495T}}.
If the position uncertainty is large, there might be many galaxies within the error box.
Sometimes, especially for SGRBs, the offset of the GRB location from the center of host galaxy
may be larger than the size of host galaxy itself, so that it is not straightforward to identify the
host galaxy without a probability argument.
It is also possible that the host galaxy of a particular GRB is too faint to be detected,
but there is a galaxy near the afterglow location by chance, so that it may be mis-identified as 
the GRB host.
Following \cite{2002AJ....123.1111B},
a chance coincidence probability $P_{\rm cc}$ is usually defined as
the possibility of a non-host galaxy identified as the host galaxy
by chance
$$P_{\rm cc}=1-{\rm exp}(-\pi r^2 \sigma(\le m_{\rm i})),$$
where $\sigma(\le m_{\rm i})$ is the surface densities of the galaxies with magnitude
$\le m_{\rm i}$, and $r$ is the effective radius, which is
a function of position uncertainty, offset and
the size of a candidate host galaxy.
In order to indicate how much the candidate host galaxy is trustable,
we list the instrument used to take the image,
and $P_{\rm cc}$ in column 3 and 4 in Table \ref{tboffset} when possible.


The basic morphological property of host galaxies 
is size, represented by the half brightness radius $R_{50}$,
which indicates the semi-major axis of the ellipse within which 
half flux of the entire galaxy is enclosed.
Sometimes the host surface brightness is fitted with the S{\'e}rsic profile
$$\Sigma(r)=\Sigma_{\rm e} {\rm exp}\{{-k_{\rm n}[(r/r_{\rm e})^{1/n}-1]}\},$$
with the effective radius $r_{\rm e}$ as the size indicator
\citep{2007ApJ...657..367W, 2010ApJ...708....9F, 2013ApJ...776...18F}.
Sometimes, the size of a host galaxy is defined as the
eighty-percent radius $R_{80}$
\citep{2006Natur.441..463F, 2010MNRAS.405...57S},
which is the major axis radius of a similar ellipse that encloses 80 percent of flux.
For these cases, we convert $R_{80}$ to $R_{50}$ by assuming that the surface 
brightness profile of the galaxy is an S{\'e}rsic profile.
Since nearly all LGRB host galaxies and most SGRB host galaxies
are disk (spiral) galaxies, $n=1$ is assumed.
This is equivalent to an exponential profile, which is 
consistent with the disk galaxy surface density profile.
For $n=1$, one has $R_{50}=R_{80}/1.79$, 
and $R_{50}=r_{\rm e}$.
The host galaxies of three SGRBs, GRB 050509B, {GRB 050724 and GRB 100117A},
are obviously elliptical galaxies with $n \sim 4$, so that
the conversion factor $R_{50}=0.968r_{\rm e}$ is applied.
If $R_{50}$ of more than one band is given,
the value for the band mostly close to optical is used
since the blue band may be affected by dust extinction.
Sometimes, no precisely defined radius is available, and
only vaguely defined ``size" or ``radius" are quoted in the literature.
In these cases, we treat them as $R_{80}$ which covers most 
flux of galaxy.
$R_{50}$ in units of arcsec and kpc are presented in column 5 and 6 of 
Table \ref{tboffset}. The parameter $n$ is also presented in column 7 when possible.
Some GRBs with angular $R_{50}$ values do not have redshift detections. For
these cases, {$z=0.5$ for SGRBs and $z=2.0$ for LGRBs} are assumed to estimate physical size of the galaxy.

Angular and physical offsets, the angular/physical separation of a GRB from the center of its 
host galaxy, are given in column 8 and 9 of Table \ref{tboffset},
in units of arcseconds and kpc, respectively.
If an offset is smaller than the positional uncertainty of the GRB or host galaxy,
an upper limit is given. 
The largest samples of LGRB offsets are from 
\cite{2002AJ....123.1111B} and \cite{2016ApJ...817..144B},
and the largest SGRB offset samples are from 
\cite{2010ApJ...708....9F} and \cite{2013ApJ...776...18F}.
For GRBs from Table 2 of \cite{2013ApJ...778..128P},
the angular distance between the afterglow and the host galaxy is used to define the offset.
{Similar to $R_{50}$, $z=0.5$ for SGRB and $z=2.0$ for LGRB are also assumed for those without redshift detections.}
In some problems, one cares more about the relative offset with respect to the size of the 
host galaxy. The normalized offset (the true offset normalized to $R_{50}$ of the host galaxy) 
is shown in column 10 of Table \ref{tboffset}.

Many GRB hosts are irregular and interacting galaxies.
For these, the size $R_{50}$, the center, and hence, the offset
of the galaxy are not well defined.
In these cases, the fraction of brightness $F_{\rm light}$, 
which is the ratio between the area of the host fainter than the GRB position
and the area of the entire host galaxy, is defined. 
It delineates how bright the GRB location is 
relative to the other regions of the host galaxy,
and reveals the local SFR, especially if the UV band image is used.
The largest LGRB $F_{\rm light}$ samples are from 
\cite{2006Natur.441..463F}, \cite{2010MNRAS.405...57S} and \cite{2016ApJ...817..144B},
and the largest SGRB $F_{\rm light}$ samples are from 
\cite{2010ApJ...708....9F} and \cite{2013ApJ...776...18F}.
Others are collected from individual papers.
The parameter $F_{\rm light}$ is given in column 11 of Table \ref{tboffset}.
$F_{\rm light} = 1$ indicates that the GRB is located in the brightest region
of the host, and $F_{\rm light} = 0$ indicates that the GRB is in the faintest region 
of the host.

\section{Distribution of properties}

With the comprehensive prompt and host galaxy properties
in Table \ref{tbbasic}, \ref{tbprompt}, \ref{tbsfr} and \ref{tboffset},
we are able to study the differences and similarities of LGRBs and SGRBs.
We present the distributions of both LGRBs and SGRBs for each property in this section.
The histograms are shown in Fig. \ref{fig1d}, and the
statistical results are presented in Table \ref{tb1dconsensus} and Table \ref{tb1dT90}.
In all the figures, LGRBs are shown in red and SGRBs in blue.
In Fig. \ref{fig1d}, the histograms of all the GRBs are presented in black.
Dotted lines show objects with redshift $z <$ 1.4, 
within which most SGRBs are located. Inspecting this sample allows one
to compare SGRBs to LGRBs in the similar redshift range,
and examine the influence of redshift on each parameter.
In the left column of all the figures, LGRBs and SGRBs are defined by the ``consensus''
criteria, i.e., GRBs with label ``S'' in Greiner's catalog
are defined as SGRBs, otherwise LGRBs.
Their statistical results are shown in Table \ref{tb1dconsensus}.
In the right column of all figures, LGRBs and SGRBs are defined by $T_{90}$ only,
i.e., GRBs with $T_{90} <$ 2 s are defined as SGRBs,
otherwise LGRBs.
Their statistical results are shown in Table \ref{tb1dT90}.
In Table \ref{tb1dconsensus} and \ref{tb1dT90}, 
the numbers of LGRBs and SGRBs with each parameter are given in column 2 and 4.
The median values and dispersion of them are given in column 3 and 5. 

In order to investigate how different the LGRB sample is from the SGRB sample,
we employ the Kolmogorov-Smirnov test (KS test),
and examine the fraction of LGRBs and SGRBs overlapping with each other.
Column 6 of Table \ref{tb1dconsensus} and \ref{tb1dT90} show 
the null probability $P_{\rm KS}$ of KS test 
between LGRBs and SGRBs for these two definitions of SGRBs and LGRBs, respectively. 
The smaller $P_{\rm KS}$ is, the more different LGRBs and SGRBs are from each other
for that particular property. The overlapping range of each parameter is shown in column 7
of both tables. 
The fractions of LGRBs and SGRBs located in the overlapping region (defined as ``overlapping fraction''
hereafter) are presented in column 8 and 9. In the following, we discuss
each property in detail.

The redshift distributions of the consensus and $T_{90}$-defined LGRBs and SGRBs
are presented in the left and right column of Fig. \ref{fig1d}, Row 1.
Photometric redshifts are not included.
It is apparent that SGRBs show a much lower redshift distribution than LGRBs,
with $z_{\rm SGRB}=0.45 \pm 0.51$ as compared with $z_{\rm LGRB}=1.64 \pm 1.30$.
The highest-redshift GRB in our sample is GRB 090423,
with $z=8.23$ obtained from absorption lines
\citep{2009GCN..9219....1T, 2009Natur.461.1258S}.
GRB 090429B has a photometric redshift $z=9.2$
\citep{2011ApJ...736....7C}
without absorption/emission lines,
and there is no host galaxy information available.
It is not included in our sample according to our primary selection criteria
given in Section 2.
The highest redshift SGRB is GRB 090426,
which is an ambiguous event with $T_{90} =1.24$ s
\citep{2009A&A...507L..45A, 2010MNRAS.401..963L}.
If it is considered as a short GRB based on the duration criterion, the overlapping fraction of 
LGRB redshift is as large as 72 \%.
If it is classified as a Type II GRB based on other information (and hence join the LGRB sample), 
the LGRB overlapping fraction in redshift is 29 \%.
Consensus SGRBs have less high redshift objects than $T_{90}$-defined SGRBs,
which results in a smaller $P_{\rm KS}$
and indicates more significant difference between the two groups.
Note that even with a redshift cut $z<1.4$,
LGRBs still show a higher median redshift than SGRBs,
due to the dominance of high redshift events over low redshift events
in this LGRBs sub-sample.

\subsection{Prompt emission properties}

Duration $T_{90}$ denotes the (observed) time scale of GRB explosions.
Physically, Type I GRBs, which have a neutron-star dense accretion torus from the debris of NS-NS
or NS-BH mergers, have a small free-fall time scale to allow short-duration GRBs. Type II GRBs,
on the other hand, having an extended stellar envelop with stellar density, have a free-fall time scale
longer than several seconds, which is natural to explain long-duration GRBs.
The distributions of $T_{90}$ for the consensus and $T_{90}$-defined LGRBs and SGRBs
are presented in the left and right panels of Figure \ref{fig1d}, Row 2, respectively.
The $T_{90}$ distribution of the entire GRB population (both LGRBs and SGRBs) (black lines)
show a peak around 50 s 
and a flat tail in the range smaller than 2 s.
The bimodality is not as significant as in the BATSE sample
\citep{1993ApJ...413L.101K}, 
due to the dominance of LGRBs.
The dominance of LGRBs is a result of the dominance of the Swift sample,
since 325/407 events in our sample are discovered by Swift,
and Swift is dominated by LGRBs due to its insensitivity to SGRBs
\citep{2011ApJS..195....2S, 2013ApJ...763...15Q}.
For $T_{90}$-defined LGRBs (red solid line) and SGRBs (blue solid line),
the $T_{90}$ criterion gives
the lowest $P_{\rm KS}$ among all the parameters 
as shown in Table \ref{tb1dconsensus} and \ref{tb1dT90}, i.e. $\sim 10^{-27}$. 
This is simply because the definitions of LGRBs and SGRBs are based on the duration criterion.
For the consensus LGRB and SGRB samples, 
the SGRB $T_{90}$ distribution extends to as long as 5.66 s, GRB 090510,
and the LGRB $T_{90}$ distribution extends to as short as 1.30 s, GRB 000926.
Such an overlap increases $P_{\rm KS}$ by one order of magnitude but still allows a
very low $P_{\rm KS}$ value, suggesting that 
the $T_{90}$ criterion is truly a good indicator to separate the two physically distinct populations.
The significant overlap in the $T_{90}$ properties (7\% in LGRBs and 20\% in SGRBs), on the other hand, suggests that 
other properties are needed to correctly place a certain GRB into the right physical category
(Type I vs. Type II).

Isotropic gamma-ray energy $E_{\rm \gamma,iso}$ gives a rough indicator of the energy budget of a GRB.
In the BH central engine scenario, the total energy budget is related to
the total material available for accretion.
Type II GRBs, having plenty of fuel from the massive star progenitor ($M > 30 M_{\sun}$ for the total mass budget), 
are expected to be 
more energetic than Type I GRBs, 
which are related to compact star mergers ($M \sim (2-3) M_{\sun}$ for the total mass budget).
In the magnetar scenario, some energy from {a} NS-NS merger may be released in the form of
gravitational {waves (GWs)} or falls into the collapsed BH, resulting in less energetic 
Type I GRBs than Type II GRBs
\citep[e.g.][]{2016PhRvD..93d4065G}.
Observationally, our sample shows that the $E_{\rm \gamma,iso}$ distribution of LGRBs 
is nearly a {Gaussian},
with an extremely low energy tail extending to $10^{47}$ erg.
The LGRBs with $E_{\rm \gamma,iso} < 10^{49}$ erg 
are usually defined as low luminosity GRBs (llGRBs), 
probably with a somewhat different physical origin from normal LGRBs
\citep{2006Natur.442.1008C, 2006Natur.442.1014S, 2007ApJ...662.1111L, 2009MNRAS.392...91V, 2011ApJ...739L..55B, 2015ApJ...812...33S}. 
Due to their rareness, the inclusion of llGRBs does not significantly influence 
the median $E_{\rm \gamma,iso}$ and the overlapping fraction of LGRBs.
Nearly all SGRBs have $E_{\rm \gamma,iso} > 10^{49}$ erg, 
so the inclusion of llGRBs does not influence the overlapping fraction with SGRBs much, either.
The median $E_{\rm \gamma,iso}$ of SGRBs is about 1.6 dex lower than the entire sample of LGRBs,
and $P_{\rm KS}$ of the $E_{\rm \gamma,iso}$ criterion is as significant as 10$^{-14}$.
However, the low redshift LGRBs shows a 0.5 dex smaller $E_{\rm \gamma,iso}$,
making $P_{\rm KS}$ eight orders of magnitude larger (but still small) if one focuses on the $z<1.4$ sample.
Due to their wide distributions ($\sigma$=1.0 dex),
the SGRBs and LGRBs show significant overlap in the $E_{\rm \gamma,iso}$ domain.
If there were no duration information, SGRBs look like the low energy tail of LGRBs,
suggesting that the $E_{\rm \gamma,iso}$ property alone is not a good criterion 
to differentiate between the two populations.

The typical peak luminosity $L_{\rm p,iso}$ of LGRBs is about 0.8 dex larger than
that of SGRBs. However, due to the large dispersion, 1.1 dex,
the difference between these two samples is not significant, either,
with $P_{\rm KS}=$ 0.007. 
LGRBs at $z<1.4$ have 0.6 dex smaller $L_{\rm p,iso}$ than 
the entire LGRB sample, making it more difficult to apply the $L_{\rm p,iso}$
criterion for classification. 
This is consistent with 
\cite{2009ApJ...703.1696Z} and \cite{2009A&A...496..585G}, 
who showed that LGRBs and SGRBs have similar $L_{\rm p, iso}$, and
their differences in $E_{\rm \gamma,iso}$ is mostly due to different durations.

It has been long known that
LGRBs have softer spectra than SGRBs.
Theoretically, such a connection is not straightforward and is model dependent
(e.g. Zhang et al. 2009 for a detailed discussion), but it may be somewhat related
to a possible higher Lorentz factor in SGRBs, {originating} from a relatively 
cleaner environment of Type I GRBs.
The hardness of a spectrum is a combination effect of
the peak energy $E_{\rm p}$, 
and the low energy photon index $\alpha$.
Consistent with previous work, LGRB $\alpha$ is $-1.01\pm0.34$, 
softer than that of SGRBs, $\alpha = -0.60\pm0.25$.
The difference between them is {moderately strong}, with $P_{\rm KS}=10^{-7}$.
The $E_{\rm p}$ center value of LGRBs is 0.49 dex smaller than SGRBs,
and the two samples have about 80\% overlaps, which 
{shows a moderately strong} difference between LGRBs and SGRBs.

The amplitude parameter $f$ shows 43\% LGRBs and 100\% SGRBs within the overlap region, 
The K-S test gives $P_{\rm KS} \sim 10^{-7}$, 
indicating a {moderately strong} difference between LGRBs and SGRBs.
As suggested by L\"u et al. (2014), the $f_{\rm eff}$ is expected to be a better indicator.
Our analysis shows $P_{\rm KS} \sim 10^{-20}$ between LGRBs and SGRBs, which is
indeed a good indicator. It is still not as significant as the $T_{90}$ criterion,
due to the smaller sample of $f_{\rm eff}$ than $T_{90}$.
{
The LGRB and SGRB overlapping fractions of $f_{\rm eff}$
are 7\% and 79\% in the consensus samples.
GRB 130427A, which has the the largest $f_{\rm eff}=4.75$ is 
an obvious outlier.
It shows an intense initial pulse with a weak tail in Swift/BAT 
while the peak is not significant in Fermi/GBM.
If excluding it from the LGRB sample,
The LGRB and SGRB overlapping fractions of $f_{\rm eff}$
are 7\% and 48\%, respectively.
}

\subsection{Host galaxy properties}

\subsubsection{Stellar mass, Star formation rate, and metallicity}

\begin{figure*}[!b]
\centering
\includegraphics[width=0.4\textwidth]{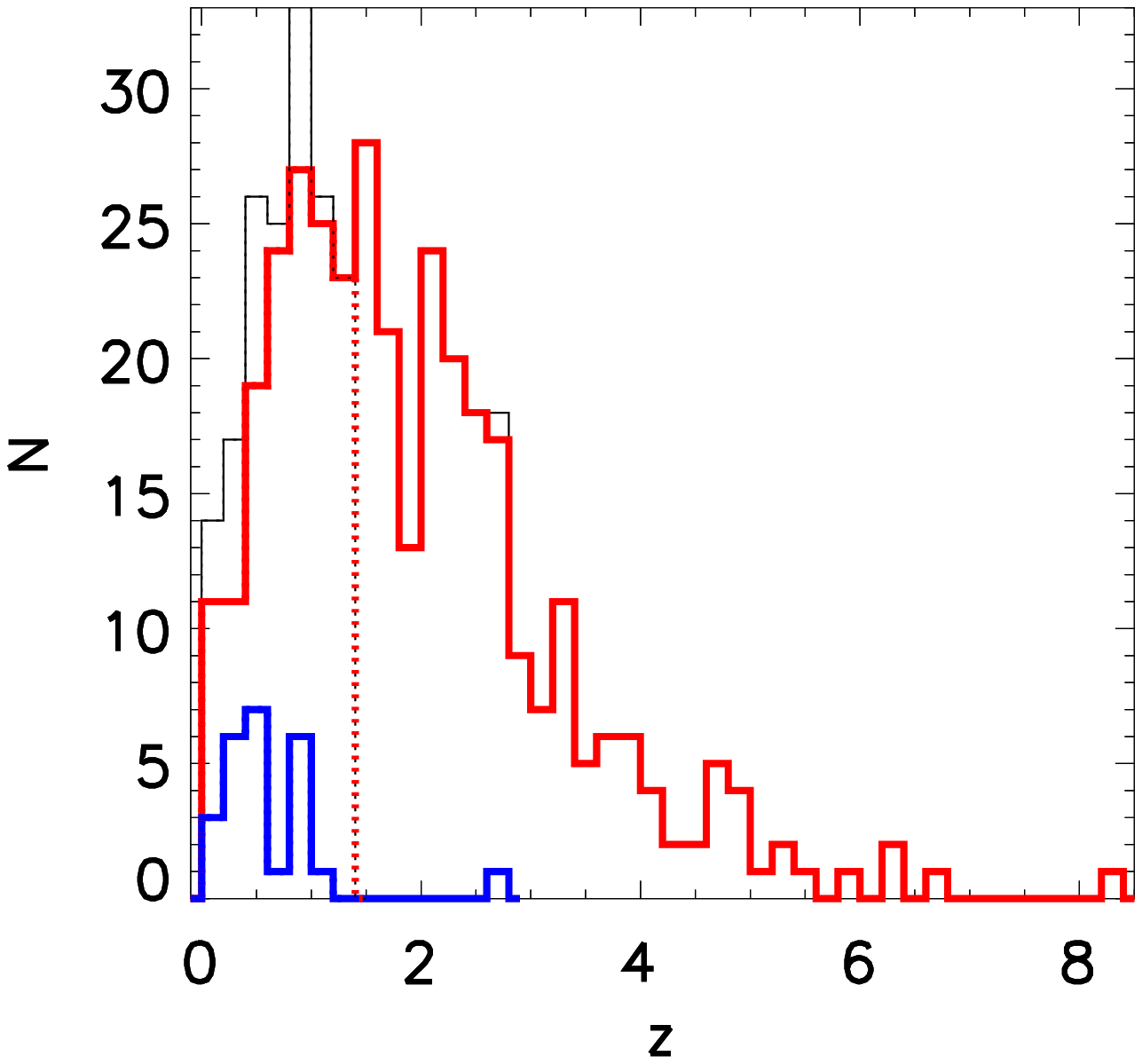}
\includegraphics[width=0.4\textwidth]{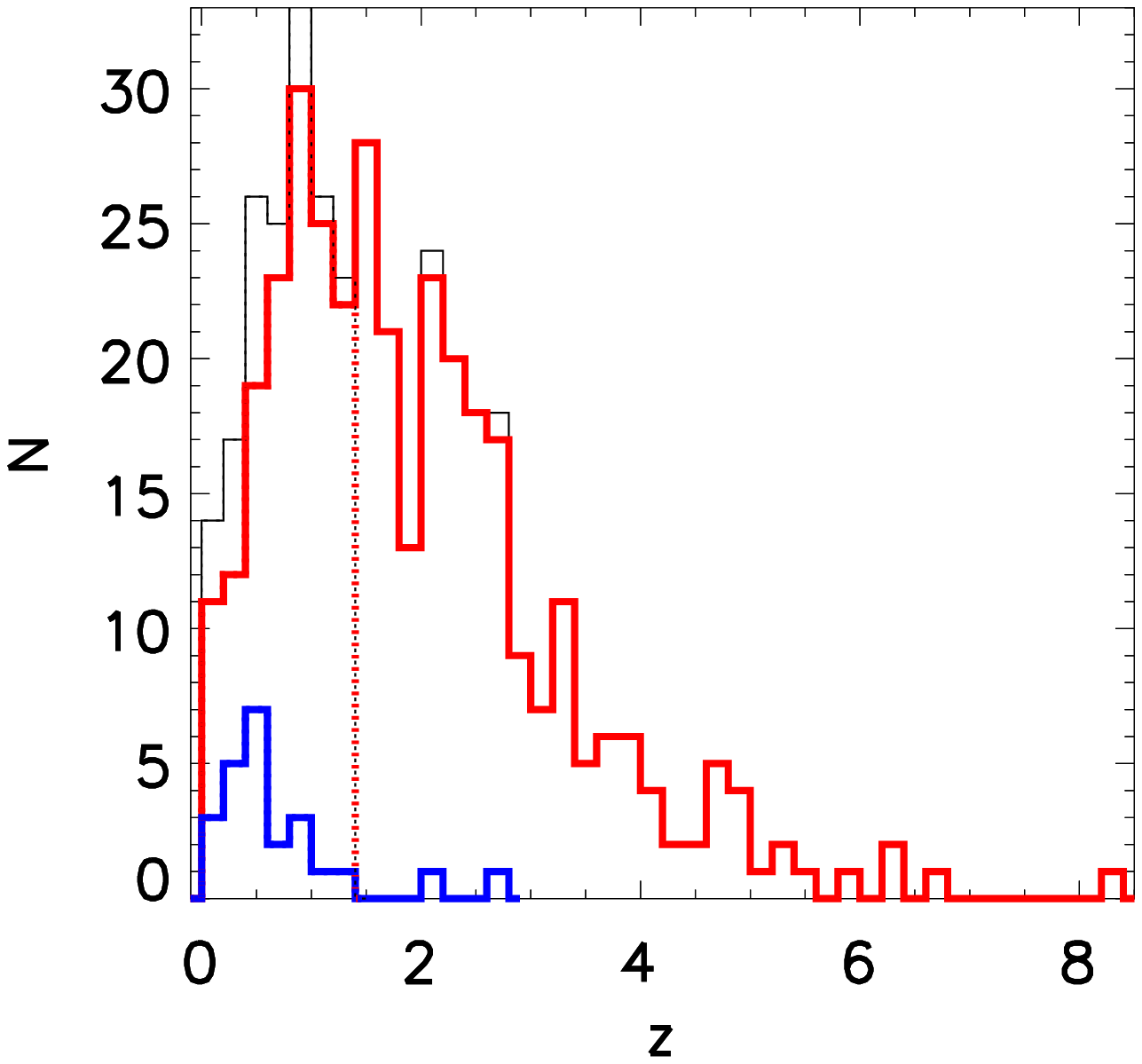}

\includegraphics[width=0.4\textwidth]{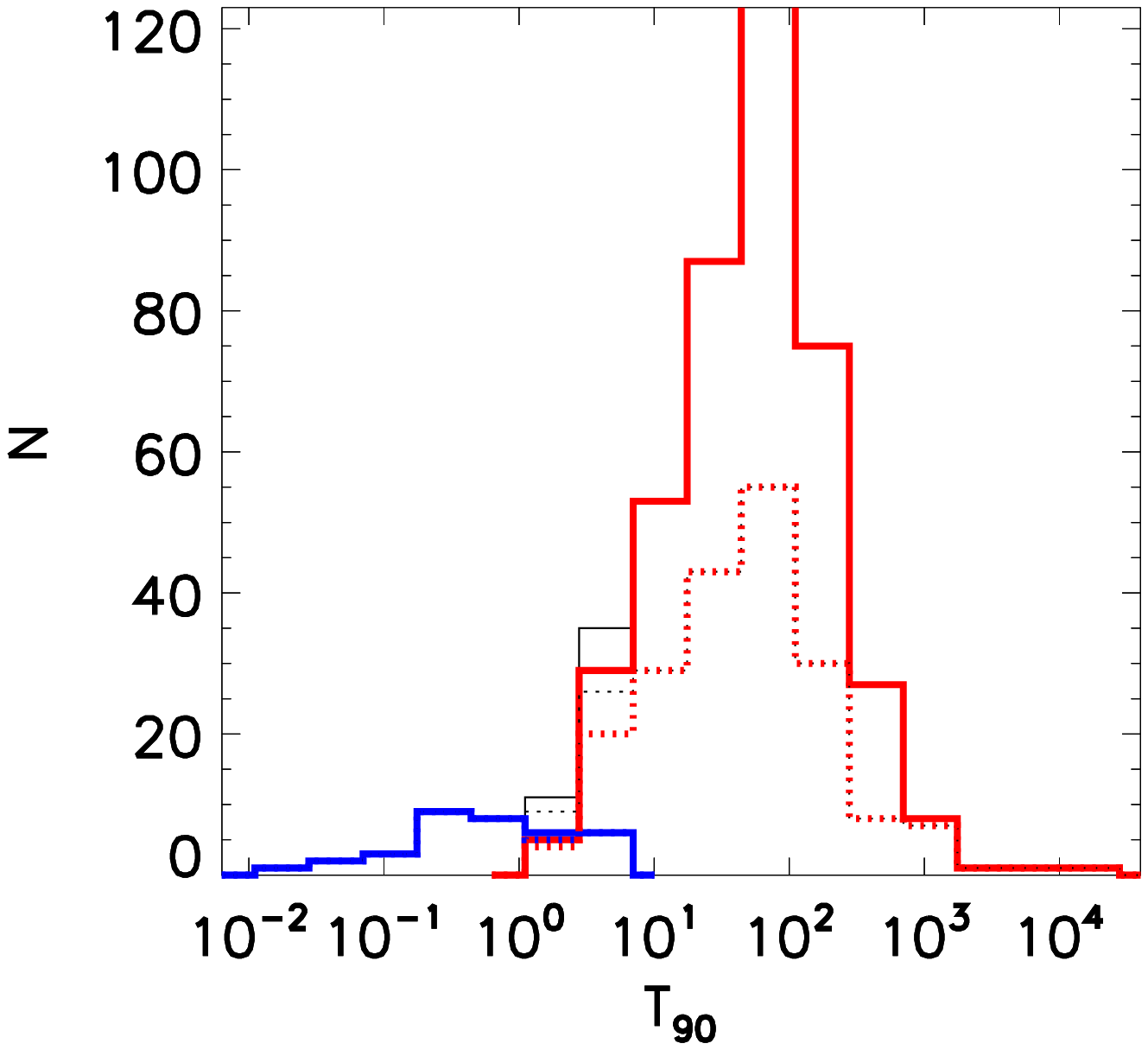}
\includegraphics[width=0.4\textwidth]{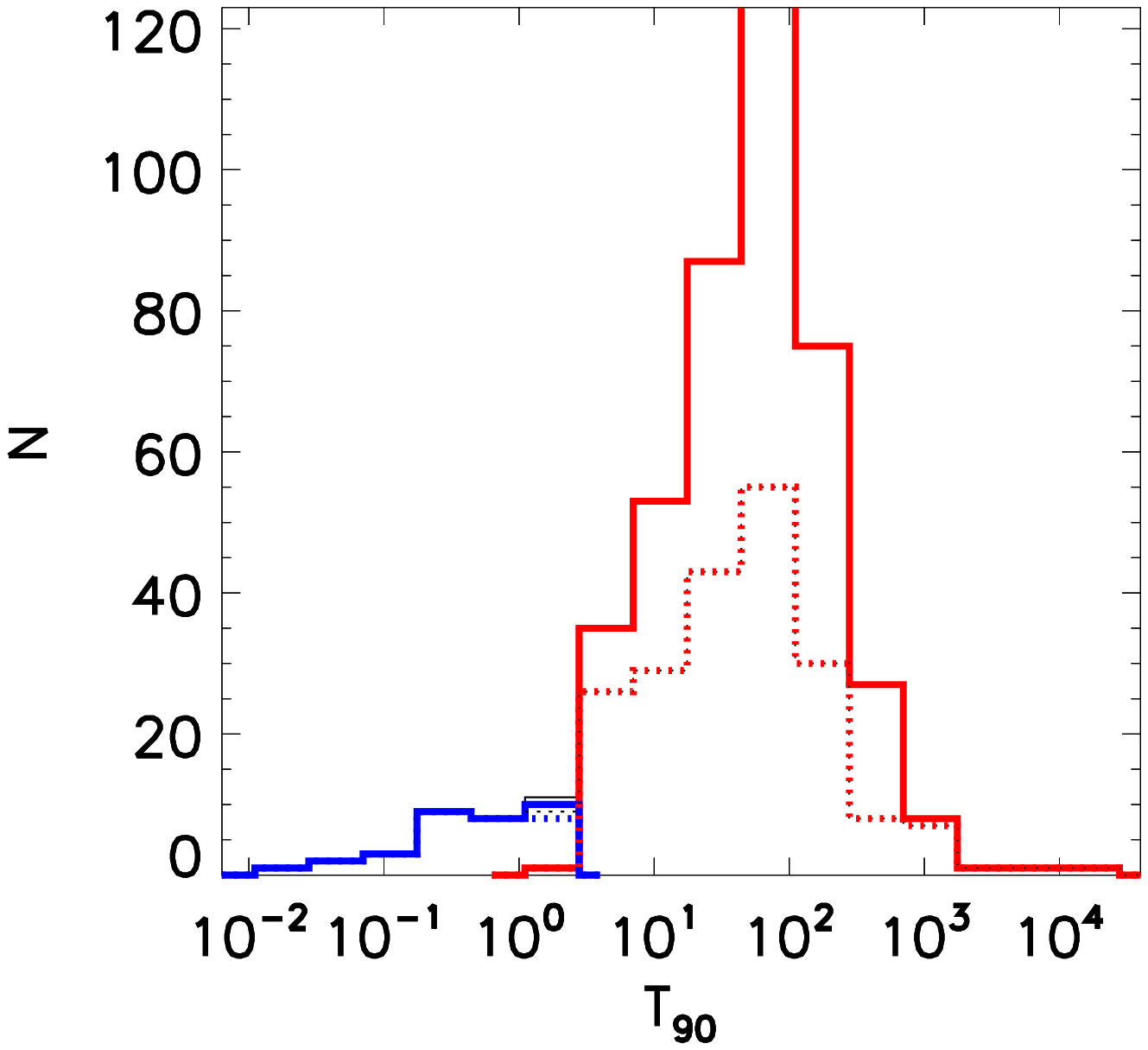}

\includegraphics[width=0.4\textwidth]{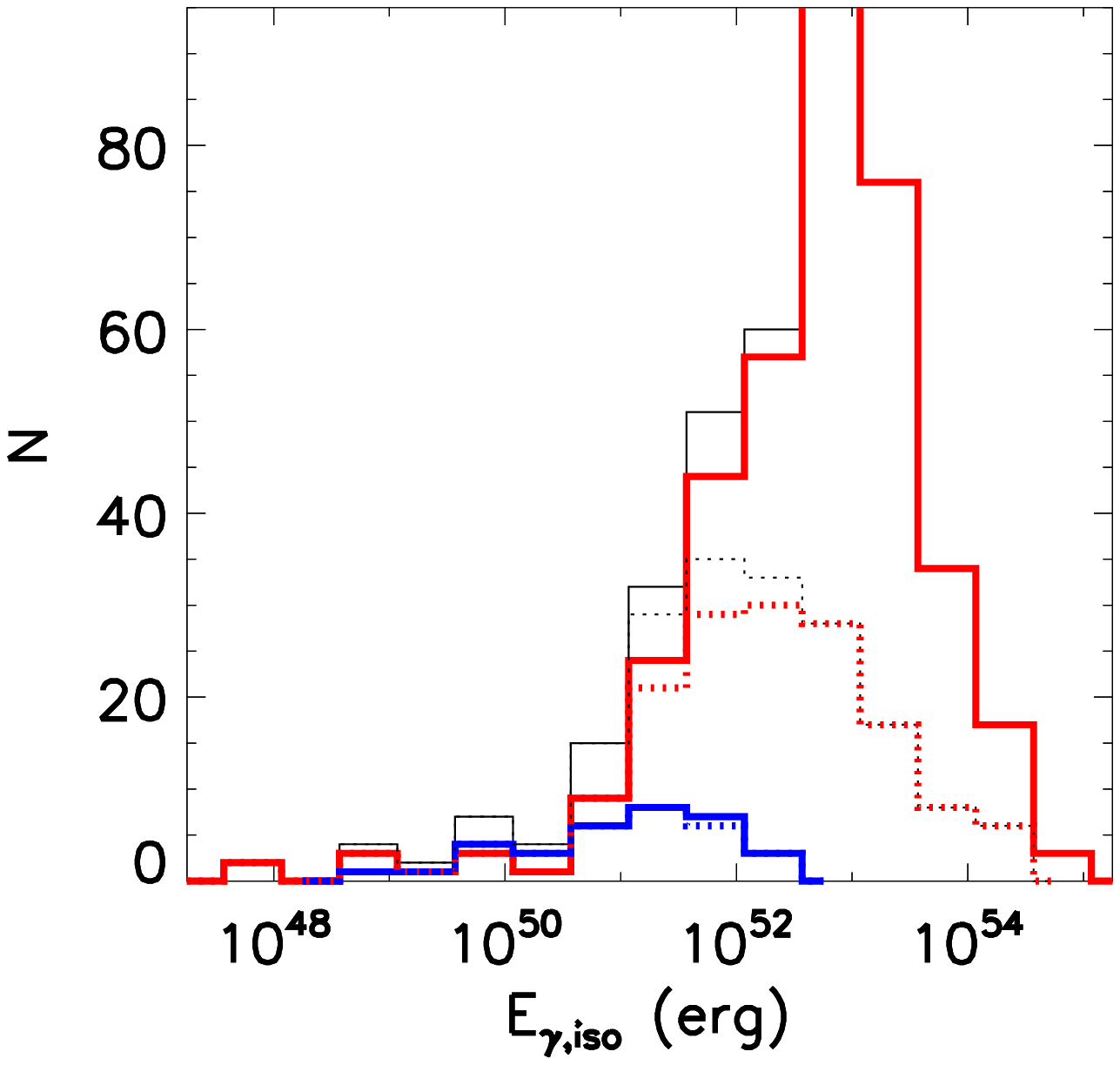}
\includegraphics[width=0.4\textwidth]{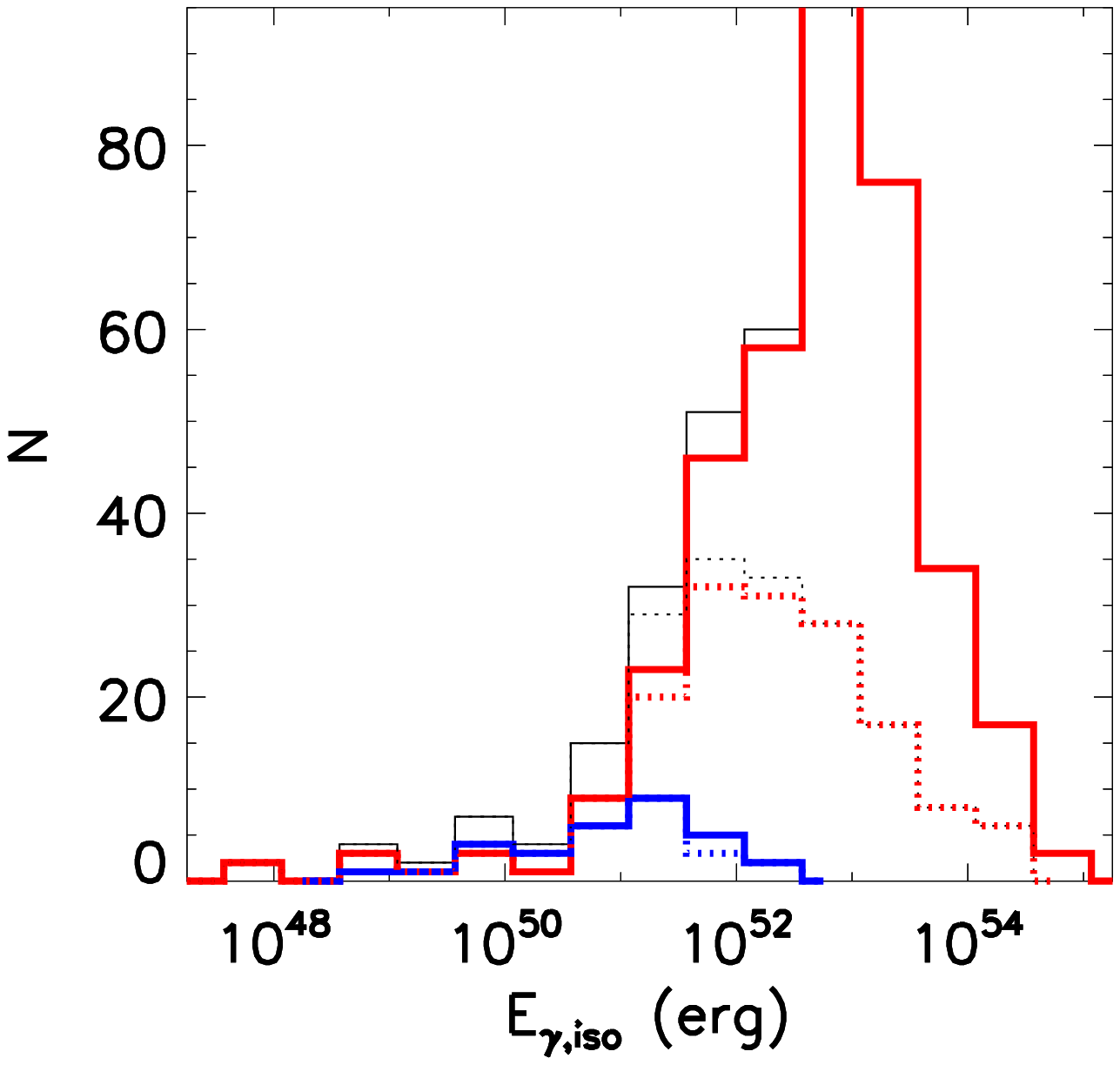}

\includegraphics[width=0.4\textwidth]{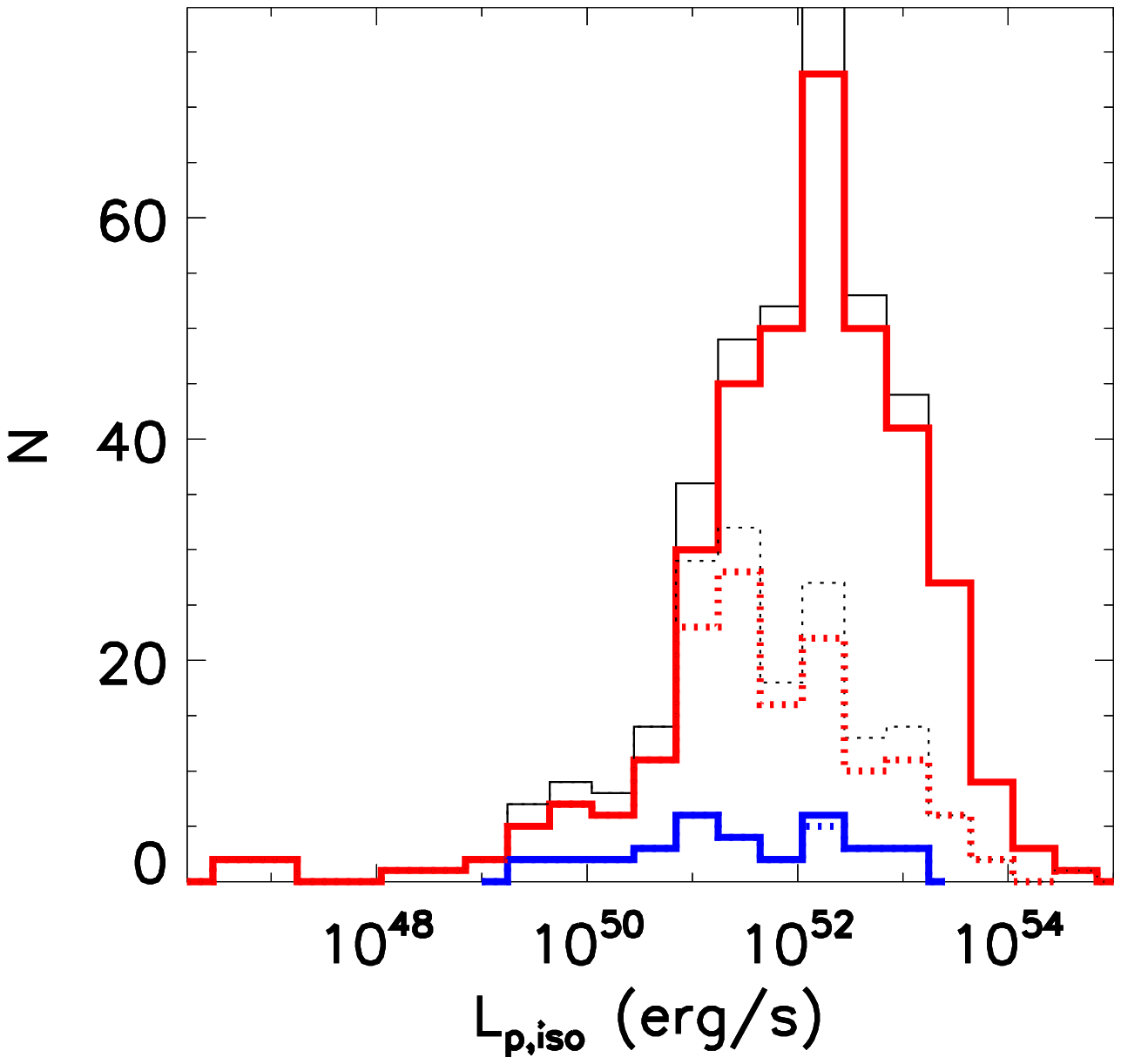}
\includegraphics[width=0.4\textwidth]{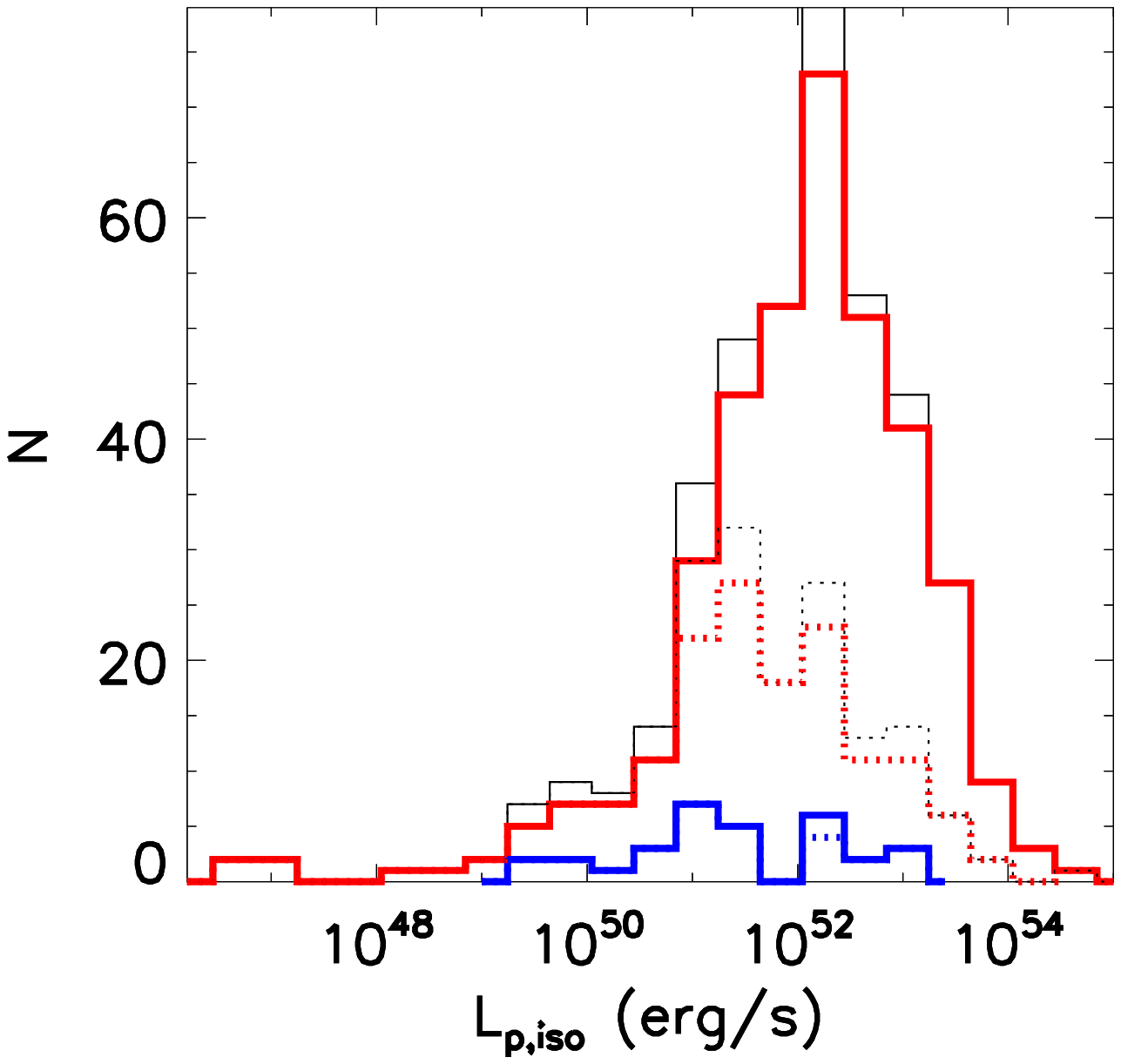}

\caption{
Distribution of prompt and host galaxy parameters of LGRBs (red lines)
and SGRBs (blue lines). 
Left panels show distributions with consensus defined LGRBs and SGRBs,
and right panels show distributions with $T_{90}$ only defined LGRBs and SGRBs.
Dotted lines show distribution of $z<1.4$ subsamples.
}
\label{fig1d}
\end{figure*}

The properties of galaxies are mainly controlled by their stellar mass $M_*$
\citep{2014ApJ...788...28V, 2015A&A...579A...2I}.
Host galaxy masses of the consensus and $T_{90}$-defined LGRBs and SGRBs
are presented in the left and right columns of Fig. \ref{fig1d}, Row 9.
In order to be consistent, only stellar masses obtained with
SED fitting are used here.
Most SGRB and LGRB host galaxies are smaller than the turnover mass of galaxies
extending to redshift 3 \citep{2015MNRAS.447....2M}.
Although the median of SGRB hosts is 0.6 dex larger than that of all LGRB hosts,
their difference is not statistically significant.
LGRB hosts with $z<1.4$ show a 0.2 dex lower stellar mass than the whole sample.
It may be an selection effect, 
since the galaxies with larger stellar masses are brighter
and easier to be observed at high redshifts.
It makes the low-$z$ LGRBs more significantly different from SGRBs.
Since the median redshift of low-$z$ LGRBs is still larger than that of SGRBs, 
a true same-redshift comparison between LGRB and SGRB host stellar masses
should show even more significant differences.
Also, it indicates that there should be more small stellar mass
host galaxies that are not been discovered yet, especially in the high redshift range.
The overlapping fractions are around 90\% for both LGRBs and SGRBs. 
The results of the consensus and $T_{90}$-defined LGRB and SGRB samples 
are consistent with each other.

The SFR represents the global star formation status of the entire galaxy.
It is expected to be large in LGRB hosts,
since LGRBs are presumed to be massive star collapsars and  
are expected to be associate with star formation.
SGRBs are believed to be related to compact star mergers,
so at least some of them are expected to be associated with the old stellar {populations}
and no recent star formation is required for the presence of SGRBs.
SFR of consensus and $T_{90}$-defined LGRBs and SGRBs 
are presented in Fig. \ref{fig1d}, Row 10.
In order to be consistent, only SFRs obtained with emission lines
are used.
The median SFR of {LGRBs} is around 0.4 dex larger than that of {SGRBs},
although their difference is not statistically significant, 
due to the large dispersion, both around 0.8 dex.
It may be also due to the generally more massive host galaxies of SGRBs,
since SFR is proportional to the stellar mass of the galaxies.
The low-$z$ LGRB hosts are similar to those of low-$z$ SGRBs.
It may be a result of the decrease of the LGRB host mass at low redshifts.
Both LGRBs and SGRBs show around 90\% overlapping fraction for SFR.
The $T_{90}$-defined samples show even less difference and larger overlaps,
suggesting the limitation of $T_{90}$ to define the physical category of GRBs.

In the sample of the consensus SGRBs, the one with an extremely large SFR is GRB 100816A.
Its $T_{90}$ is reported to be 2.9 s in the Swift GRB table and 
1.99 s in \cite{2013EAS....61..345P}. 
With a small spectral lag 10 $\pm$ 25 ms\citep{2010GCN..11113...1N},
it is suggested to be a SGRB in Greiner's catalog.
Considering its high SFR 
\citep{2015A&A...581A.125K}
and possible interacting nature of the host galaxy
\citep{2010GCN..11123...1T}, 
we would suggest it to be still a Type II GRB. 
We still keep it in the consensus SGRB sample based on our sample selection criterion.
Changing it to the consensus LGRB sample 
makes the median log(SFR) of SGRB to be 0.08, i.e., 1.2 $M_{\sun}$ yr$^{-1}$,
with a dispersion 0.71,
and results in a lower $P_{\rm KS}$ 0.04 for this criterion.

For the bursts with both SFR and stellar mass $M_*$, 
specific SFR (${\rm sSFR} = {\rm SFR}/M_*$) of the host galaxy is available.
Since sSFR describes SFR per unit stellar mass,
it is {a} more relevant parameter to describe 
the star formation status in the GRB location.
The distributions of sSFRs are presented in Fig. \ref{fig1d}, Row 11.
sSFR shows a more significant difference between LGRBs and SGRBs than SFR.
In general, the sSFR of LGRBs is 0.5 dex higher than SGRBs. 
The redshift evolution of the sSFR for the LGRB host is not significant, 
even though the redshift evolution of sSFR of the entire universe is apparent,
with a peak at $z=2-3$. 
This may indicate that sSFR is directly related to the LGRB rate.
Another factor might be the selection effects. 
Since the massive hosts with less sSFRs are {easier detected},
high redshift samples should on average show smaller sSFRs relative to 
the true distribution.
The $T_{90}$-defined sample shows a 0.3 dex less difference than the consensus sample,
again indicating the limitation of the $T_{90}$-only criterion.

In the consensus LGRB sample, the GRB with the lowest sSFR, 0.006 Gyr$^{-1}$, is GRB {050219A}
\citep{2014A&A...572A..47R}.
Its host was discovered by GROND and confirmed by VLT. 
No HST image is available. 
It is an elliptical galaxy 4.6" away from the GRB XRT location, with a
1.9" positional uncertainty.
The estimated chance coincide probability is $P_{\rm cc}=0.8\%$. 
If we exclude GRB 050219, 
the LGRB sSFR becomes $0.05\pm 0.62$ Gyr$^{-1}$ and $P_{\rm KS}=0.01$.
The overlap range becomes (-1.17, 1.05) and
the overlapping fraction becomes 90\% and 78\% 
for the consensus LGRBs and SGRBs, respectively.

LGRB progenitor models prefer a low metallicity environment,
since it would keep enough angular momentum in the core the star to launch a jet.
On the other hand, no metallicity limitation is required for SGRBs.
The distributions of metallicity [X/H] of the consensus and $T_{90}$-defined
LGRBs and SGRBs are presented in the left and right columns of Fig. \ref{fig1d}, Row 12.
If one event has double values, 
the average value is plotted.
For the consensus samples, SGRBs show a 0.5 dex richer metallicity than LGRBs,
and are consistent with the highest end of the consensus LGRBs.
The $P_{\rm KS}$ value is 0.0007, indicating a relatively significant difference.
The overlapping fractions are 47\% and 100\% for LGRBs and SGRBs, respectively, which 
is as low as that of the amplitude parameter $f$. 
However, the $z<1.4$ LGRB sample is much metal richer than the whole LGRB sample.
In this redshift range, the difference between median LGRBs and SGRBs becomes 0.3 dex, 
and the overlapping fraction of LGRBs is increases to 73\%.
Since the average redshift of low-$z$ {LGRBs} is still higher than {SGRBs},
and metallicity from \cite{2009ApJ...690..231B} may overestimate the SGRB metallicity,
the difference between SGRBs and LGRBs in the same redshift {bin} may be even milder.
This makes metallicity not a good indicator of the physical origin of individual GRBs. 

\subsubsection{Morphological properties: galaxy size and Offset}

Galaxy size is correlated with stellar mass,
according to the galaxy types
\citep{2014ApJ...788...28V}.
The $R_{50}$ distributions of
the consensus and $T_{90}$-defined LGRB and SGRB samples are
presented in Fig. \ref{fig1d}, Row 13.
The LGRB host size is typically 0.31 dex smaller than that of SGRBs 
in the consensus samples, with a small $P_{\rm KS}=4\times10^{-4}$.
The overlapping fraction of LGRBs is $\sim$ 69\%.
For $z<1.4$, the $R_{50}$ distribution of LGRB hosts is consistent with
that of the whole sample, only 0.06 index larger.
However, due to shrinkage of the sample size,  
The $P_{\rm KS}$ value is one order of magnitude larger.
The $T_{90}$ defined SGRB sample includes more small size hosts,
again indicating the limitation of the $T_{90}$ criterion.

{SGRB} offsets are expected to be larger than LGRBs,
since the explosion of SNe that formed the NSs and BHs in the merger systems
would have given the system two kicks, so that the system may have a large
offset from the original birth location in the host galaxy. 
The cumulative offset distribution of SGRBs indeed differ from that of LGRBs
\citep{2010ApJ...708....9F, 2013ApJ...776...18F, 2014ARA&A..52...43B}. 
Our analysis shows that the typical physical offset of SGRBs, in units of kpc, 
is 0.77 dex larger 
than that of LGRBs. The KS test gives $P_{\rm KS} = 10^{-4}$.
However, the overlapping fractions of both LGRBs and SGRBs are as large as 80\%.
Only 5 of the 28 SGRBs show offsets larger than all LGRBs.
The redshift evolution of the offsets is not significant. 
At $z<1.4$, LGRBs have the same median physical offset as the whole sample.

The offset normalized to the host size $R_{50}$ is a more physical parameter to delineate 
the location of a SGRB within the host galaxy.
Also, normalized offset does not require the measurements of the absolute values 
of the offset and host size, so one can include events without redshift measurements as well.
The normalized offset distributions are presented in Fig. \ref{fig1d}, Row 15. 
In general, SGRBs are 0.27 dex larger than LGRBs,
and mildly different with $P_{\rm KS}$=0.02.
No redshift evolution is seen.
Only 3 SGRBs have normalized offset larger than all LGRBs, 
and only 8 LGRBs have normalized offset smaller than all SGRBs.
The overlapping fractions are as high as 90\%.

The surface brightness fraction $F_{\rm light}$ is expected to be large 
for LGRBs since they are believed to be associated with the highest local SFR
in the galaxy. 
The $F_{\rm light}$ of SGRBs is expected to be small since compact star mergers usually
are expected to be kicked from the star forming regions {by the time the} merger happens.
It is also a parameter that does not require a redshift measurement.
They are presented in the last row of Fig. \ref{fig1d}.
It can be seen that SGRBs tend to be located in the faint regions of their hosts
and LGRBs tend to be located in the bright regions of their hosts.
A SGRB within the brightest region of its host is the ambiguous GRB 090426,
which {has} $F_{\rm light}=0.82$.
Although the numbers of both consensus SGRBs and LGRBs with $F_{\rm light}$ 
measurement are relatively small, 
$F_{\rm light}$ shows the most significant difference between the two types of GRBs
in the host galaxy properties, with
$P_{\rm KS} = 7\times 10^{-5}$.
The regions where LGRBs are located are 60\% fraction brighter than 
the {regions} where SGRBs reside.
Excluding GRB 090426, the overlapping fractions become
48\% and 100\% for LGRB and SGRBs.
At $z<1.4$, LGRBs are located in even brighter regions of their hosts, 
and SGRBs are located 
in even fainter regions due to the exclusion of 
ambiguous GRB 090426.
It makes the difference between LGRBs and SGRBs even more significant, and 
the overlapping fractions are as low as 40\%. 
It is one of the best physical origin indicator candidates.
Similar to other parameters, $T_{90}$-defined samples show 
less difference between LGRBs and SGRBs.

\subsection{Simulated 1-D distribution}

The KS test provides a statistical judgement 
about how different two groups of data are.
By definition, it is sensitive to the sample size, the number of objects within each group.
$T_{90}$ has the largest sample size among all the tested properties, with $403$ in total,
so it is easier to show more significant differences, with very small $P_{\rm KS}$ values.
Physically, however, we want to
examine how efficient each property of GRB is for
distinguishing SGRBs from LGRBs. 
It is a fair comparison only if we use equal sample size for each property.
We then simulate 400 GRBs (which is roughly the $T_{90}$ sample size)
for each property, based on the observational sample we already have.


The simulated numbers of LGRBs and {SGRBs}, median and dispersion of each property,
and null probability of the KS test are presented in {Table} \ref{tb1dfake}.\footnote{Although 
the absolute $P_{\rm KS}$ value depends on    
the seed of random generator, the relative significance of difference physical parameters do not change.}
For each property, the sum of the LGRB and SGRB numbers are 400.
The median and dispersion of each property are generally the same as
the observed sample.
According to $P_{\rm KS}$, 
$f_{\rm eff}$ shows the most significant difference between LGRB and SGRBs,
$P_{\rm KS} \sim 10^{-38}$.
This suggests that $f_{\rm eff}$ is the most efficient criterion 
for LGRB and SGRB classification, even better than $T_{90}$.
Besides $T_{90}$ and $f_{\rm eff}$, two prompt emission properties,
the host galaxy property $F_{\rm light}$ shows $P_{\rm KS}=4 \times 10^{-19}$, 
In the $z<1.4$ sample, 
$F_{\rm light}$ shows an even more significant difference between LGRBs and SGRBs,
with $P_{\rm KS}=4 \times 10^{-32}$. 
It suggests that $F_{\rm light}$, as a {representative} of the host galaxy properties, 
is also a good indicator of LGRB and SGRB classification.
Besides these, $f$, $\alpha$, $E_{\rm \gamma, iso}$, physical offsets, and size of the host galaxy $R_{50}$ also show
significant differences between LGRBs and SGRBs.
However, opposite to common sense,
SFR does not show a significant difference between the two classes. 
This may be due to the generally larger mass of SGRB host galaxies, which compensate 
their relatively low sSFR.
On the other hand, the 8\% to 100\% overlapping fractions of each parameter do not change
with the sample size.
As a result, multiple parameters are always needed to tell the physical categories of GRBs.


\section{2-D distributions of the properties}

Two-dimensional distributions of properties play an important role in classifications
of astronomical objects. A famous example is the Hertzsprung - Russell diagram
for stars. In GRBs, the duration - hardness ratio plot played an important role of
defining LGRBs and SGRBs (e.g. Kouveliotou et al. 1993).


Since in this paper we perform a joint analysis between prompt emission properties
and host galaxy properties of GRBs, it is interesting to investigate these two
types of properties in pairs in 2-D distribution plots. This would allow us to
investigate whether there are distinct 2-D distribution plots that can clearly
separate two physical classes of GRBs.
In the following, we examine the difference between LGRBs and SGRBs 
in different combinations of prompt emission properties vs. host galaxy properties.
Redshift vs. host galaxy property plots are also presented,
in order to study the selection effects and possible redshift evolution.
Since the eventual goal is to investigate the differences between Type I and Type II GRBs
using these plots, we use the consensus SGRB and LGRB samples (which
already considered multiple criteria other than $T_{90}$) in the analysis.
All the 2-D distribution plots are presented in Figure \ref{fig2d}, 
and the statistical results are presented in Table \ref{tb2d}. 
The numbers of LGRBs and SGRBs for each pair of parameters are shown in column
2,7,12,17 of Table \ref{tb2d}.

Since the standard 2-D KS test only works well for samples without correlations,
and since some of our 2-D plots show mild to significant correlations,
we perform a rotated KS test to investigate 
how different LGRB and SGRB samples are from each other.
For each 2-D plot, we rotate the axis with 180 trial angles from $0-180$ degrees
and calculate the $P_{\rm KS}$ along the new $x$-axis for each angle.
We then choose the lowest $P_{\rm KS}$ as the $P_{\rm KS}$ of that particular 2-D plot.
The angle with the lowest $P_{\rm KS}$ and the corresponding $P_{\rm KS}$ value
are presented in Table \ref{tb2d} for each plot.
The black line segment in the circle at the lower left corner of each plot shows the
direction of the $x$-axis with the lowest $P_{\rm KS}$. 
We also test the possible correlation among LGRB sample between each parameter pair 
with the Spearman correlation.
The Spearman correlation $\rho_{\rm s}$ and 
the null probability of the Spearman correlation $P_{\rm S}$
are also presented in Table \ref{tb2d}.

Since $T_{90}$ defines LGRBs and SGRBs, plots related to $T_{90}$ have LGRBs and
SGRBs are well separated and show low $P_{\rm KS}$ values,
with the lowest $P_{\rm KS}$ angle at about 90$^{\circ}$,
especially if the host galaxy parameters {do} not show significant difference between 
LGRBs and SGRBs.
There is generally no correlation between $T_{90}$ and the host galaxy parameters.

In the plots with gamma-ray spectral parameters $\alpha$ and $E_{\rm p}$,
in general SGRBs have harder $\alpha$, larger $E_{\rm p}$
and show mild difference from LGRBs.
However, the overlap is still significant, and one cannot 
clearly distinguish the two classes by any of these plots.

In the plots with $E_{\rm \gamma, iso}$,
LGRBs and SGRBs show obvious differences but still overlap with each other.
The $P_{\rm KS}$ value is not as significant as the 1D plots
due to about two thirds reduction of the total number.
The best separated plot is $E_{\rm \gamma, iso}$ vs. offset.
Plots with $L_{\rm p,iso}$ do not show as significant difference
between LGRBs and SGRBs as $E_{\rm \gamma, iso}$ plots.
There are some correlations for LGRBs
in $E_{\rm \gamma, iso}$/$L_{\rm p,iso}$ vs $M_*$/SFR plots,
as is also shown in \cite{2010ApJ...709..664R}.
Some anticorrelations are also shown
in $E_{\rm \gamma, iso}$/$L_{\rm p,iso}$ vs [X/H] plots.

The 2D plots involving $f$ and $f_{\rm eff}$ are similar to those involving $T_{90}$.
In particular, those involving $f_{\rm eff}$ show significant differences between LGRBs and SGRBs,
even though significant overlapping is observed.

In several plots involving redshift, LGRBs show apparent correlations between $z$ and other
parameters. Most of these correlations may be partially attributed to observational selection effects. 
In the plots of log $M_*$ vs. $z$ and SFR vs. $z$, 
LGRB hosts with higher redshifts
generally have larger stellar masses and larger SFRs. 
This is likely due to a selection effect, since galaxies with larger stellar masses are usually 
brighter and therefore detectable at higher redshifts, and since SFR is mainly determined by stellar mass.
However, this selection effect is not obvious for SGRBs.
SGRB hosts are generally more massive than LGRB hosts at a same
redshift, while having nearly the same SFR.
This results in a smaller sSFR for SGRB hosts,
as shown in the plot of sSFR vs. $z$, as expected.
In the plot of sSFR vs. $z$, the influence of stellar mass on SFR
is generally removed and no significant redshift evolution is shown.
A strong evolution of metallicity can be seen in the [X/H] vs. $z$ plot,
which also shows higher metallicity of SGRB hosts relative to the 
LGRB hosts at the same redshift.

The host sizes $R_{50}$ of SGRBs are generally larger than those of LGRBs {at} all redshifts,
even though much overlap is seen.
A mild, negative correlation between galaxy size $R_{50}$ and redshift
can be noticed.
Since selection effects may {create} a positive correlation,
this negative correlation should be intrinsic, even though it is not significant.
There is also a mild, negative correlation between $F_{\rm light}$
and redshift. 
$F_{\rm light}$ of SGRBs shows a tentative positive correlation with 
redshift despite of a wide spread. 
SGRBs and LGRBs are generally more separated at $z<1$ than at $z>1$.

\section{Conclusions and Discussion}

In this paper, we present a sample of 407 GRBs detected before June 30th, 2014, with both prompt emission 
and host galaxy properties. Most GRBs (375) have spectroscopic redshift measurements. The other 32 bursts are included
because of their host galaxy information. The prompt emission properties include
duration $T_{90}$, spectral peak energy $E_{\rm p}$, low energy photon index $\alpha$,
isotropic $\gamma$-ray energy $E_{\rm \gamma, iso}$, peak luminosity $L_{\rm p, iso}$,
amplitude parameters $f$ and $f_{\rm eff}$.
The host galaxy properties include
star formation rate SFR, stellar mass $M_*$, specific star formation rate sSFR,
metallicity [X/H], galaxy size $R_{50}$, physical offsets of GRBs from the center of the host $R_{\rm off}$,
normalized offset $r_{\rm off}=R_{\rm off}/R_{50}$, and brightness fraction $F_{\rm light}$.
We pay special attention to the comparison between $T_{90}$-defined SGRBs and LGRBs,
and more importantly, the physically defined Type I vs. Type II GRBs. For the latter, we compare
the ``consensus'' samples of SGRBs and LGRBs as listed in Jochen Greiner's catalog, in which
the definition of each SGRB was based on multiple criteria, with some of them having $T_{90}$
longer than 2 s. For both definitions of SGRB/LGRB samples, we present the 
one-dimensional (1D) histograms of the two types, compare their distributions,
and quantify their overlapping fractions. For the consensus samples, we further presented
{series} two-dimensional (2D) scatter plots between prompt emission properties and
host galaxy properties, aiming at identifying good parameters to separate the two
types of bursts. Our results can be summarized as follows.


1. In 1D diagrams, all the prompt emission properties
and host galaxy properties show more or less overlaps between SGRBs and LGRBs. 
No property shows a clear separation between consensus SGRBs and LGRBs.
The duration $T_{90}$ and the effective amplitude parameter $f_{\rm eff}$
are two parameter that have the lowest overlaps. 
{The overlapping fractions for
the $f_{\rm eff}$ histograms are 7\% for LGRBs and 20\% for SGRBs.}
The overlapping fractions for
the $f_{\rm eff}$ histograms are 7\% for LGRBs and 79\%
\footnote{48\% if the outlier GRB 130427A is excluded.}
 for SGRBs, respectively.
Other parameters have much larger overlapping fractions, typically 50\%-80\% for LGRBs 
and 80\%-100\% for SGRBs. This suggests that no single parameter alone is good 
enough to place a particular burst into the right physical category.

2. The $T_{90}$-defined LGRB and SGRB samples show more overlaps than
the consensus LGRBs and SGRBs in most properties other than $T_{90}$,
especially in host galaxy properties.
This indicates that {the} $T_{90}$-only criterion mis-classifies some GRBs.
Other properties are needed as supplementary criteria to classify GRBs physically.

3. None of the 2D prompt emission vs host galaxy property plots
show a clear separation between the consensus LGRBs and SGRBs.
It suggests that simple 2D plots are not good enough for 
Type I and Type II GRB classifications.

4. The three best parameters to classify GRBs are
the effective amplitude $f_{\rm eff}$, $T_{90}$, and the brightness fraction $F_{\rm light}$.
They show the smallest overlapping fractions and the smallest null probability 
$P_{\rm KS}$ in the simulated 1-D distributions.

5. 
Some correlations between prompt emission properties and host galaxy properties 
are found in some 2D plots, such as $L_{\rm p, iso}$/$E_{\rm \gamma, iso}$ vs. $M_*$, 
$L_{\rm p, iso}$/$E_{\rm \gamma, iso}$ vs. SFR,
$L_{\rm p, iso}$/$E_{\rm \gamma, iso}$ vs. [X/H], etc.
(see Fig. \ref{fig2d} and Table \ref{tb2d}).
However, all these parameters show even more significant correlation 
with redshift, indicating that the correlations may be significantly subject to
observational selection effects.

The significant overlapping nature of the observed properties {suggests} that it is not
always easy to identify the correct physical category of GRBs. Multiple observational
criteria are needed to give more robust judgement, as suggested by Zhang et al.
(2009). This first paper in a series {presents} all the observational data and 1-D
and 2-D overlapping properties. In a follow-up paper, we will develop a quantitative
method to apply the multiple observational criteria to classify GRBs into the
Type I vs. Type II physical categories. 


\acknowledgments
This work is partially supported by NASA through grants NNX14AF85G and NNX15AK85G.
We thank the anonymous referee for detailed review and very useful suggestions, and
Antonion Cucchiara, Wenfai Fong, T. Kr{\"u}hler, Sandra Savaglio and Qiang Yuan for helpful discussion.
We also acknowledge the public data available at the Swift catalog (http://swift.gsfc.nasa.gov/archive/grb\_table/), 
and the SIMBAD database,
operated at CDS, Strasbourg, France.

\bibliographystyle{apj}
\bibliography{refs}

\clearpage
\setlength{\voffset}{-25mm}

\clearpage
\begin{figure*}

\includegraphics[width=0.4\textwidth]{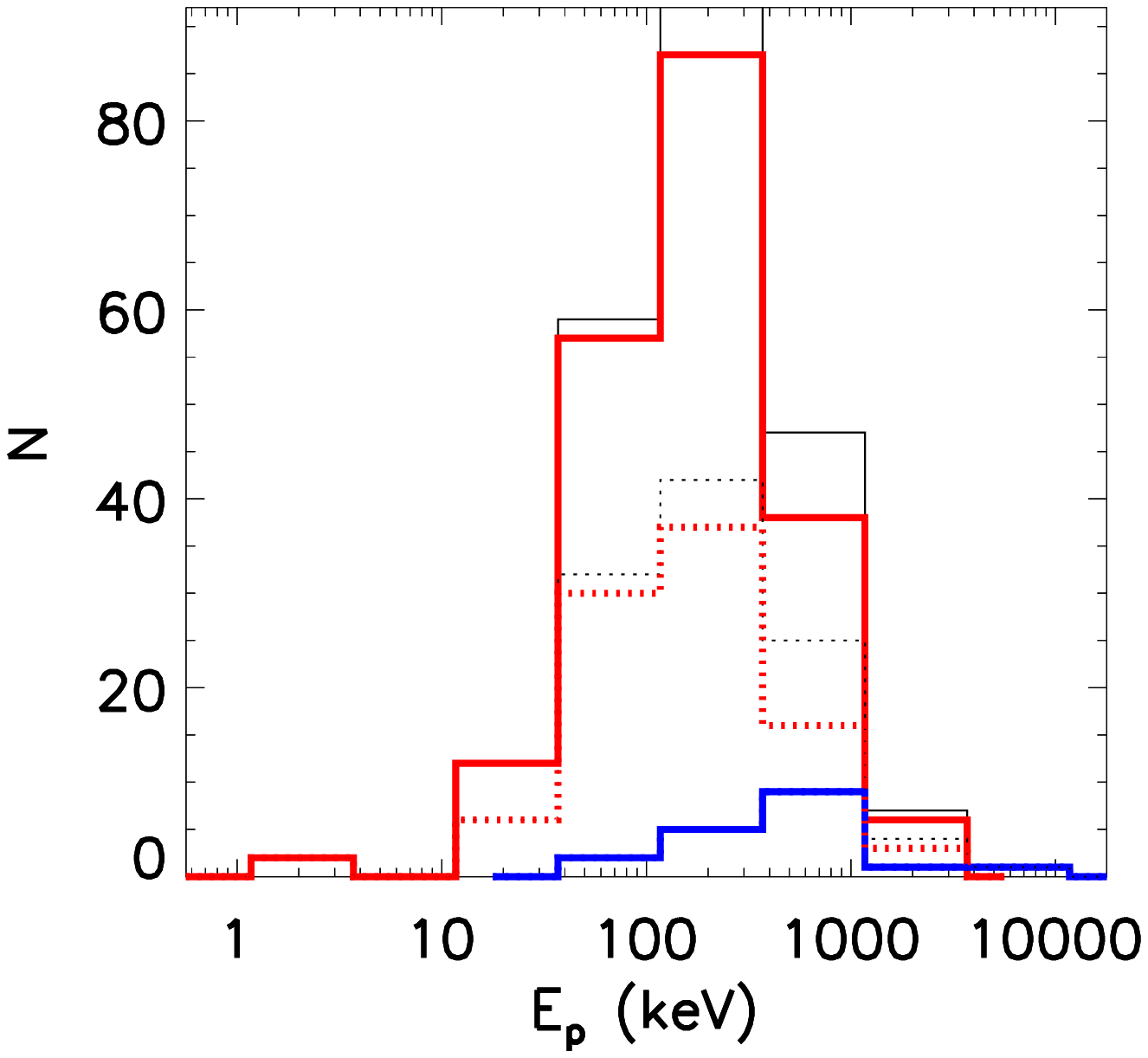}
\includegraphics[width=0.4\textwidth]{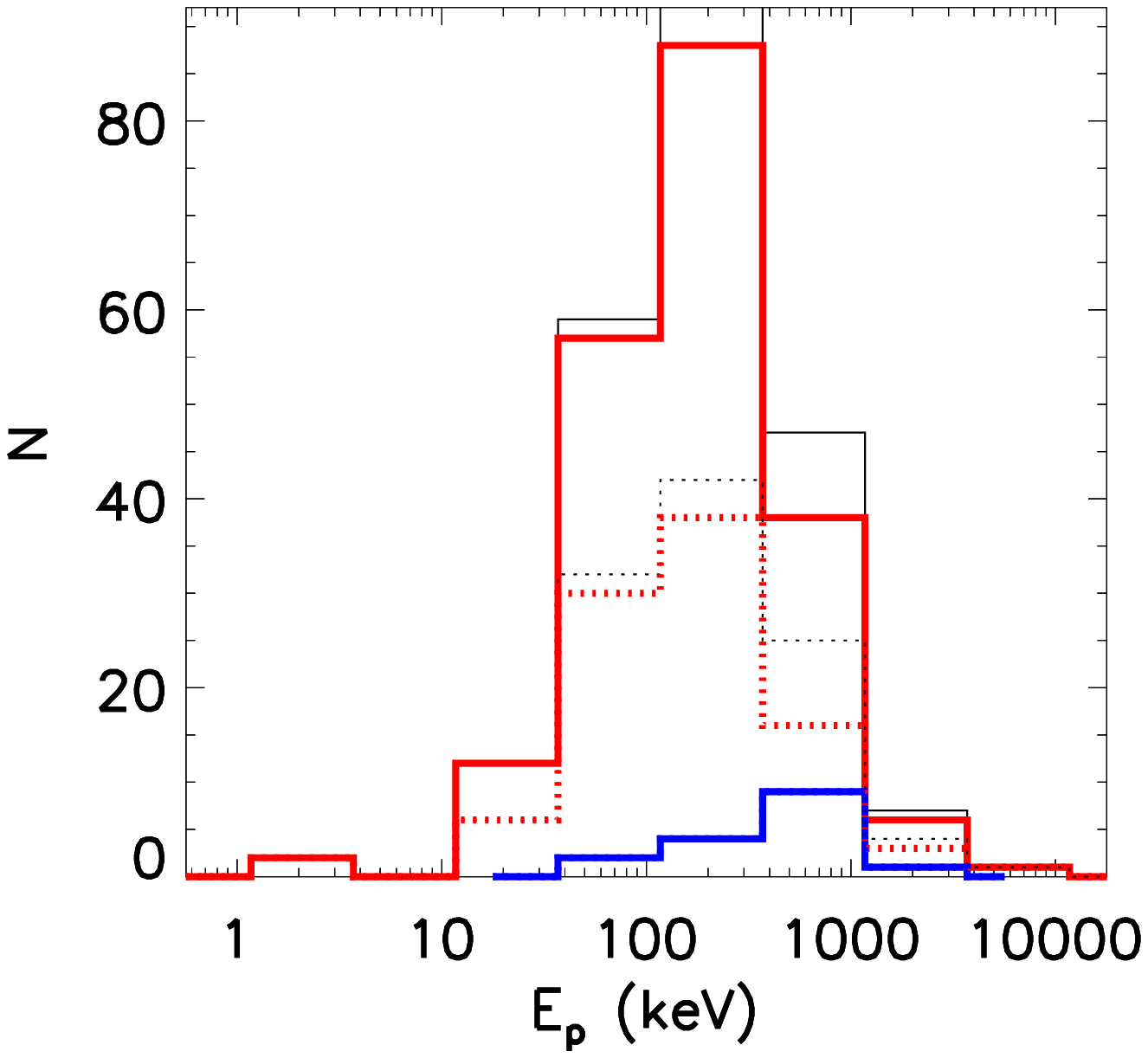}

\includegraphics[width=0.4\textwidth]{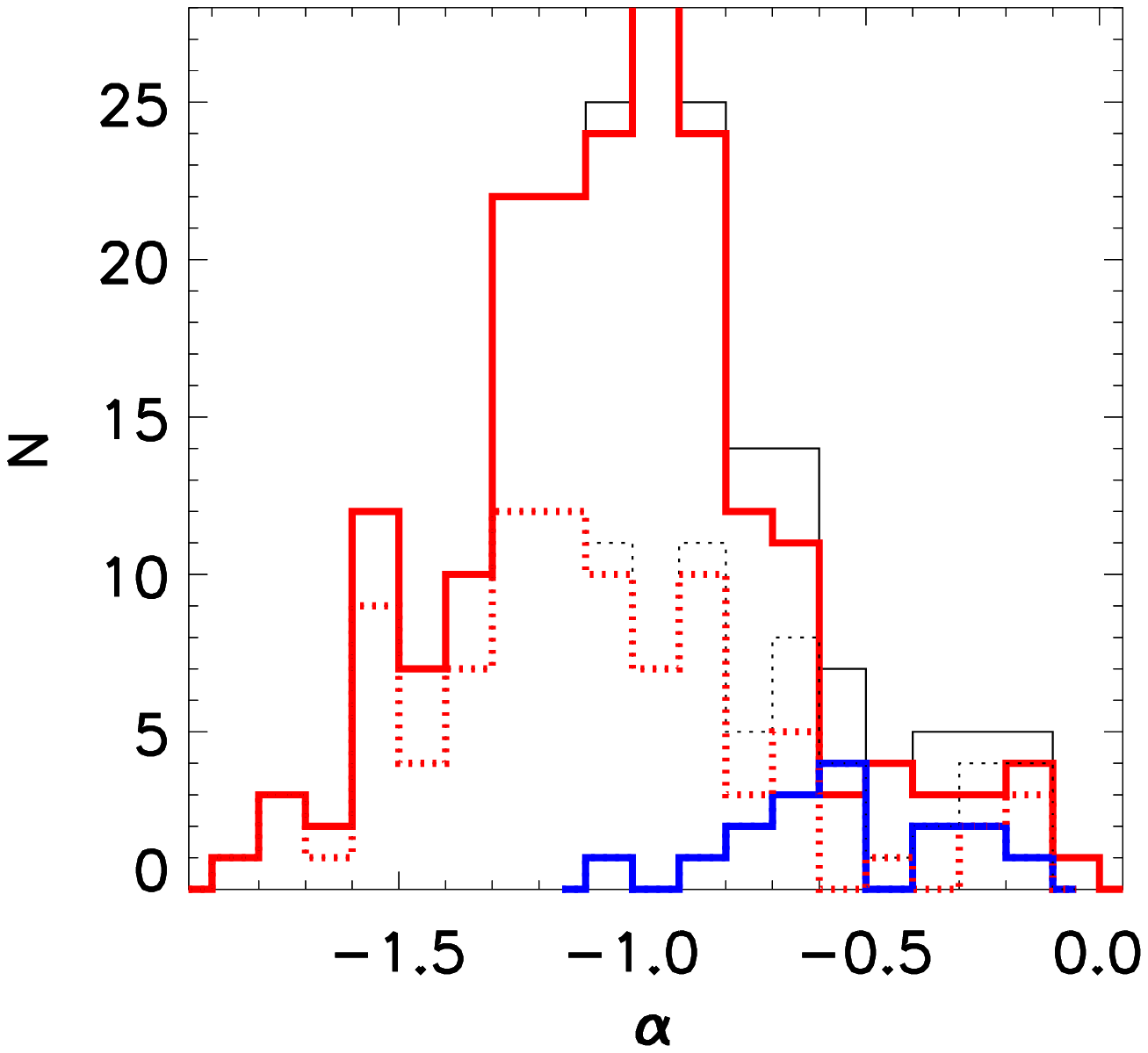}
\includegraphics[width=0.4\textwidth]{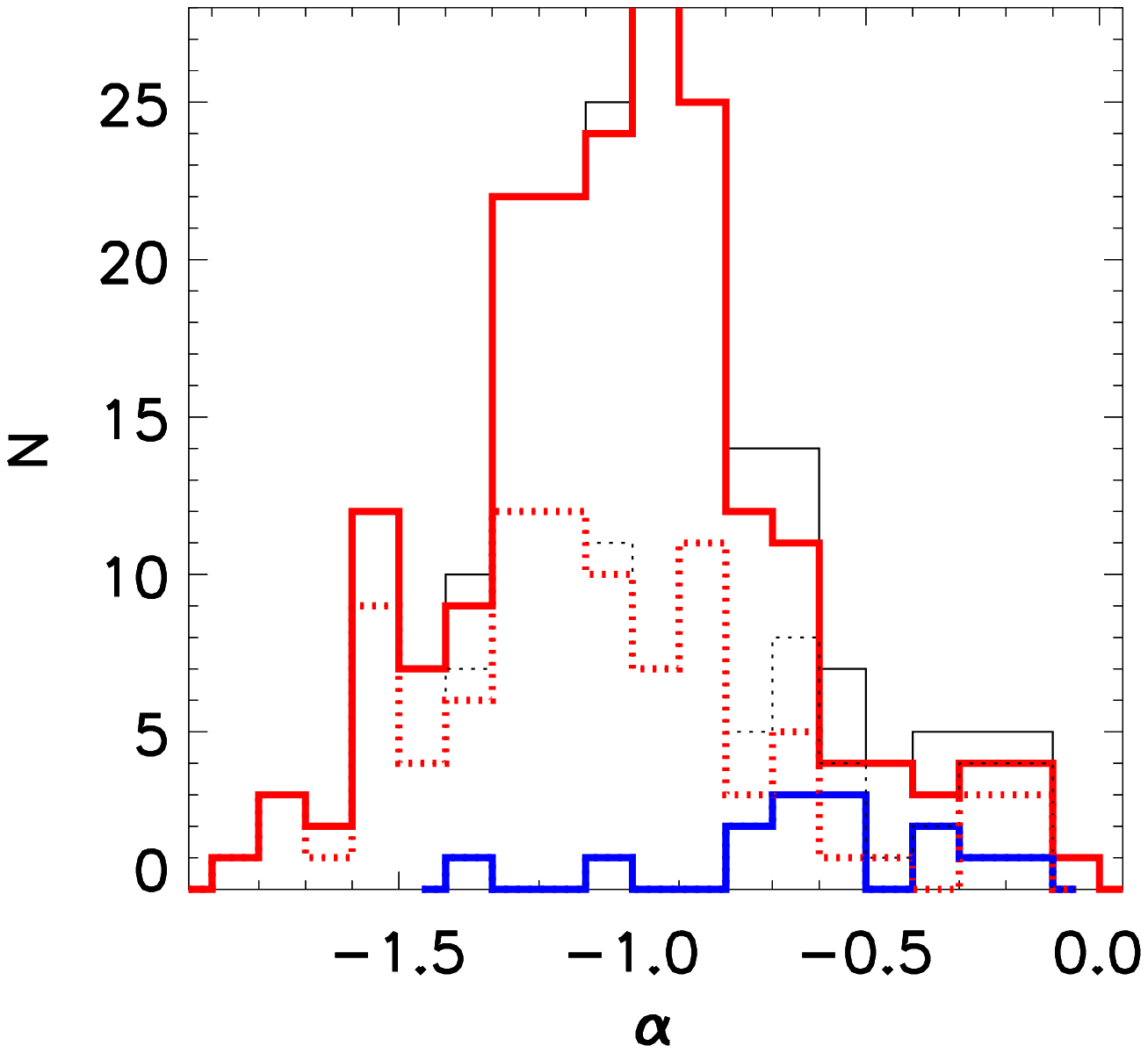}

\includegraphics[width=0.4\textwidth]{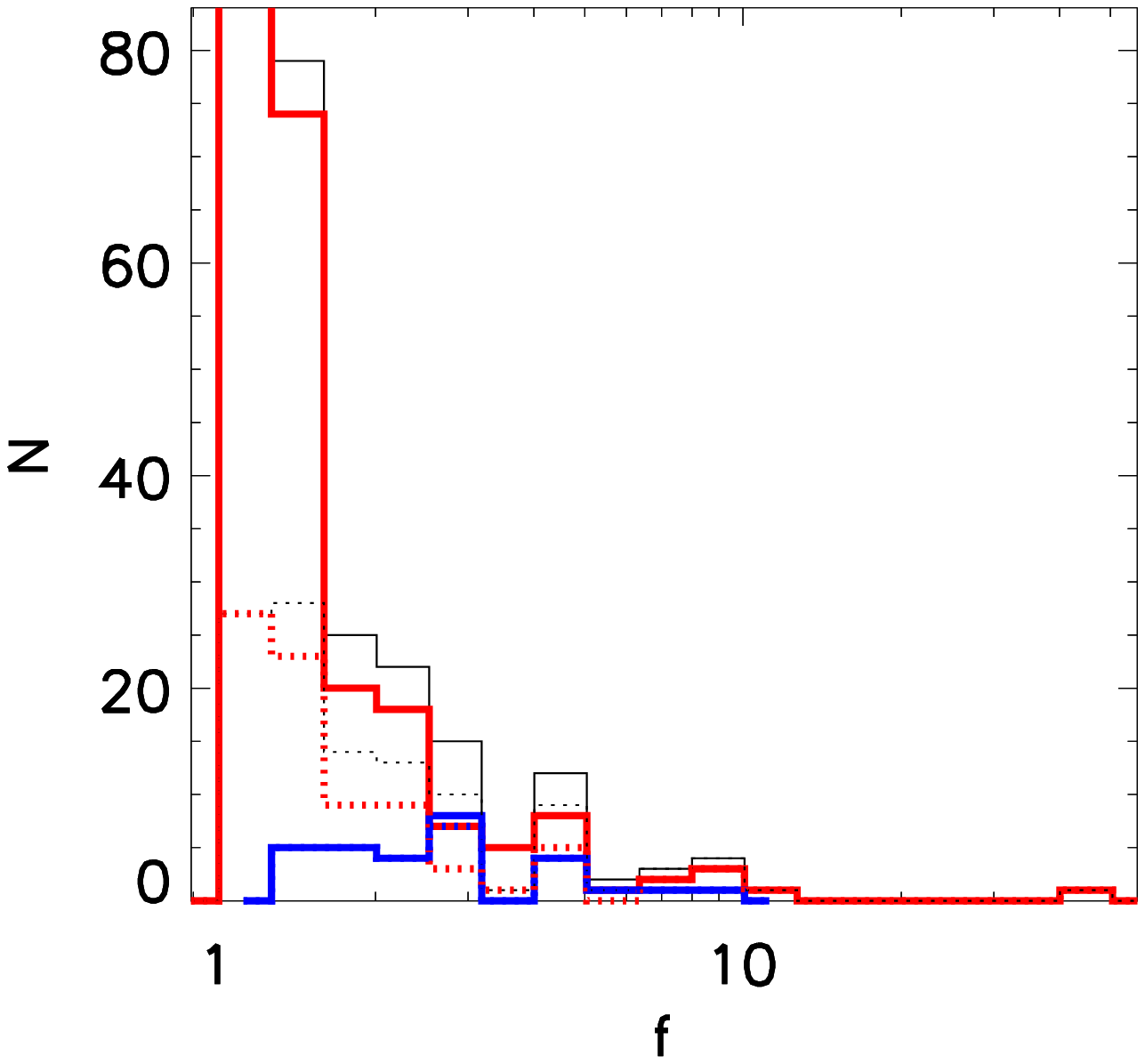}
\includegraphics[width=0.4\textwidth]{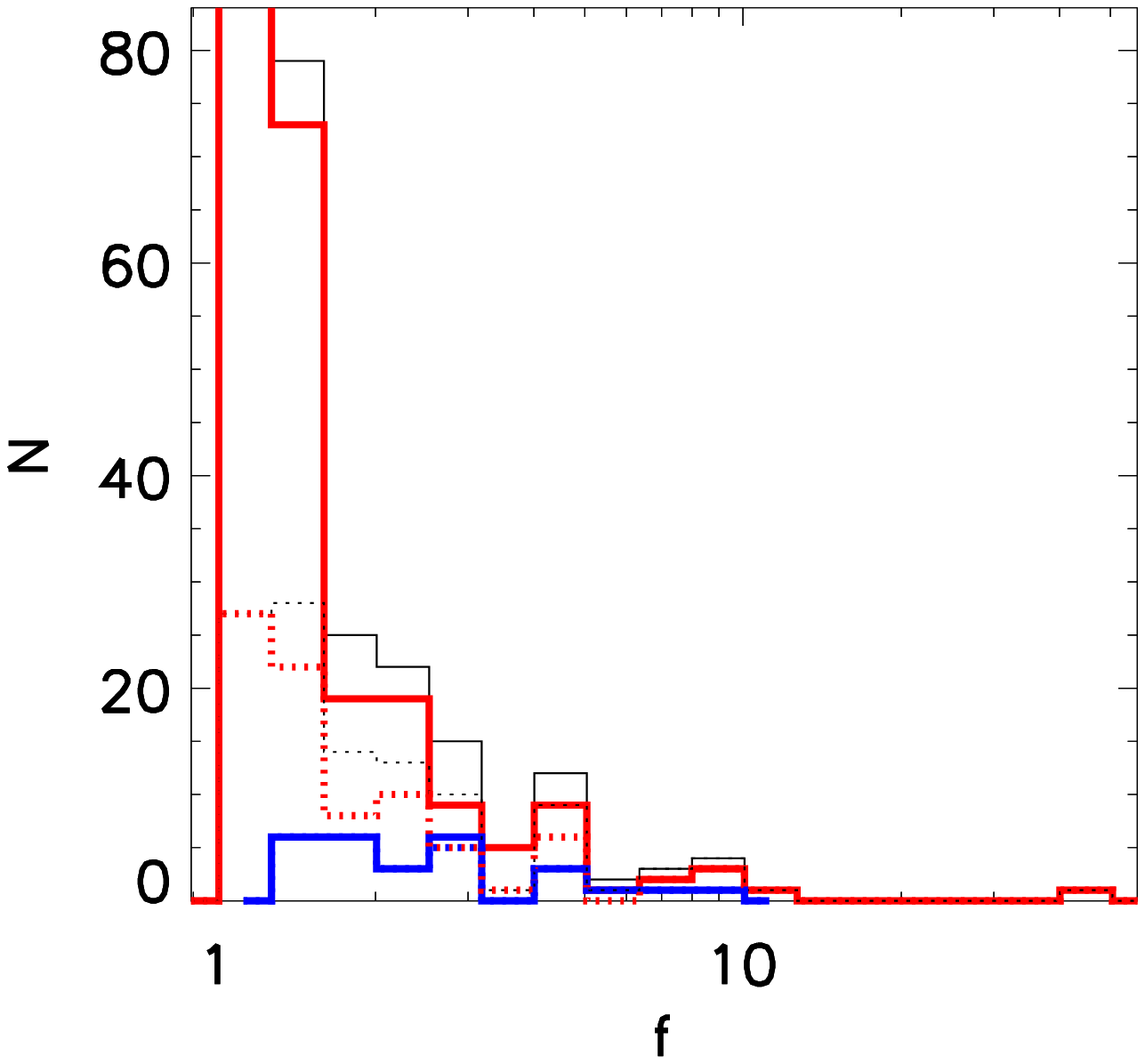}

\includegraphics[width=0.4\textwidth]{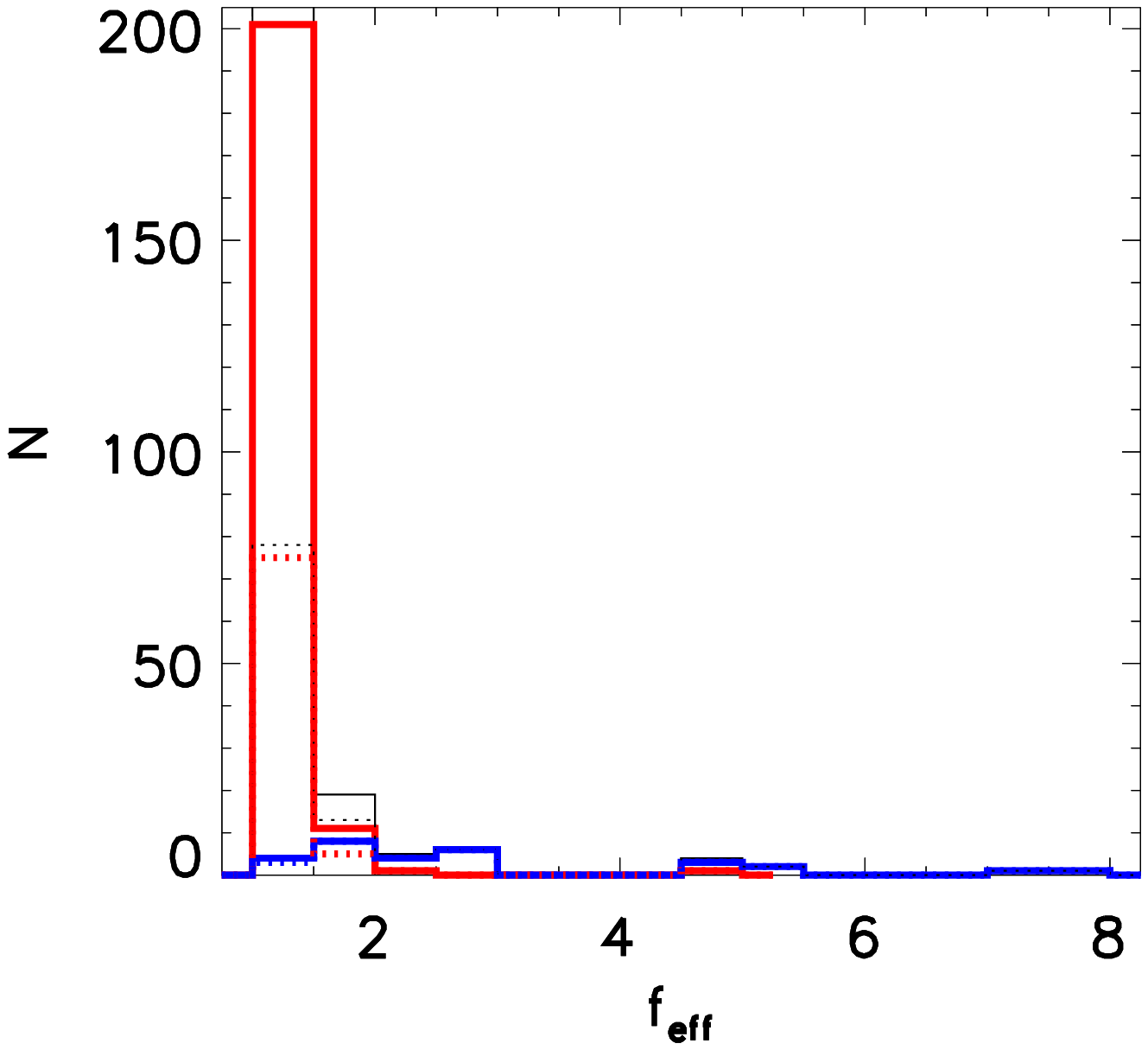}
\includegraphics[width=0.4\textwidth]{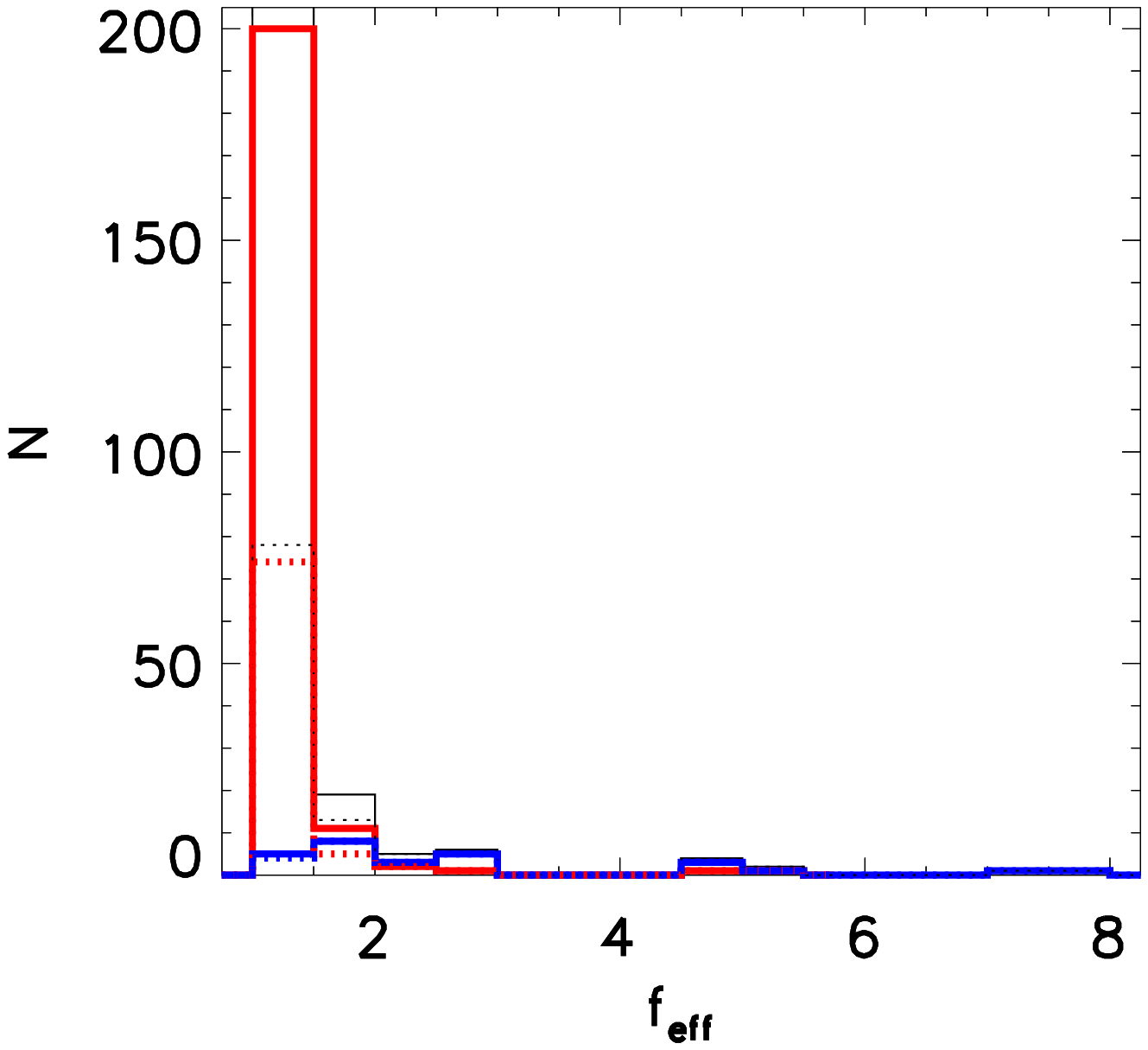}

\center{Fig. \ref{fig1d}---Continued}
\end{figure*}


\clearpage
\begin{figure*}

\includegraphics[width=0.4\textwidth]{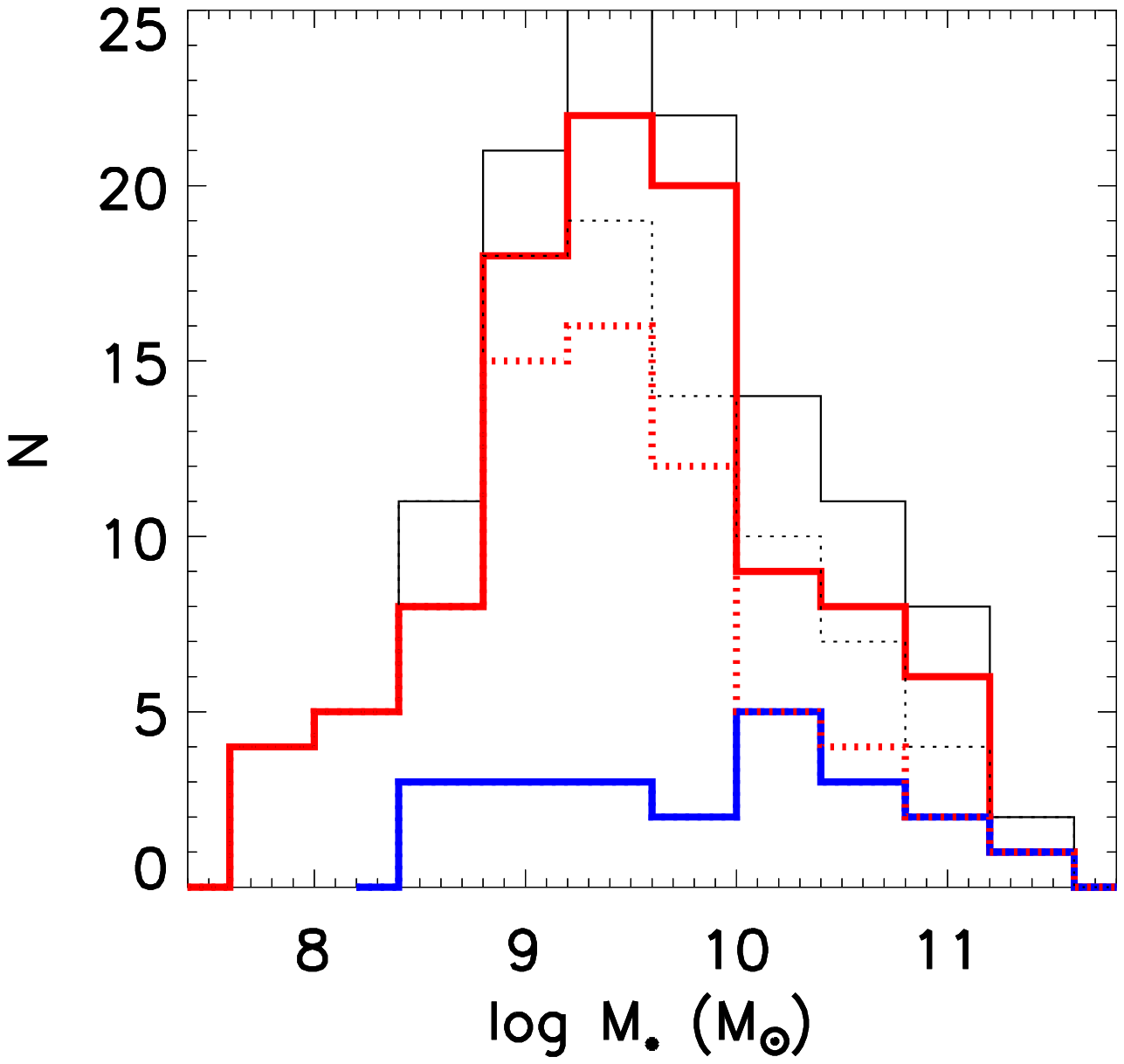}
\includegraphics[width=0.4\textwidth]{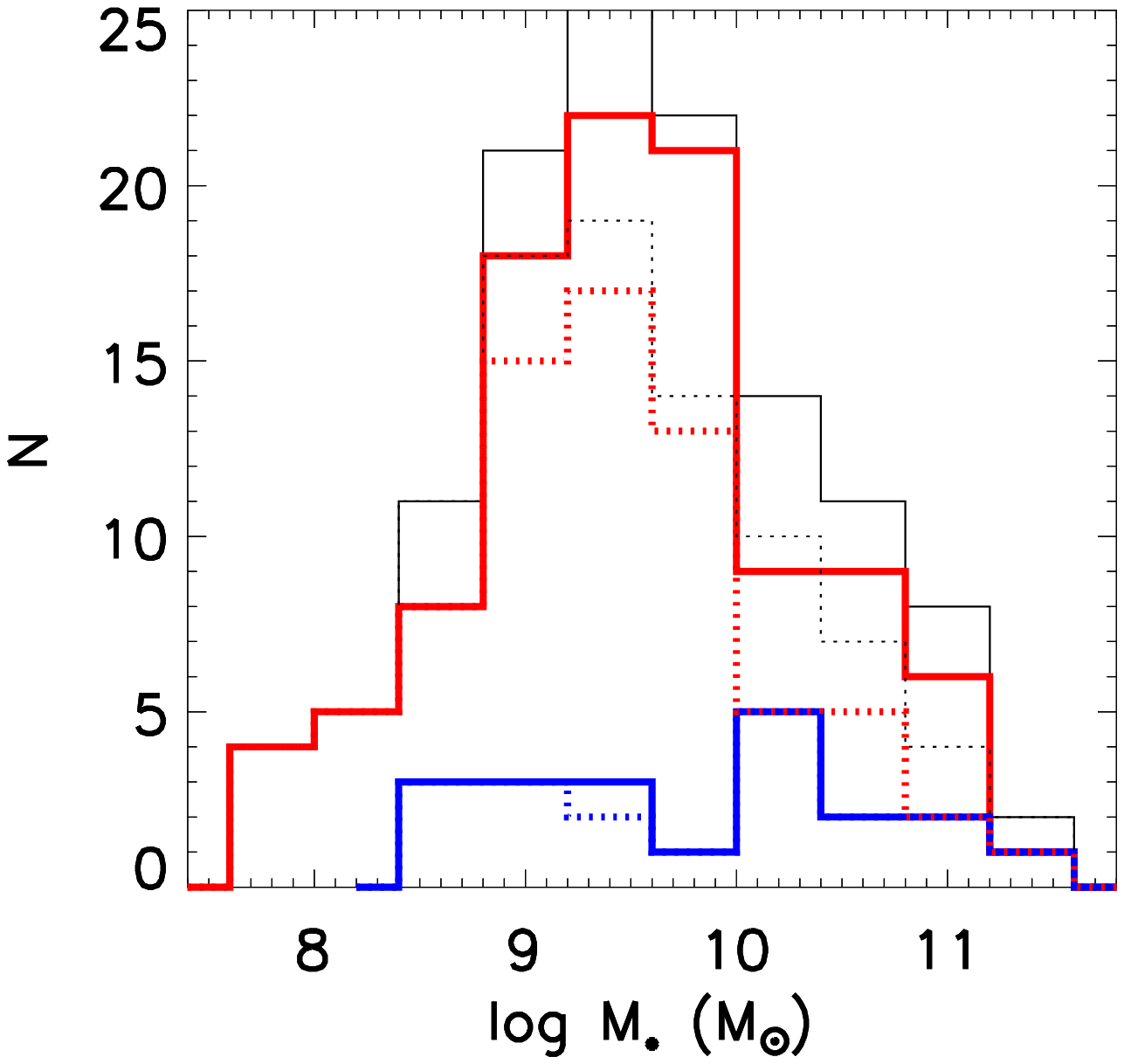}

\includegraphics[width=0.4\textwidth]{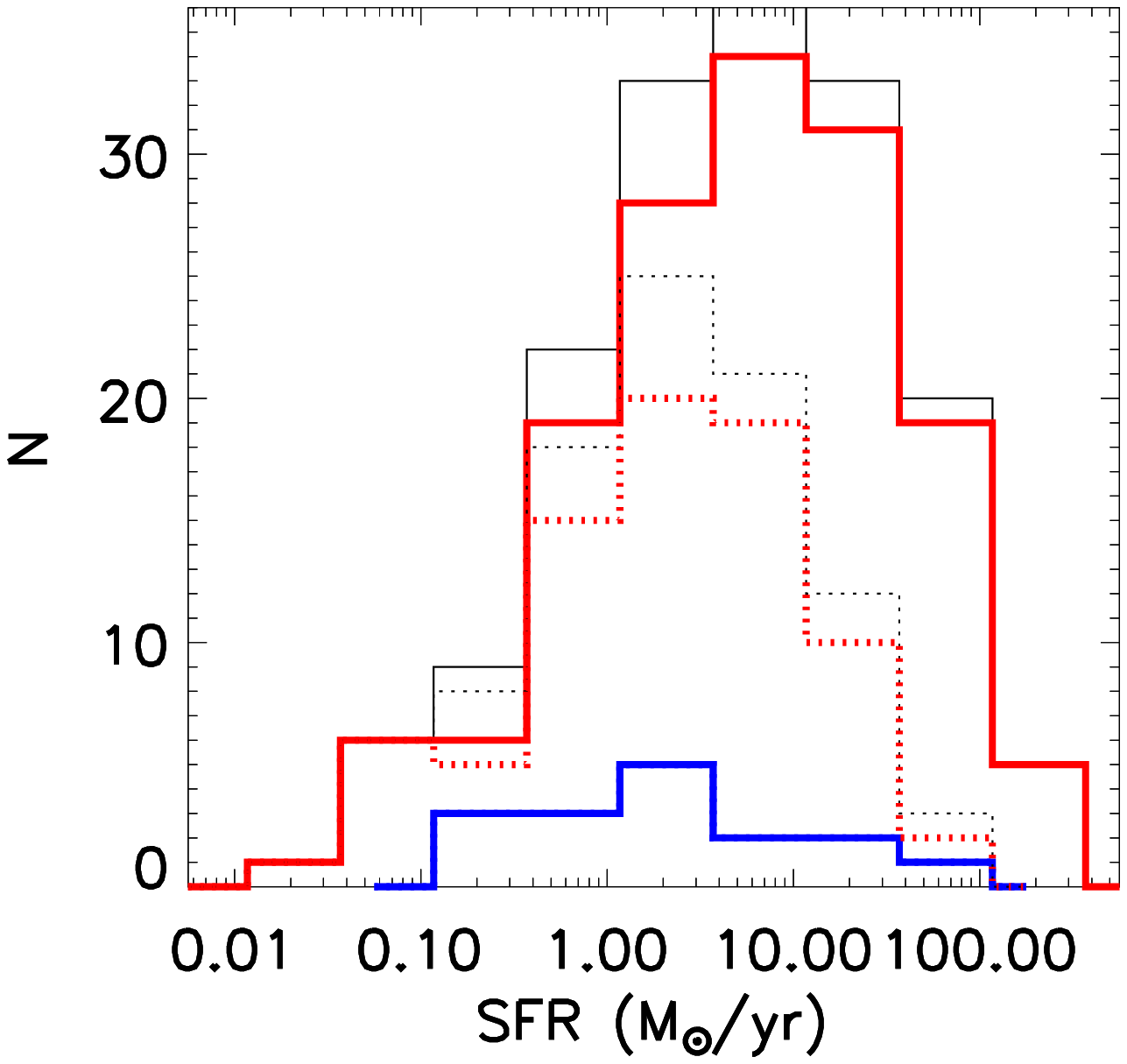}
\includegraphics[width=0.4\textwidth]{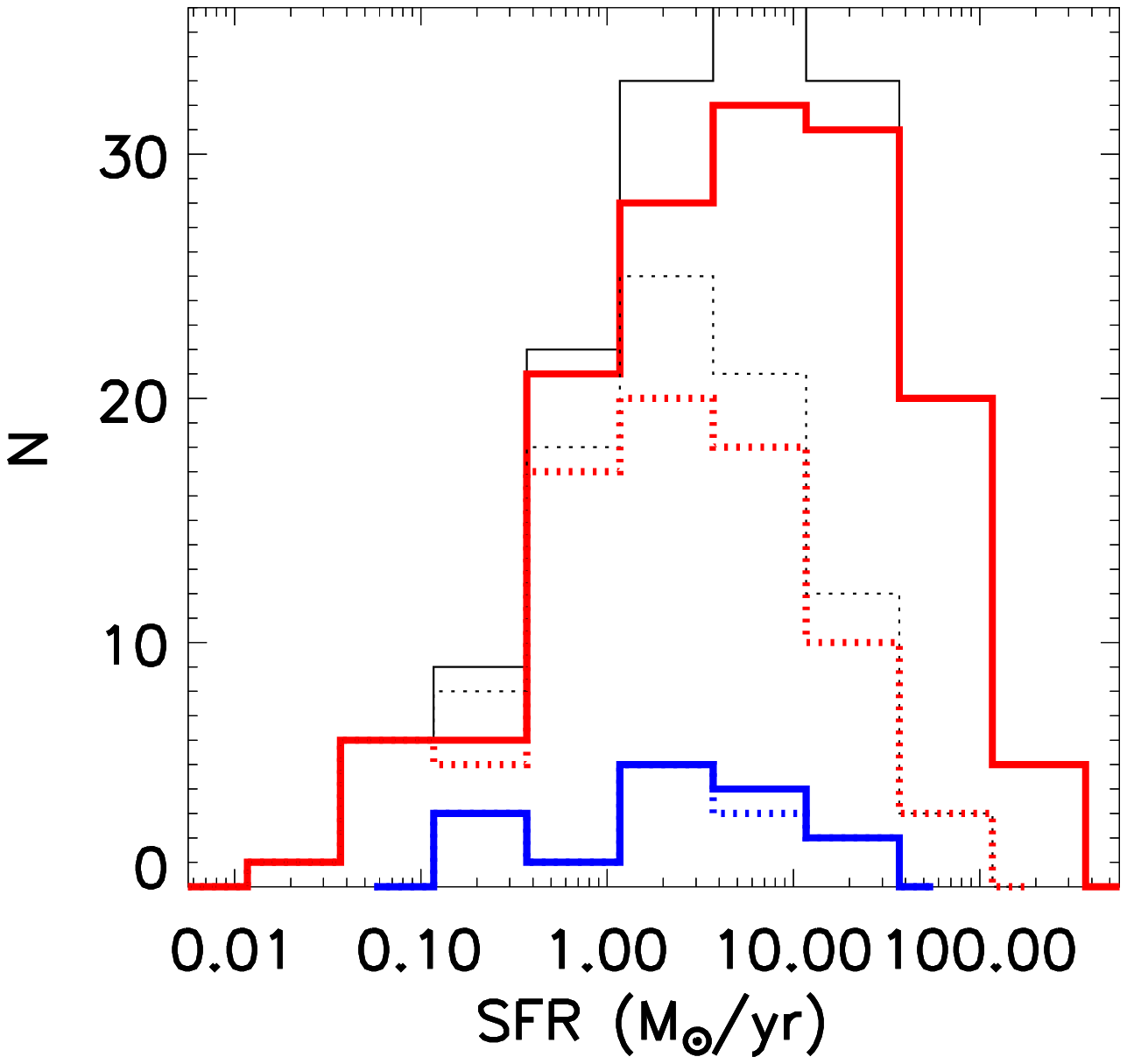}

\includegraphics[width=0.4\textwidth]{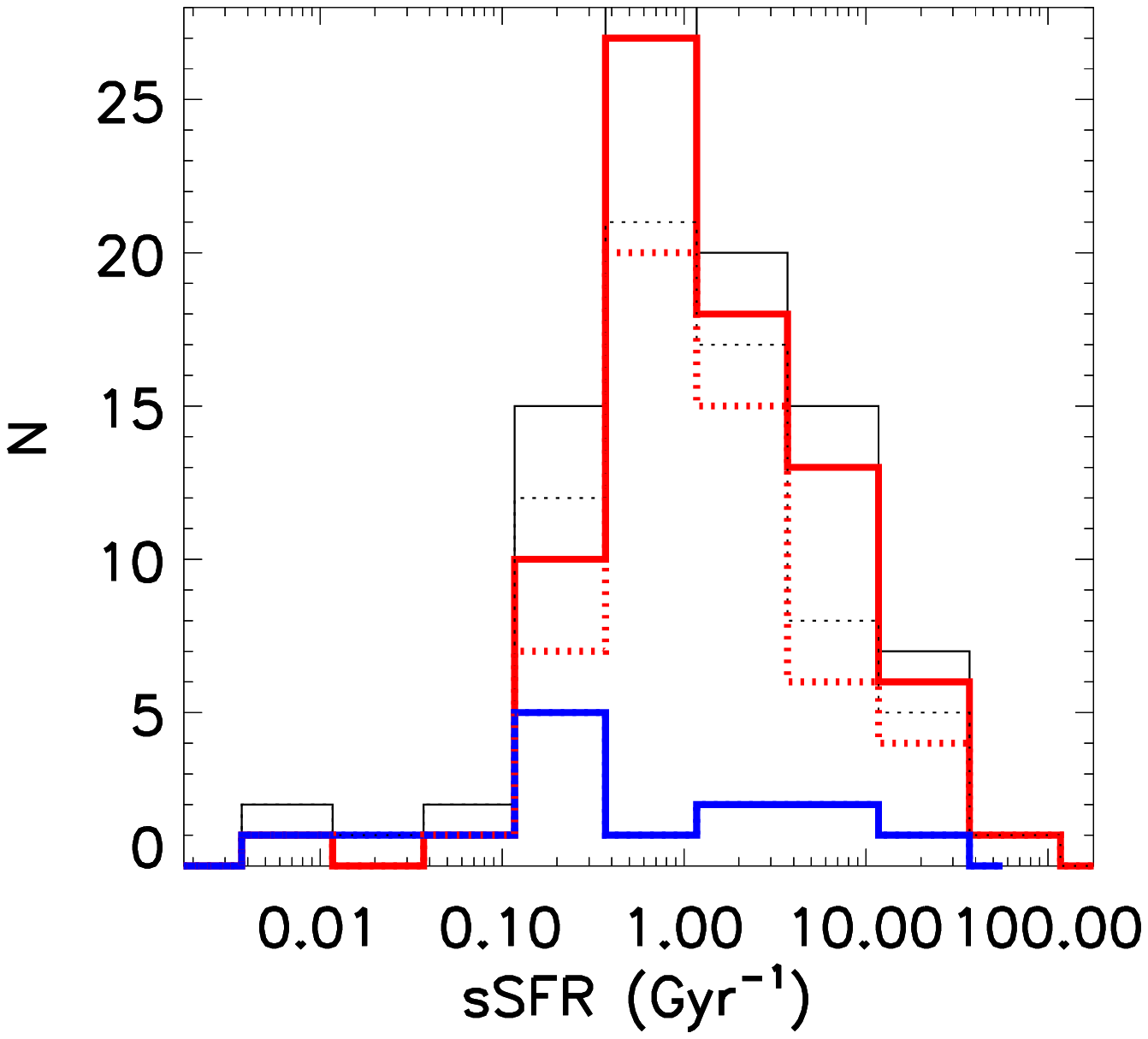}
\includegraphics[width=0.4\textwidth]{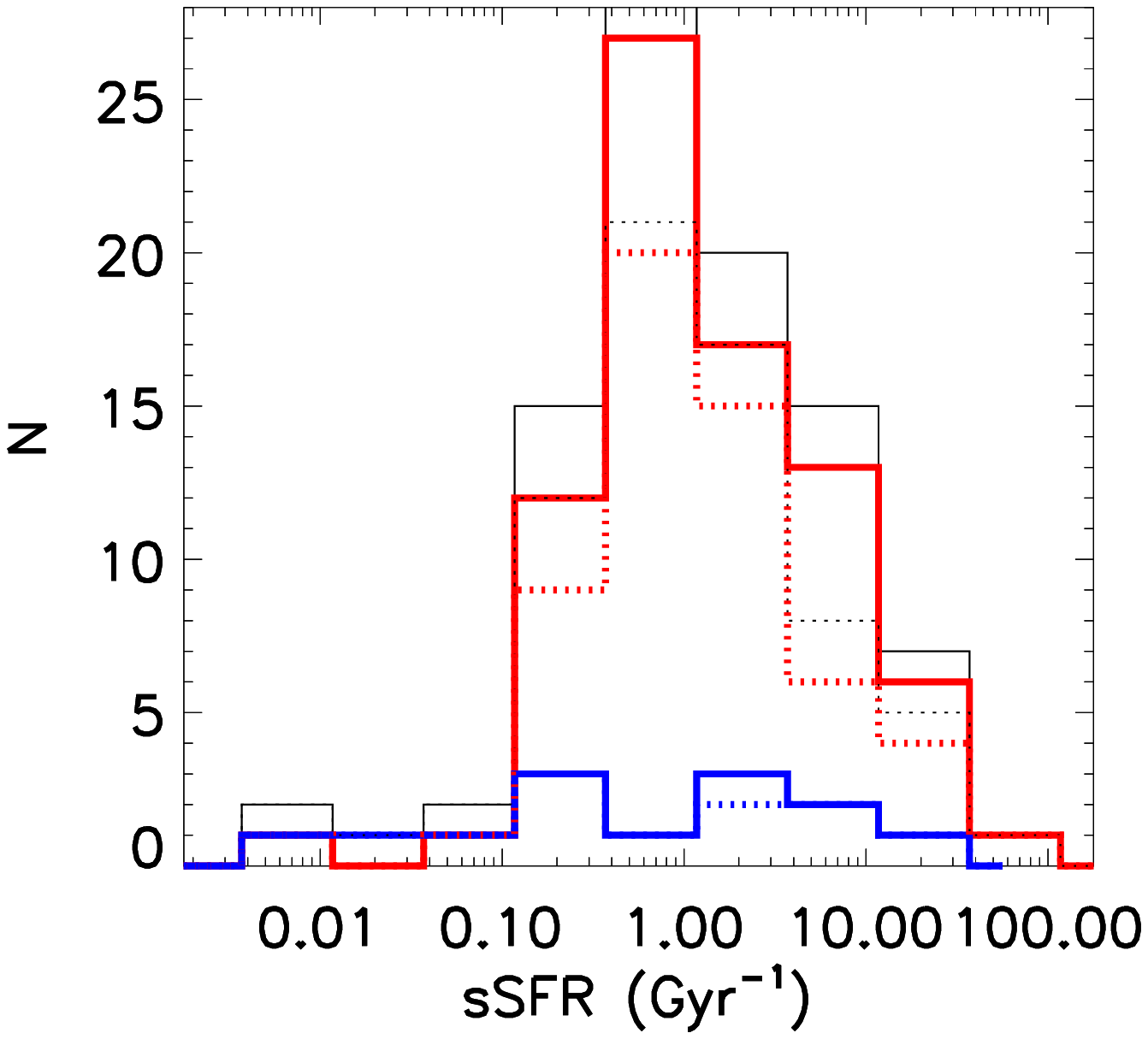}

\includegraphics[width=0.4\textwidth]{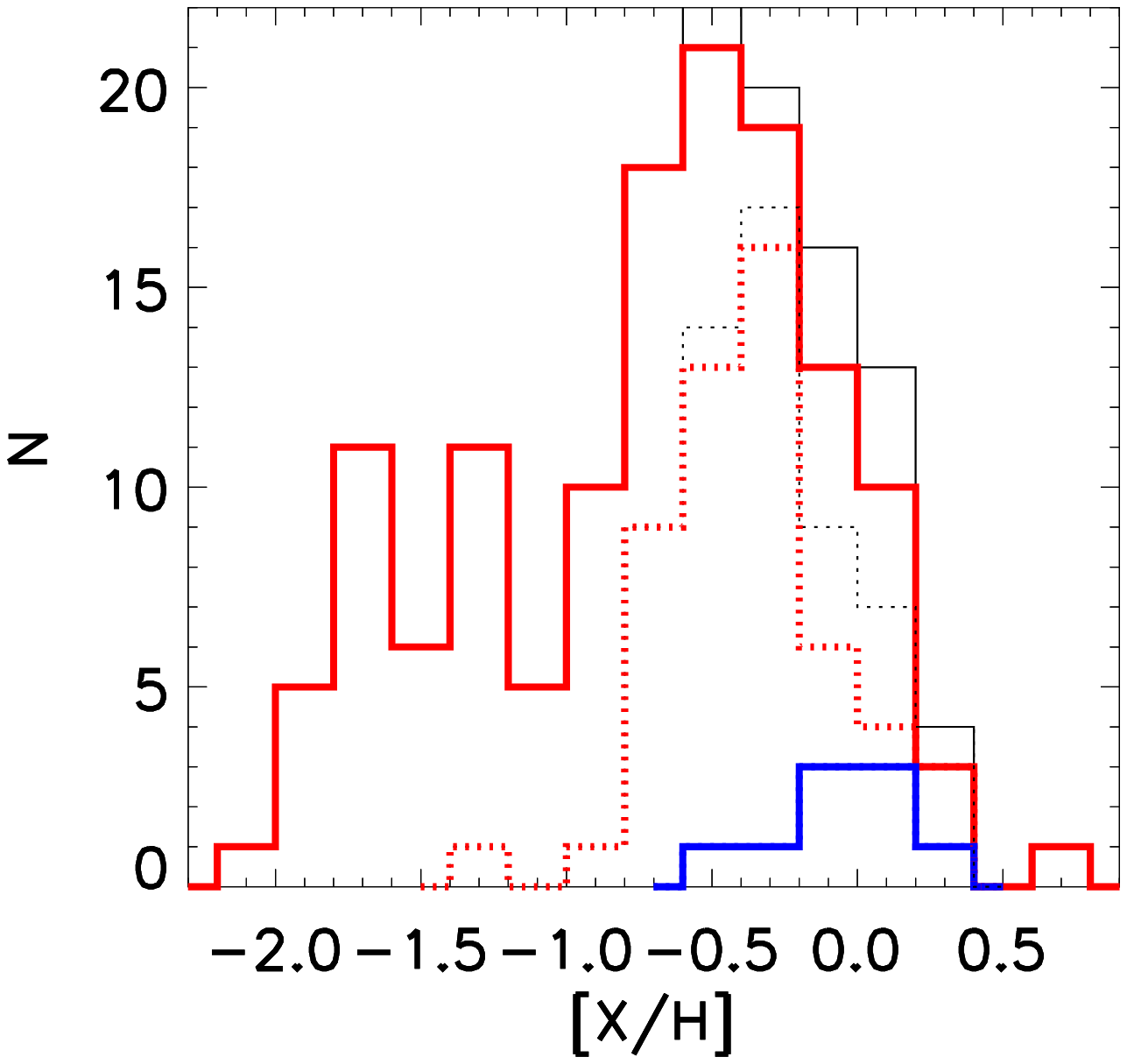}
\includegraphics[width=0.4\textwidth]{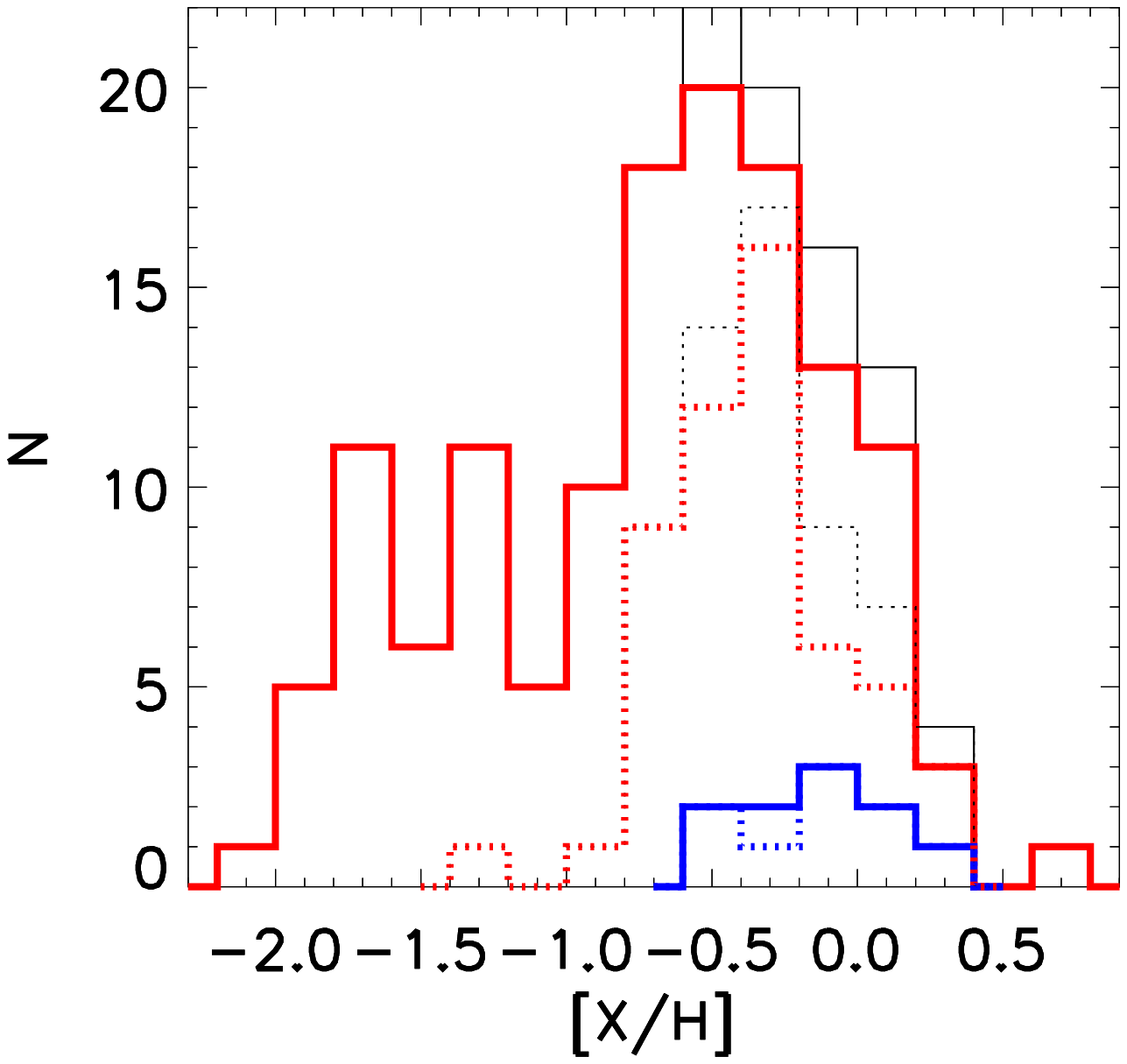}

\center{Fig. \ref{fig1d}---Continued}
\end{figure*}


\clearpage
\begin{figure*}

\includegraphics[width=0.4\textwidth]{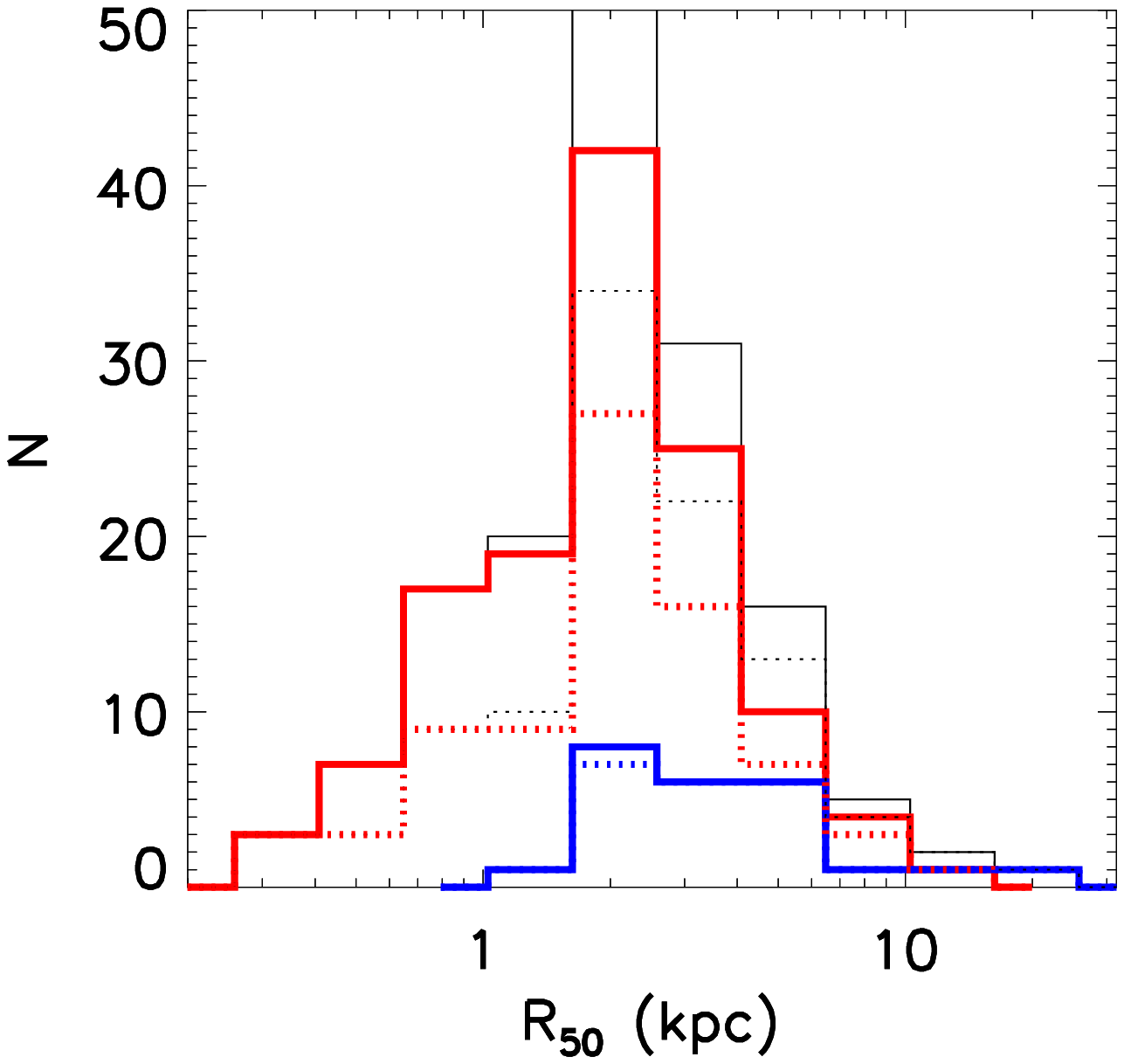}
\includegraphics[width=0.4\textwidth]{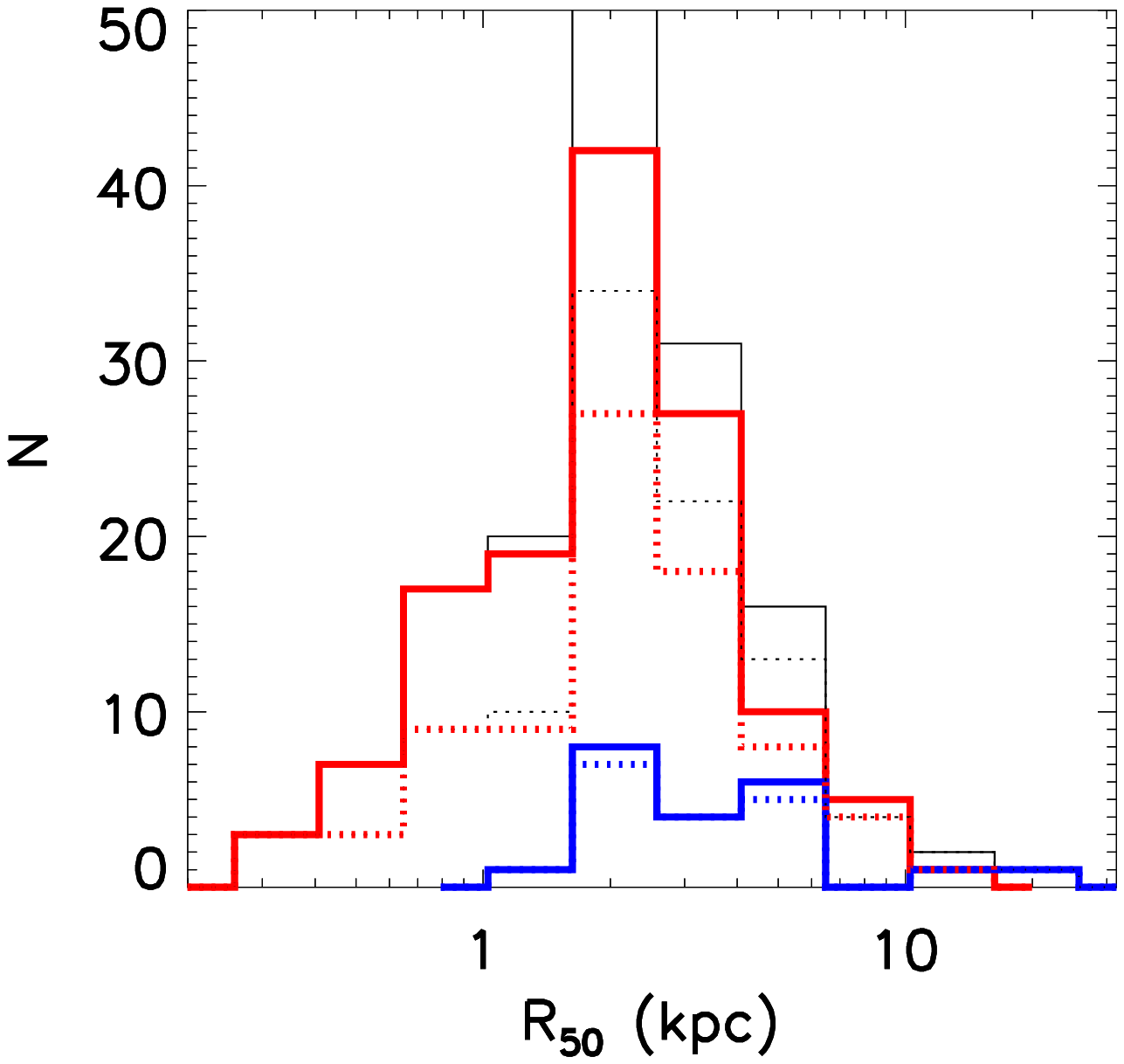}

\includegraphics[width=0.4\textwidth]{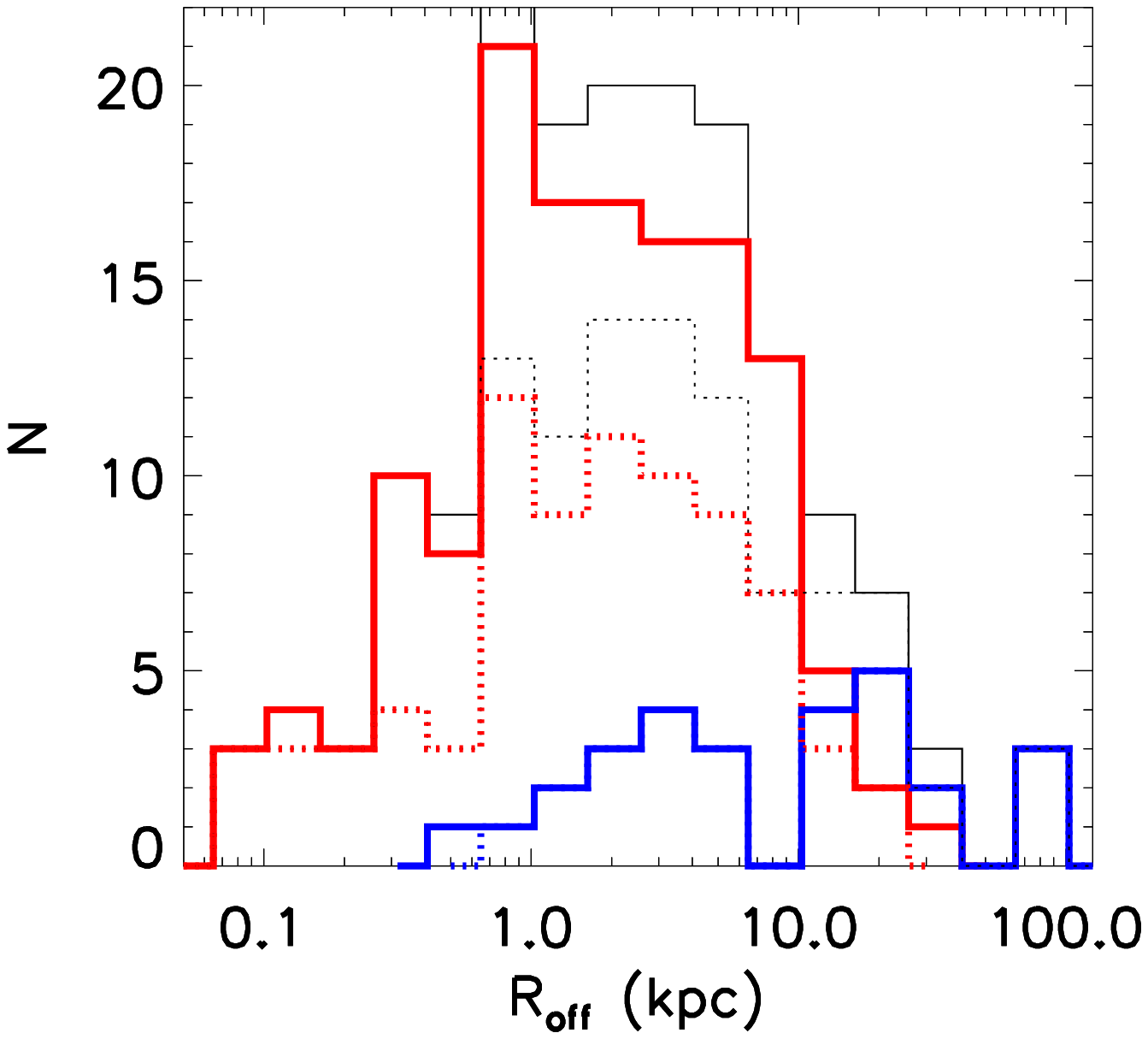}
\includegraphics[width=0.4\textwidth]{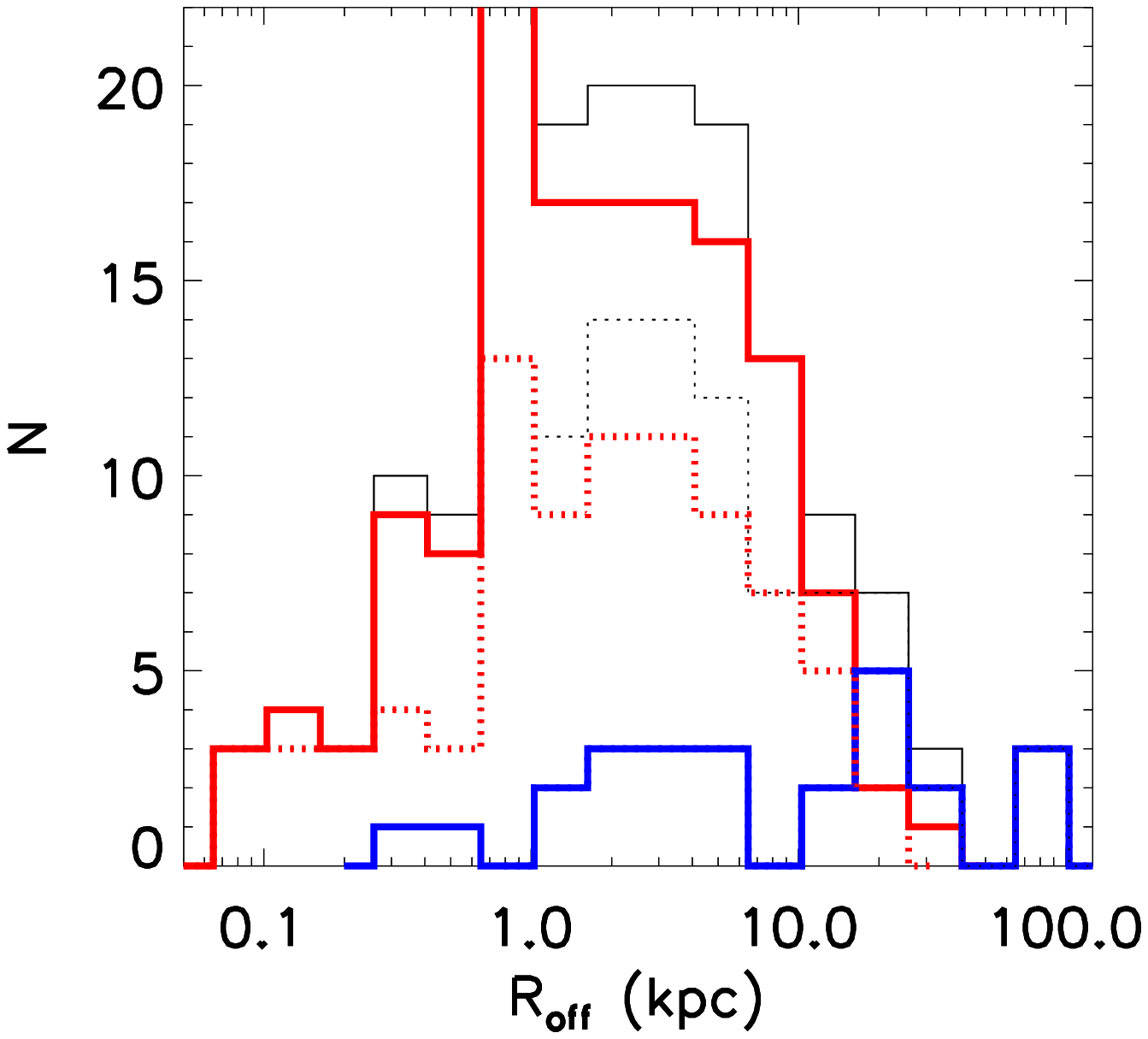}

\includegraphics[width=0.4\textwidth]{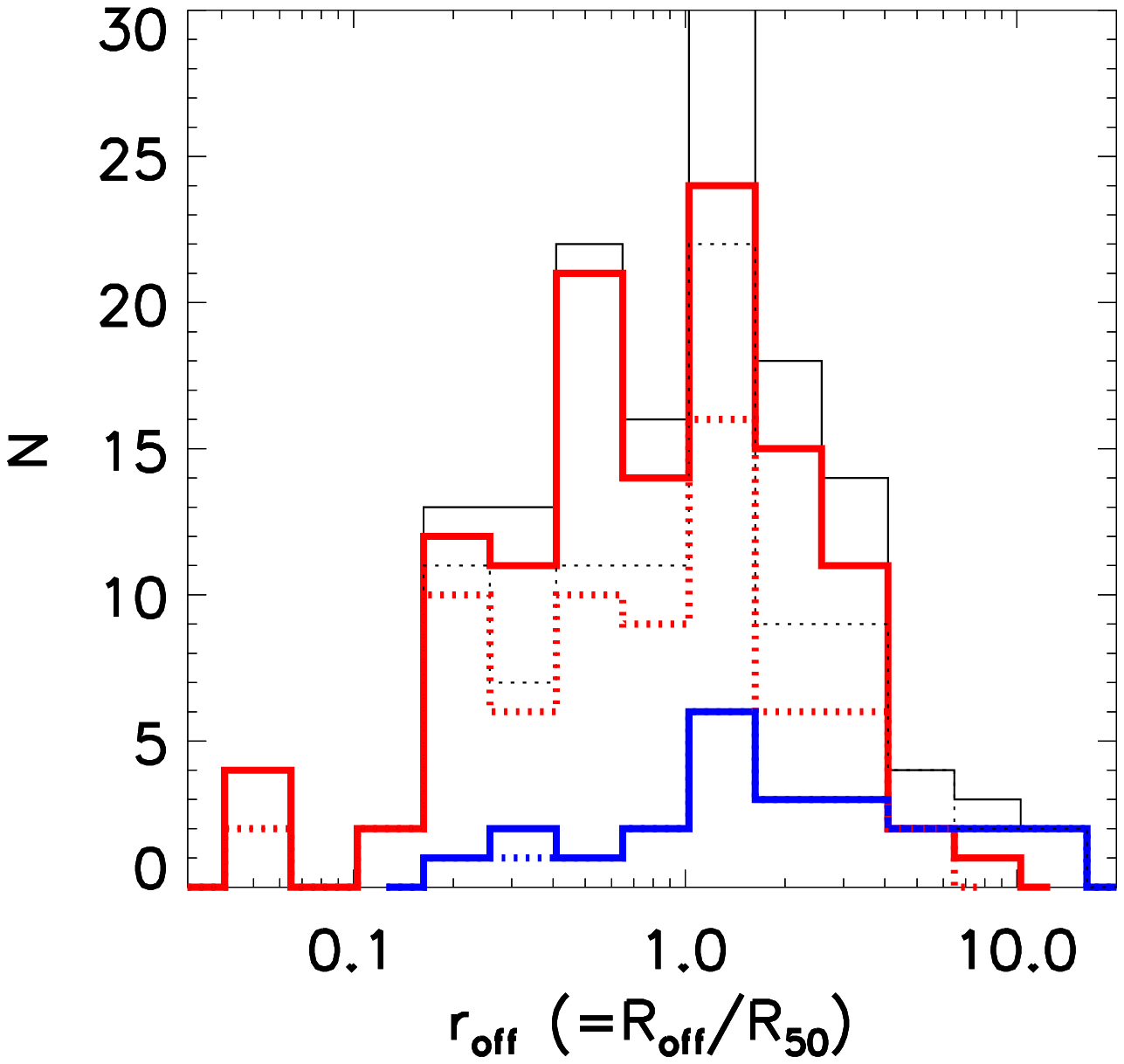}
\includegraphics[width=0.4\textwidth]{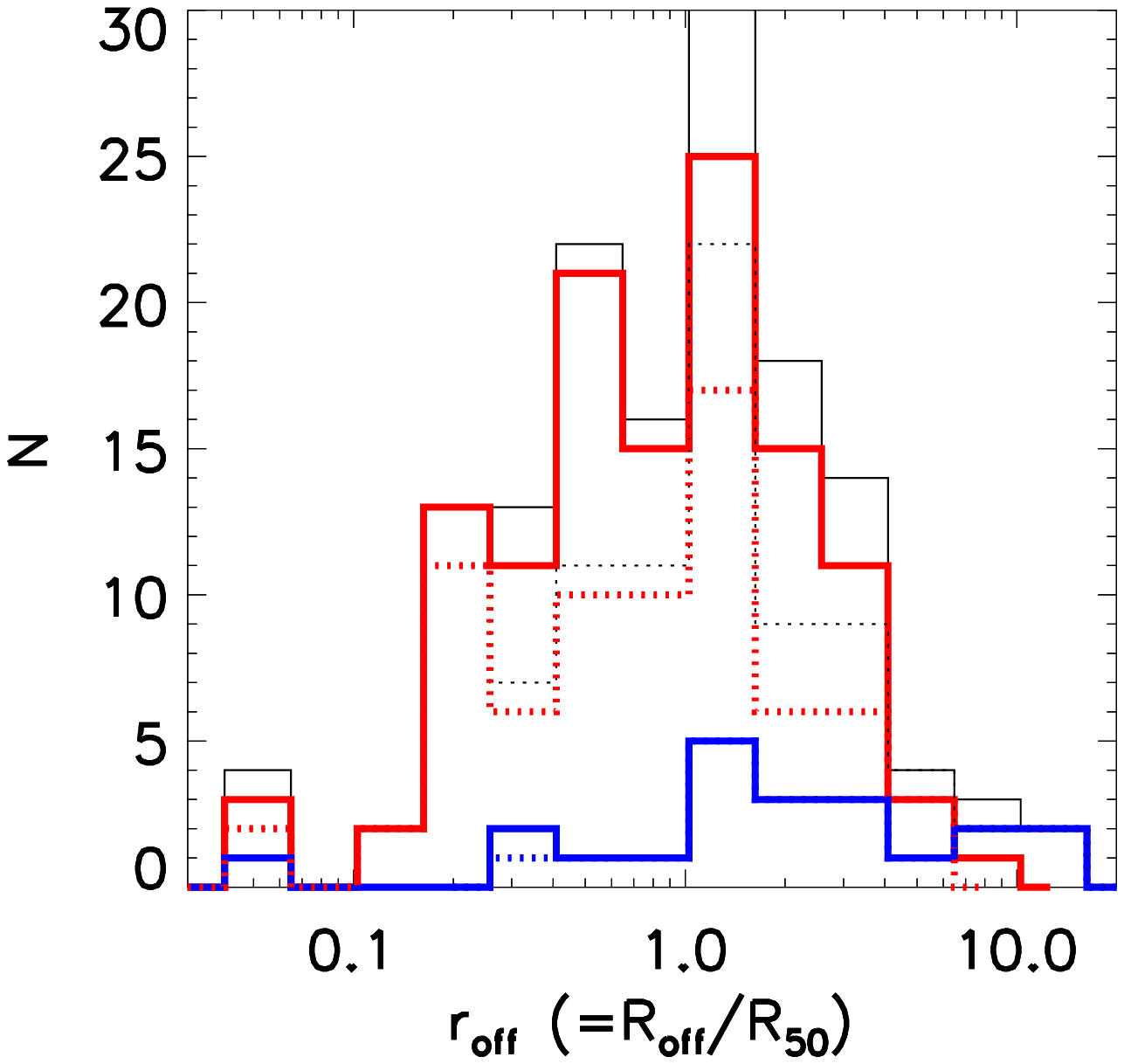}

\includegraphics[width=0.4\textwidth]{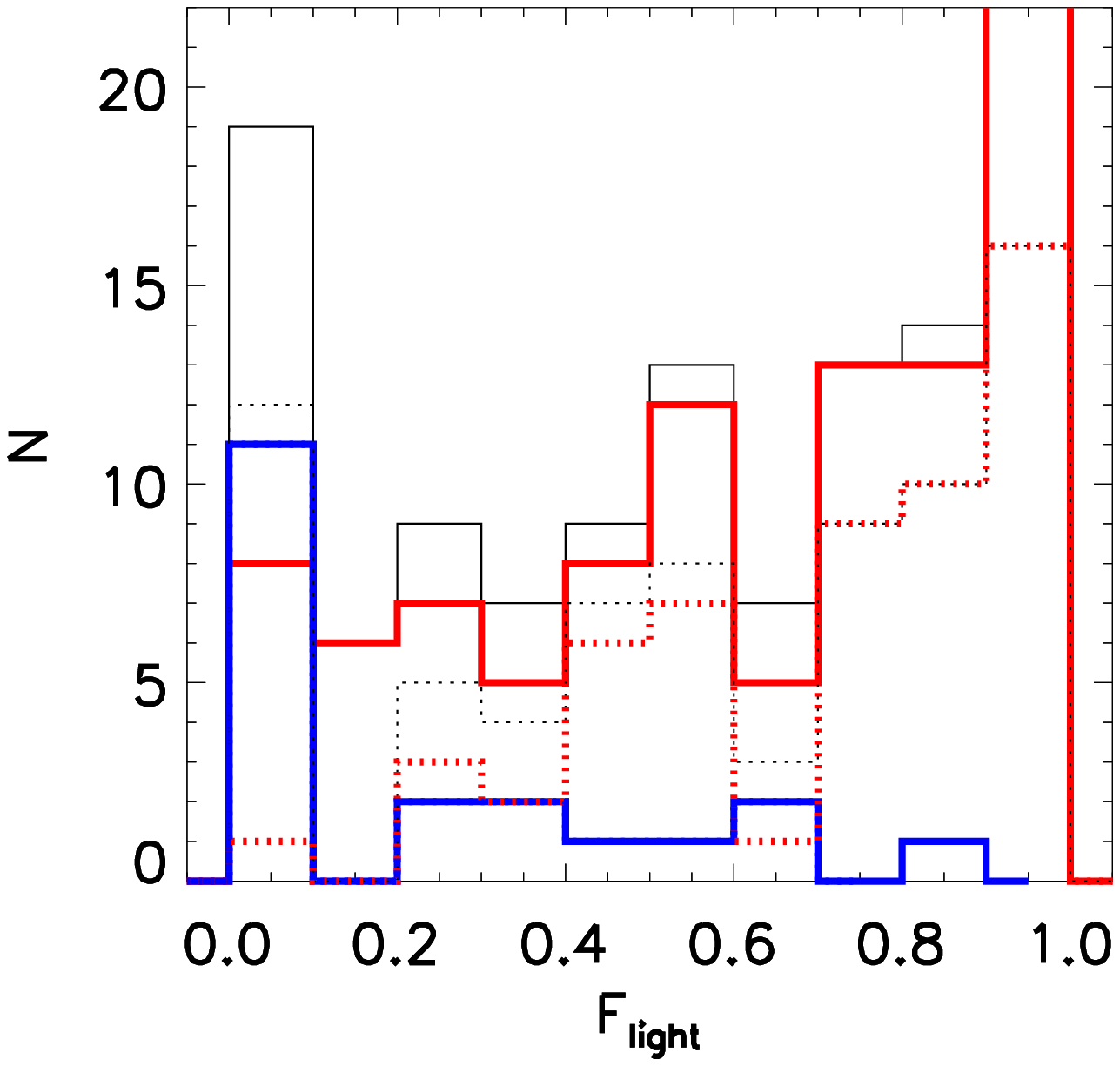}
\includegraphics[width=0.4\textwidth]{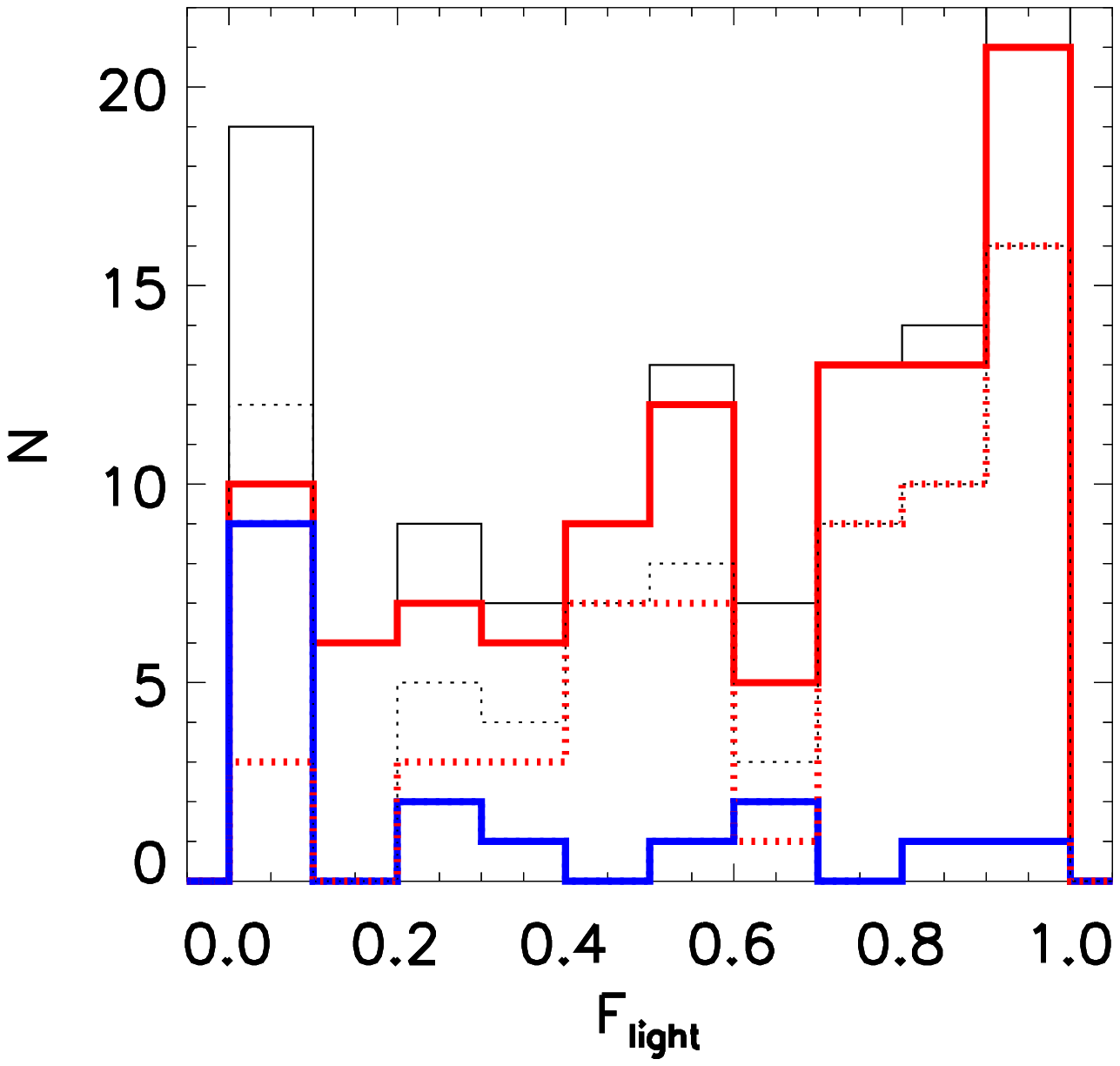}

\center{Fig. \ref{fig1d}---Continued}
\end{figure*}

\clearpage

\begin{figure*}[!b]
\centering
\includegraphics[width=0.4\textwidth]{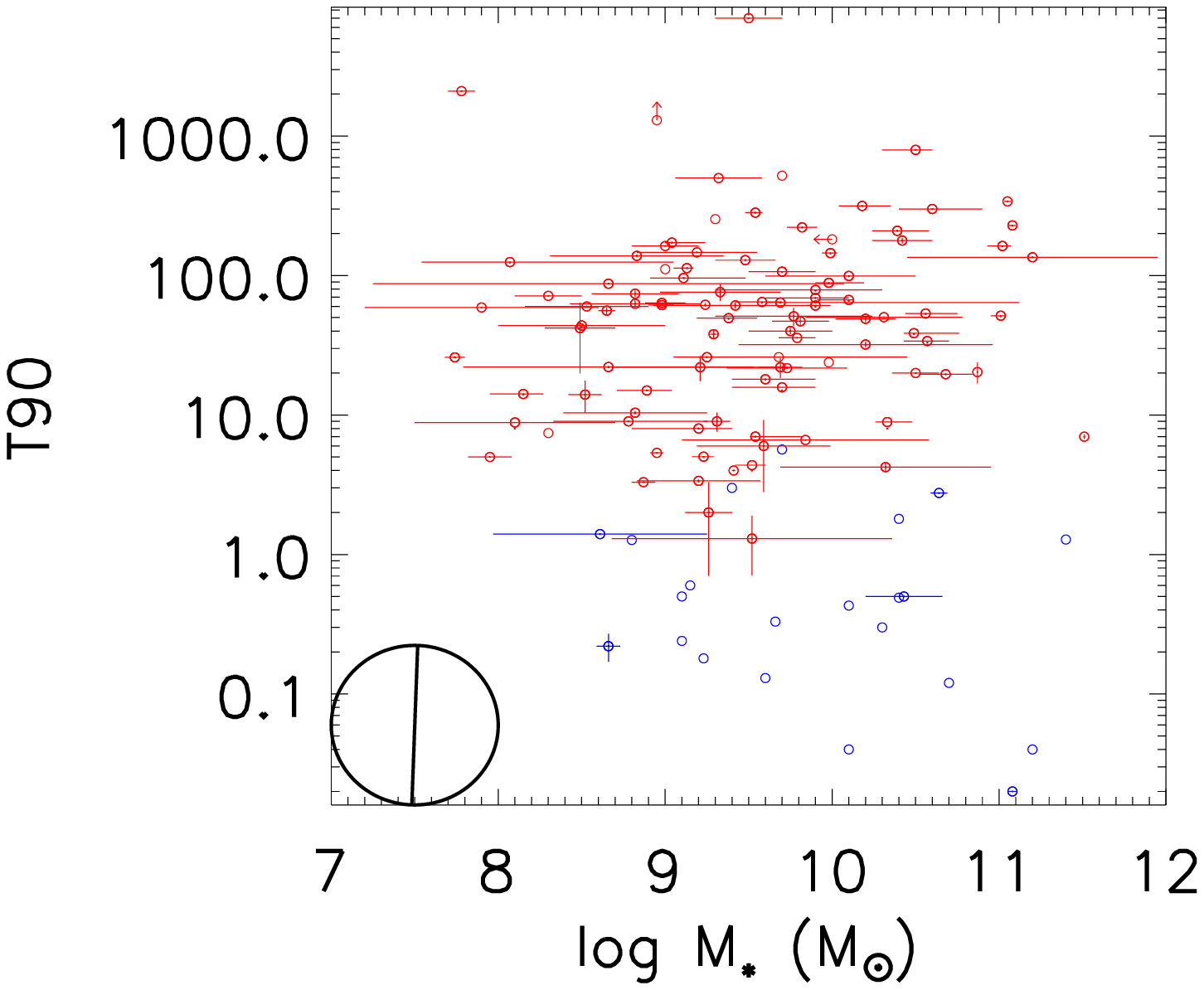}
\includegraphics[width=0.4\textwidth]{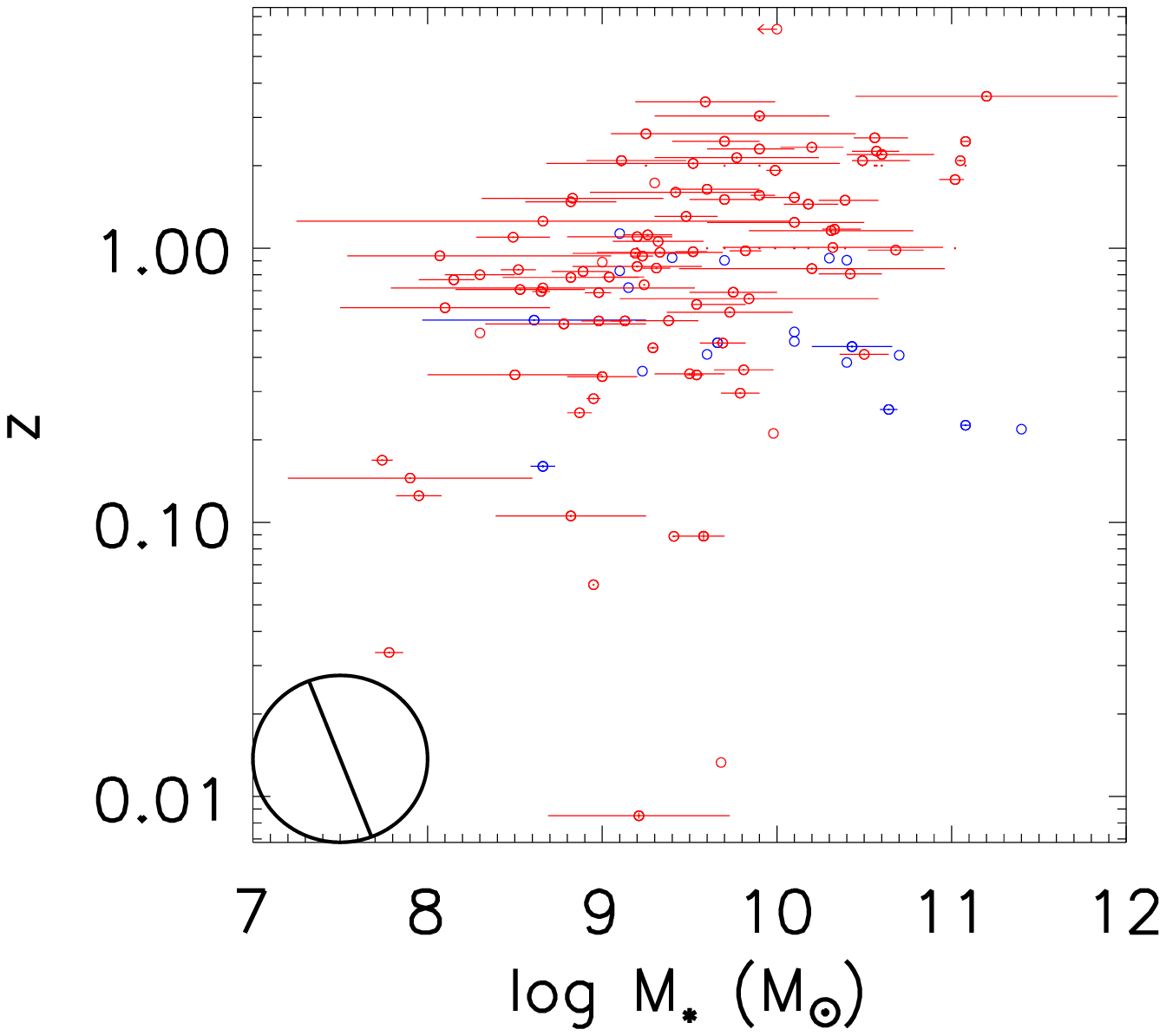}

\includegraphics[width=0.4\textwidth]{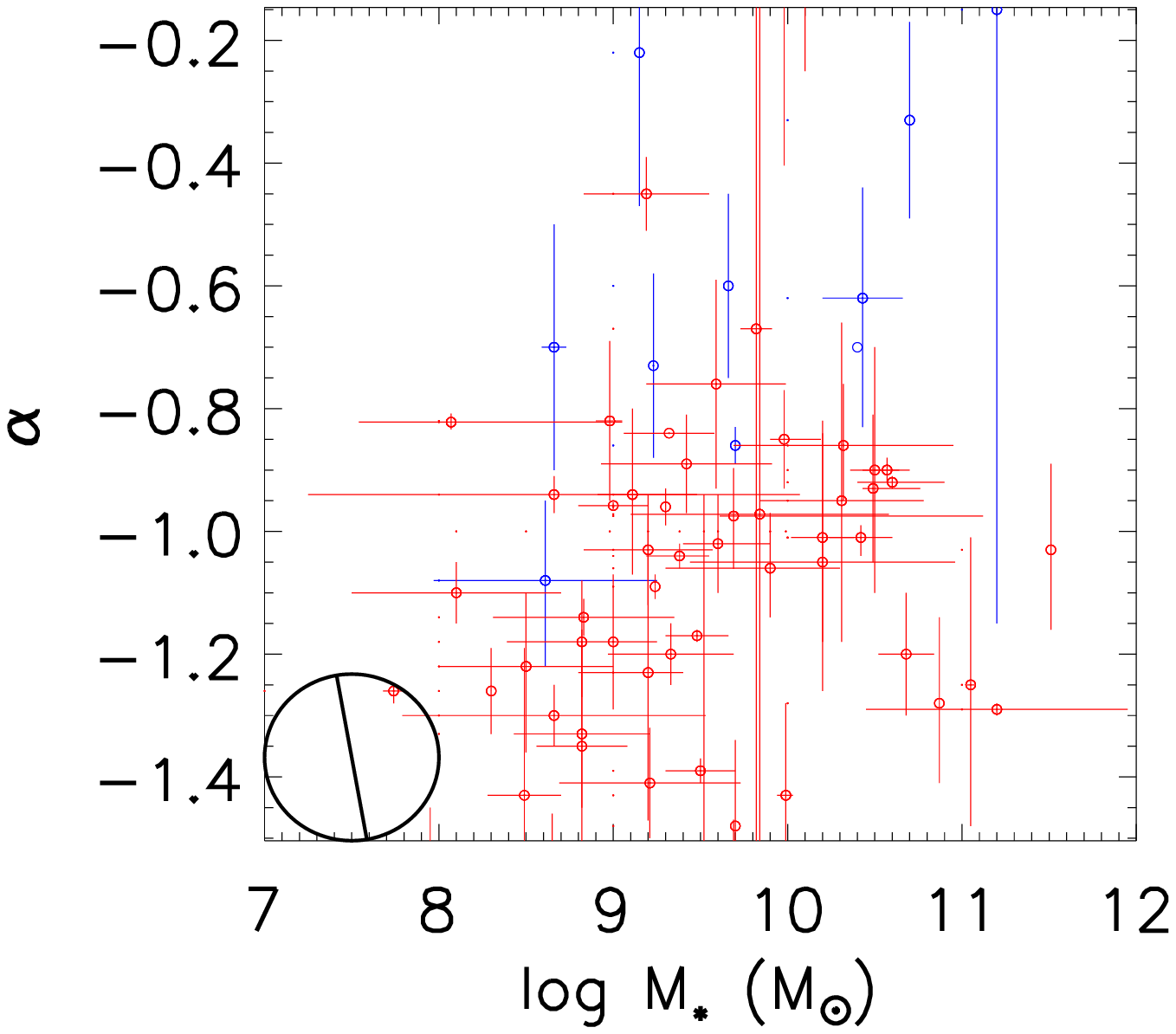}
\includegraphics[width=0.4\textwidth]{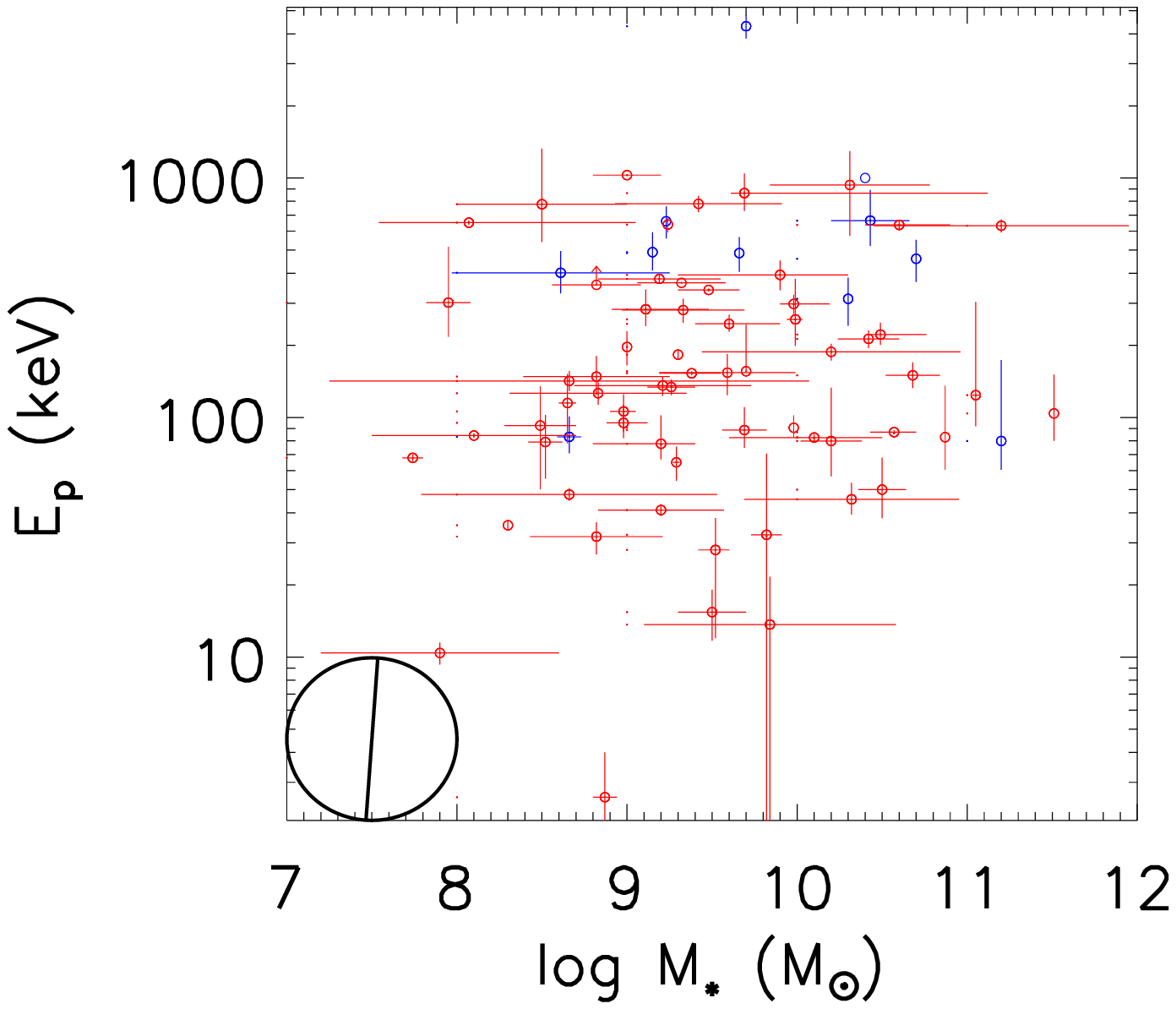}

\includegraphics[width=0.4\textwidth]{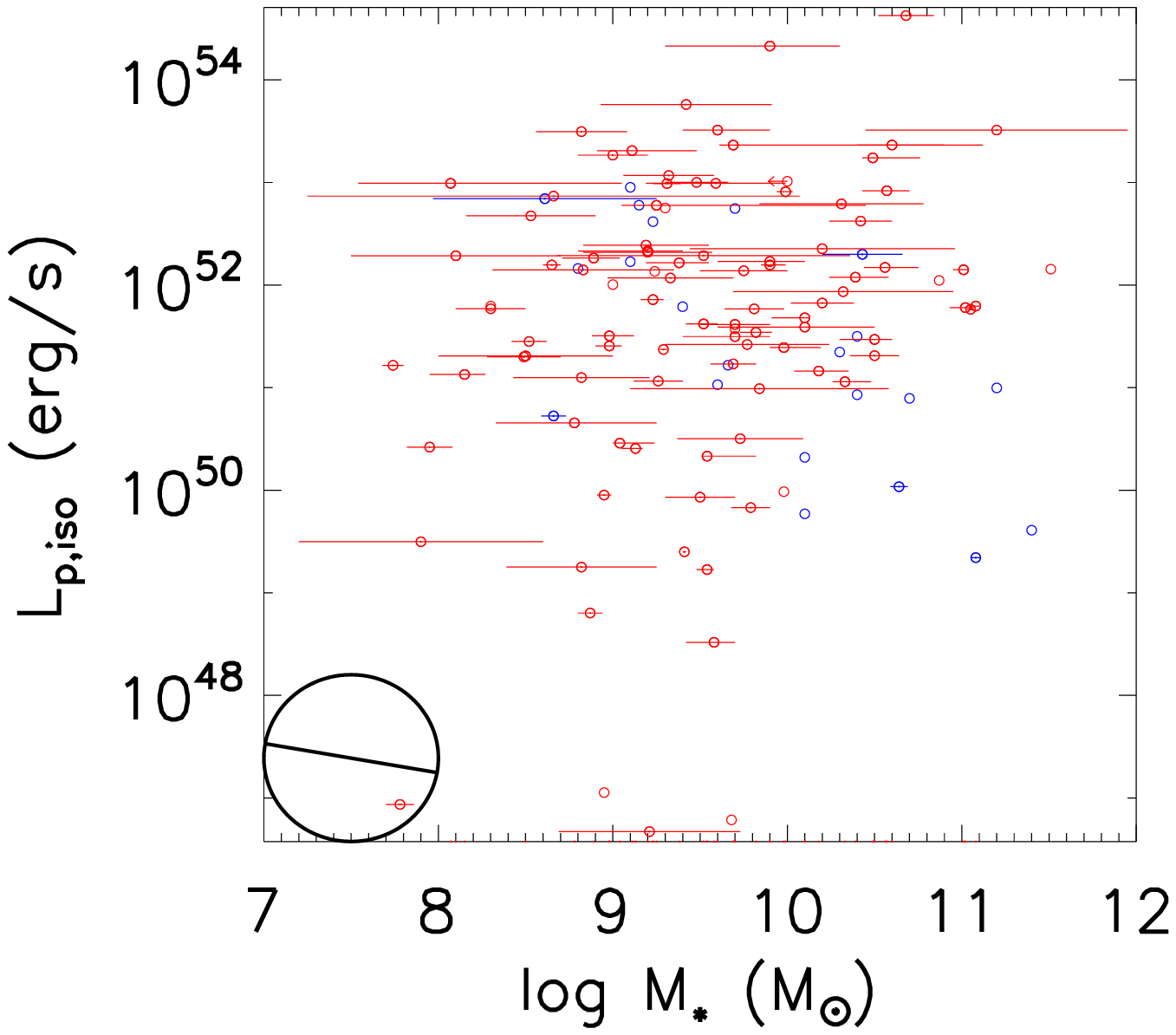}
\includegraphics[width=0.4\textwidth]{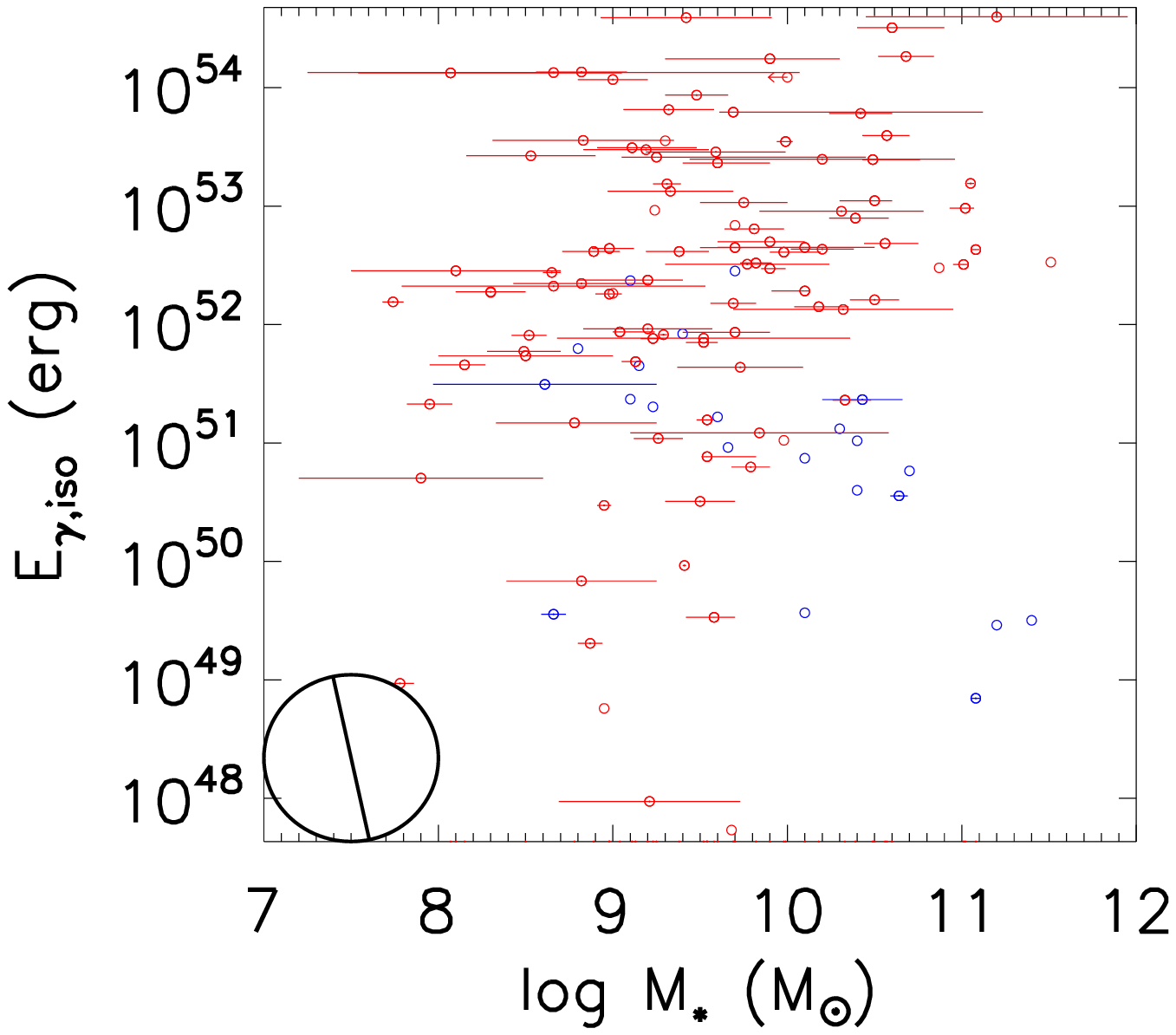}

\includegraphics[width=0.4\textwidth]{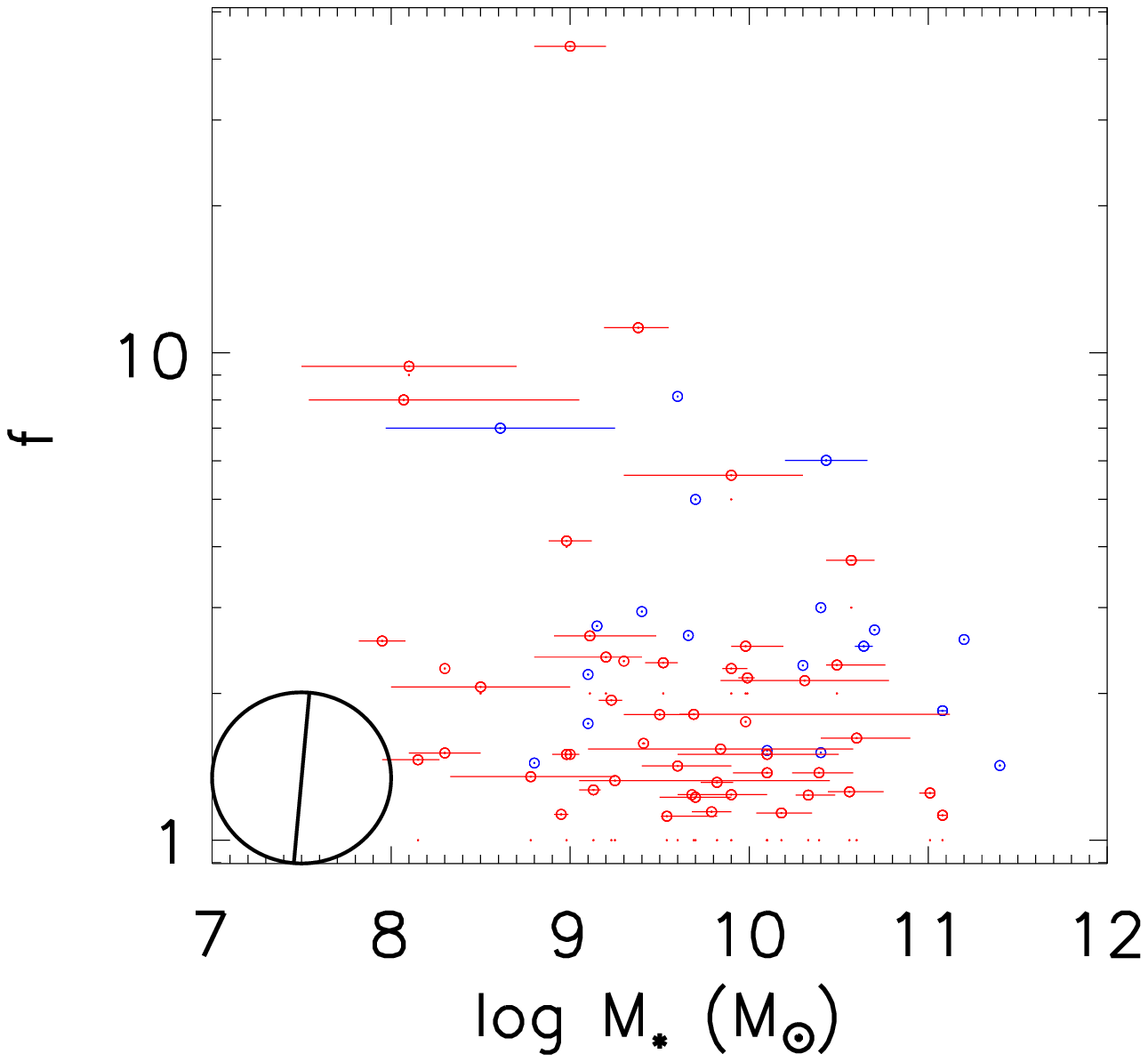}
\includegraphics[width=0.4\textwidth]{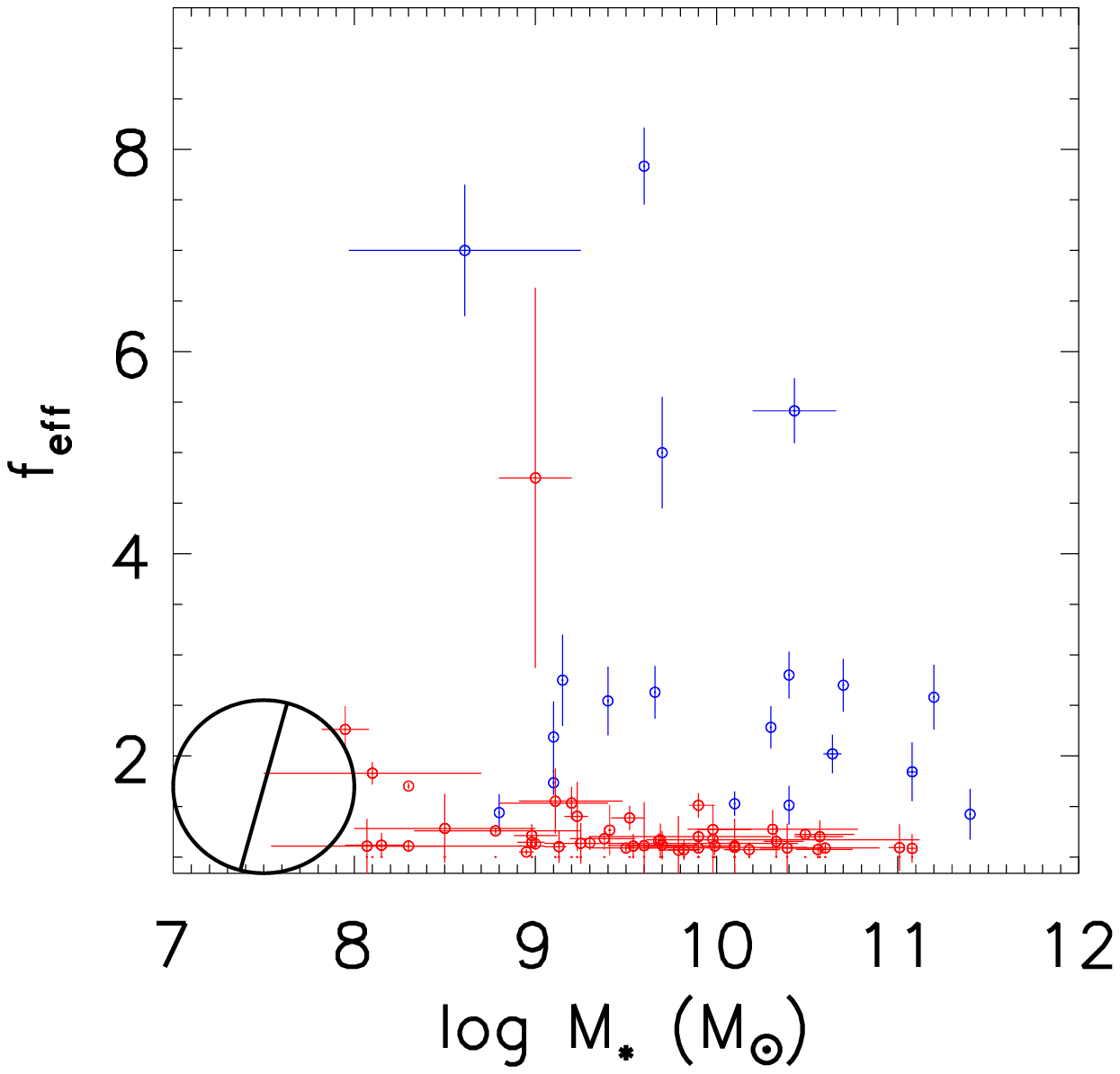}

\caption{
Prompt emission VS host galaxy property 2D plots of LGRBs (red dots)
and SGRBs (blue dots). Black lines show the rotated new x-axis for the 
lowest $P_{\rm KS}$.
}
\label{fig2d}
\end{figure*}

\clearpage
\begin{figure*}

\includegraphics[width=0.4\textwidth]{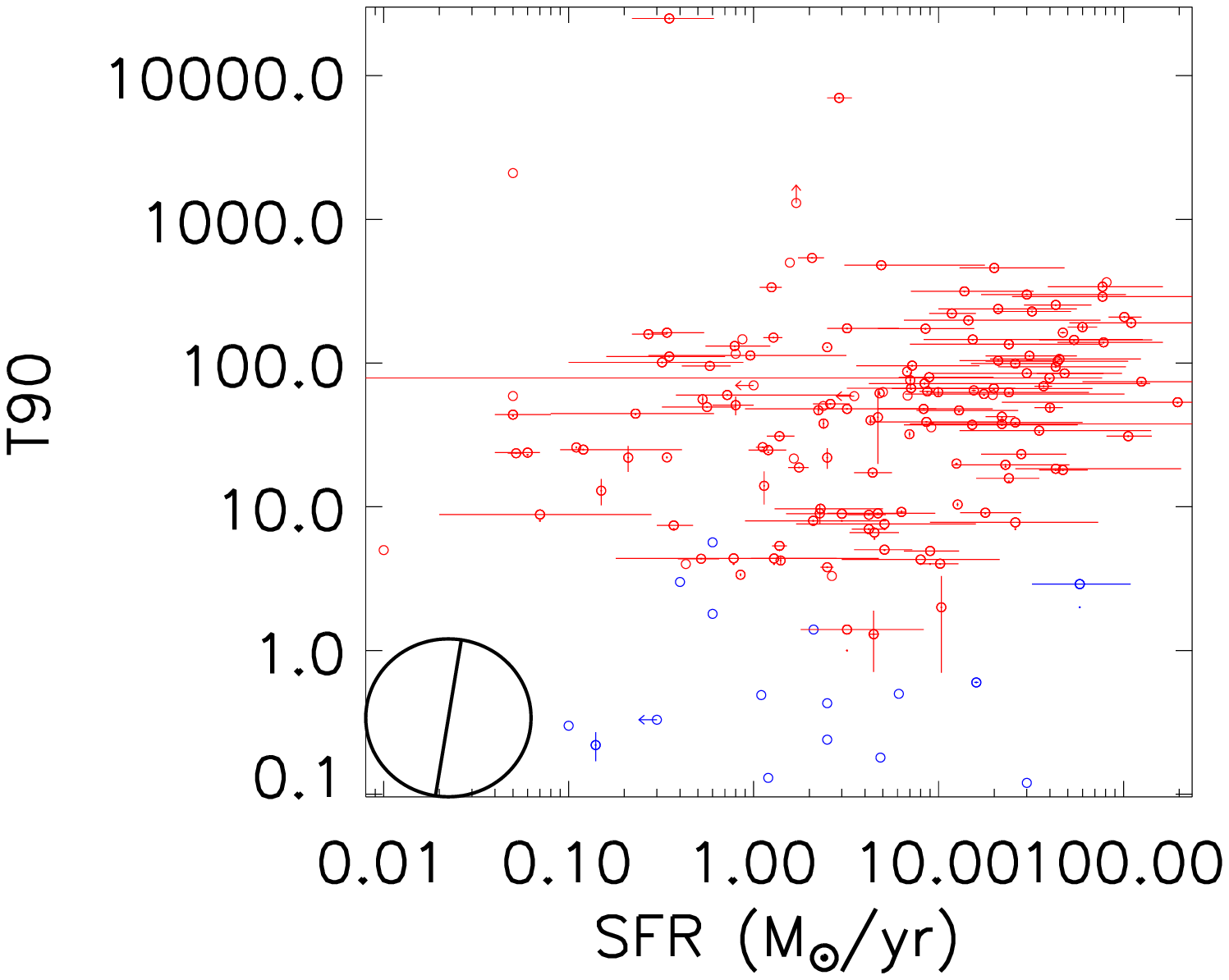}
\includegraphics[width=0.4\textwidth]{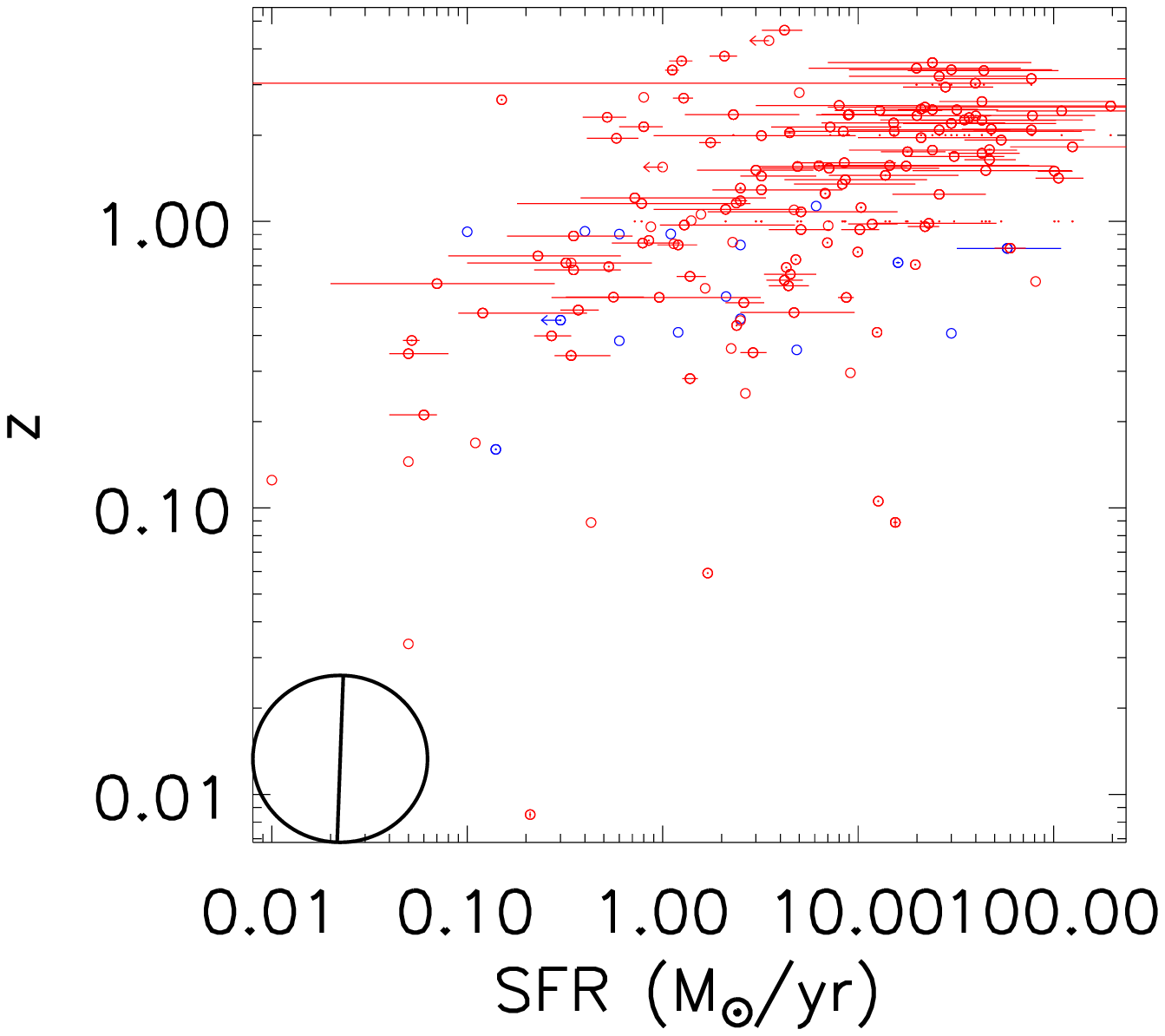}

\includegraphics[width=0.4\textwidth]{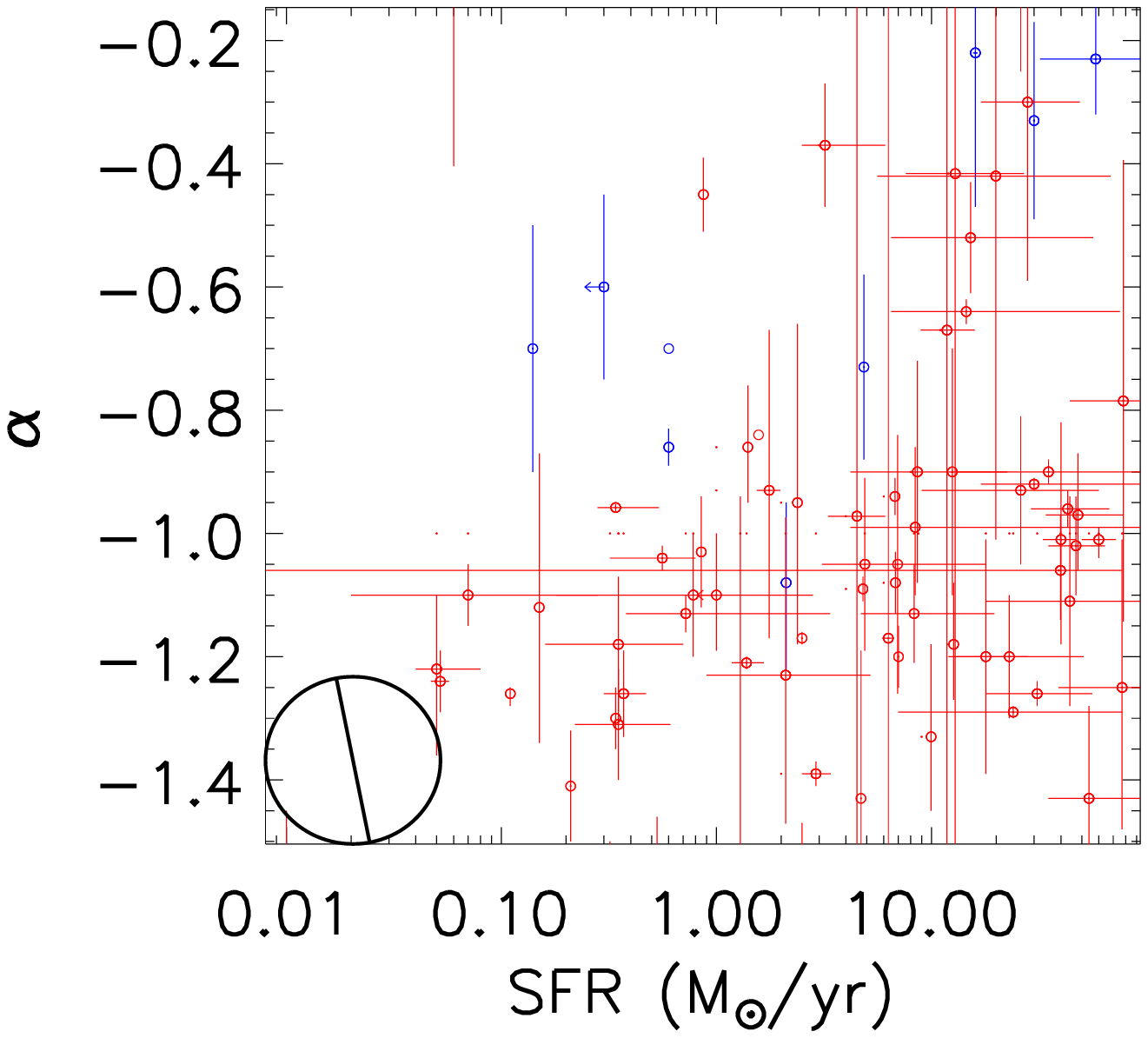}
\includegraphics[width=0.4\textwidth]{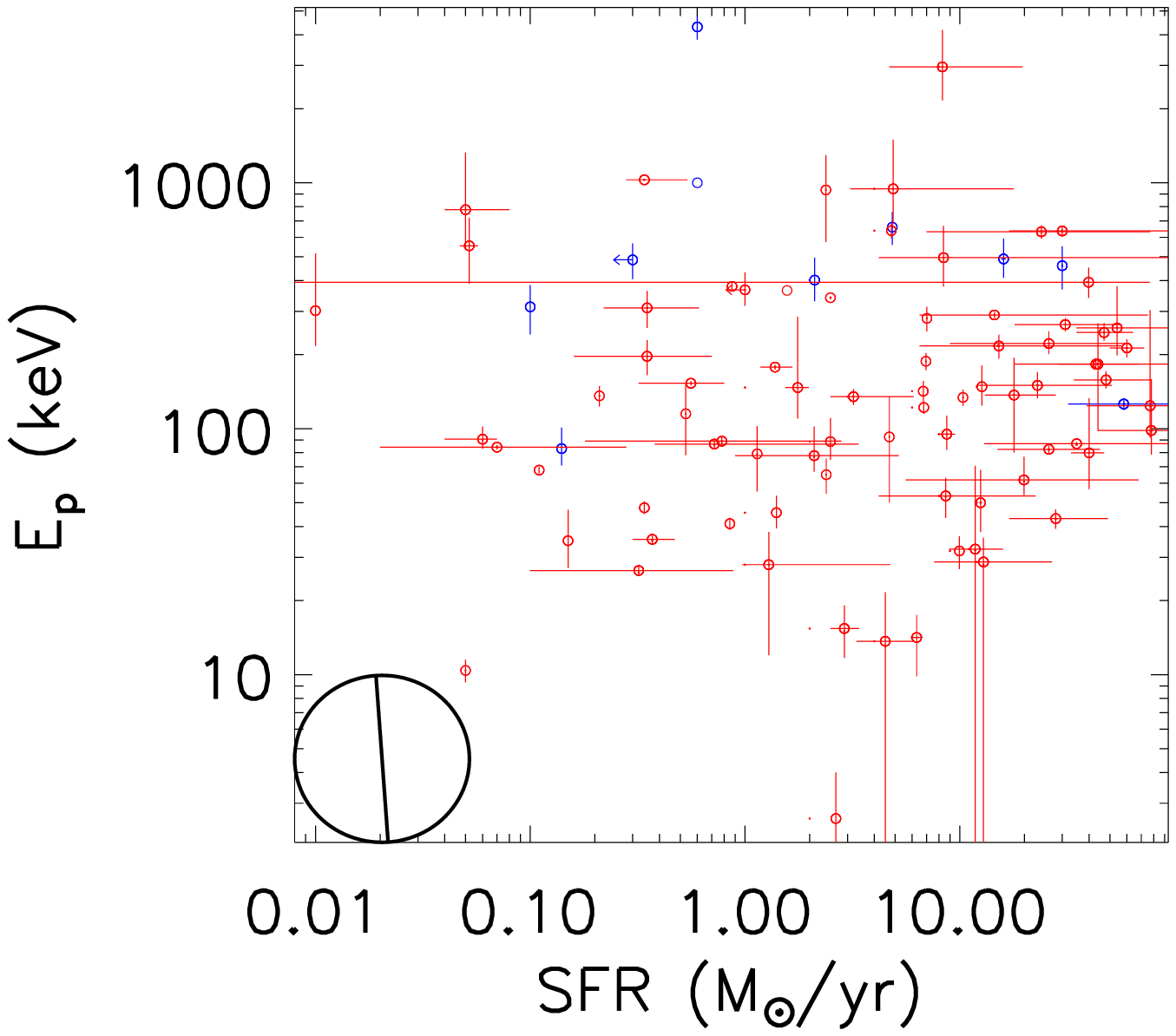}

\includegraphics[width=0.4\textwidth]{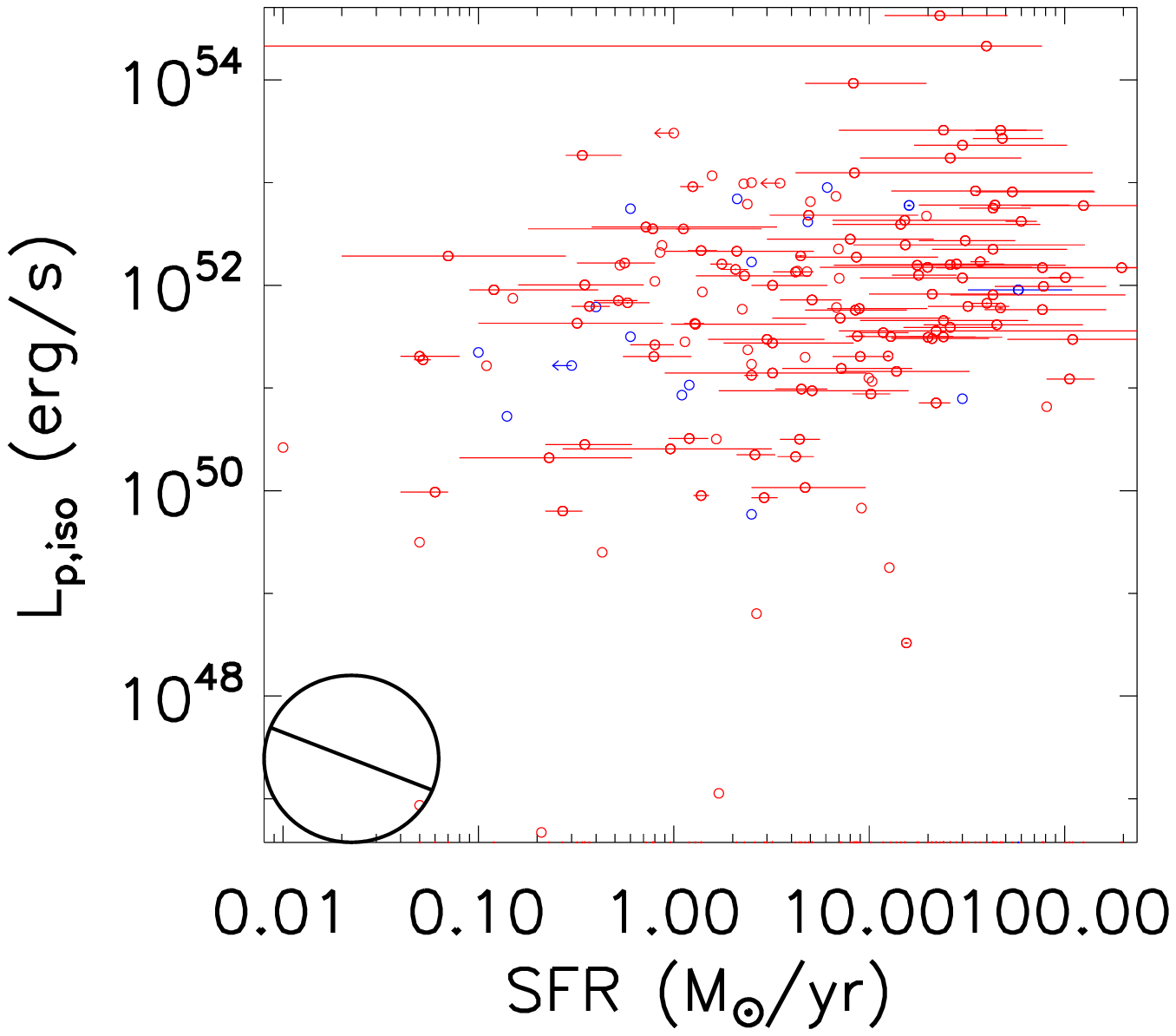}
\includegraphics[width=0.4\textwidth]{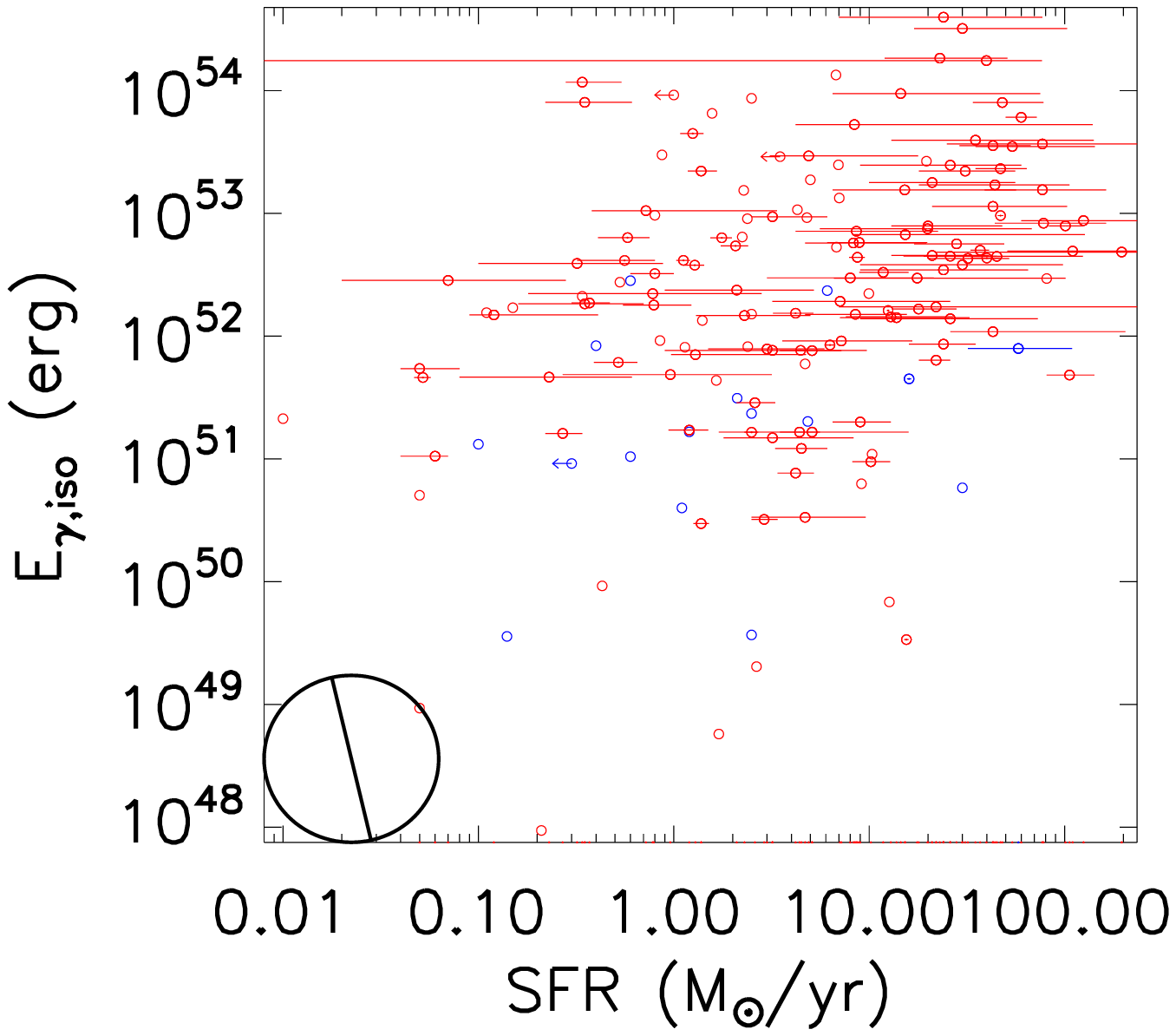}

\includegraphics[width=0.4\textwidth]{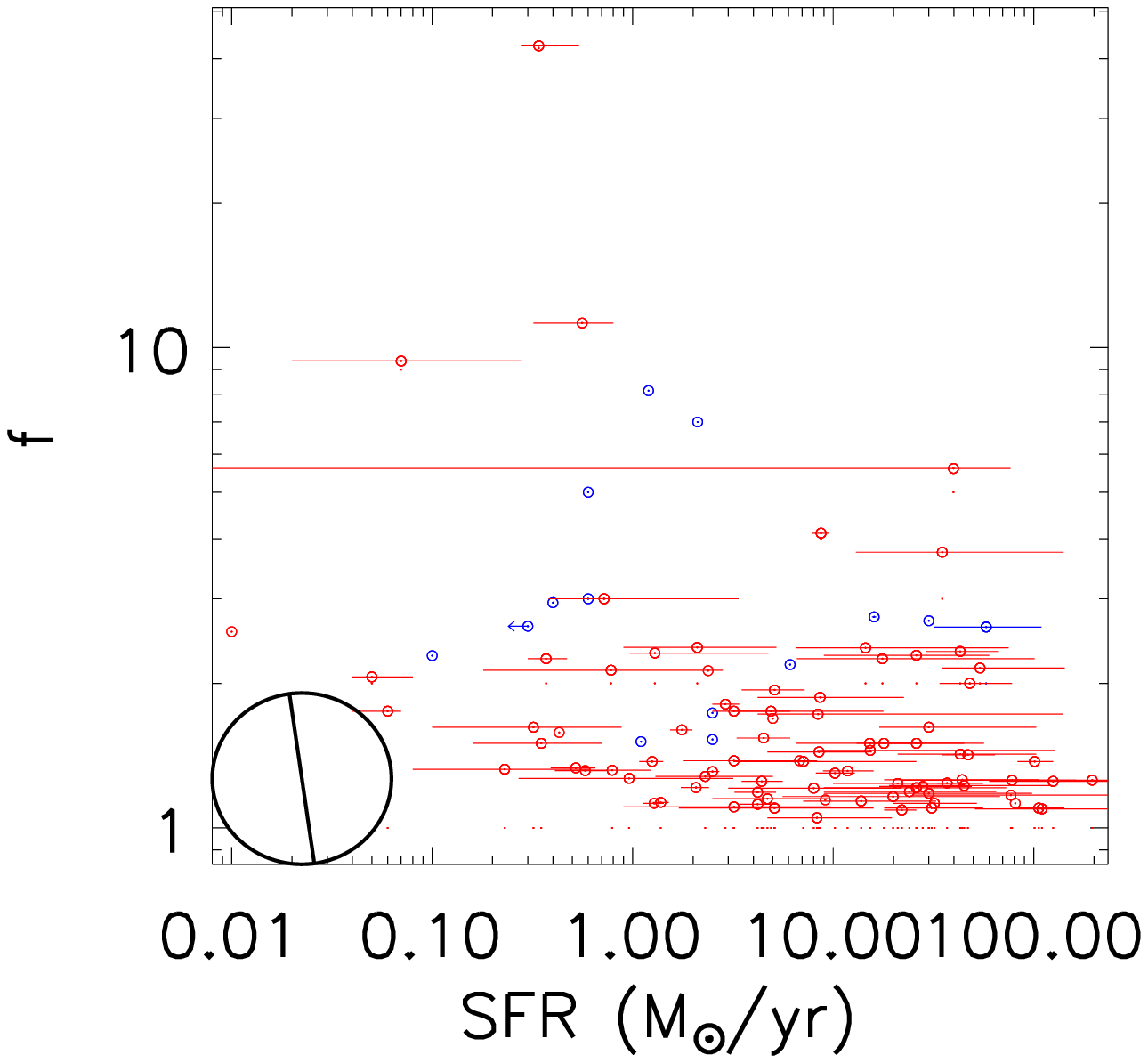}
\includegraphics[width=0.4\textwidth]{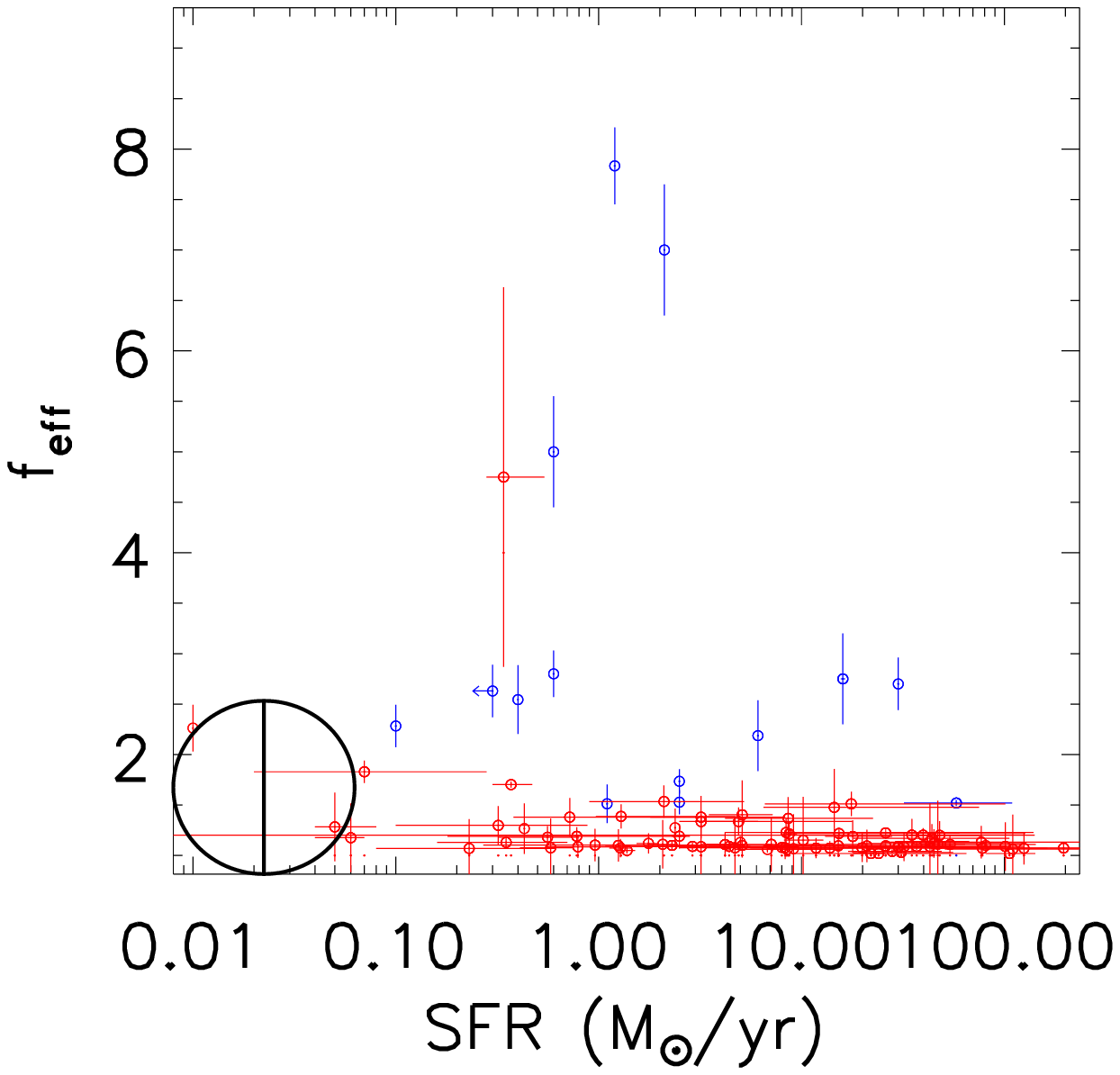}

\center{Fig. \ref{fig2d}---Continued}
\end{figure*}


\clearpage
\begin{figure*}

\includegraphics[width=0.4\textwidth]{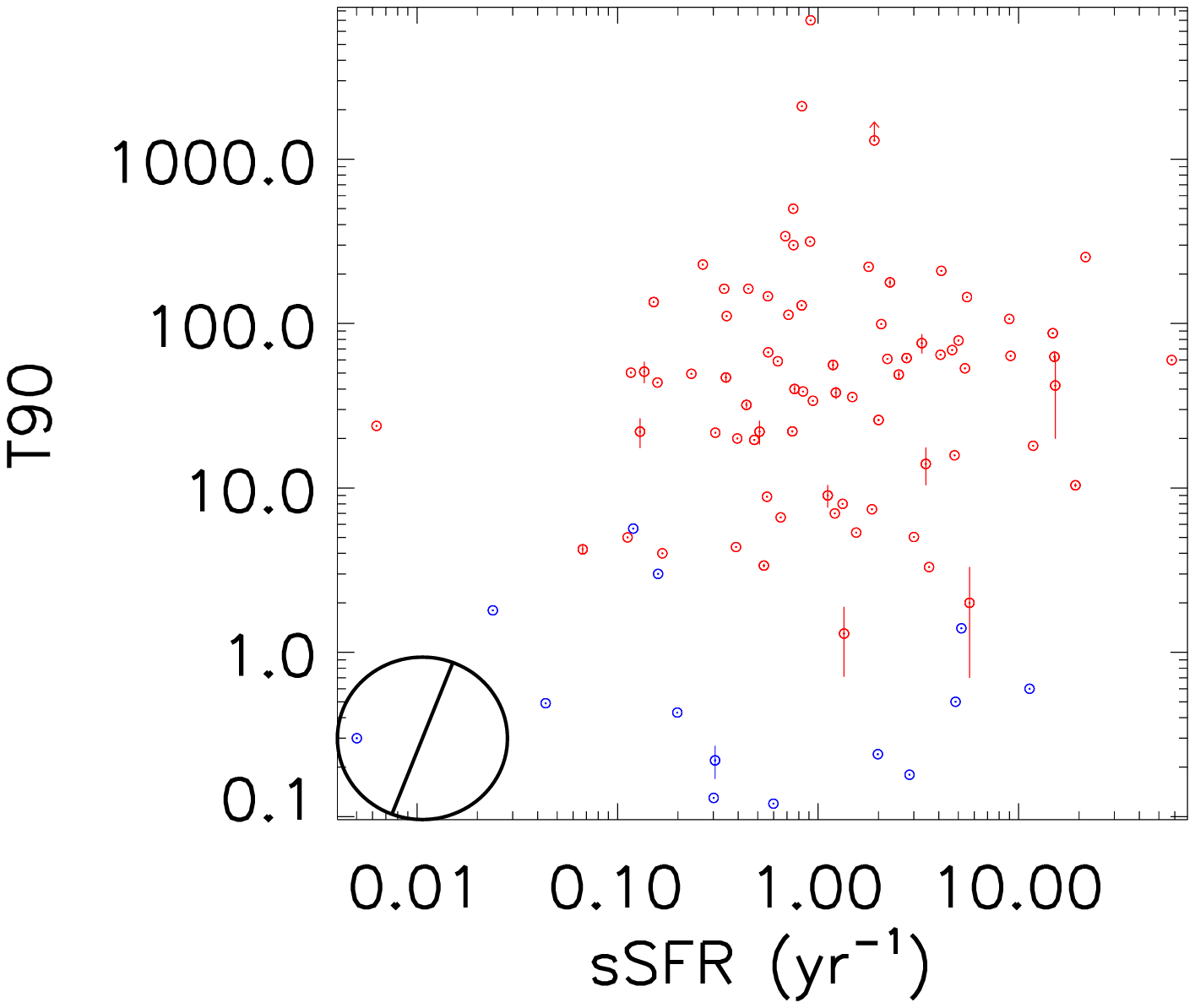}
\includegraphics[width=0.4\textwidth]{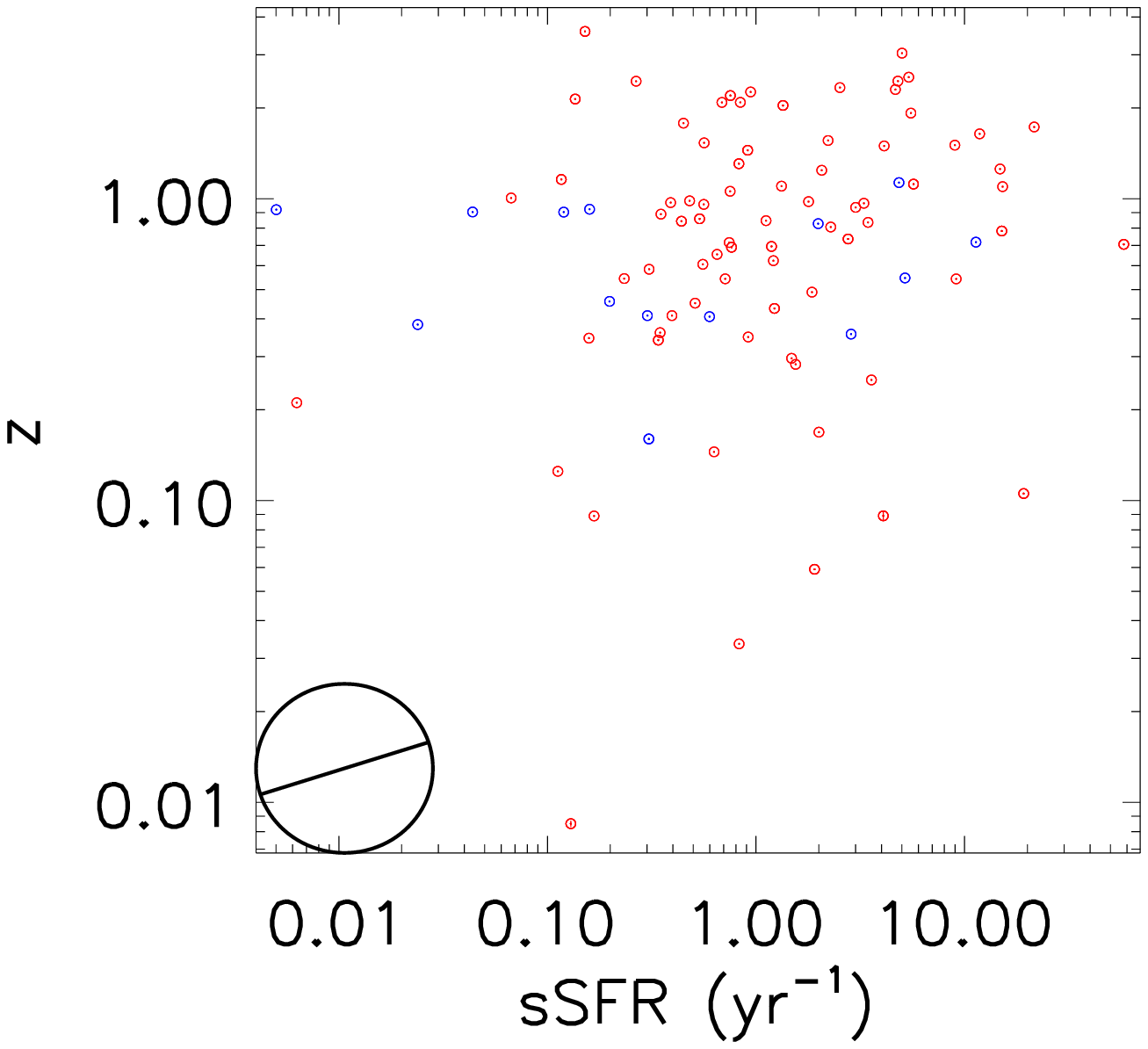}

\includegraphics[width=0.4\textwidth]{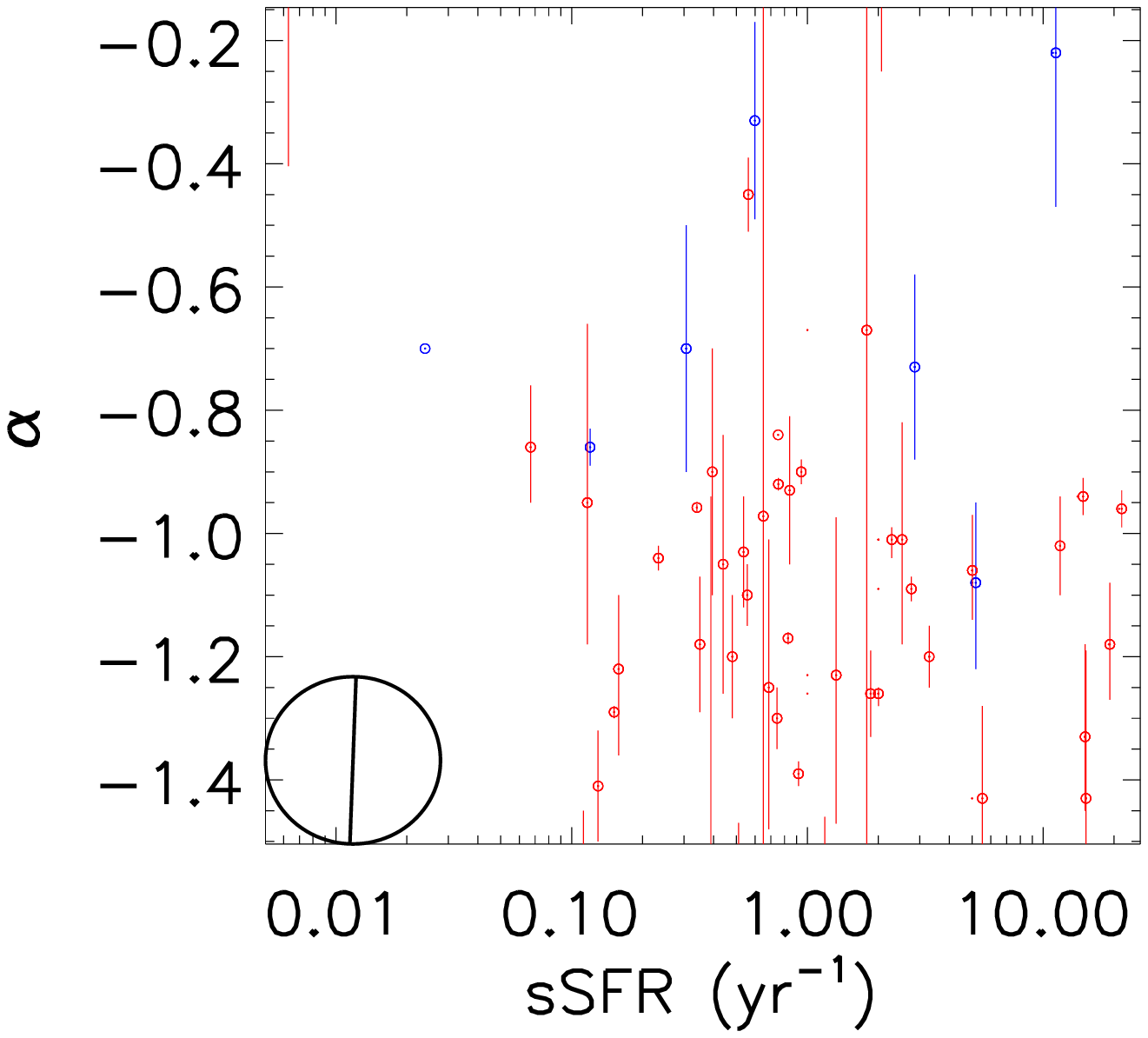}
\includegraphics[width=0.4\textwidth]{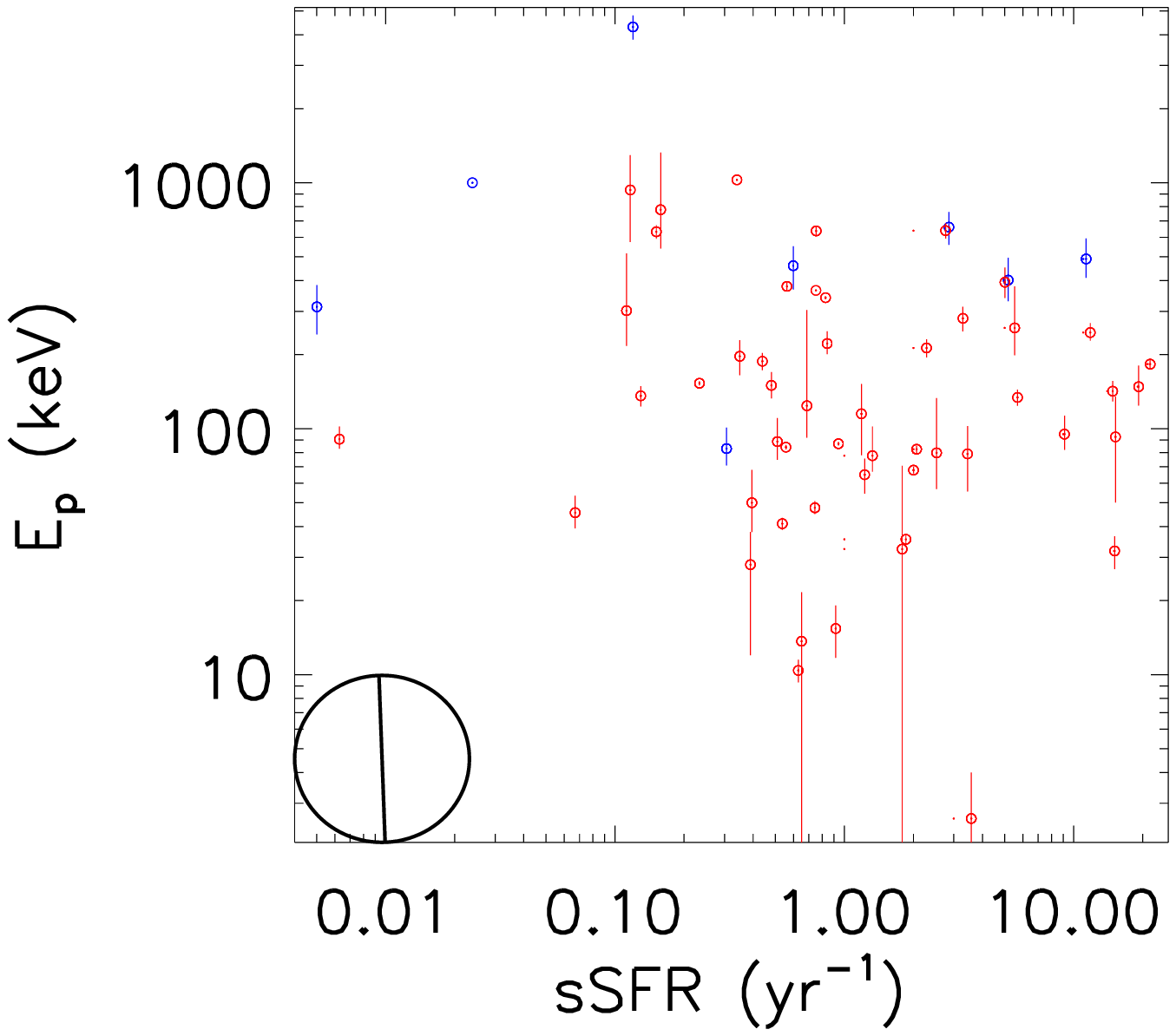}

\includegraphics[width=0.4\textwidth]{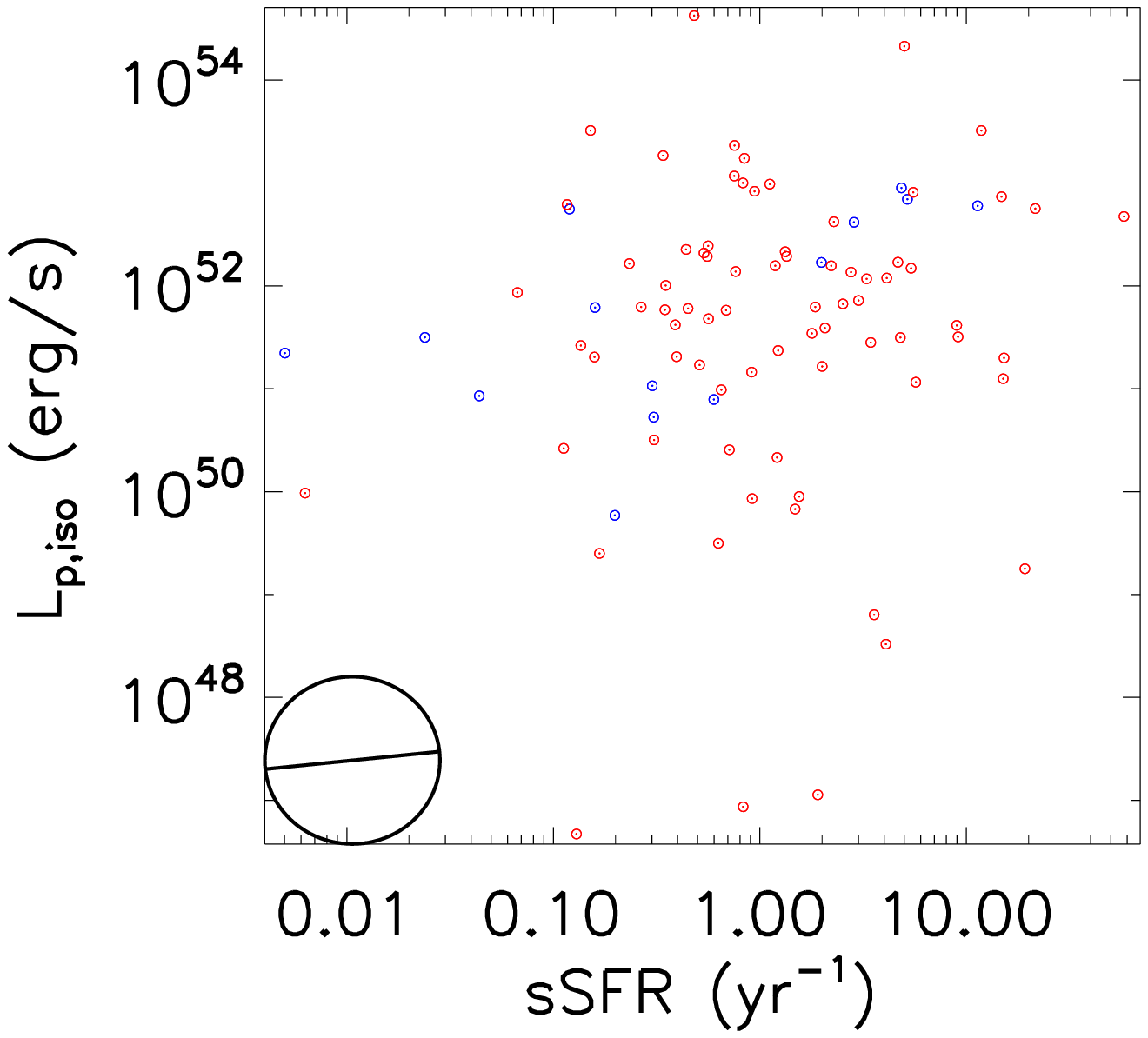}
\includegraphics[width=0.4\textwidth]{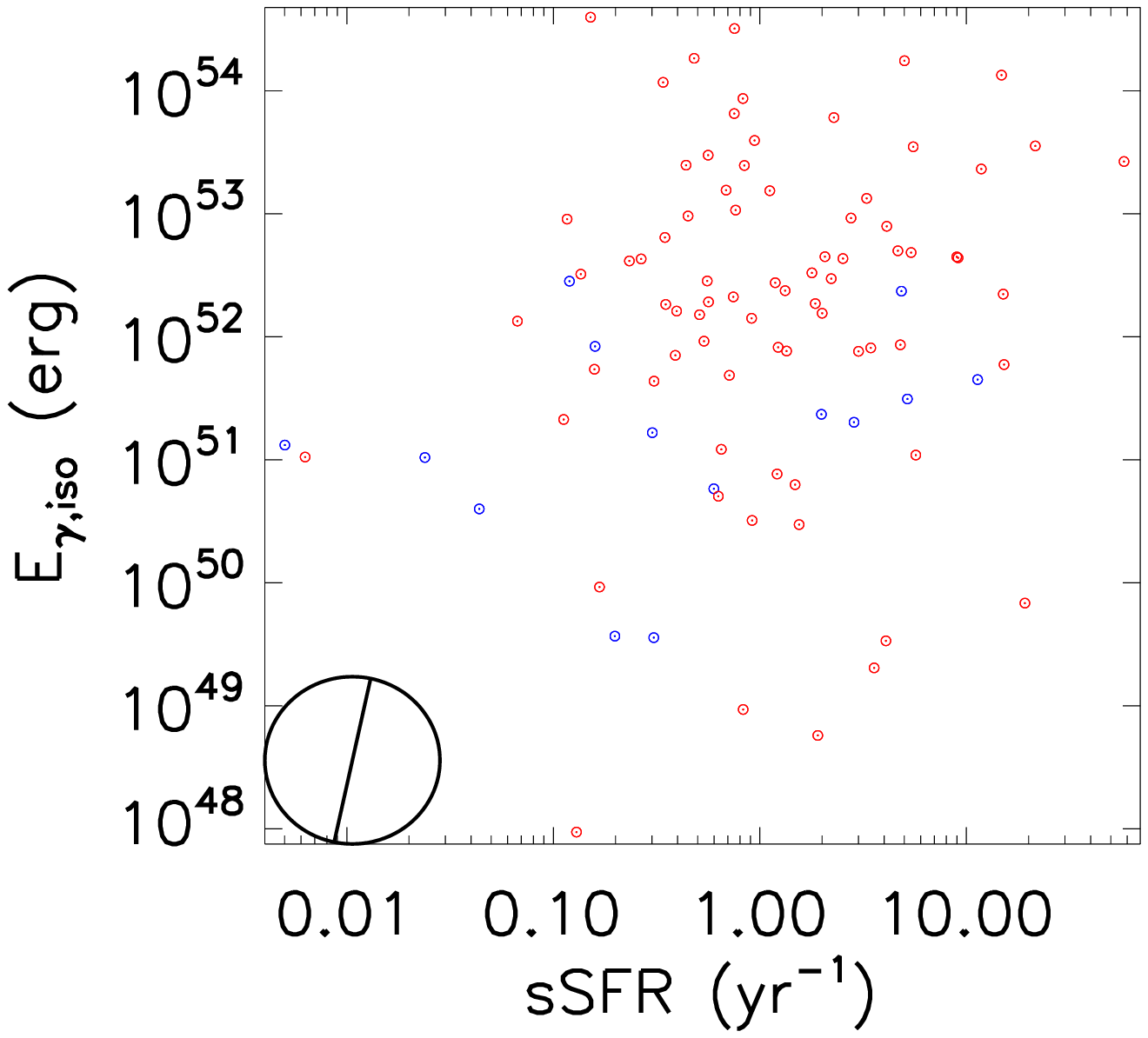}

\includegraphics[width=0.4\textwidth]{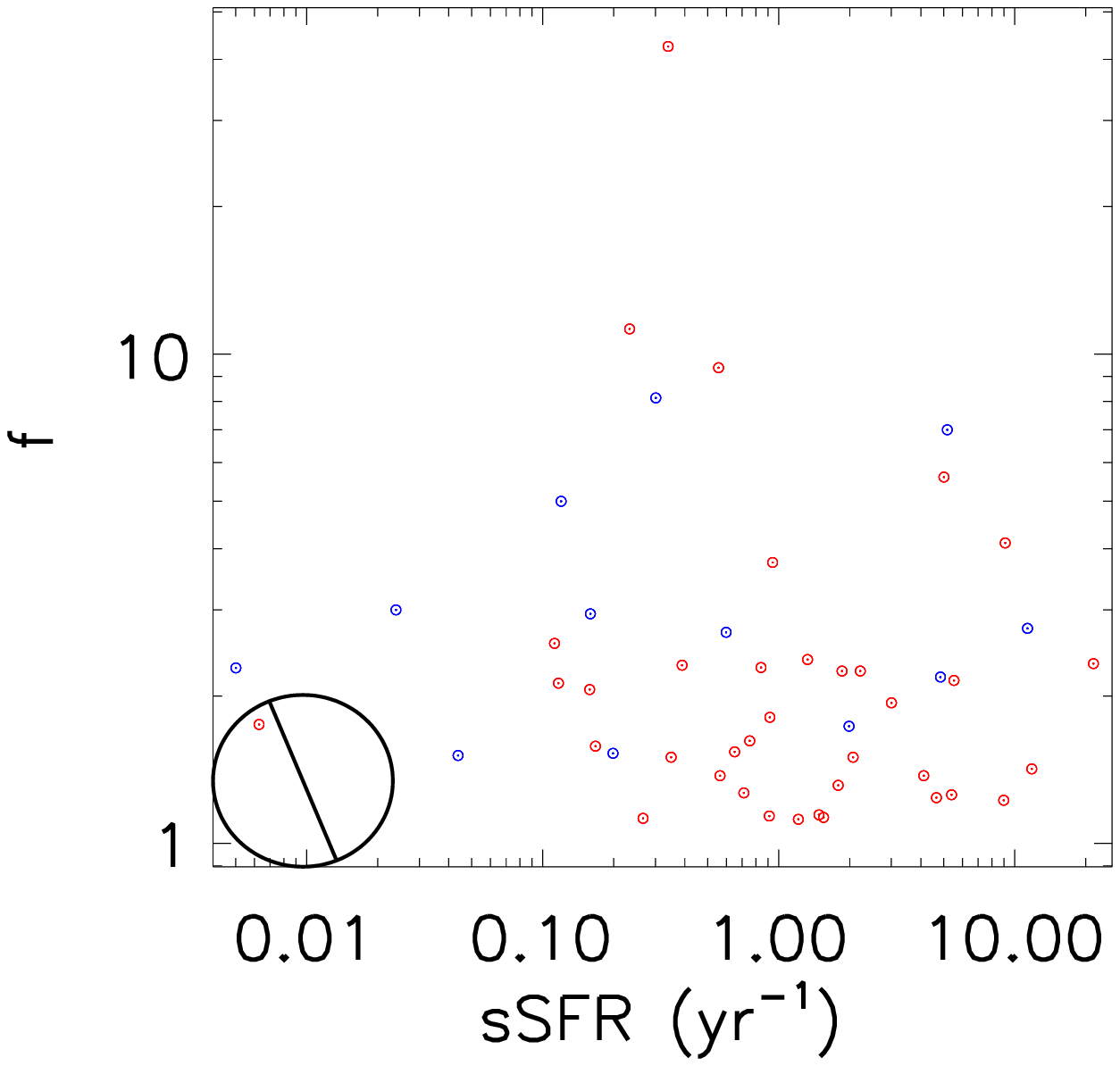}
\includegraphics[width=0.4\textwidth]{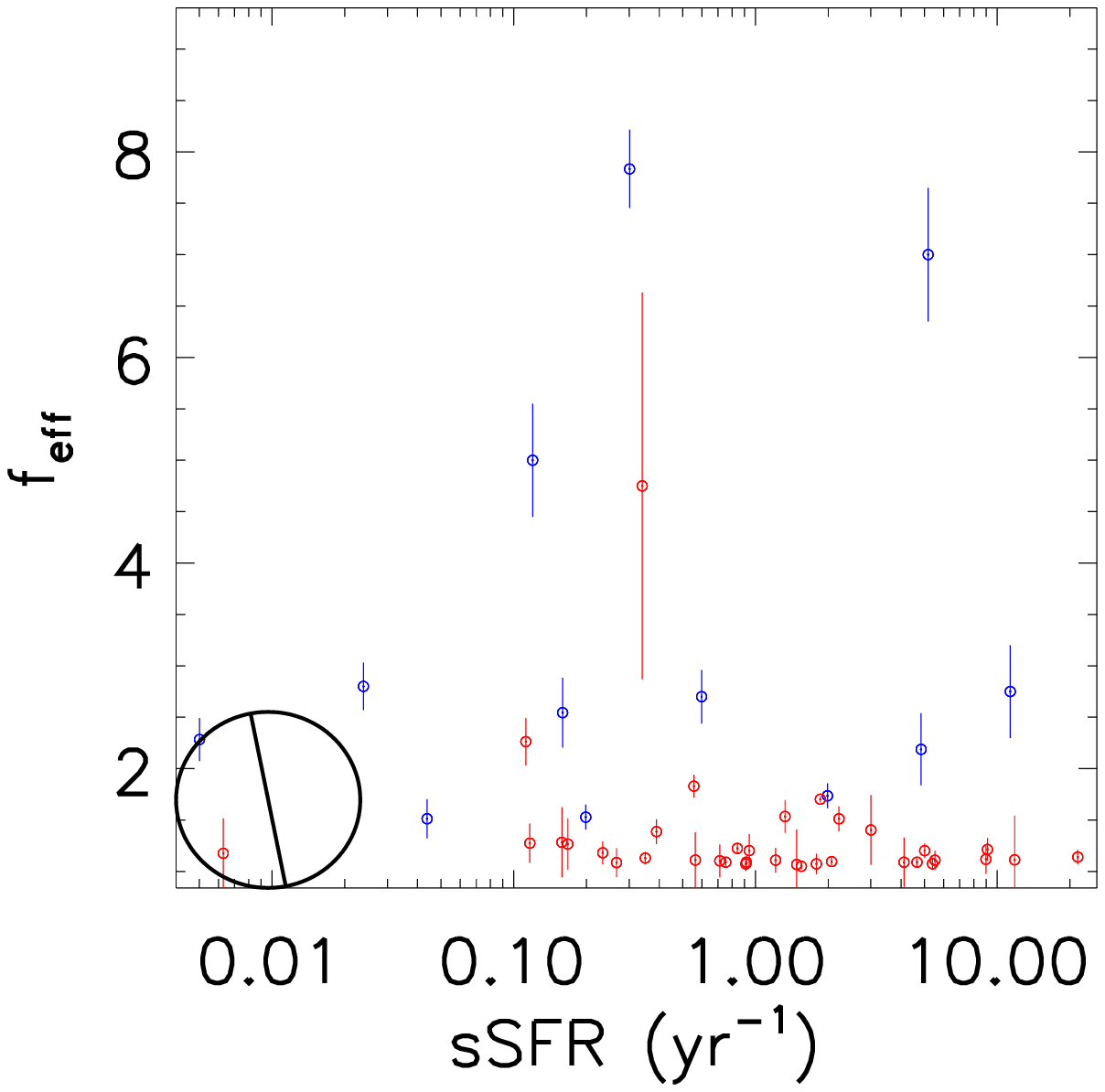}

\center{Fig. \ref{fig2d}---Continued}
\end{figure*}


\clearpage
\begin{figure*}

\includegraphics[width=0.4\textwidth]{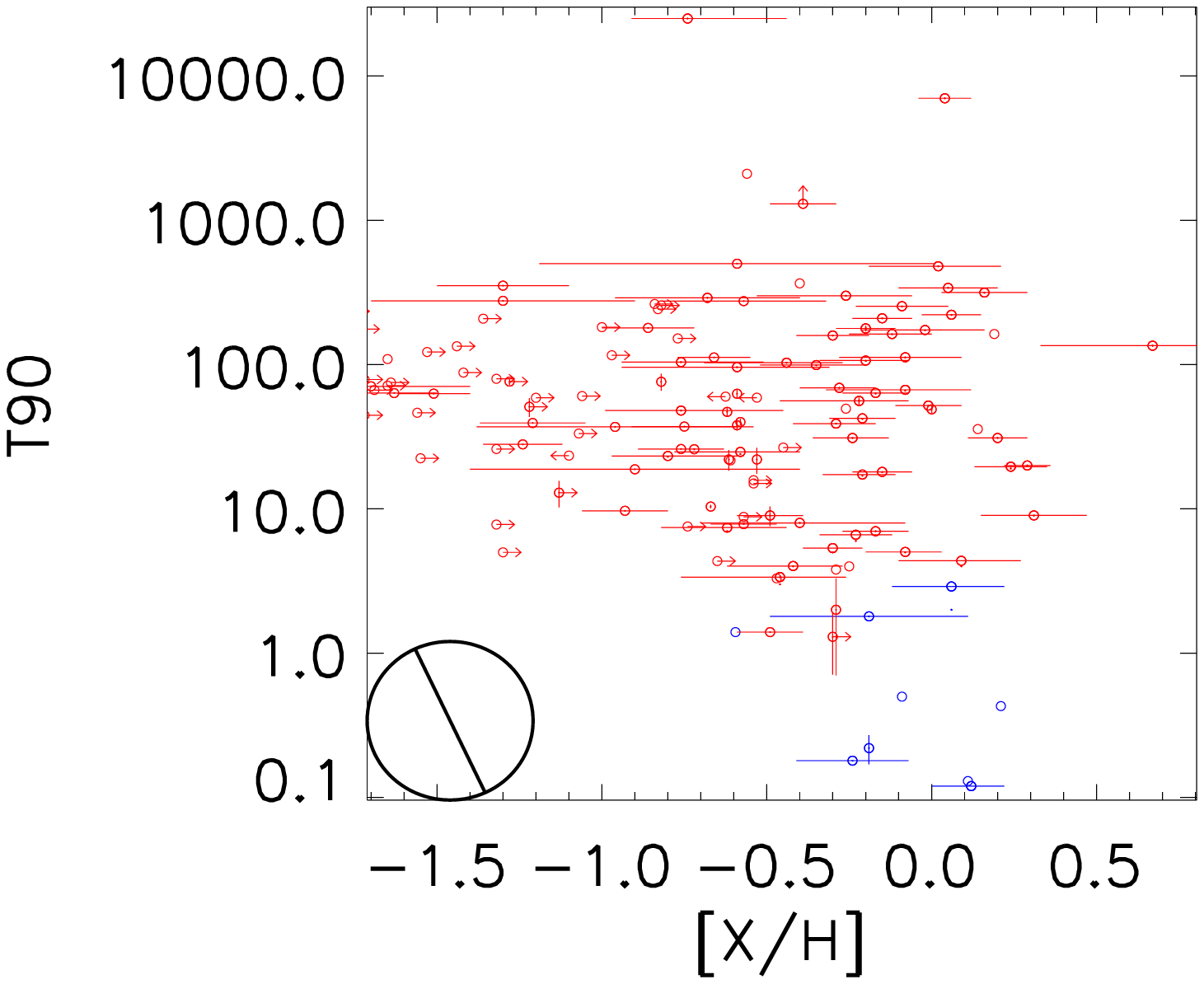}
\includegraphics[width=0.4\textwidth]{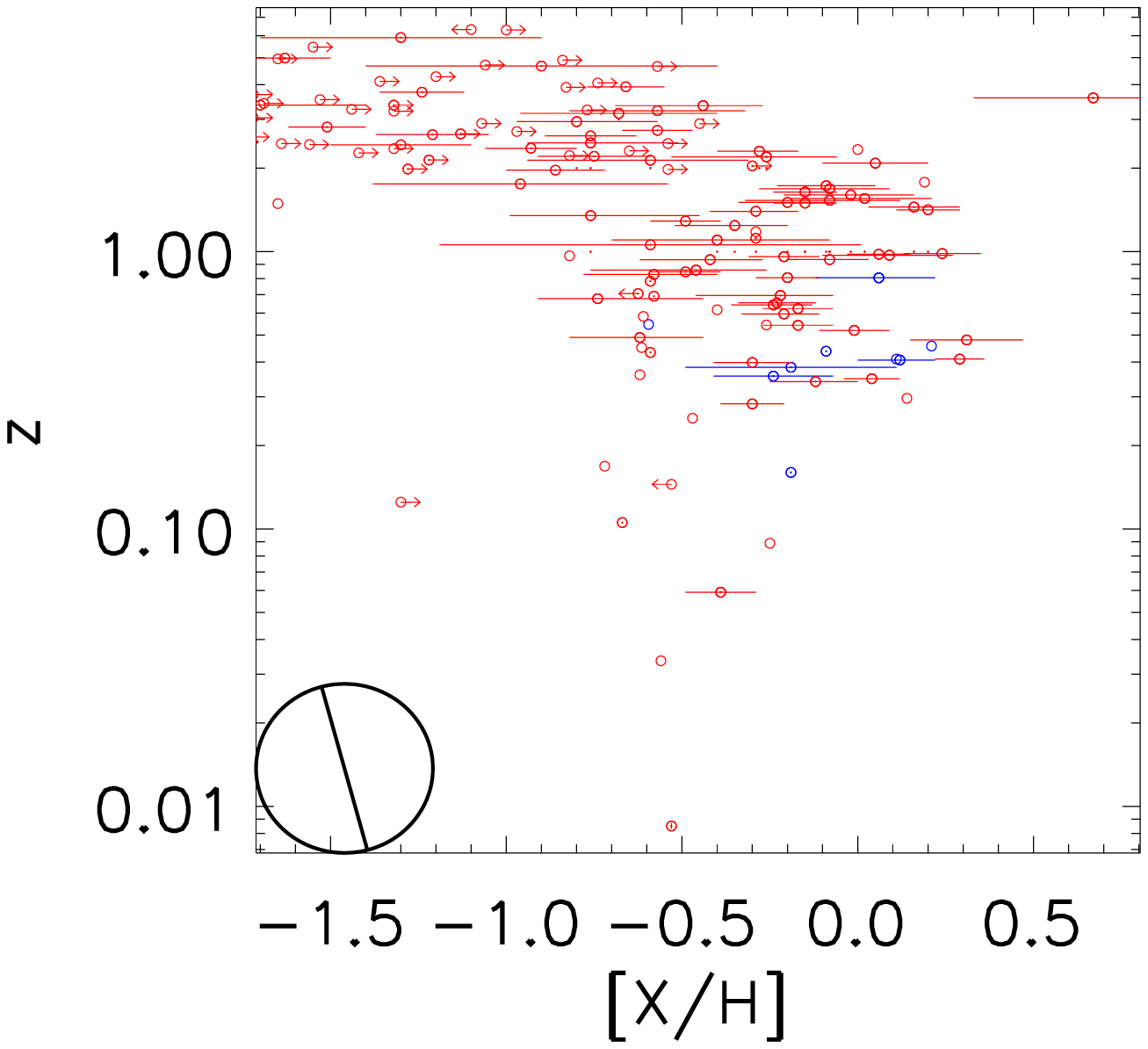}

\includegraphics[width=0.4\textwidth]{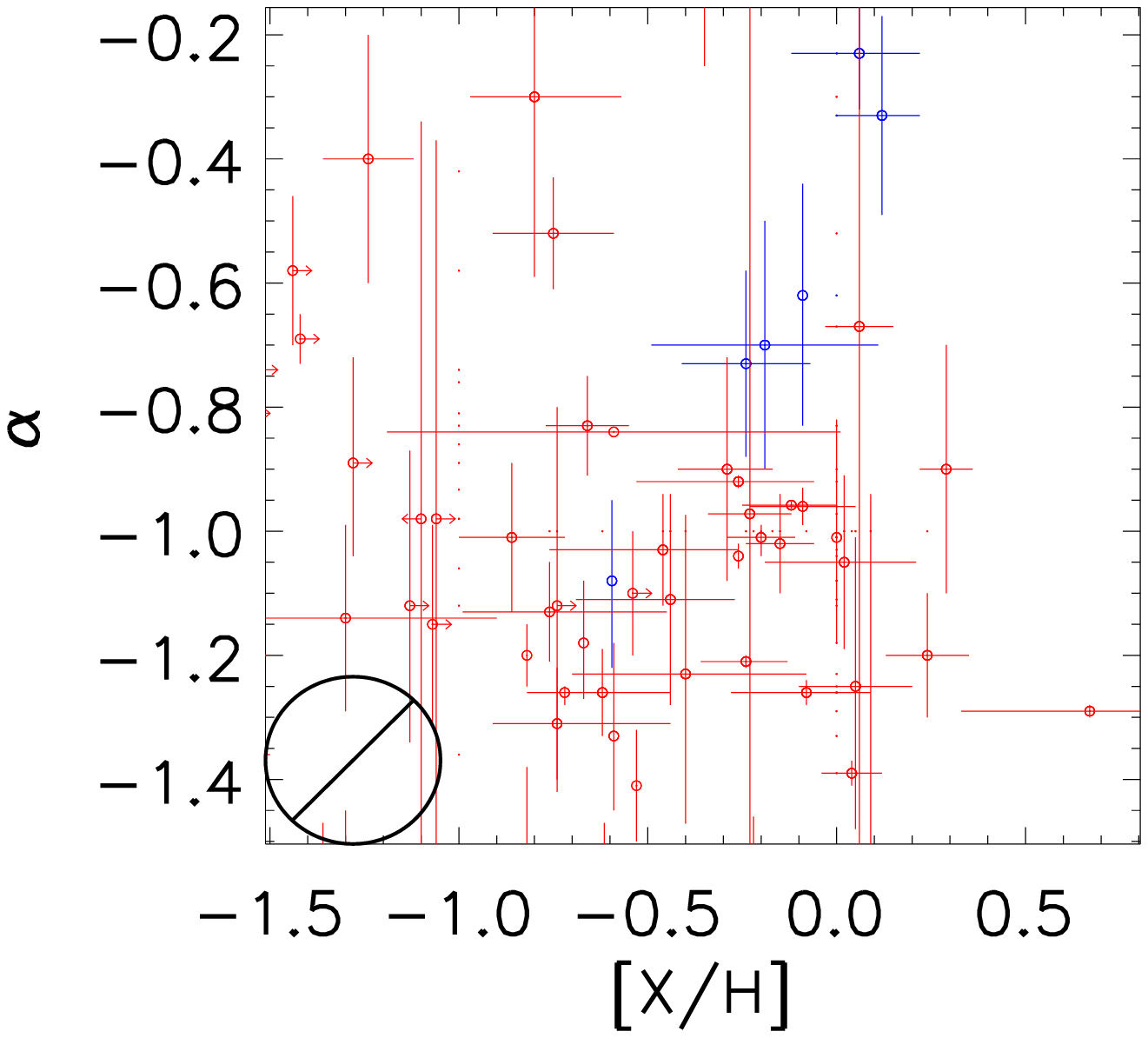}
\includegraphics[width=0.4\textwidth]{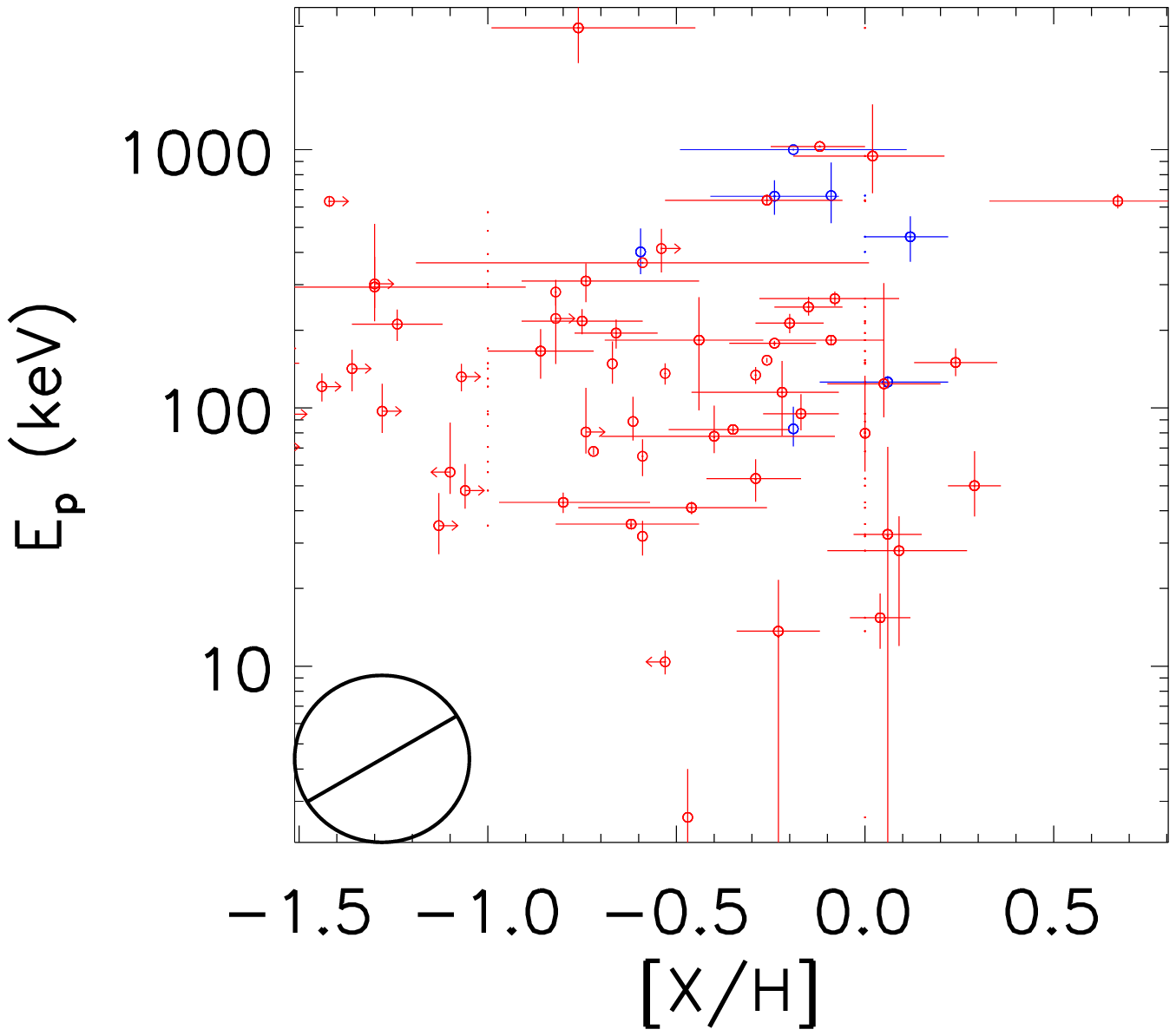}

\includegraphics[width=0.4\textwidth]{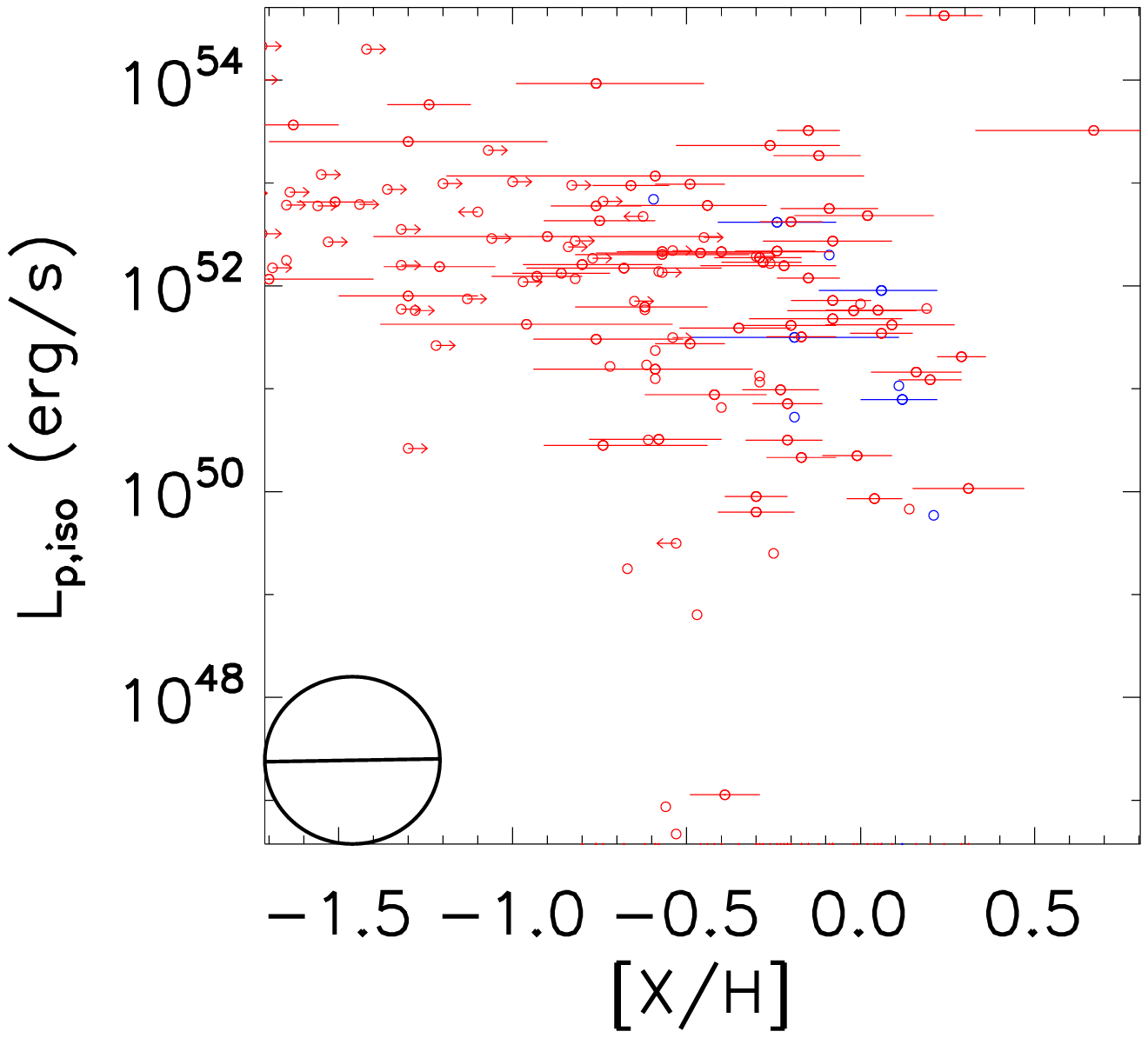}
\includegraphics[width=0.4\textwidth]{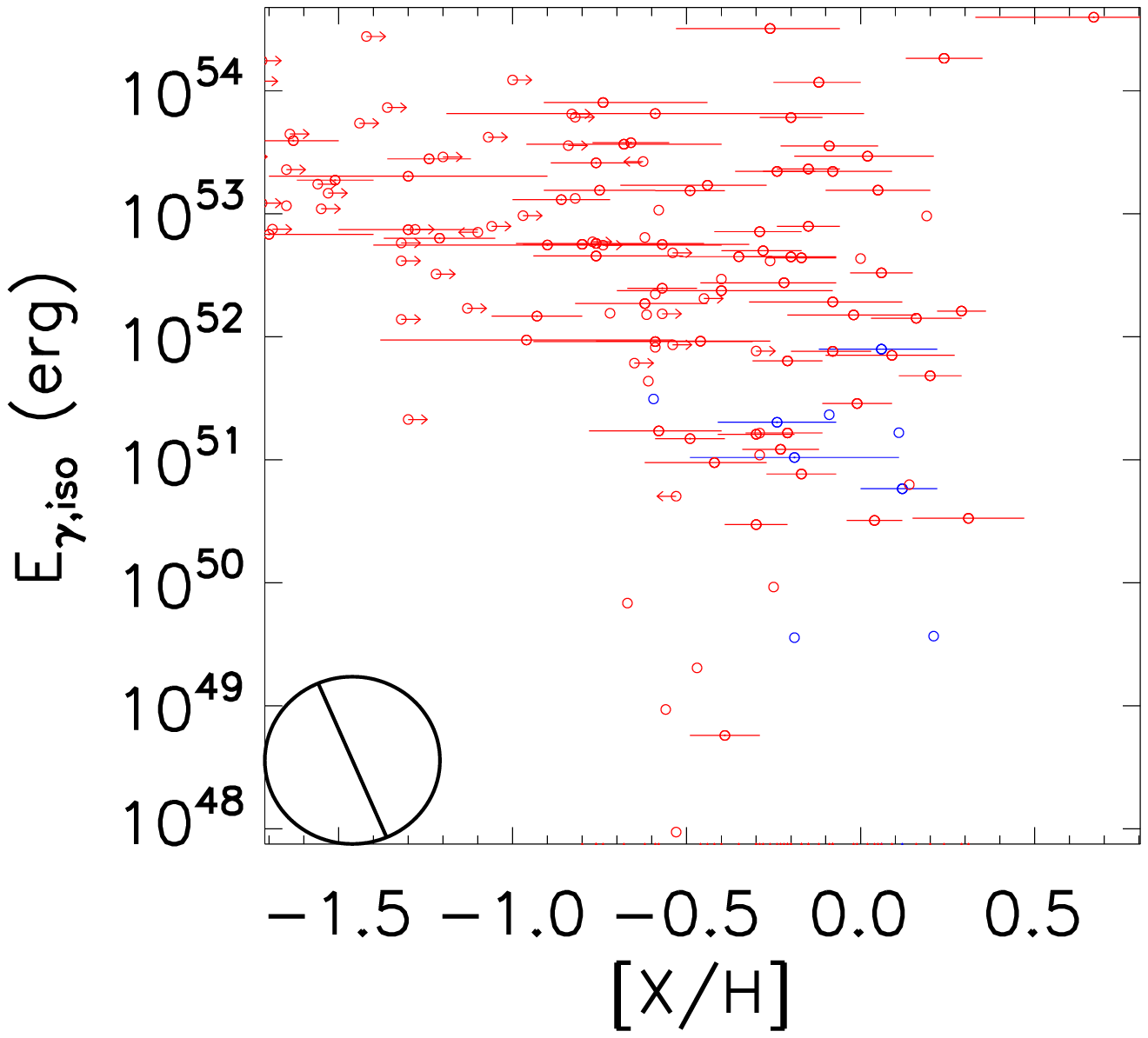}

\includegraphics[width=0.4\textwidth]{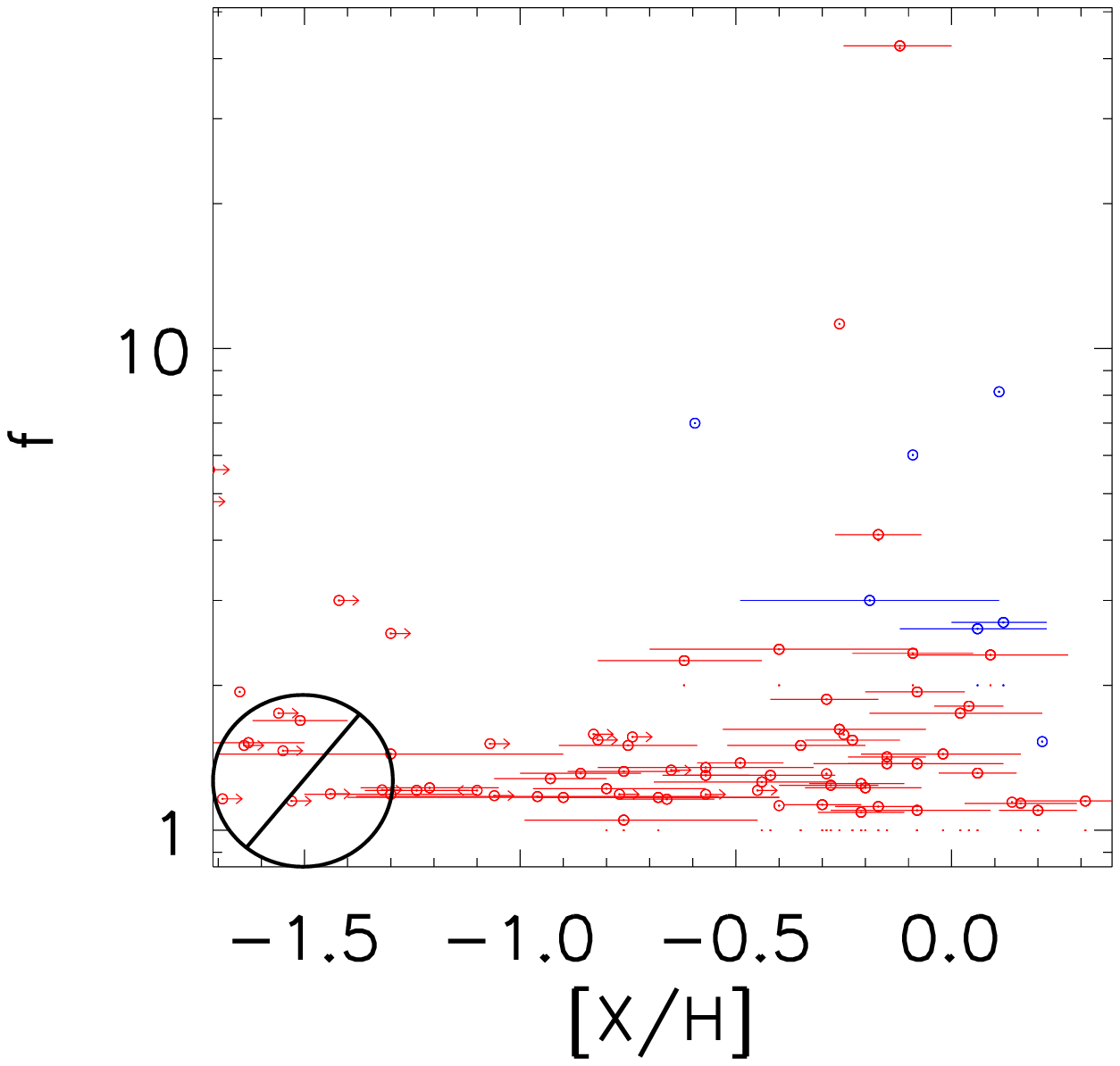}
\includegraphics[width=0.4\textwidth]{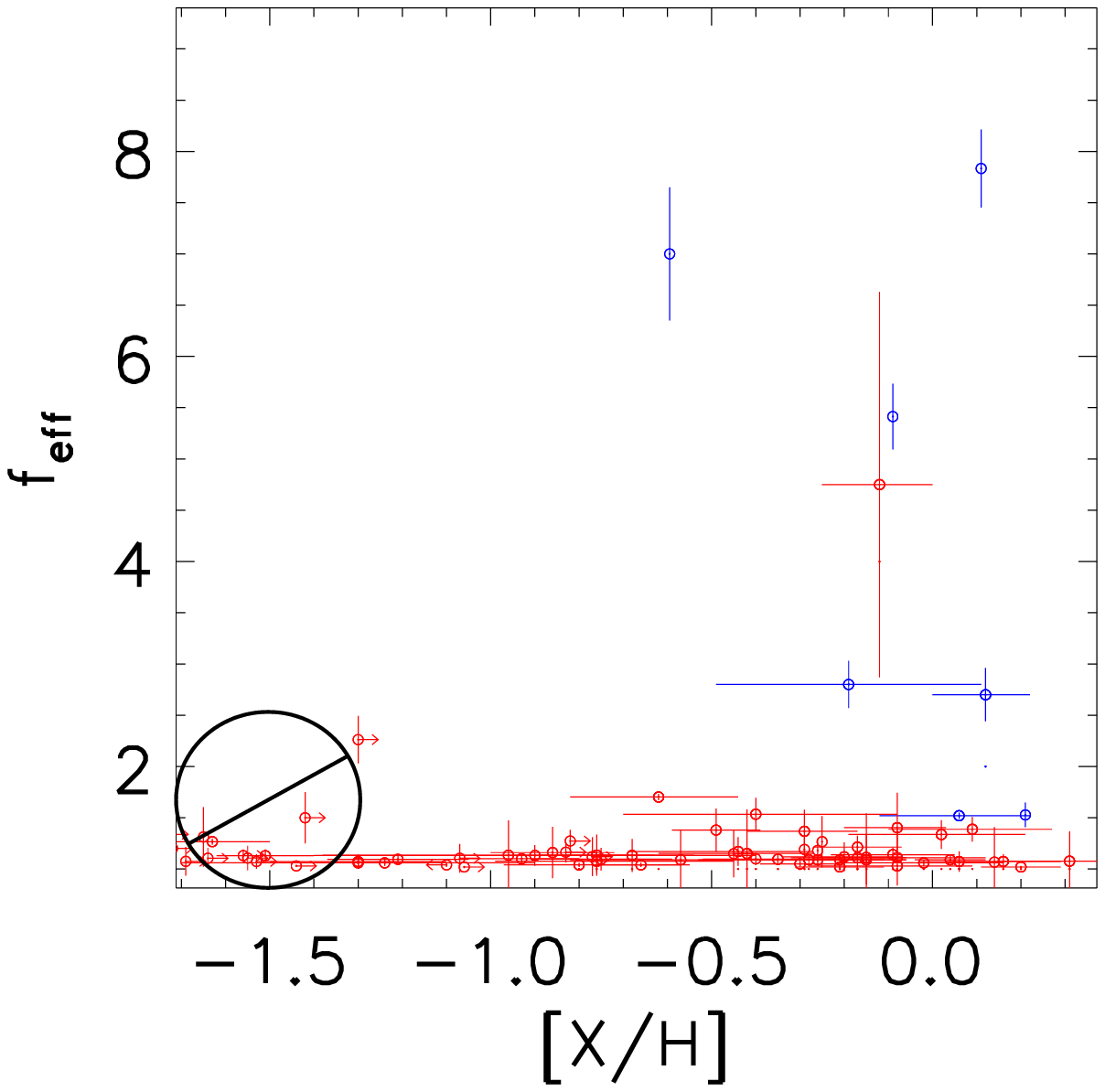}

\center{Fig. \ref{fig2d}---Continued}
\end{figure*}


\clearpage
\begin{figure*}

\includegraphics[width=0.4\textwidth]{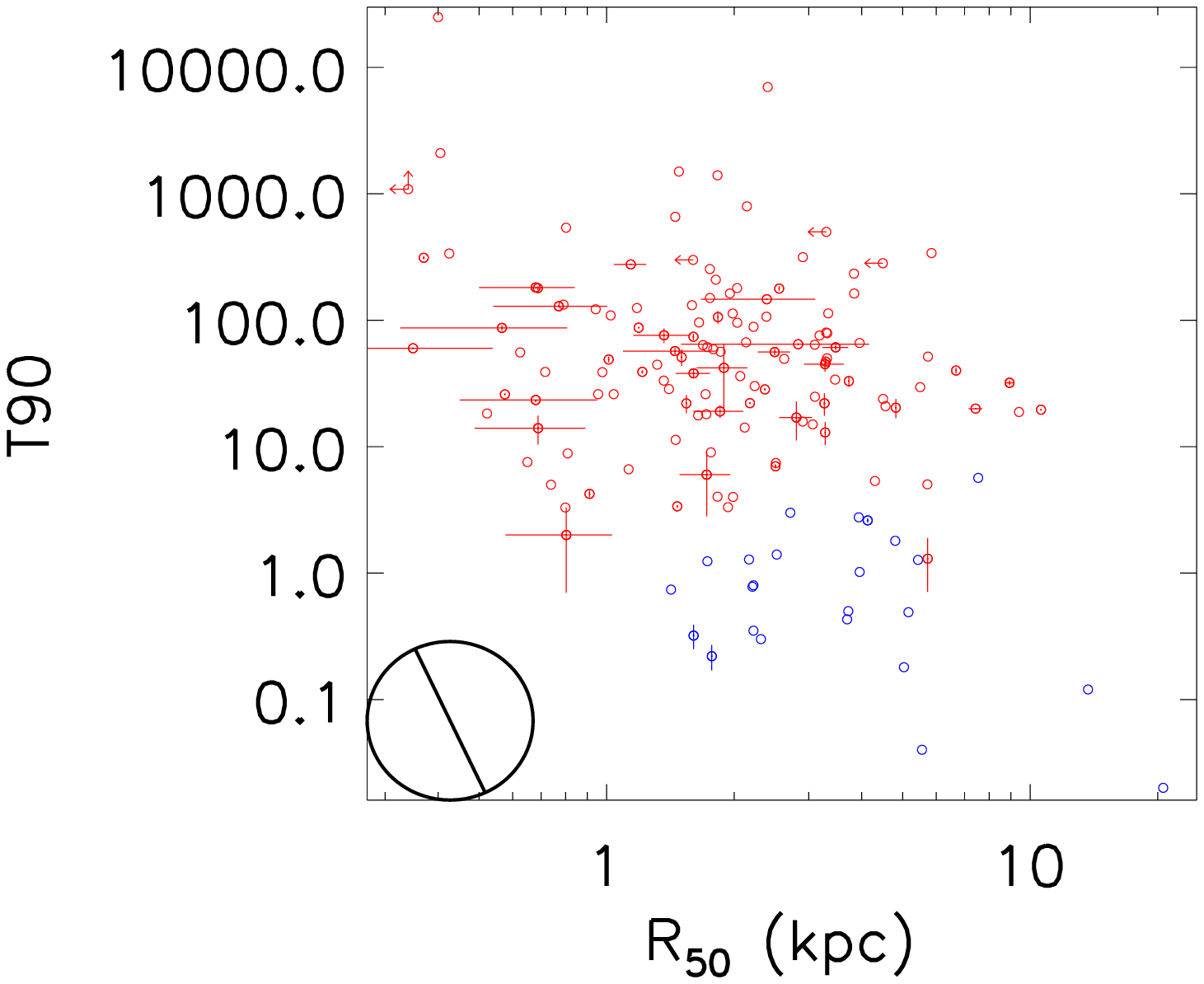}
\includegraphics[width=0.4\textwidth]{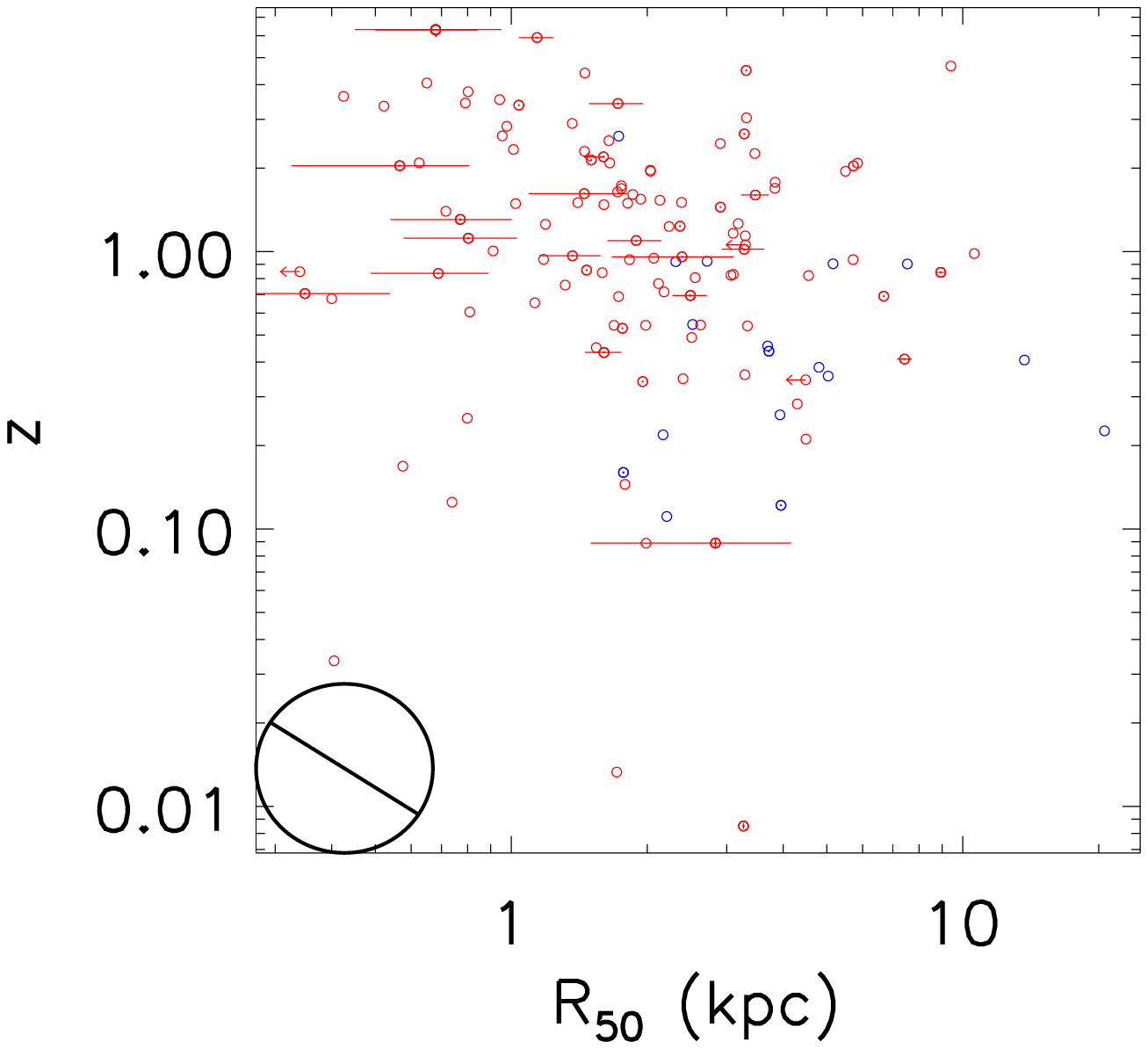}

\includegraphics[width=0.4\textwidth]{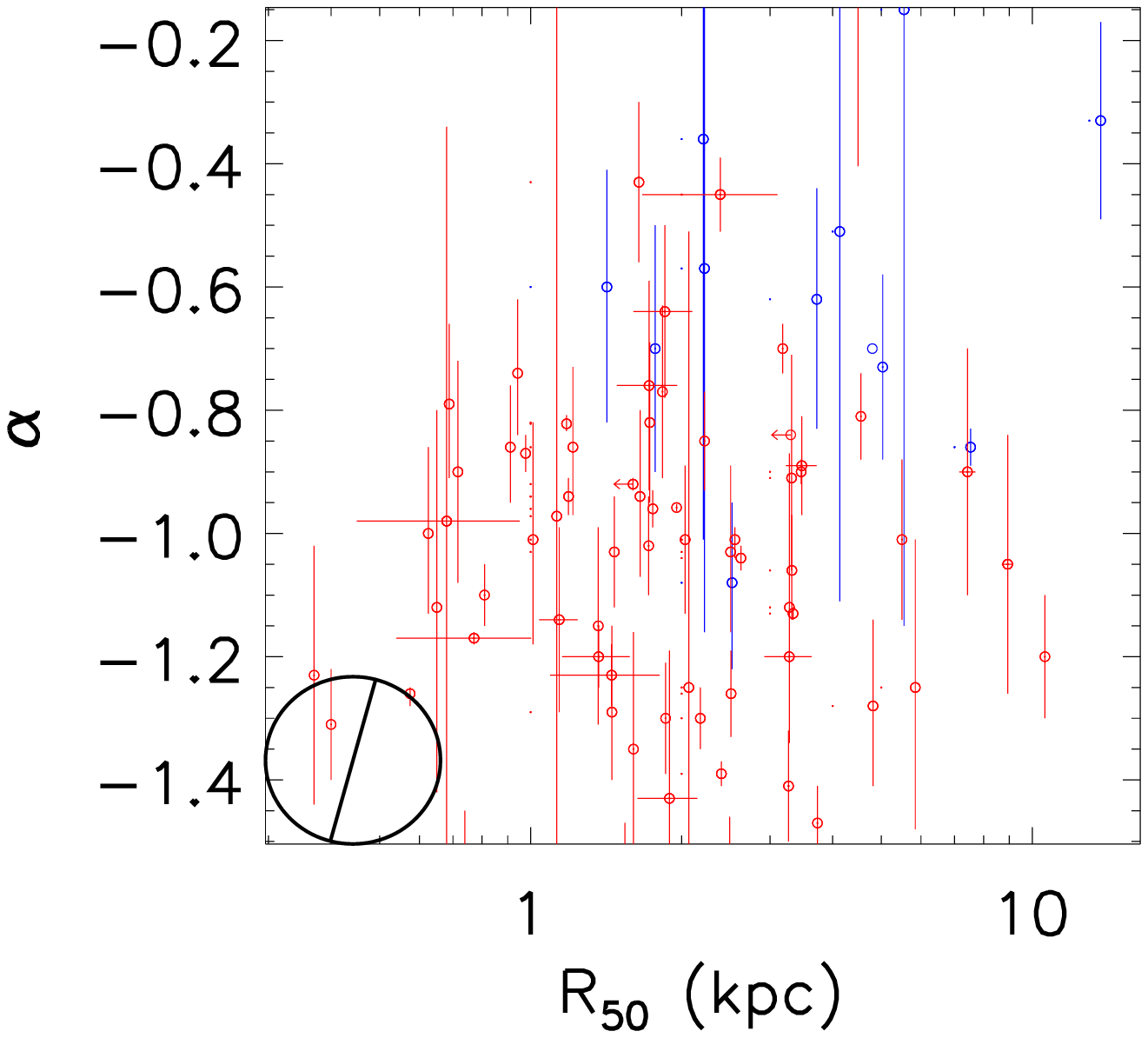}
\includegraphics[width=0.4\textwidth]{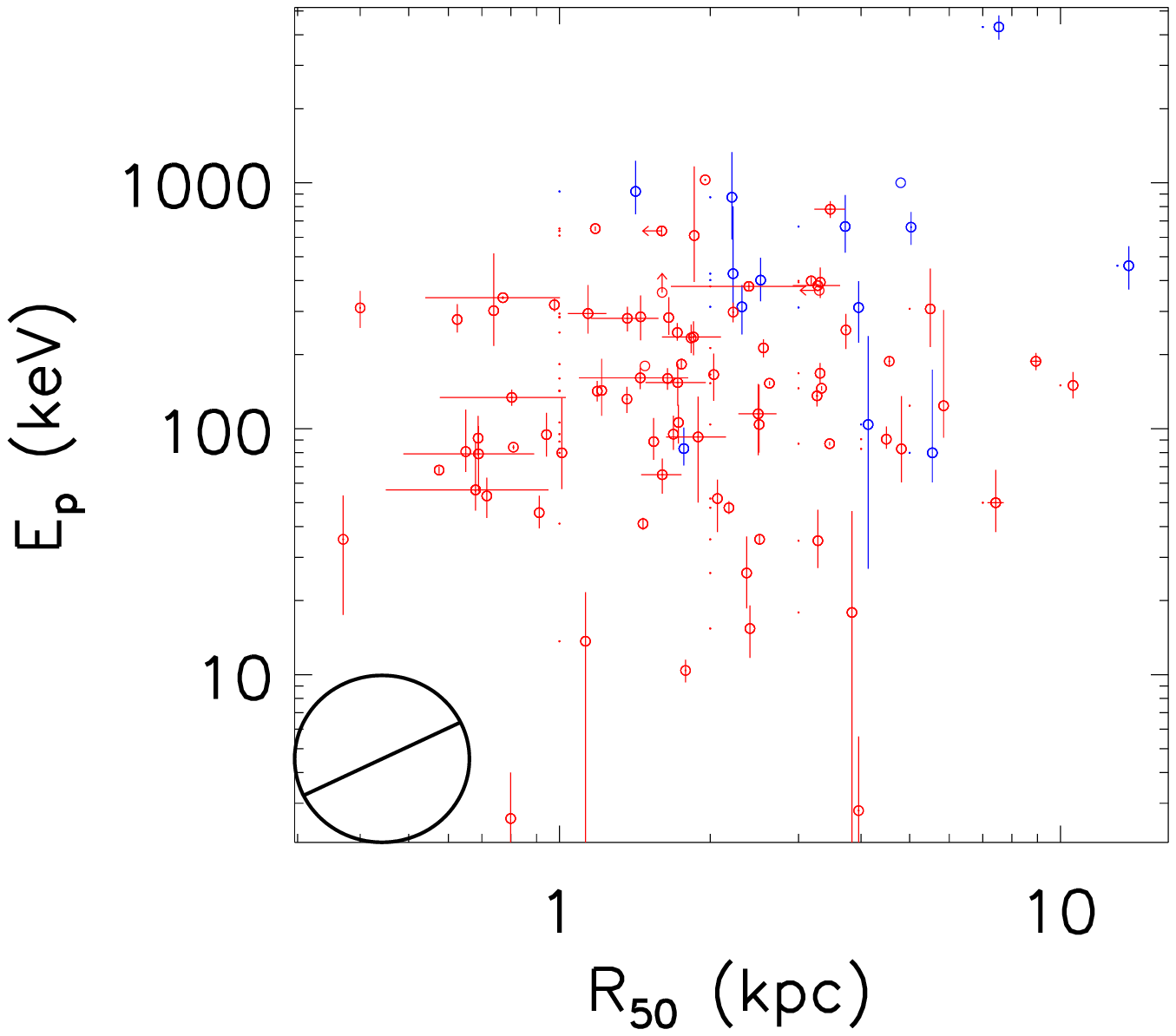}

\includegraphics[width=0.4\textwidth]{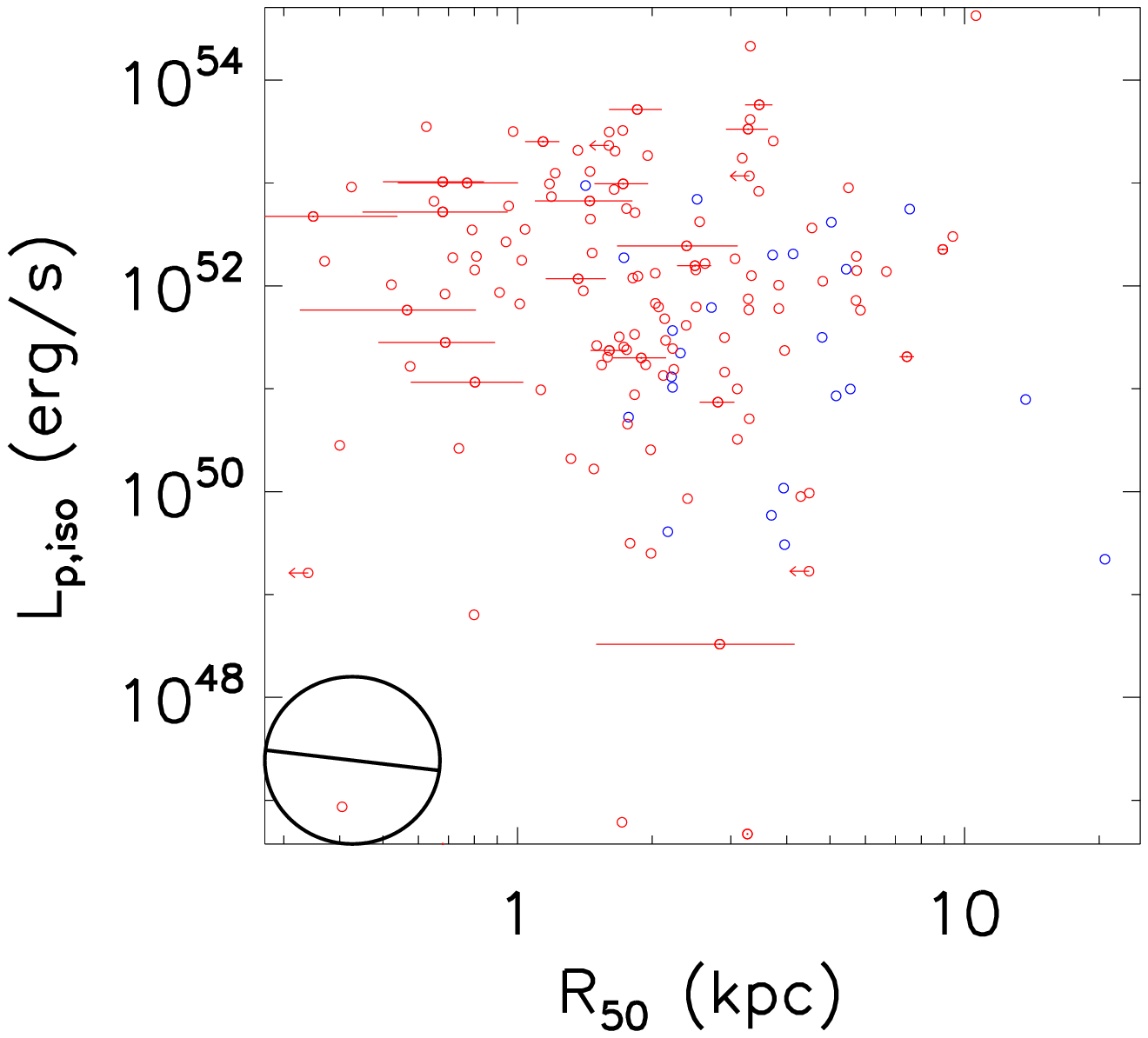}
\includegraphics[width=0.4\textwidth]{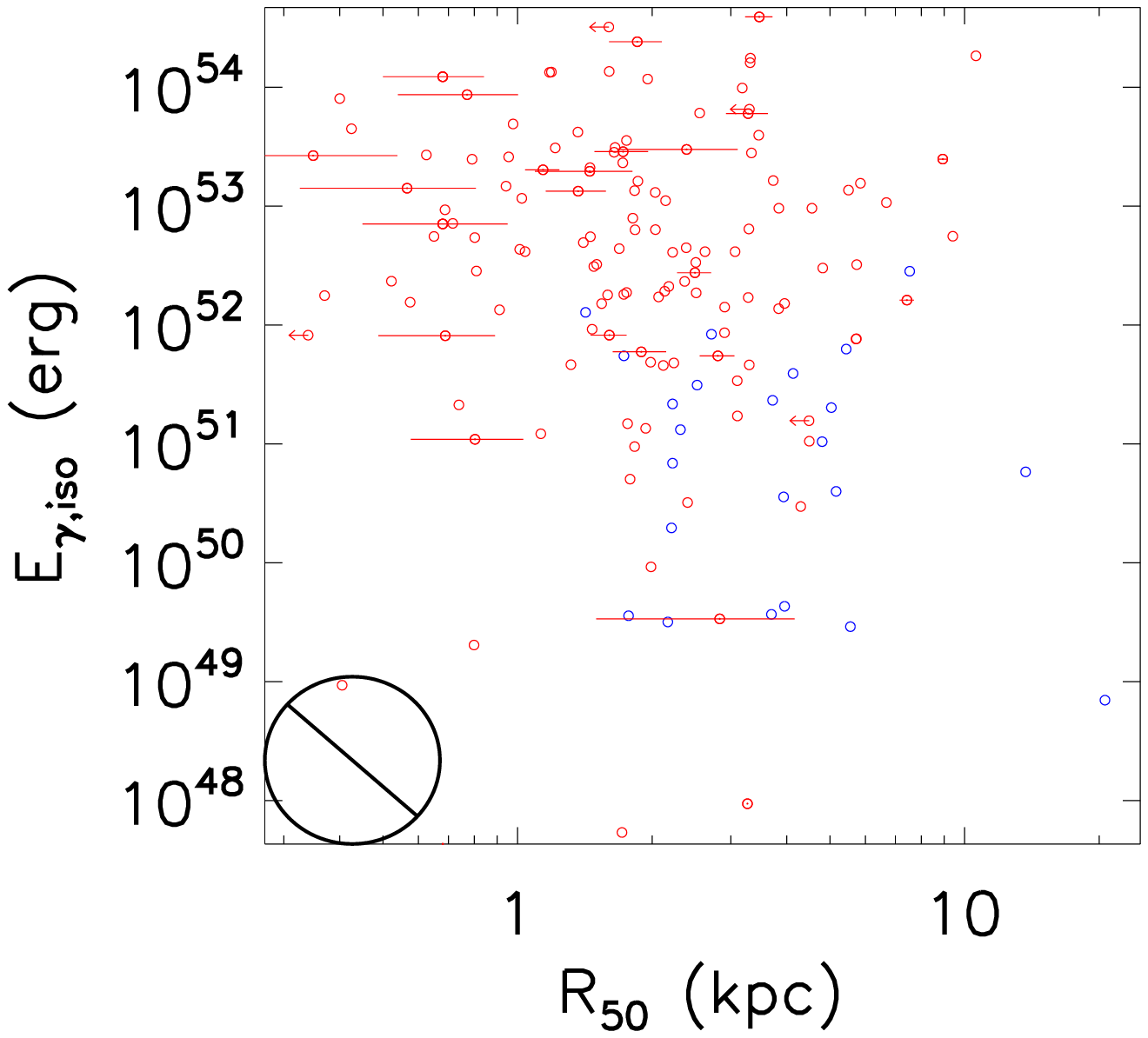}

\includegraphics[width=0.4\textwidth]{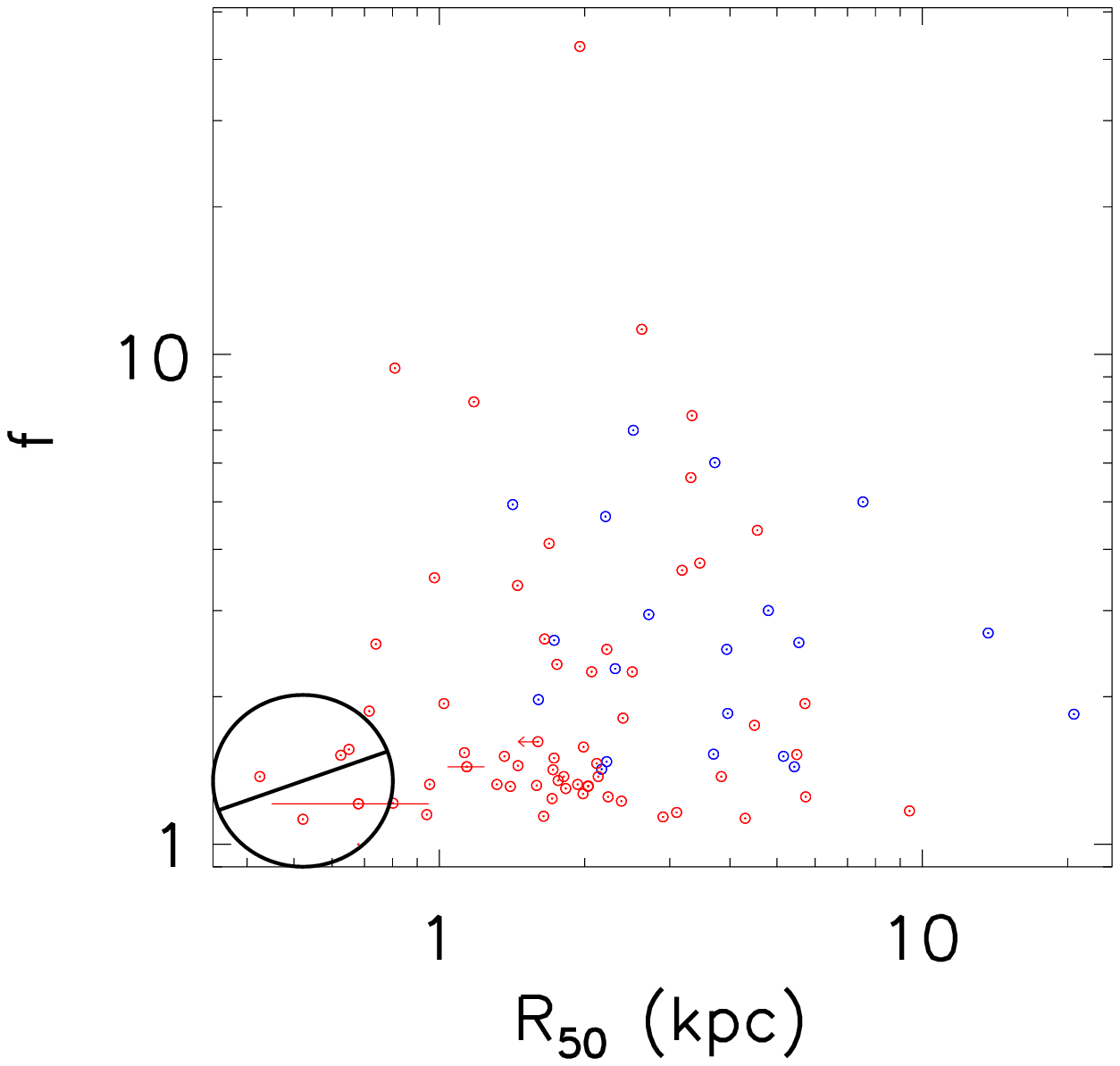}
\includegraphics[width=0.4\textwidth]{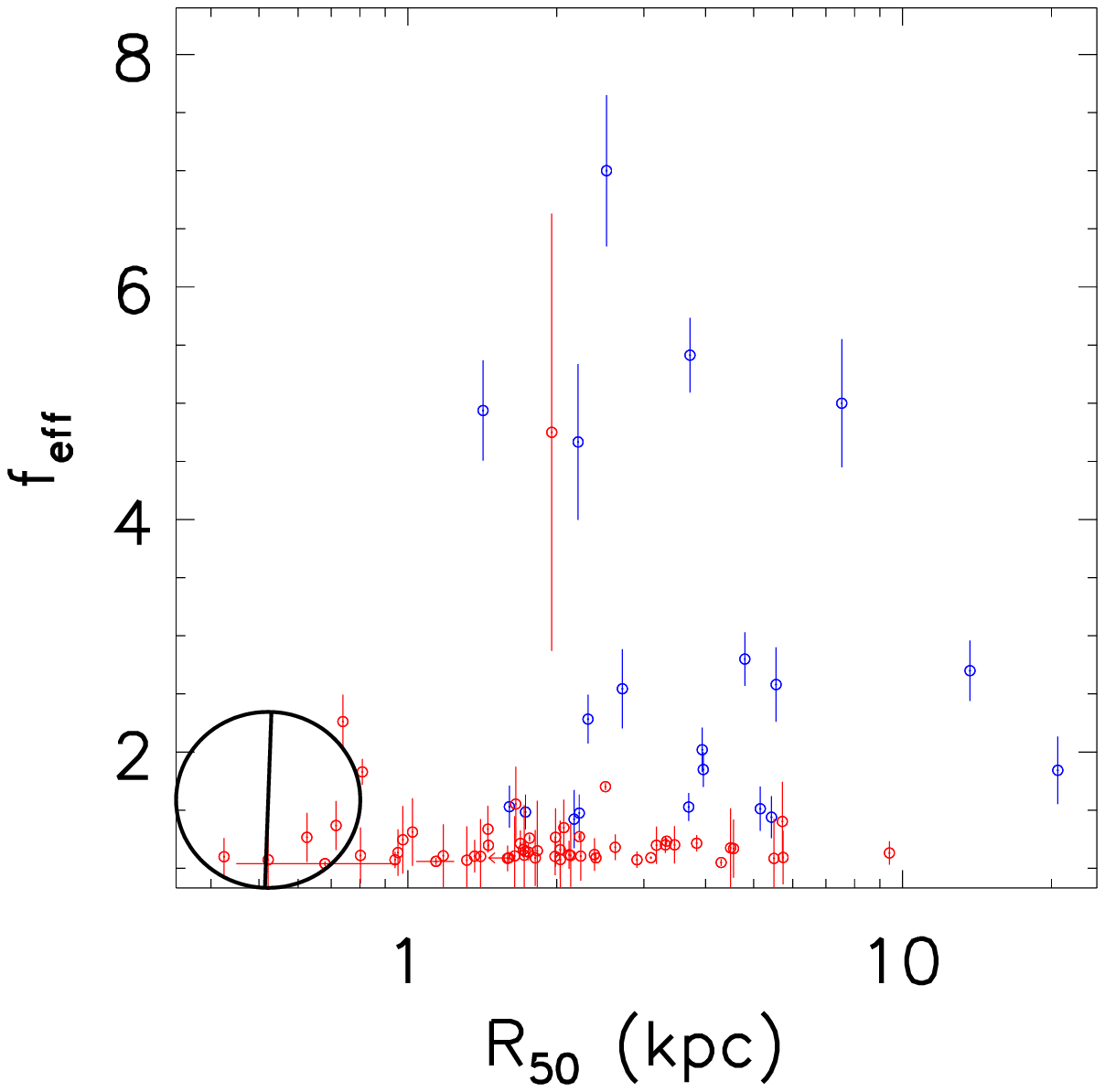}

\center{Fig. \ref{fig2d}---Continued}
\end{figure*}


\clearpage
\begin{figure*}

\includegraphics[width=0.4\textwidth]{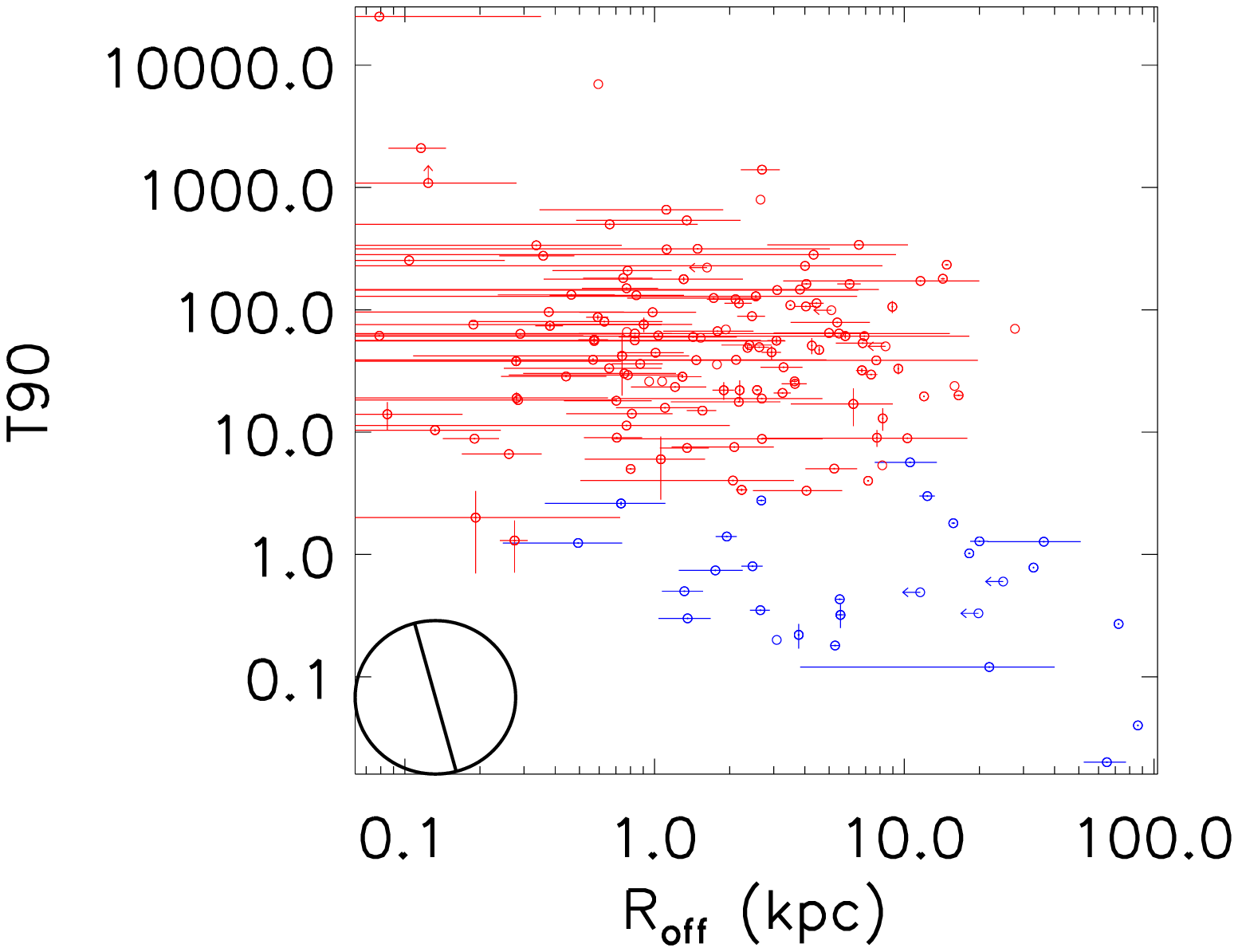}
\includegraphics[width=0.4\textwidth]{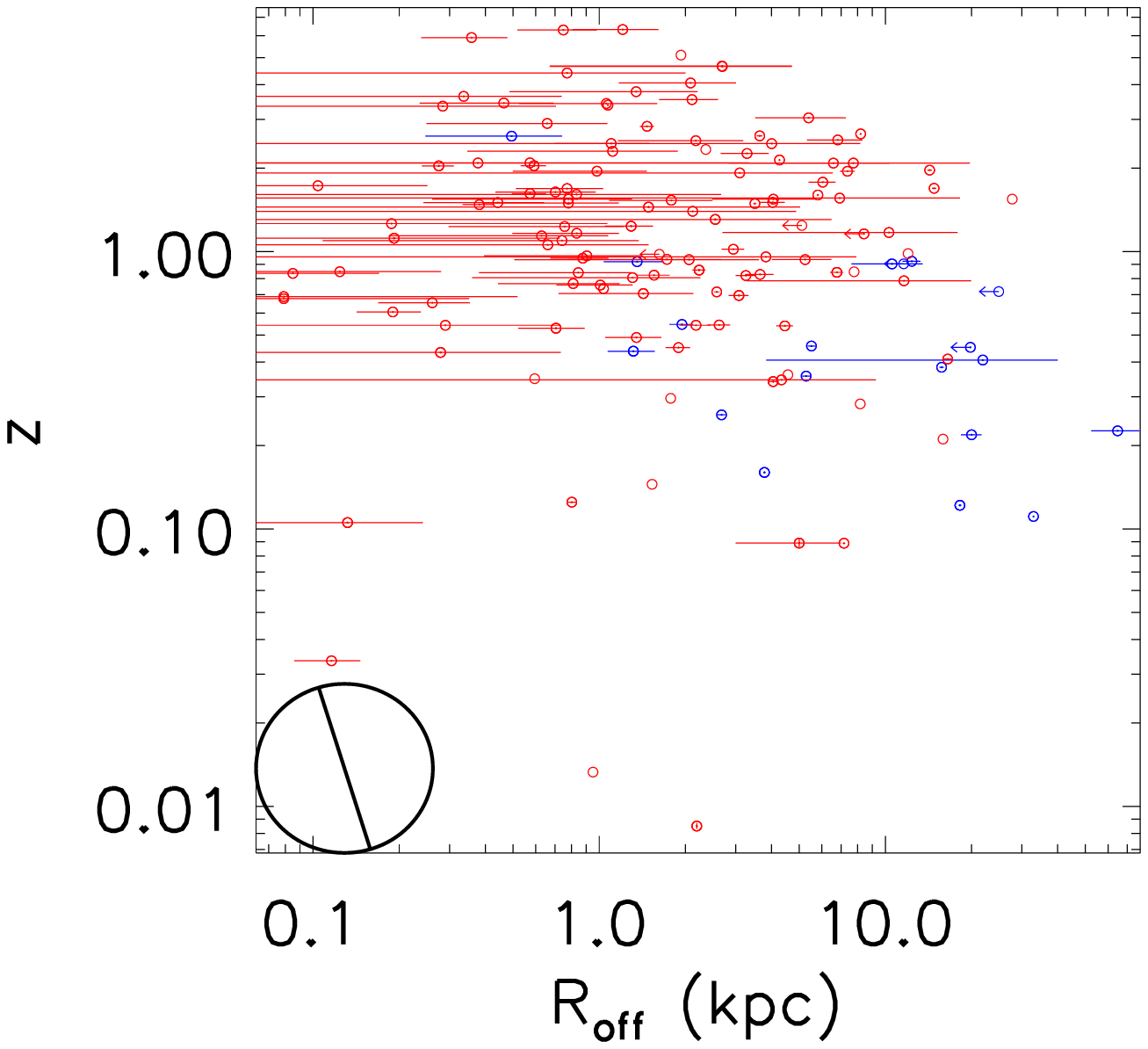}

\includegraphics[width=0.4\textwidth]{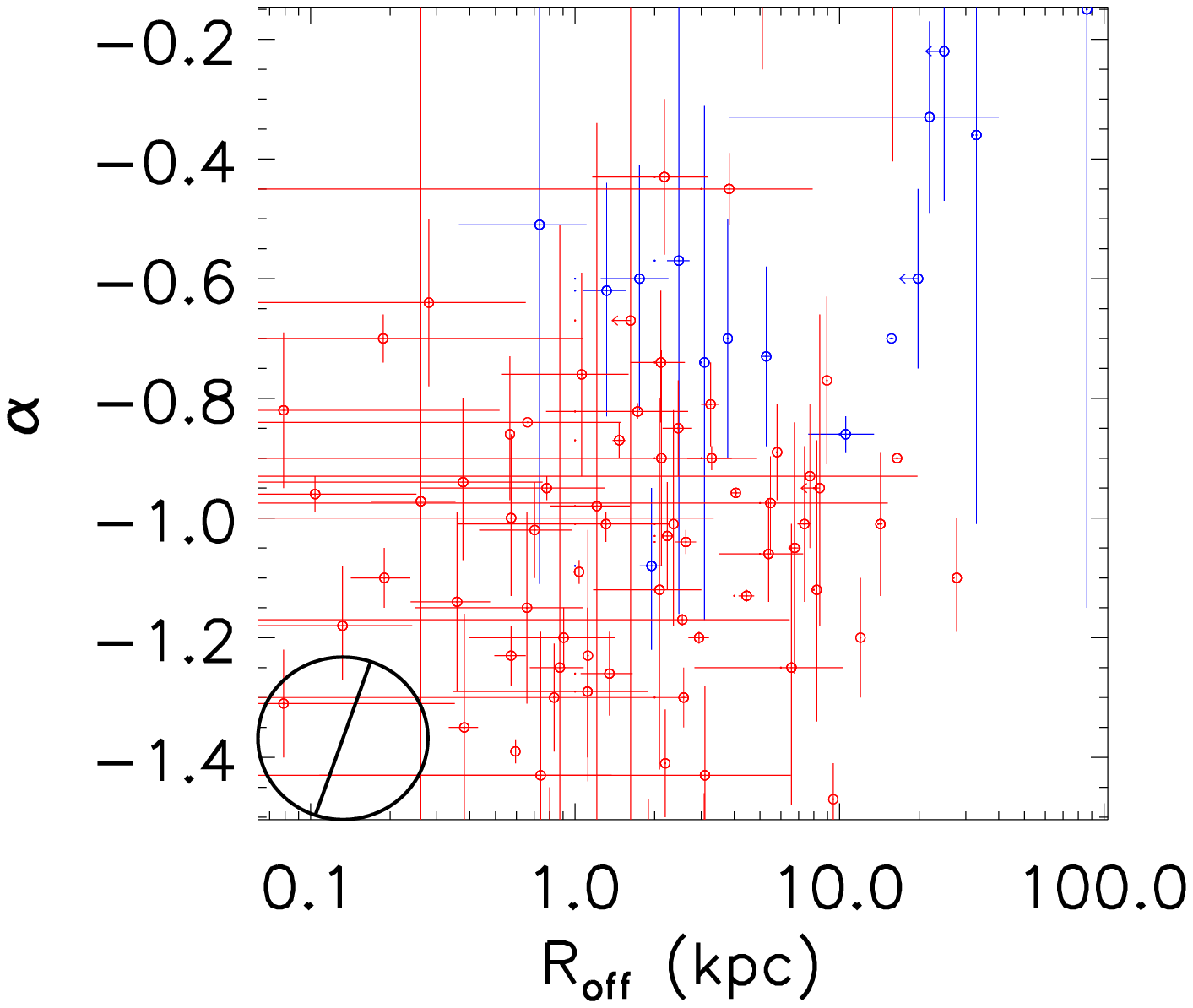}
\includegraphics[width=0.4\textwidth]{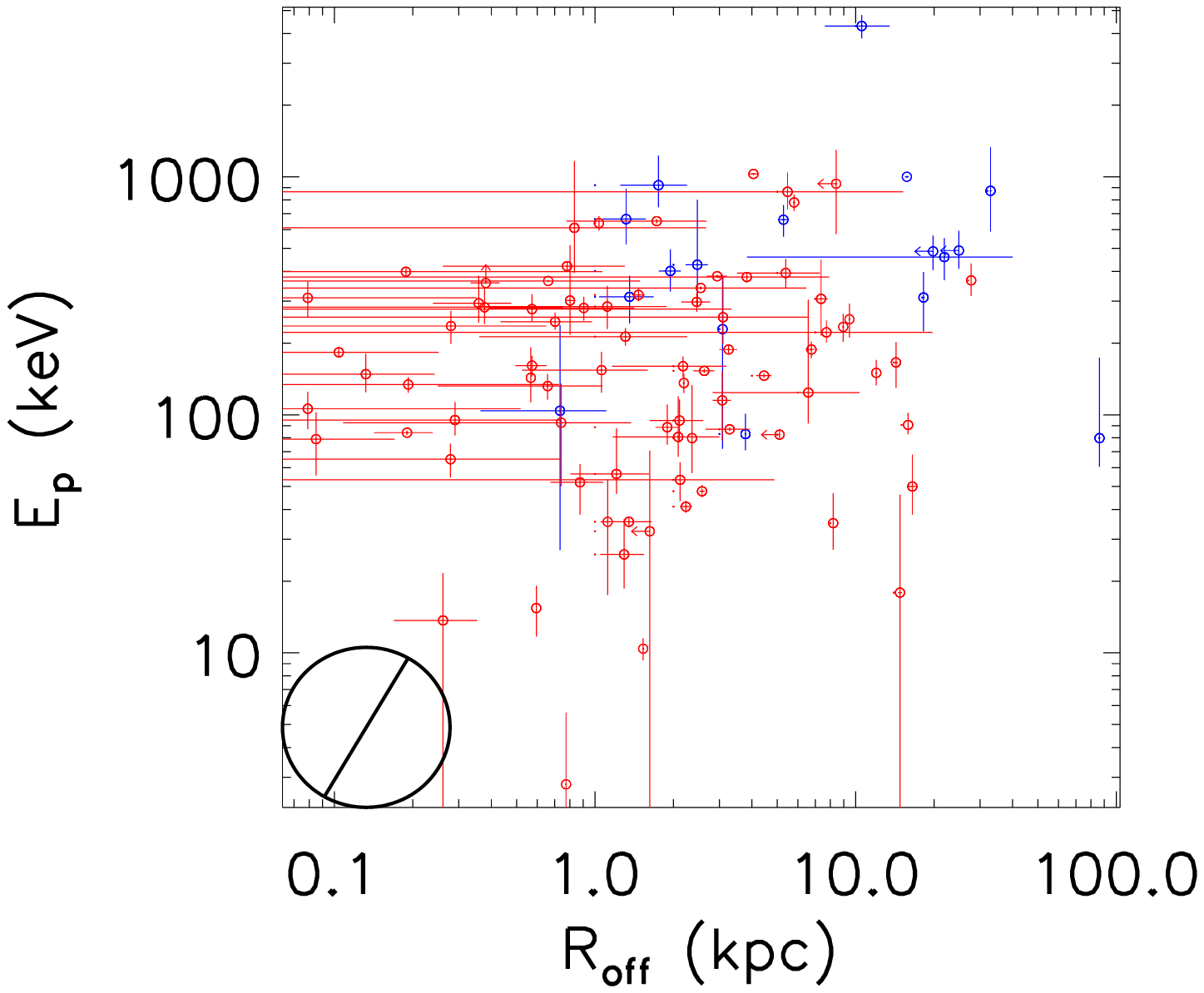}

\includegraphics[width=0.4\textwidth]{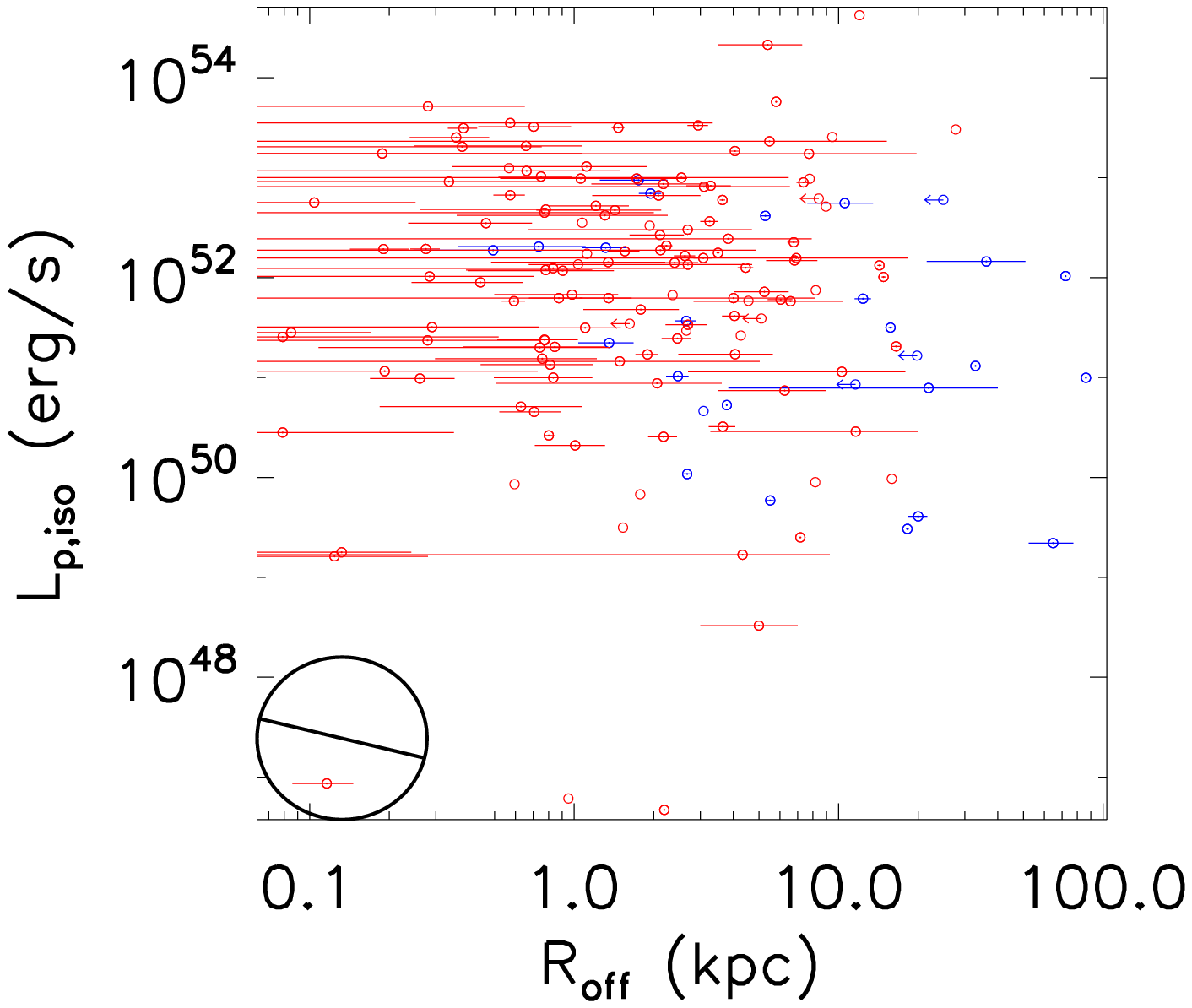}
\includegraphics[width=0.4\textwidth]{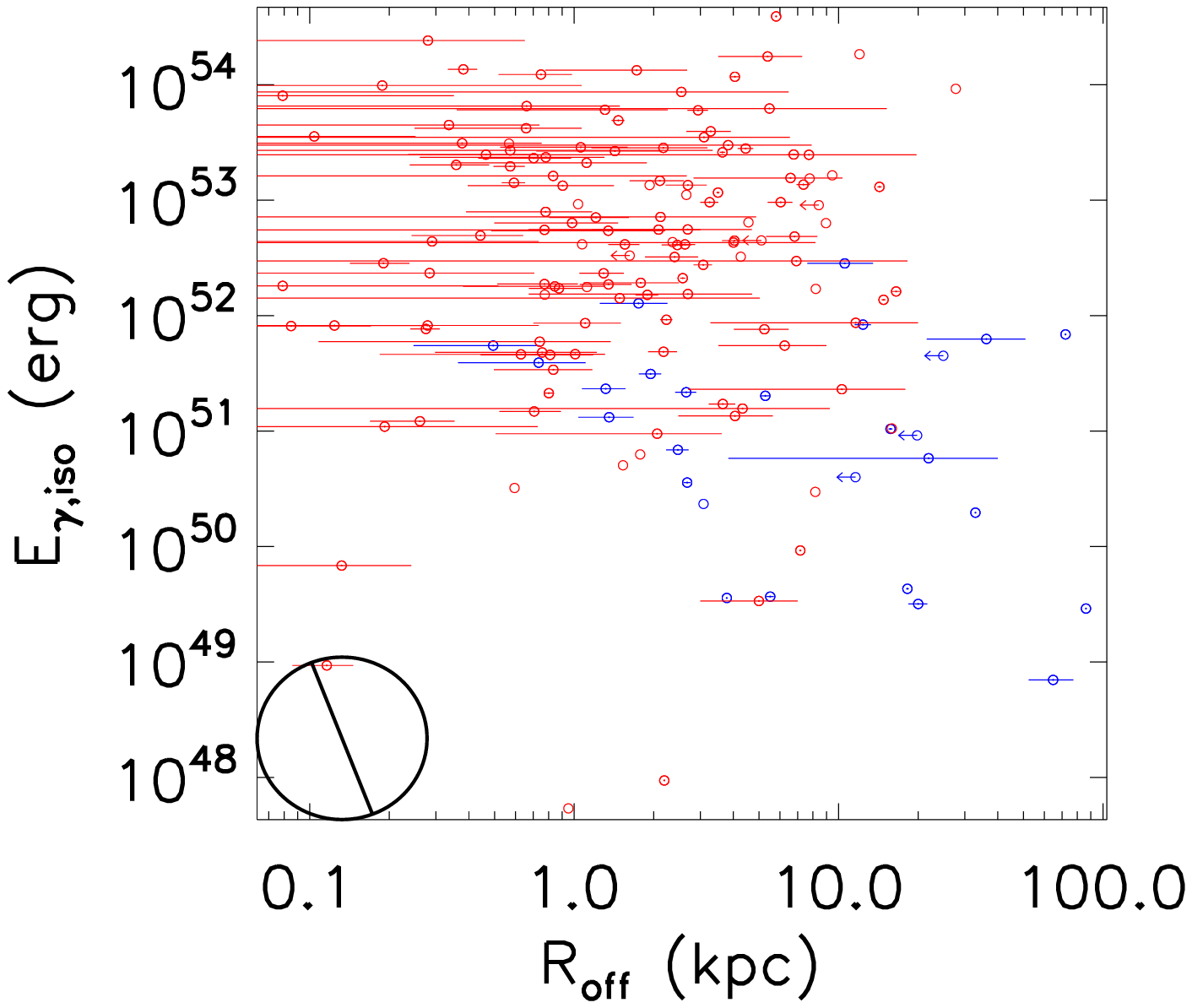}

\includegraphics[width=0.4\textwidth]{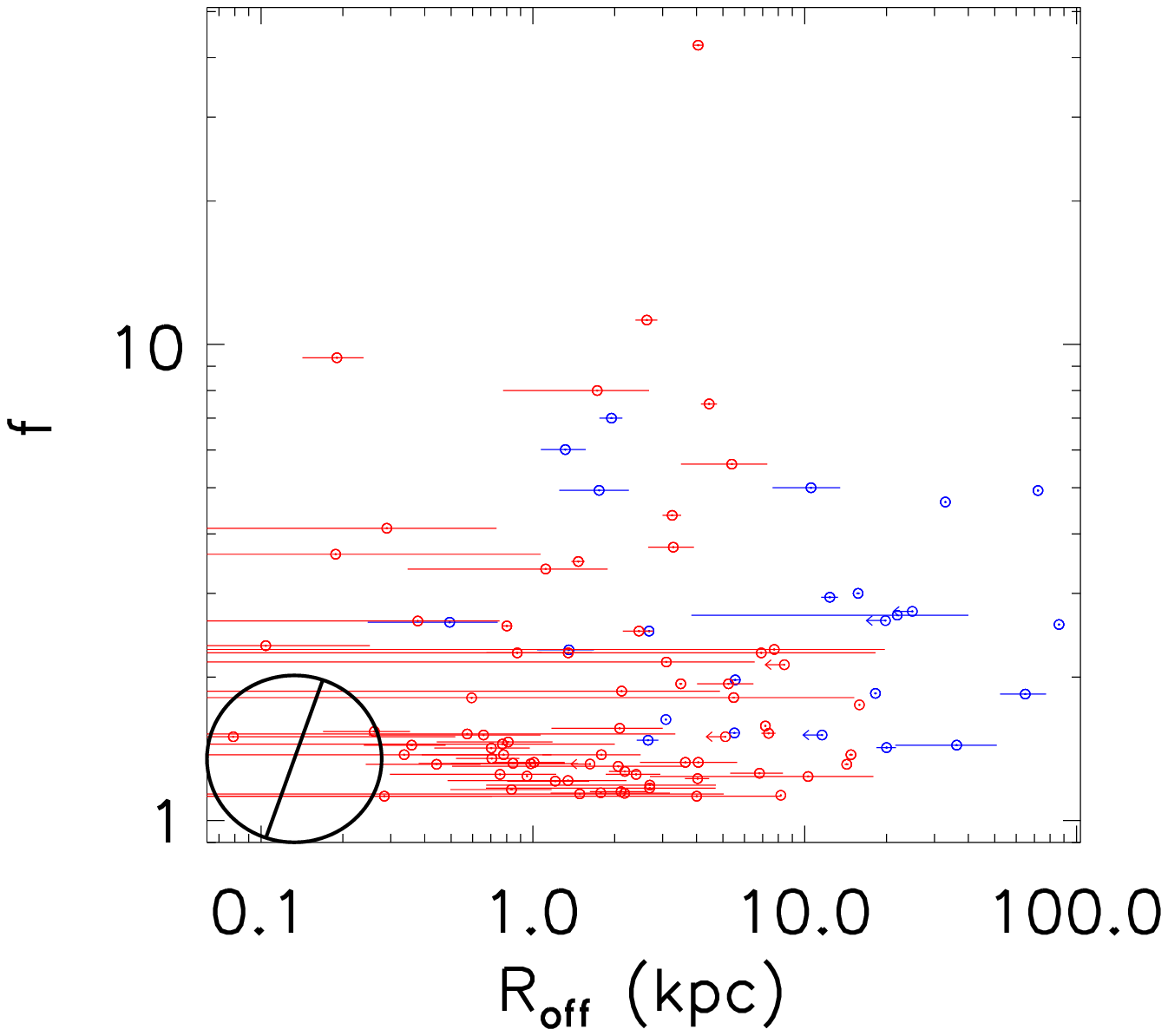}
\includegraphics[width=0.4\textwidth]{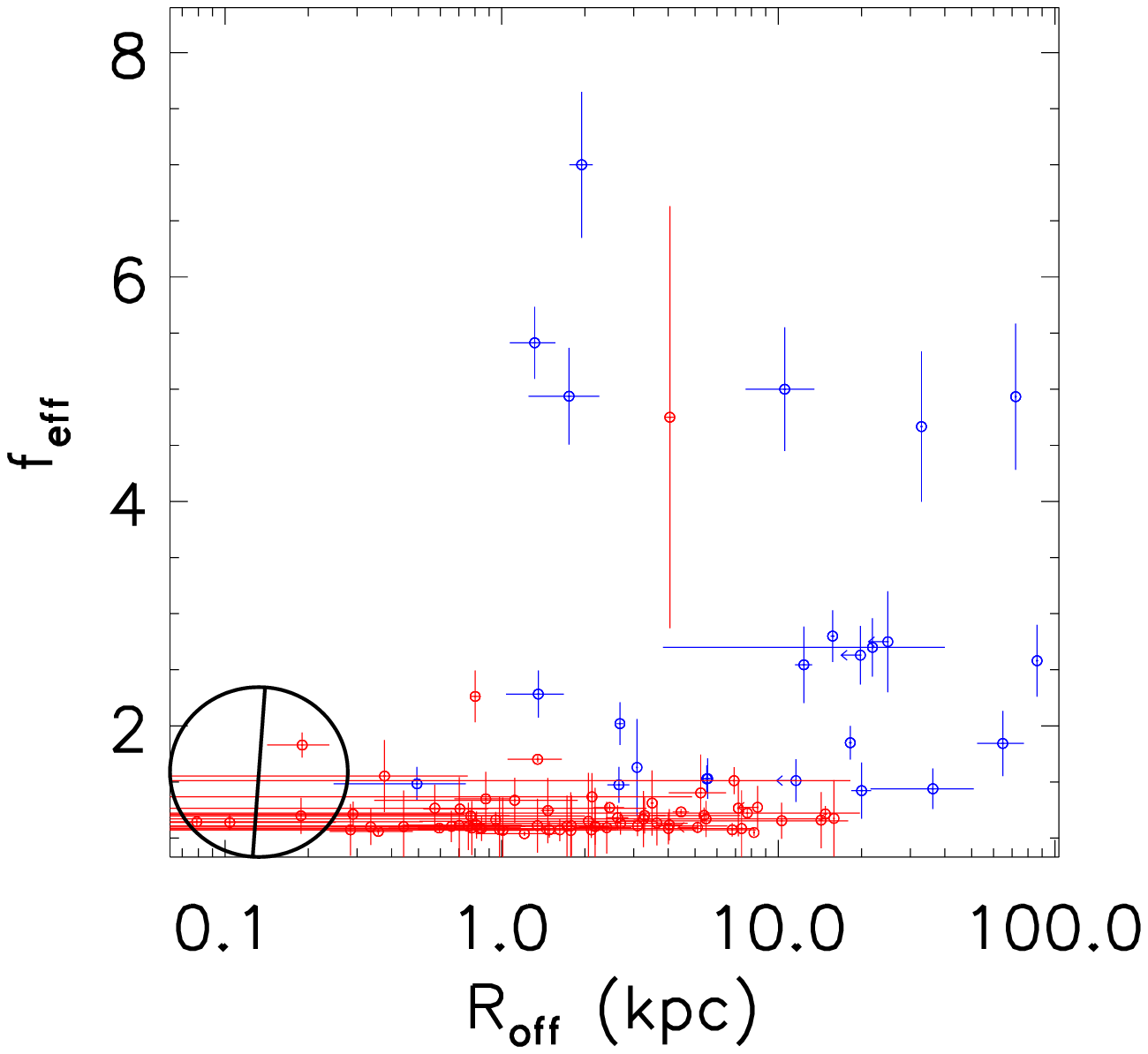}

\center{Fig. \ref{fig2d}---Continued}
\end{figure*}


\clearpage
\begin{figure*}

\includegraphics[width=0.4\textwidth]{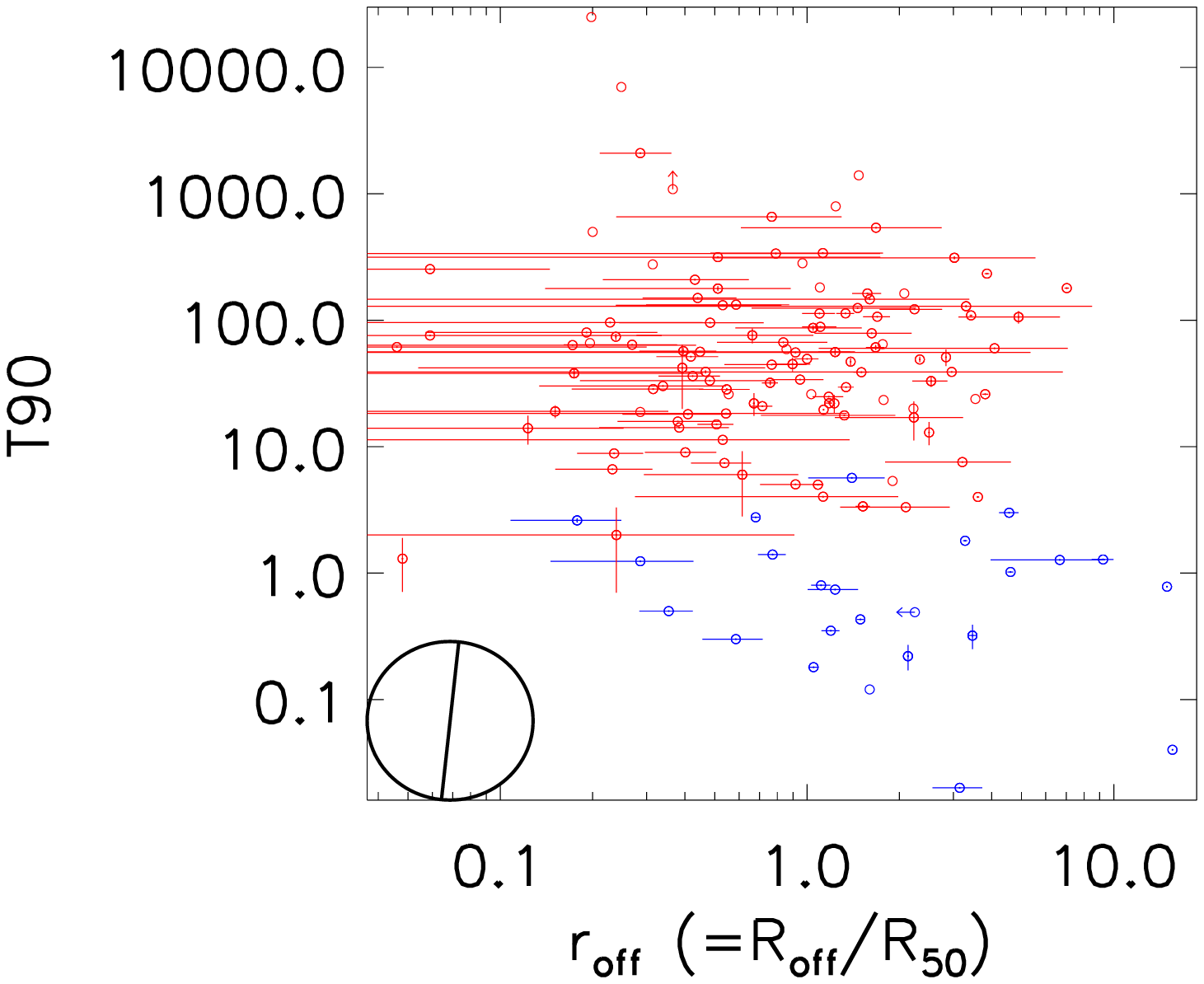}
\includegraphics[width=0.4\textwidth]{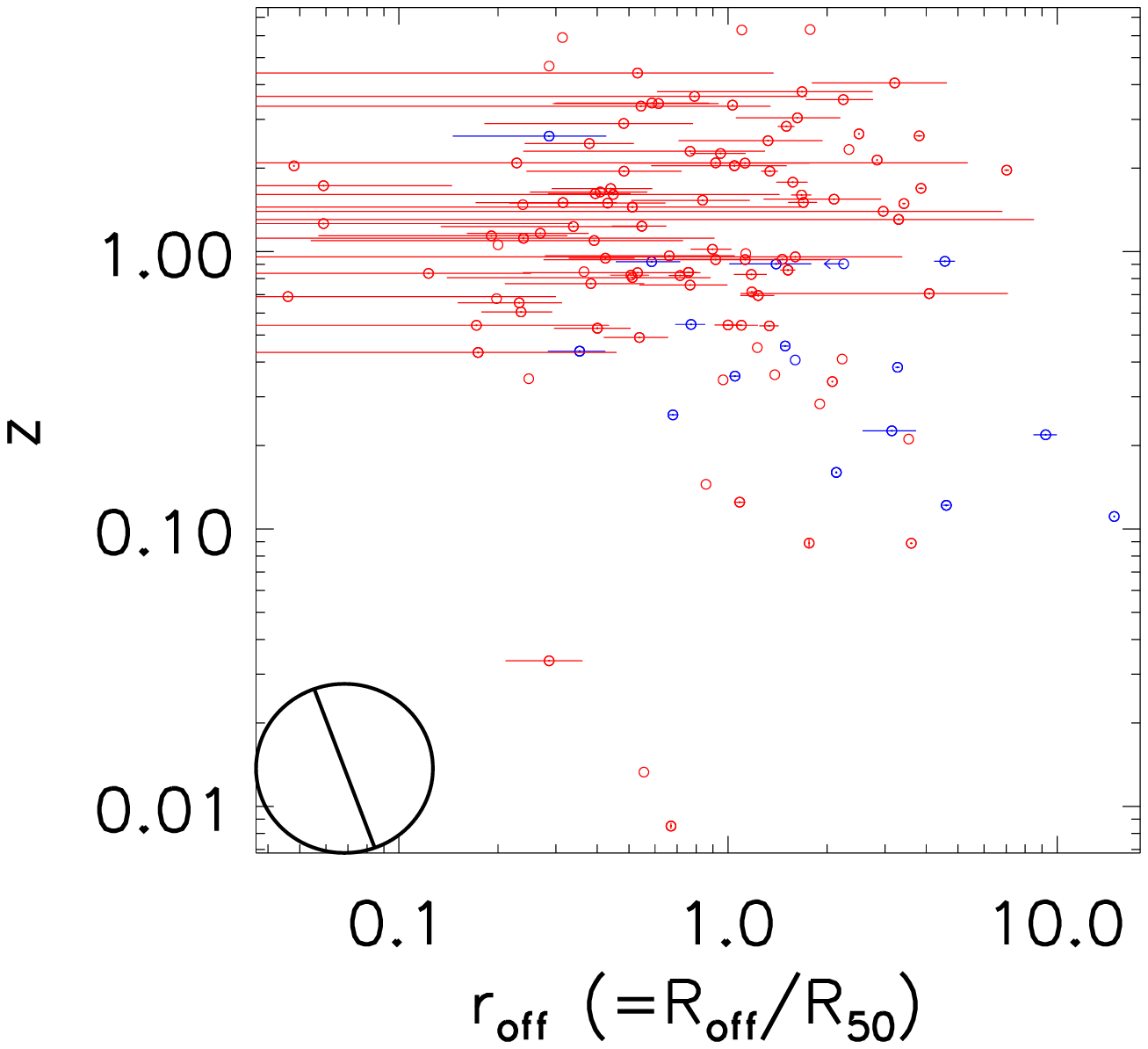}

\includegraphics[width=0.4\textwidth]{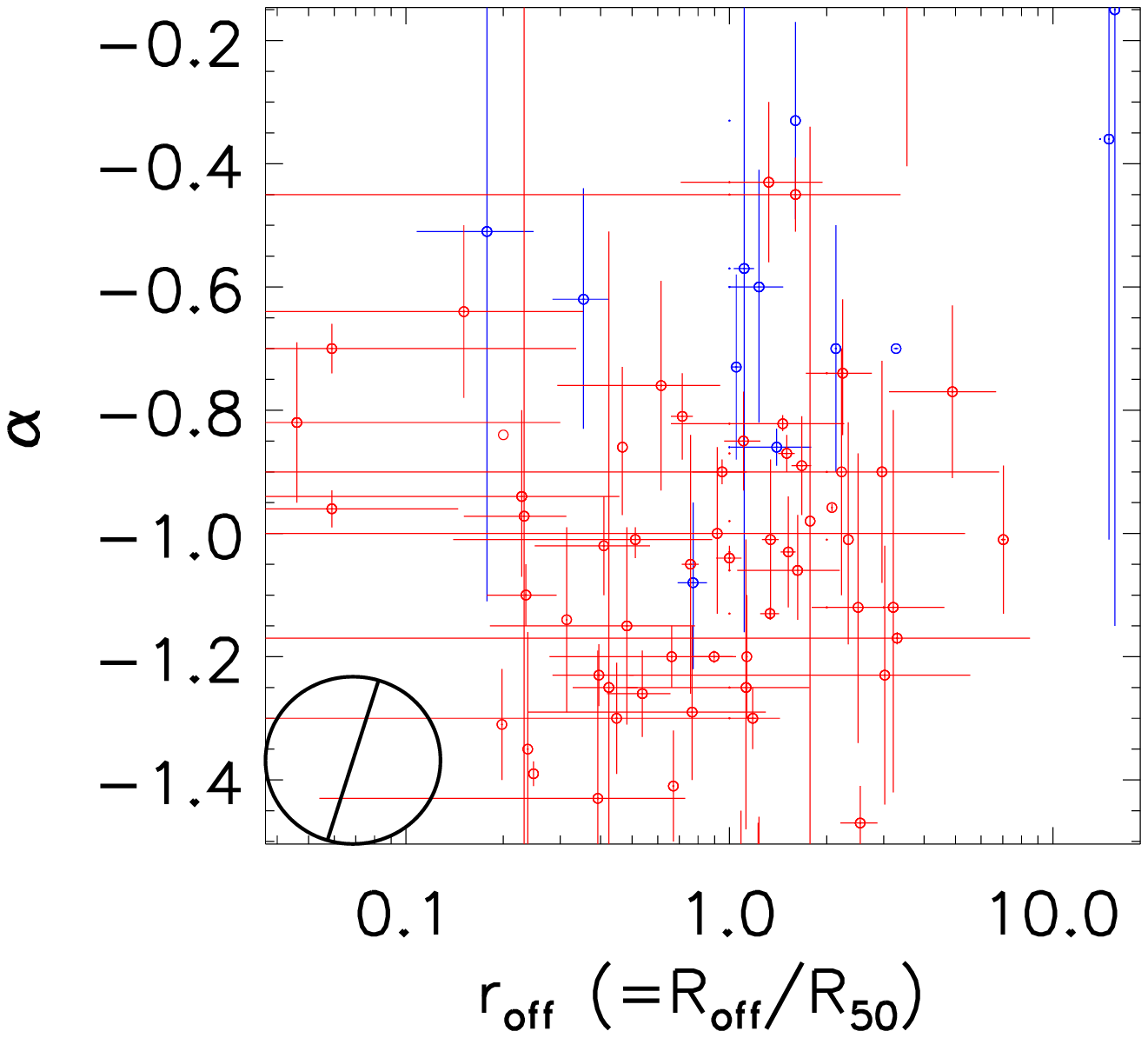}
\includegraphics[width=0.4\textwidth]{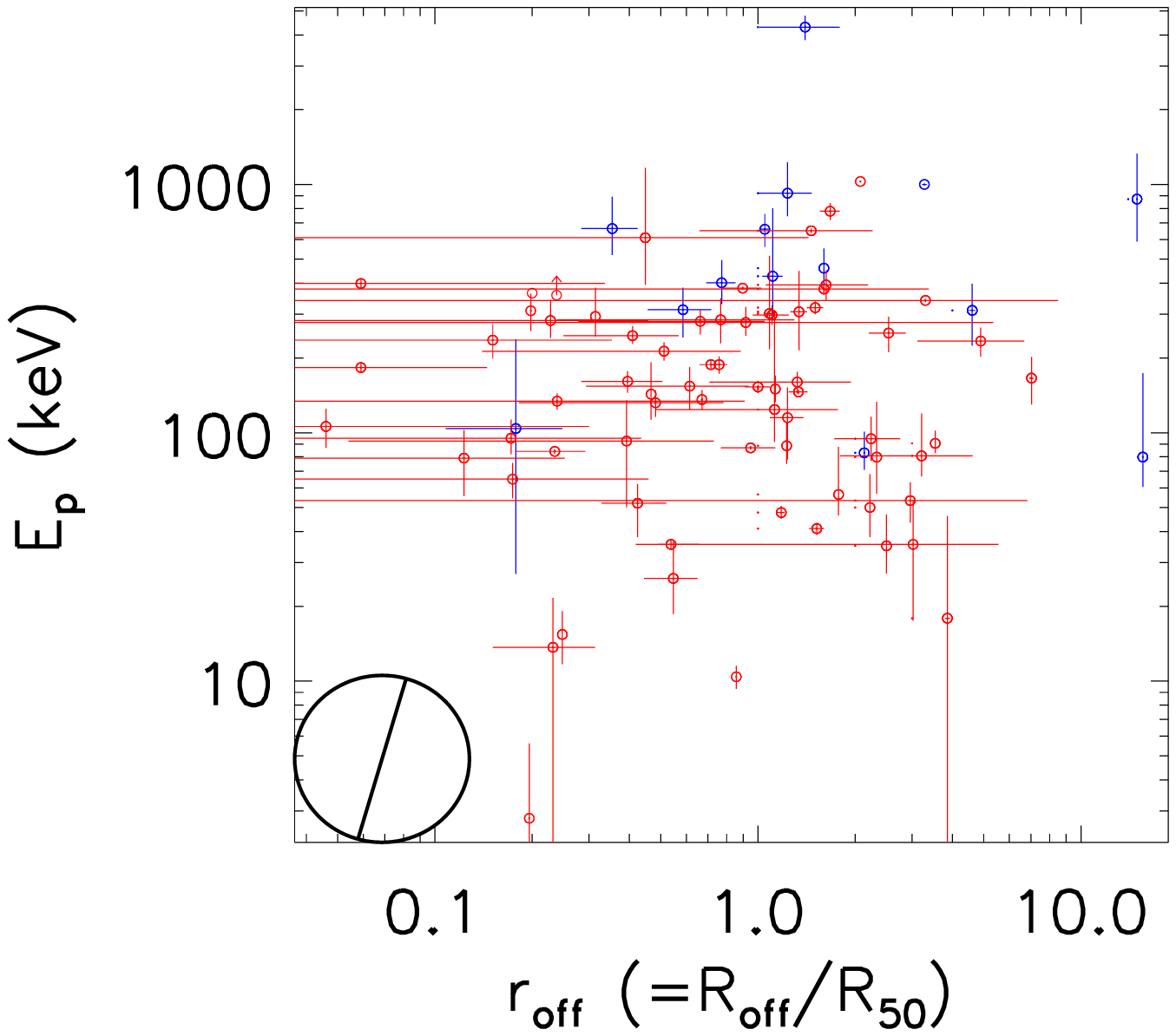}

\includegraphics[width=0.4\textwidth]{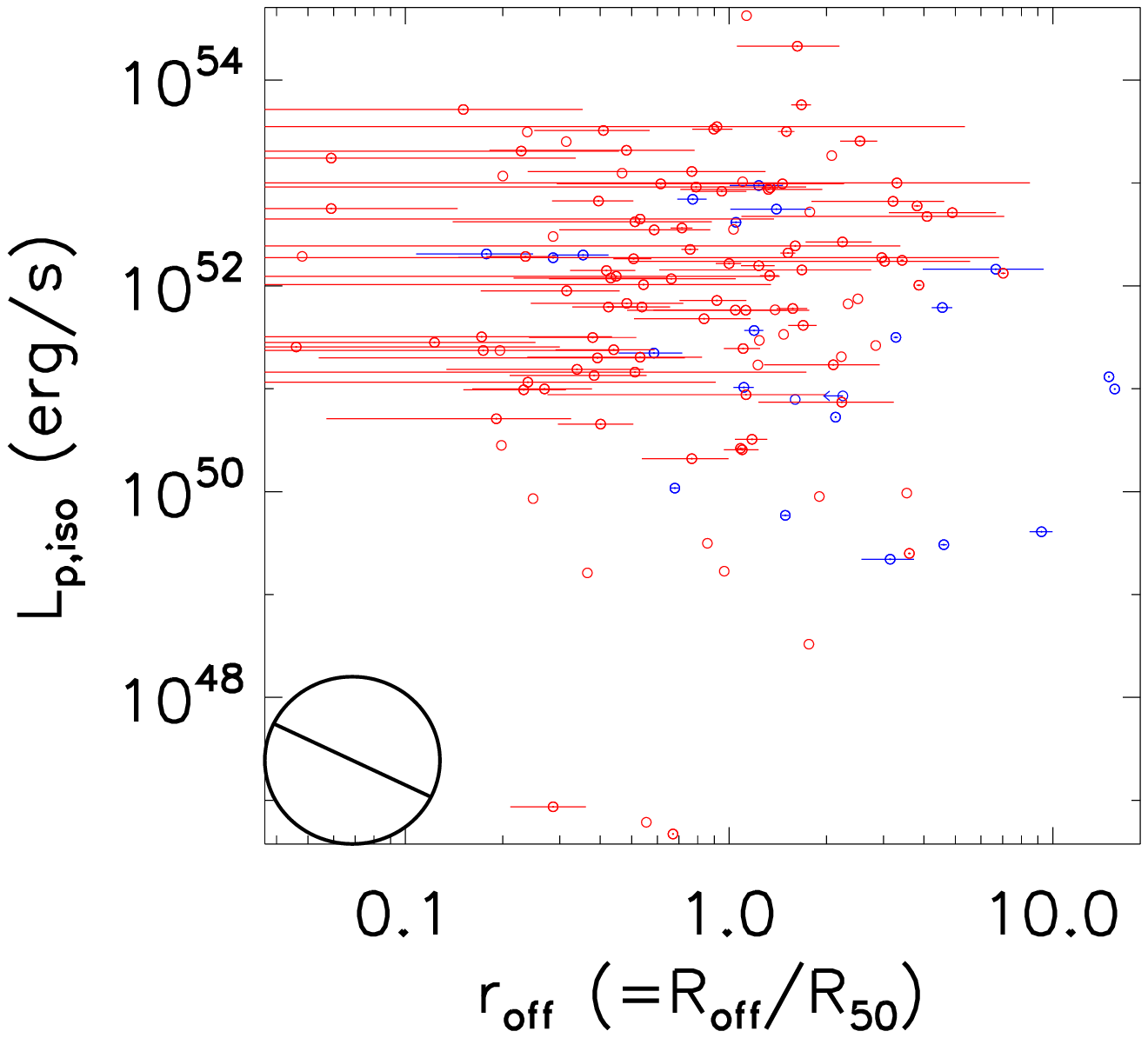}
\includegraphics[width=0.4\textwidth]{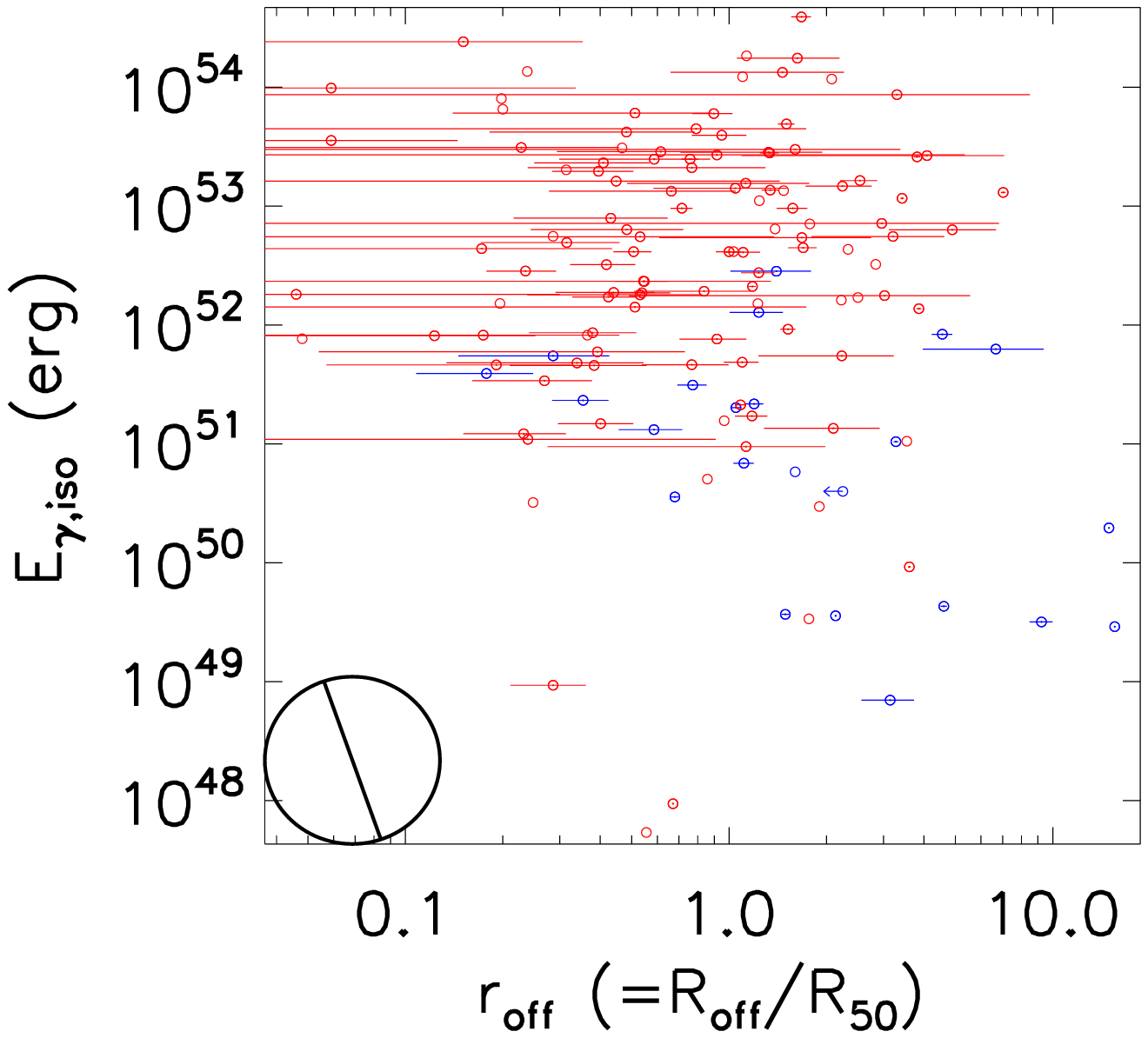}

\includegraphics[width=0.4\textwidth]{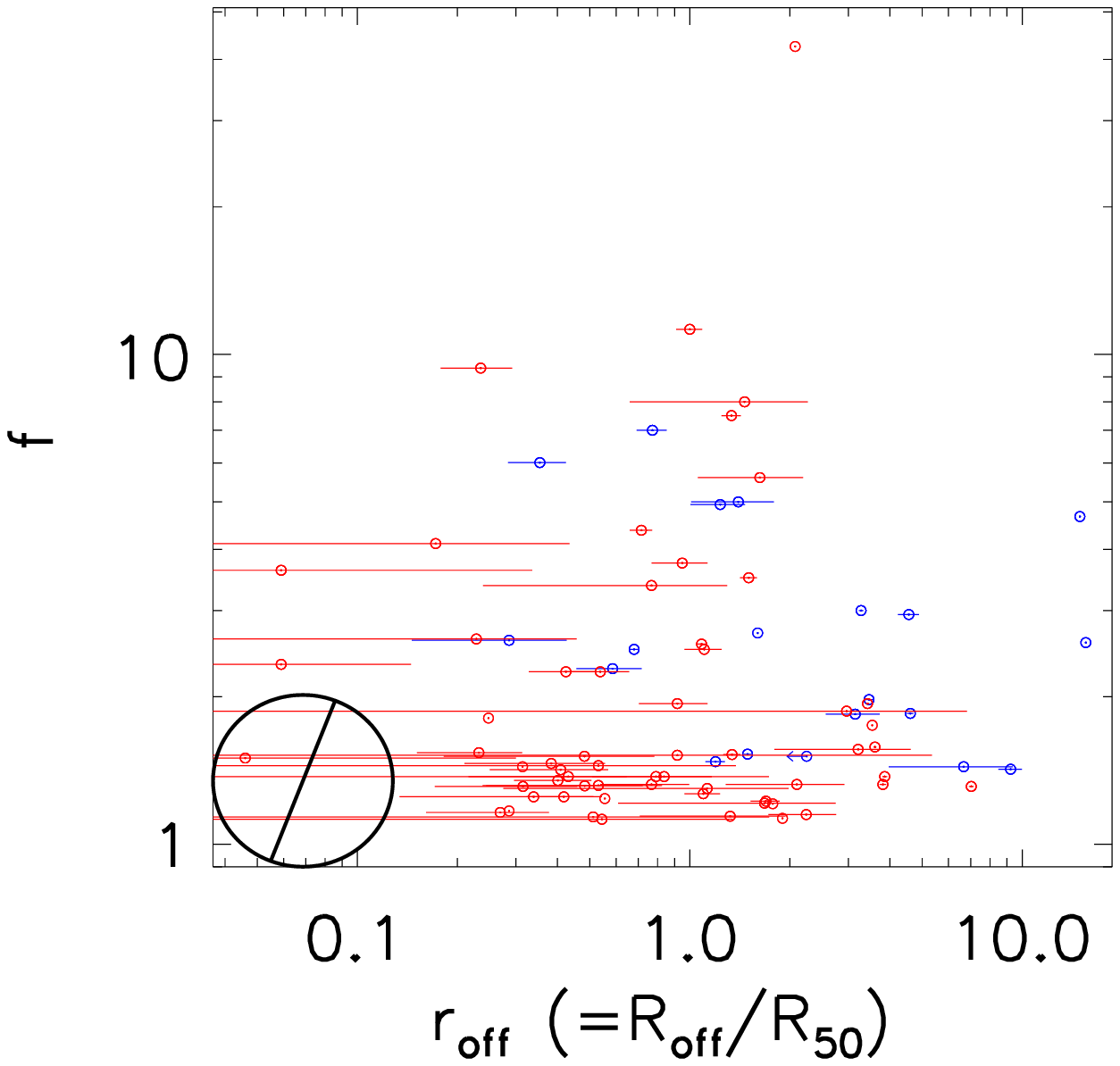}
\includegraphics[width=0.4\textwidth]{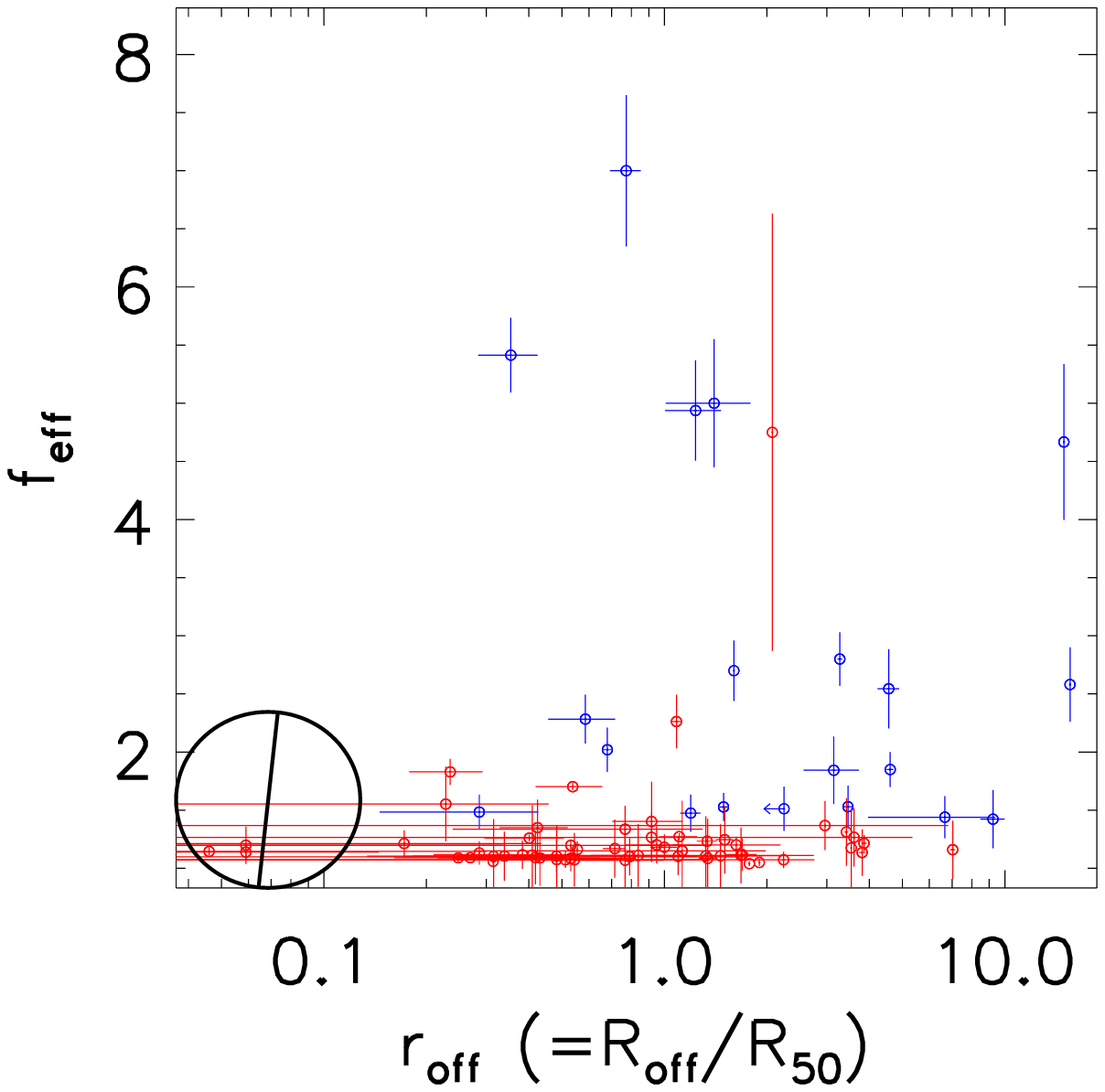}

\center{Fig. \ref{fig2d}---Continued}
\end{figure*}


\clearpage
\begin{figure*}

\includegraphics[width=0.4\textwidth]{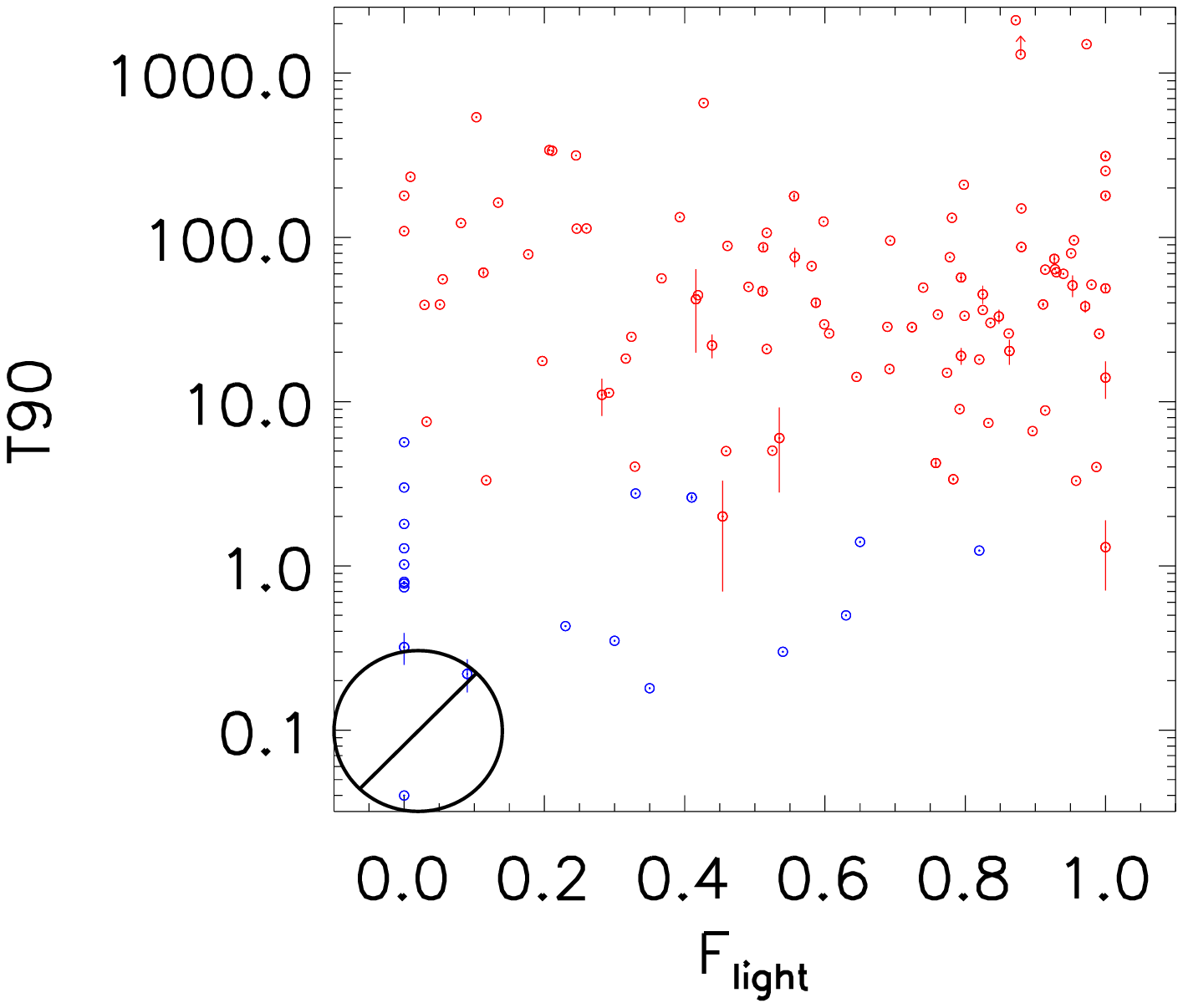}
\includegraphics[width=0.4\textwidth]{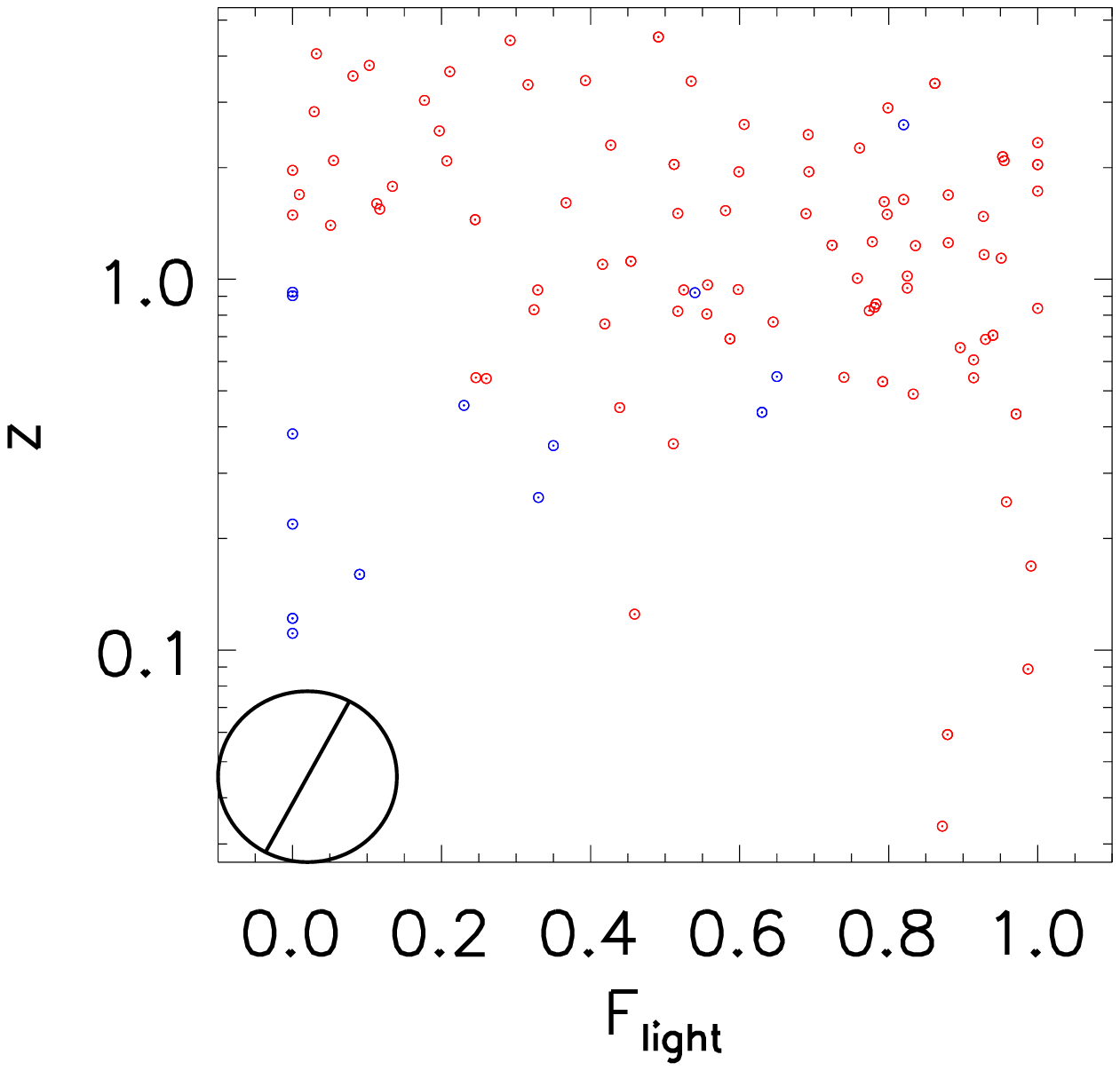}

\includegraphics[width=0.4\textwidth]{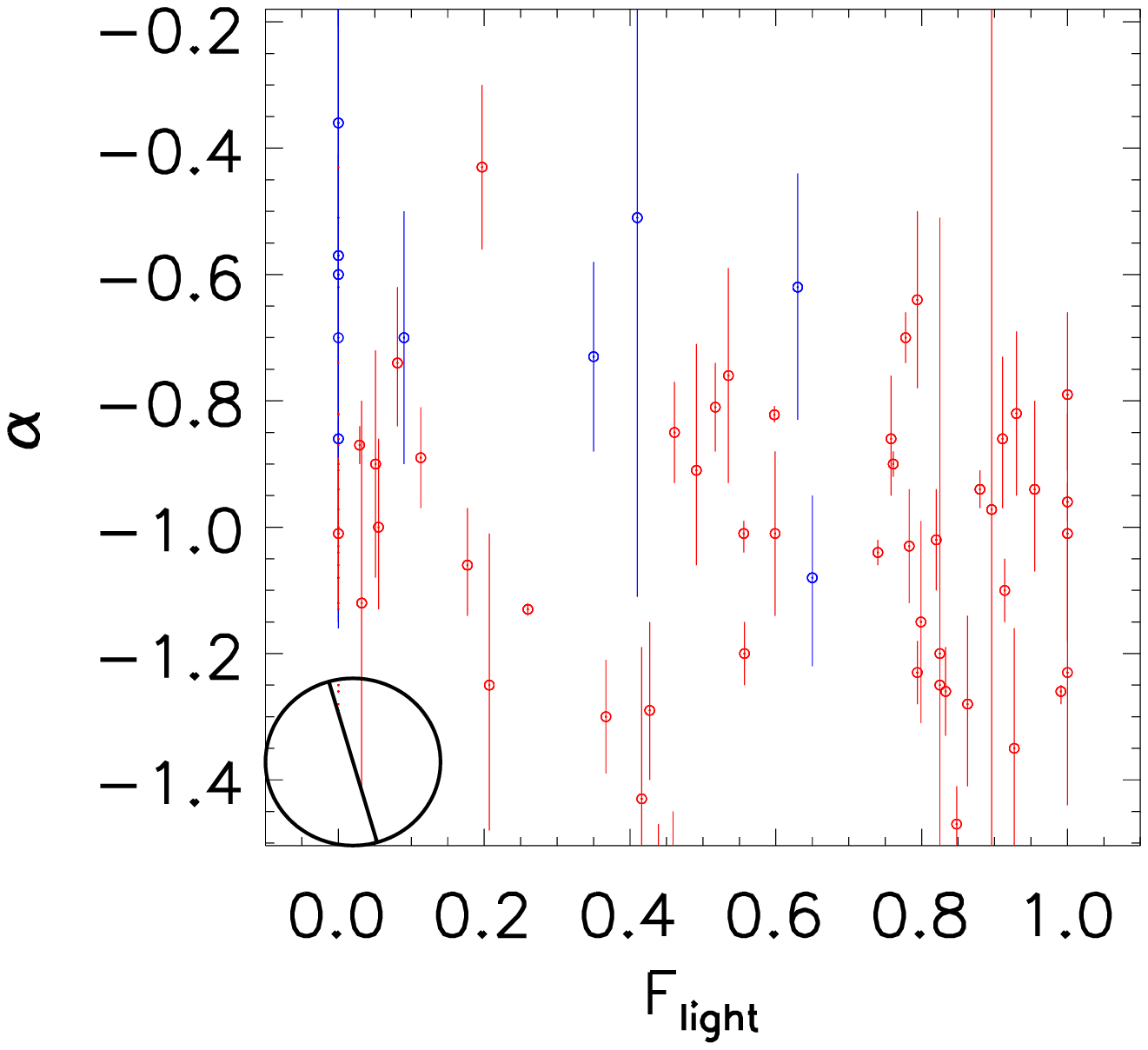}
\includegraphics[width=0.4\textwidth]{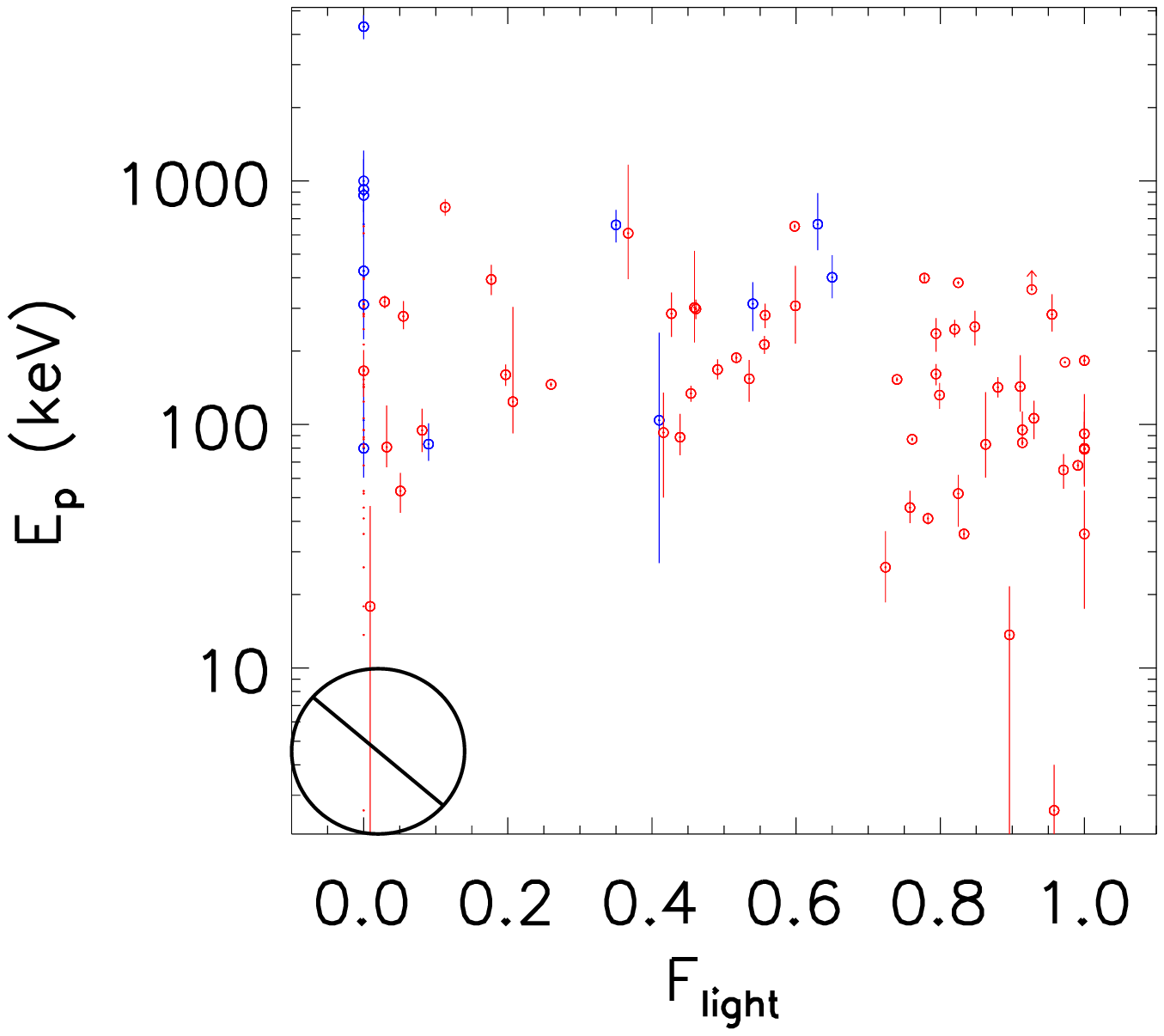}

\includegraphics[width=0.4\textwidth]{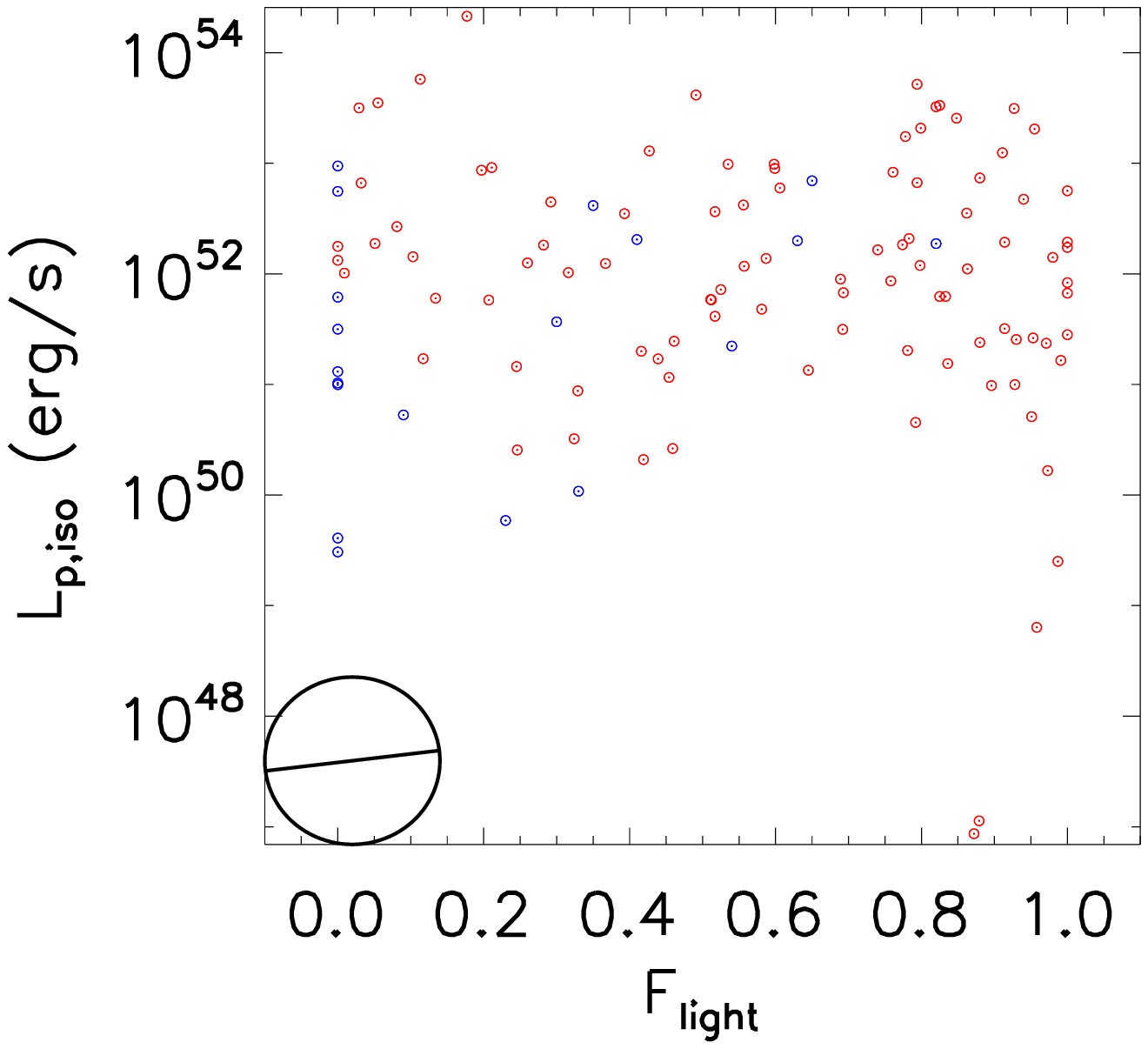}
\includegraphics[width=0.4\textwidth]{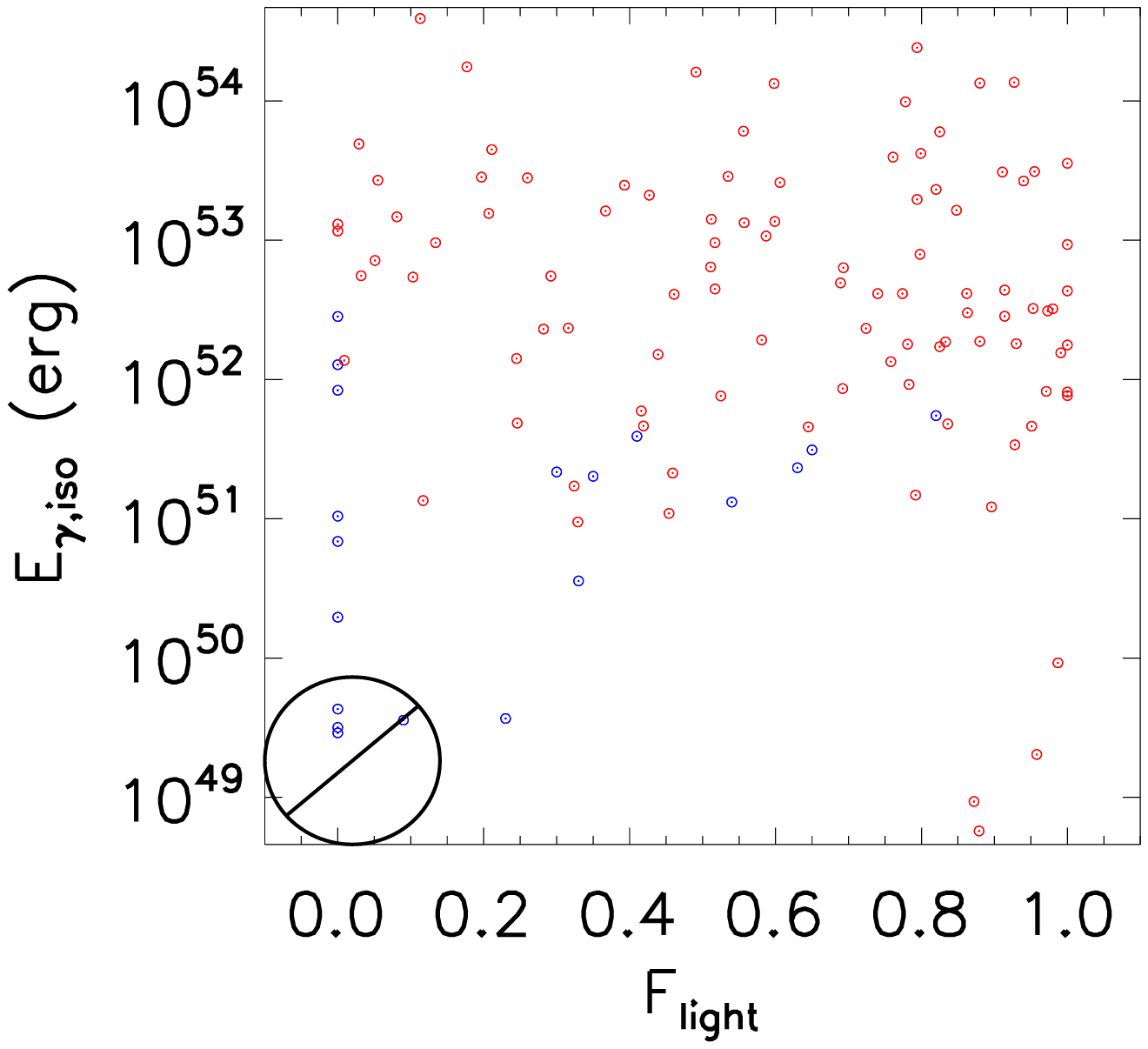}

\includegraphics[width=0.4\textwidth]{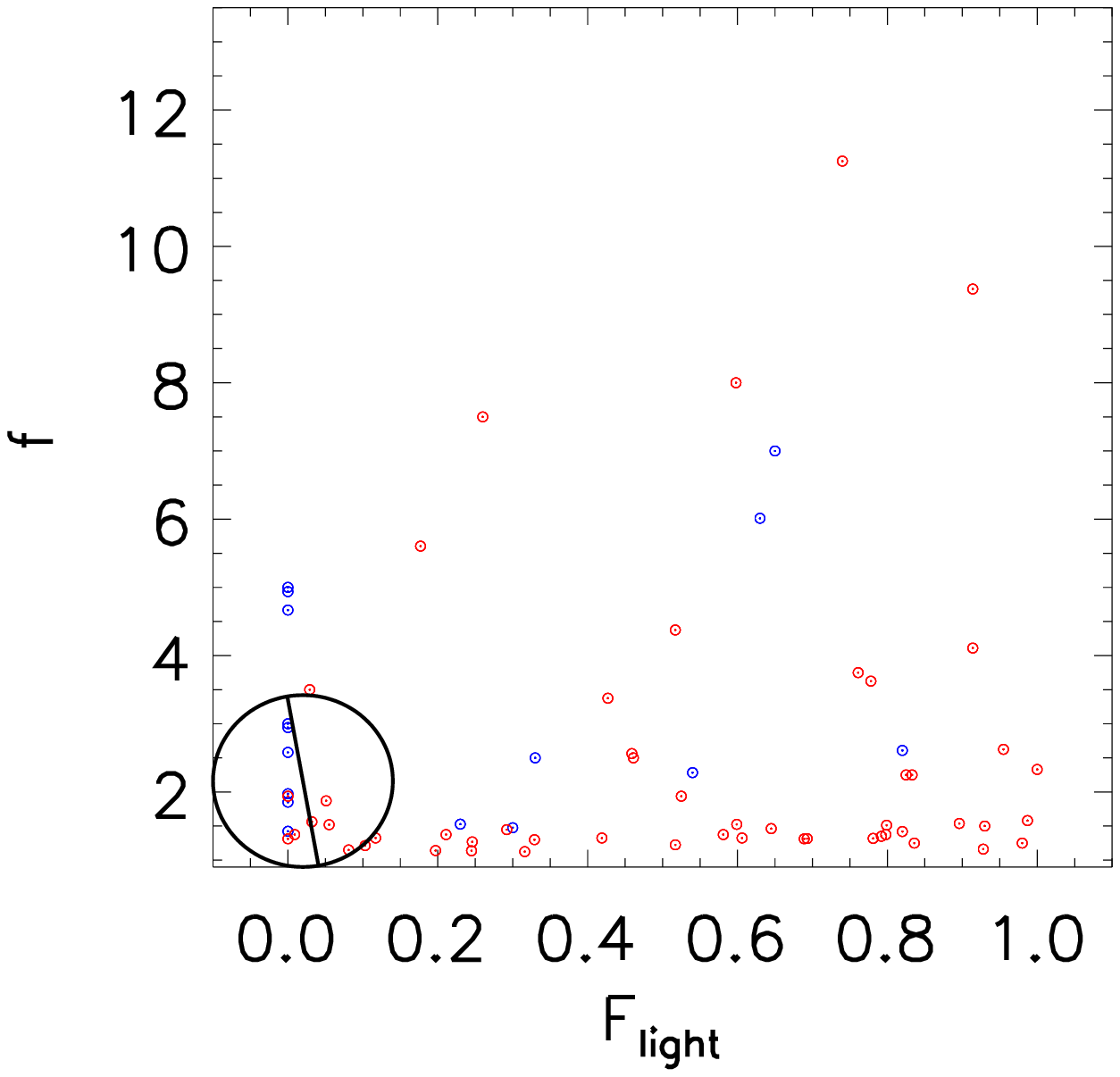}
\includegraphics[width=0.4\textwidth]{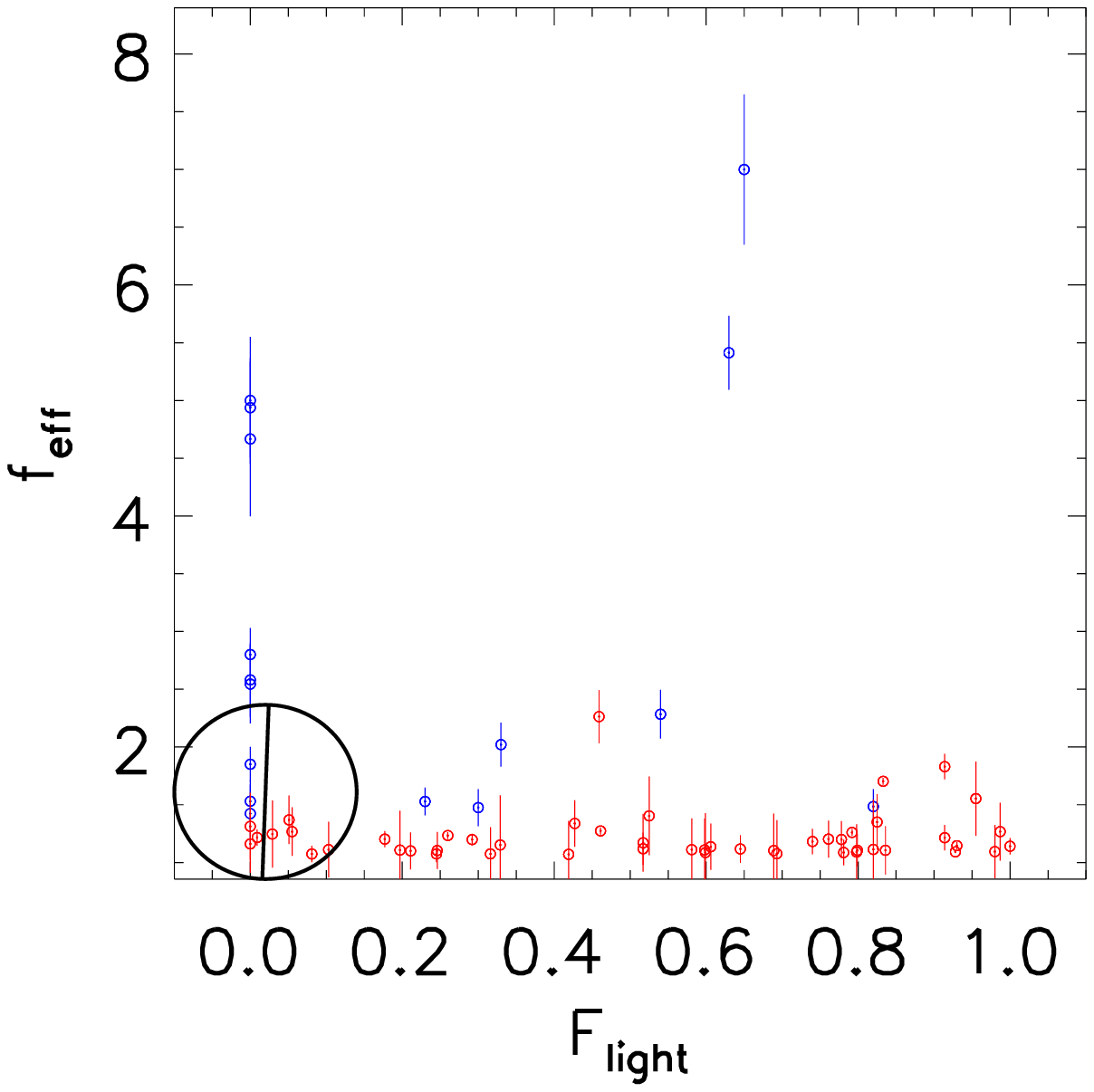}

\center{Fig. \ref{fig2d}---Continued}
\end{figure*}

\begin{deluxetable}{lcccccccccccc}
\tabletypesize{\scriptsize}
\tablecolumns{12}
\tablewidth{0pc}
\tablecaption{basic \label{tbbasic}}
\tablehead {
\colhead{GRB}        &
\multicolumn{3}{c}{redshift} &
\colhead{} & 
\multicolumn{3}{c}{$T_{90}$} & 
\colhead{$E_{\gamma, \rm iso}$}      &
\colhead{$L_{\rm p,iso}$}      &
\colhead{$f$} &
\colhead{$f_{\rm eff}$} &\\\cline{2-4}\cline{6-8}

\colhead{}        &
\colhead{$z$} &
\colhead{method}      &
\colhead{Ref}      &
\colhead{} &
\colhead{Detector} &
\colhead{[s]} &
\colhead{Ref}      &
\colhead{[${\rm erg\ cm^{-2}}$] }      &
\colhead{[${\rm erg\ s^{-1}\ cm^{-2}}$]}      &
\colhead{} &
\colhead{} &\\

\colhead{(1)}        &
\colhead{(2)} &
\colhead{(3)}      &
\colhead{(4)}      &
\colhead{} &
\colhead{(5)} &
\colhead{(6)} &
\colhead{(7)}      &
\colhead{(8)}      &
\colhead{(9)}      &
\colhead{(10)} &
\colhead{(11)} &
}

\startdata

140606B& $0.384$ & E & $279$ & $ $ & Fermi & $23.6$ & $39$ & $4.60\times 10^{51}$ & $1.88\times 10^{51}$ & $...$ & $...$ \\
140518A& $4.707$ & A & $272$ & $ $ & Swift & $60.5$ & $42$ & $7.92\times 10^{52}$ & $2.89\times 10^{52}$ & $    1.18$ & $1.02\pm0.01$ \\
140515A& $6.32$ & A & $271$ & $ $ & Swift & $23.4$ & $42$ & $7.08\times 10^{52}$ & $5.23\times 10^{52}$ & $    1.21$ & $1.04\pm0.02$ \\
140512A& $0.725$ & A & $273$ & $ $ & Swift & $154.8$ & $42$ & $5.63\times 10^{52}$ & $6.71\times 10^{51}$ & $    1.78$ & $1.25\pm0.13$ \\
140508A& $1.027$ & A & $270$ & $ $ & Fermi & $44.3$ & $38$ & $2.26\times 10^{53}$ & $6.74\times 10^{52}$ & $...$ & $...$ \\
140506A& $0.8893$ & E & $275$ & $ $ & Swift & $111.1$ & $42$ & $1.83\times 10^{52}$ & $1.01\times 10^{52}$ & $    1.50$ & $1.13\pm0.06$ \\
140430A& $1.6019$ & E & $275$ & $ $ & Swift & $173.6$ & $42$ & $1.51\times 10^{52}$ & $5.73\times 10^{51}$ & $    1.44$ & $1.06\pm0.03$ \\
140428A& $4.7$ & A & $269$ & $ $ & Swift & $17.42$ & $42$ & $3.69\times 10^{52}$ & $3.10\times 10^{52}$ & $...$ & $...$ \\
\enddata
\end{deluxetable}

\begin{landscape}
\begin{deluxetable}{lcccccccccccc}
\setlength{\voffset}{10mm}
\tabletypesize{\scriptsize}
\tablecolumns{13}
\tablewidth{0pc}
\tablecaption{prompt \label{tbprompt}}
\tablehead {
\colhead{GRB}        &
\colhead{Detector} &
\colhead{$\Gamma$} &
\colhead{$\alpha$} &
\colhead{$\beta$}       &
\colhead{$E_{\rm p}$} &
\multicolumn{2}{c}{fluence $S_{\gamma}$} &
\colhead{} &
\multicolumn{2}{c}{peak flux $F_{\rm p}$$(P_{\rm p})$} &
\colhead{reference}      &\\\cline{7-8}\cline{10-11}

\colhead{}        &
\colhead{fluence/flux} &
\colhead{} &
\colhead{} &
\colhead{}       &
\colhead{[keV]} &
\colhead{[$10^{-7}\ \rm erg\ cm^{-2}$]}      &
\colhead{Band} &
\colhead{} &
\colhead{} &
\colhead{Band} &
\colhead{}      &\\


\colhead{(1)}        &
\colhead{(2)} &
\colhead{(3)} &
\colhead{(4)} &
\colhead{(5)}       &
\colhead{(6)} &
\colhead{(7)}      &
\colhead{(8)} &
\colhead{} &
\colhead{(9)} &
\colhead{(10)}      &
\colhead{(11)} &

}

\startdata

140606B& Fermi & $...$ & $-1.24^{+0.05}_{-0.05}$ & $-2.20^{+0.52}_{-0.52}$ & $554^{+165}_{-165}$ & $75.9\pm0.4$ & $10- 1000$ & $ $ & $13.2\pm0.3^{\rm p}$ & $10- 1000$ & $103$ \\
140518A& Swift & $-1.89$ & $-0.98^{+0.61}_{-0.53}$ & $...$ & $47.9^{+12.7}_{-7.1}$ & $10\pm1$ & $15-150$ & $ $ & $1.0\pm0.1^{\rm p}$ & $15-150$ & $107$ \\
140515A& Swift & $1.78\pm0.13$ & $-0.98^{+0.64}_{-0.55}$ & $...$ & $56.4^{+31.3}_{-9.9}$ & $5.9\pm0.6$ & $15-150$ & $ $ & $0.9\pm0.1^{\rm p}$ & $15-150$ & $107$ \\
140512A& Fermi & $1.45\pm0.04$ & $-1.22^{+0.02}_{-0.02}$ & $-3.2^{+1.6}_{-1.6}$ & $682^{+70}_{-70}$ & $293\pm0$ & $10- 1000$ & $ $ & $11.0\pm0.3^{\rm p}$ & $10- 1000$ & $103,107$ \\
140508A& Fermi & $...$ & $-1.19^{+0.02}_{-0.02}$ & $-2.36^{+0.10}_{-0.10}$ & $263^{+14}_{-14}$ & $614\pm1$ & $10- 1000$ & $ $ & $66.8\pm1.0^{\rm p}$ & $10- 1000$ & $103$ \\
140506A& Fermi & $1.68\pm0.16$ & $-1.18^{+0.11}_{-0.11}$ & $-2.3^{+0.4}_{-7.7}$ & $197^{+32}_{-32}$ & $65.9\pm1.2$ & $10- 1000$ & $ $ & $15.7\pm0.7^{\rm p}$ & $10- 1000$ & $103,107$ \\
140430A& Swift & $2.00\pm0.22$ & $...$ & $...$ & $...$ & $11\pm2$ & $15-150$ & $ $ & $2.5\pm0.2^{\rm p}$ & $15-150$ & $107$ \\
140428A& Swift & $1.54\pm0.26$ & $...$ & $...$ & $...$ & $3.4\pm0.6$ & $15-150$ & $ $ & $0.6\pm0.2^{\rm p}$ & $15-150$ & $107$ \\
\enddata
\end{deluxetable}
\end{landscape}

\begin{landscape}
\begin{deluxetable}{lcccccccccccccccc}
\setlength{\voffset}{10mm}
\tabletypesize{\scriptsize}
\tablecolumns{15}
\tablewidth{0pc}
\tablecaption{sfr \label{tbsfr}}
\tablehead {
\colhead{GRB}        &
\colhead{z}        &
\colhead{Instrument} &
\multicolumn{2}{c}{log $M_*$} &
\colhead{} &
\multicolumn{2}{c}{SFR} &
\colhead{sSFR}       &
\multicolumn{2}{c}{[X/H]} &
\colhead{} &
\multicolumn{2}{c}{$A_{\rm V}$} &
\colhead{reference}      &\\\cline{4-5}\cline{7-8}\cline{10-11}\cline{13-14}

\colhead{}        &
\colhead{}        &
\colhead{(Spectrum)} &
\colhead{[$\rm M_{\odot}$]} &
\colhead{method} &
\colhead{} &
\colhead{[$\rm M_{\odot}\ yr^{-1}$]}       &
\colhead{method} &
\colhead{[Gyr$^{-1}$]} &
\colhead{} &
\colhead{method} &
\colhead{} &
\colhead{[mag]}      &
\colhead{method}      &
\colhead{}      &\\

\colhead{(1)}        &
\colhead{(2)} &
\colhead{(3)} &
\colhead{(4)} &
\colhead{(5)}       &
\colhead{} &
\colhead{(6)} &
\colhead{(7)}      &
\colhead{(8)} &
\colhead{(9)} &
\colhead{(10)}      &
\colhead{} &
\colhead{(11)} &
\colhead{(12)} &
\colhead{(13)} &
}

\startdata

140606B&$0.384$&GTC/OSIRIS&$...$&...& &$0.052\pm0.005$&H$\alpha$&...&$...$&...& & $0.3^{+0.89}_{-0.30}$&Balmer&$87$\\
140518A&$4.707$&Gemini/GMOS&$...$&...& &$...$&...&...&$>-1.06$&A/S& & $...$&...&$79$\\
140515A&$6.32$&GTC/OSIRIS&$...$&...& &$...$&...&...&$<-1.1$&A/O& & $0.11\pm0.02$&AG-SED&$75$\\
140512A&$0.725$&NOT/ALFOSC&$...$&...& &$...$&...&...&$...$&...& & $...$&...&$0$\\
140508A&$1.027$&NOT/ALFOSC&$...$&...& &$...$&...&...&$...$&...& & $...$&...&$0$\\
140506A&$0.8893$&VLT/X-Shooter&$9$&SED& &$0.35^{+0.35}_{-0.19}$&H$\alpha$&0.35&$...$&...& & $...$&...&$68,77$\\
140430A&$1.6019$&VLT/X-Shooter&$...$&...& &$8.5^{+7.1}_{-3.8}$&H$\alpha$&...&$-0.020^{+0.18}_{-0.19}$&E& & $...$&...&$77$\\
140428A&$4.7$&Keck I/LRIS&$...$&...& &$...$&...&...&$...$&...& & $...$&...&$0$\\
\enddata
\end{deluxetable}
\end{landscape}

\begin{landscape}
\begin{deluxetable}{lcccccccccccccc}
\setlength{\voffset}{10mm}
\tabletypesize{\scriptsize}
\tablecolumns{10}
\tablewidth{0pc}
\tablecaption{offset \label{tboffset}}
\tablehead {
\colhead{GRB}        &
\colhead{z}        &
\colhead{Instrument} &
\colhead{$P_{cc}$} &
\colhead{$R_{50}$} &
\colhead{$R_{50}$}       &
\colhead{n} &
\colhead{$R_{\rm off}$}      &
\colhead{$R_{\rm off}$}      &
\colhead{$r_{\rm off}$}       &
\colhead{$F_{\rm light}$}       &
\colhead{reference}      &\\

\colhead{}        &
\colhead{}        &
\colhead{(Image)} &
\colhead{} &
\colhead{["]} &
\colhead{[kpc]}       &
\colhead{} &
\colhead{["]}      &
\colhead{[kpc]}      &
\colhead{$[R_{50}]$}       &
\colhead{}       &
\colhead{}      &\\

\colhead{(1)}        &
\colhead{(2)}        &
\colhead{(3)} &
\colhead{(4)} &
\colhead{(5)} &
\colhead{(6)}       &
\colhead{(7)} &
\colhead{(8)}      &
\colhead{(9)}      &
\colhead{(10)}       &
\colhead{(11)}       &
\colhead{(12)}      &

}

\startdata

140606B&$0.384$&...&$...$&$...$&$...$&$...$&$...$&$...$&$...$&$...$&$0$\\
140518A&$4.707$&...&$...$&$...$&$...$&$...$&$...$&$...$&$...$&$...$&$0$\\
140515A&$6.32$&HST&$0.014$&$0.12^{+0.05}_{-0.04}$&$0.680^{+0.27}_{-0.23}$&$...$&$0.210\pm0.070$&$1.21\pm0.40$&$1.776$&$...$&$47$\\
140512A&$0.725$&...&$...$&$...$&$...$&$...$&$...$&$...$&$...$&$...$&$0$\\
140508A&$1.027$&...&$...$&$...$&$...$&$...$&$...$&$...$&$...$&$...$&$0$\\
140506A&$0.8893$&...&$...$&$...$&$...$&$...$&$...$&$...$&$...$&$...$&$0$\\
140430A&$1.6019$&...&$...$&$...$&$...$&$...$&$...$&$...$&$...$&$...$&$0$\\
140428A&$4.7$&...&$...$&$...$&$...$&$...$&$...$&$...$&$...$&$...$&$0$\\
\enddata
\end{deluxetable}
\end{landscape}

\begin{landscape} 
\setlength{\voffset}{10mm}
\begin{table*}[!htb] \scriptsize
\begin{center}
\caption{Statistical results of properties for consensus LGRB/SGRB definition. 
\label{tb1dconsensus} }
\begin{tabular}{l|ccccc|cccccc}
\hline\hline

 &                \multicolumn{8}{c}{Consensus L/S} \\\hline
Name & No. & value & No. & value &$P_{\rm KS}$ & Overlap & LGRB in & SGRB in \\
     & LGRB &      & SGRB &       &             & range   & overlap (\%) & overlap (\%) \\\hline

log $T_{90}$ (s)&         369&$1.70\pm0.60$&          34&$-0.30\pm0.57$&3.e-26&(0.11,0.75)&           7&          20&\\
z&         351&$1.64\pm1.30$&          25&$0.45\pm0.51$&8.e-11&(0.11,2.61)&          72&         100&\\
log $E_{\rm iso}$ (erg)&         369&$52.7\pm1.0$&          33&$51.1\pm0.9$&5.e-14&(48.8,52.5)&          36&         100&\\
log $L_{\rm iso}$ (erg s$^{-1}$)&         366&$52.1\pm1.1$&          33&$51.3\pm1.1$&7.e-03&(49.3,53.0)&          78&         100&\\
$\alpha$&         196&$-1.01\pm0.34$&          16&$-0.60\pm0.25$&2.e-07&(-1.08,-0.15)&          55&         100&\\
\hline
\end{tabular}
\end{center}
\end{table*}
\end{landscape}

\begin{landscape} 
\setlength{\voffset}{10mm}
\begin{table*}[!htb] \scriptsize
\begin{center}
\caption{Statistical results of properties for T90 defined LGRB/SGRB. 
\label{tb1dT90} }
\begin{tabular}{l|ccccc|cccccc}
\hline\hline

 &                \multicolumn{8}{c}{T90 L/S} \\\hline
Name & No. & value & No. & value &$P_{\rm KS}$ & Overlap & LGRB in & SGRB in \\
     & LGRB &      & SGRB &       &             & range   & overlap (\%) & overlap (\%) \\\hline

log $T_{90}$ (s)&         371&$1.70\pm0.60$&          32&$-0.31\pm0.50$&2.e-27&(0.30,0.26)&       0&       0\\
z&         352&$1.63\pm1.31$&          24&$0.46\pm0.60$&6.e-08&(0.11,2.61)&          73&         100\\
log $E_{\rm iso}$ (erg)&         371&$52.7\pm1.0$&          31&$51.0\pm0.9$&5.e-15&(48.8,52.4)&          35&         100\\
log $L_{\rm iso}$ (erg s$^{-1}$)&         368&$52.1\pm1.1$&          31&$51.2\pm1.0$&1.e-03&(49.3,53.0)&          80&         100\\
$\alpha$&         198&$-1.01\pm0.34$&          14&$-0.60\pm0.32$&3.e-06&(-1.36,-0.15)&          84&          92\\
\hline
\end{tabular}
\end{center}
\end{table*}
\end{landscape}

\begin{landscape} 
\setlength{\voffset}{0mm}
\begin{table*}[!htb] \scriptsize
\begin{center}
\caption{Statistical results of Simulated 1D distributions.}
\label{tb1dfake}
\begin{tabular}{l|ccccc|cccccccc}
\hline\hline

 &                \multicolumn{5}{c}{Consensus L/S} & \multicolumn{5}{|c}{T90 L/S} \\\hline
Name & N(LGRB) & value & N(SGRB) & value &$P_{\rm KS}$ & N(LGRB) & value & N(SGRB) & value &$P_{\rm KS}$ \\\hline

log $T_{90}$ (s)&         359&$1.67\pm0.57$&          41&$-0.13\pm0.61$&5.e-30&         365&$1.67\pm0.57$&          35&$-0.30\pm0.50$&1.e-29\\
z&         380&$1.75\pm1.38$&          20&$0.44\pm0.31$&5.e-09&         378&$1.75\pm1.38$&          22&$0.46\pm0.46$&6.e-08\\
log $E_{\rm iso}$ (erg)&         363&$52.6\pm1.0$&          37&$51.0\pm1.0$&7.e-14&         370&$52.6\pm1.0$&          30&$50.9\pm0.9$&4.e-16\\
log $L_{\rm iso}$ (erg s$^{-1}$)&         368&$52.2\pm1.1$&          32&$51.2\pm1.2$&3.e-04&         370&$52.2\pm1.1$&          30&$51.2\pm1.2$&3.e-04\\
$\alpha$&         367&$-1.03\pm0.33$&          33&$-0.60\pm0.17$&7.e-15&         369&$-1.01\pm0.33$&          31&$-0.62\pm0.30$&2.e-12\\
log $E_{\rm p}$ (keV)&         362&$2.19\pm0.49$&          38&$2.69\pm0.41$&4.e-10&         365&$2.19\pm0.50$&          35&$2.66\pm0.33$&1.e-09\\
\hline
\end{tabular}
\end{center}
\end{table*}
\end{landscape}

\begin{landscape} 
\setlength{\voffset}{20mm}
\begin{table*}[!htb] \scriptsize
\begin{center}
\caption{Statistical results of 2D.}
\label{tb2d}
\begin{tabular}{l|ccccc|ccccc|ccccc|ccccc}
\hline\hline

 & \multicolumn{5}{c}{T90} & \multicolumn{5}{|c}{z} & \multicolumn{5}{|c}{$\alpha$} & \multicolumn{5}{|c}{$E_{\rm p}$ (keV)} \\
& No. & angle & $P_{\rm KS}$ & $\rho_{\rm s}$ & $P_{\rm s}$ & No. & angle & $P_{\rm KS}$ & $\rho_{\rm s}$ & $P_{\rm s}$ & No. & angle & $P_{\rm KS}$ & $\rho_{\rm s}$ & $P_{\rm s}$ & No. & angle & $P_{\rm KS}$ & $\rho_{\rm s}$ & $P_{\rm s}$ \\
\hline

$M_* (M_{\odot})$ & 101/22 & 87 &  4.e-15 &  2.e-01 &  1.e-01 & 94/20 & 111 &  1.e-04 &  5.e-01 &  3.e-07 & 60/10 & 100 &  2.e-05 &  3.e-01 &  2.e-02 & 63/11 & 86 &  6.e-04 &  1.e-01 &  4.e-01\\
SFR ($M_{\odot}$ yr$^{-1}$) & 149/16 & 80 &  7.e-12 &  3.e-01 &  2.e-03 & 149/16 & 87 &  4.e-05 &  5.e-01 &  4.e-11 & 74/9 & 101 &  4.e-05 &  3.e-01 &  5.e-03 & 77/10 & 94 &  8.e-04 &  1.e-01 &  3.e-01\\
sSFR (yr$^{-1}$) & 77/14 & 68 &  2.e-10 &  1.e-01 &  3.e-01 & 77/14 & 18 &  1.e-02 &  2.e-01 &  7.e-02 & 48/7 & 87 &  8.e-04 & -1.e-01 &  4.e-01 & 51/8 & 92 &  4.e-04 & -1.e-01 &  4.e-01\\
$\rm [X/H]$ & 134/9 & 115 &  3.e-08 & -8.e-02 &  4.e-01 & 134/9 & 104 &  1.e-05 & -6.e-01 &  7.e-14 & 65/7 & 46 &  2.e-04 & -3.e-01 &  4.e-02 & 68/7 & 31 &  2.e-03 & -1.e-01 &  3.e-01\\
$R_{50}$ (kpc) & 128/24 & 115 &  8.e-19 & -1.e-01 &  2.e-01 & 113/17 & 147 &  3.e-06 & -2.e-01 &  7.e-02 & 74/12 & 75 &  2.e-07 &  5.e-02 &  7.e-01 & 79/14 & 26 &  6.e-06 &  8.e-02 &  5.e-01\\
$R_{\rm off}$ (kpc) & 136/28 & 104 &  1.e-20 & -5.e-02 &  6.e-01 & 124/19 & 107 &  2.e-08 & -1.e-02 &  9.e-01 & 75/15 & 71 &  5.e-09 &  2.e-01 &  1.e-01 & 79/17 & 60 &  5.e-05 &  8.e-02 &  5.e-01\\
\hline
\end{tabular}
\end{center}
\end{table*}
\end{landscape}


%
%

\end{document}